\documentclass[12pt]{article}
\pdfoutput=1
\usepackage{jheppub, bm, wrapfig}
\numberwithin{equation}{section}
\graphicspath{{./figures-reducedset/}}

\renewcommand{\v}{{\mathbf v}}

\newcommand{\HAT}{\hat}

\newcommand{\bea}{\begin{eqnarray}}
\newcommand{\eea}{\end{eqnarray}}
\newcommand{\be}{\begin{equation}}
\newcommand{\ee}{\end{equation}}
\newcommand{\bse}{\begin{subequations}}
\newcommand{\ese}{\end{subequations}}
\newcommand{\mb}{\mathbf}

\newcommand{\wt}{\widetilde}

\newcommand{\ol}{\overline}
\newcommand{\ds}{\displaystyle}

\newcommand{\eg}{\emph{e.g.}}
\newcommand{\ie}{\emph{i.e.}}
\newcommand{\cf}{\emph{cf.}}

\newcommand{\Z}{{\mathbb Z}}
\newcommand{\R}{{\mathbb R}}
\newcommand{\C}{{\mathbb C}}

\renewcommand{\S}{{\mathbb S}}
\newcommand{\X}{{\mathbb X}}
\newcommand{\Y}{{\mathbb Y}}

\newcommand{\cp}{{\mathbb{CP}}}

\newcommand{\Tr}{{\rm Tr \,}}
\renewcommand{\Re}{{\rm Re}}
\renewcommand{\Im}{{\rm Im}}

\newcommand{\pd}{\partial}

\newcommand{\CA}{\mathcal{A}}
\newcommand{\CB}{\mathcal{B}}
\newcommand{\CC}{\mathcal{C}}
\newcommand{\CD}{\mathcal{D}}

\newcommand{\CF}{\mathcal{F}}
\newcommand{\CG}{\mathcal{G}}
\newcommand{\CH}{\mathcal{H}}
\newcommand{\CI}{\mathcal{I}}

\newcommand{\CL}{\mathcal{L}}
\newcommand{\CM}{\mathcal{M}}
\newcommand{\CN}{\mathcal{N}}
\newcommand{\CO}{\mathcal{O}}

\newcommand{\CR}{\mathcal{R}}
\newcommand{\CS}{\mathcal{S}}
\newcommand{\CT}{\mathcal{T}}

\newcommand{\CW}{\mathcal{W}}

\newcommand{\mtwod}{{\tilde d}}

\title
{Boundaries, Mirror Symmetry, and Symplectic Duality in 3d $\CN=4$ Gauge Theory}

\author[1,2]{Mathew Bullimore}
\author[2,3,4]{Tudor Dimofte}
\author[3]{Davide Gaiotto}
\author[5]{Justin Hilburn}

\affiliation[1]{Mathematical Institute, University of Oxford, Woodstock Road, Oxford, OX2 6GG, UK}
\affiliation[2]{School of Natural Sciences, Institute for Advanced Study, Princeton, NJ 08540, USA}
\affiliation[3]{Perimeter Institute for Theoretical Physics, Waterloo, Ontario, Canada N2L 2Y5}
\affiliation[4]{Department of Mathematics and Center for Quantum Mathematics and Physics, UC Davis, Davis, CA 95616, USA}
\affiliation[5]{Department of Mathematics, University of Oregon, Eugene, Oregon 97403, USA}

\emailAdd{mathew.bullimore@maths.ox.ac.uk}
\emailAdd{tudor@math.ucdavis.edu}
\emailAdd{dgaiotto@perimeterinstitute.ca}
\emailAdd{jhilburn@math.upenn.edu}

\abstract{We introduce several families of $\CN=(2,2)$ UV boundary conditions in 3d $\mathcal N=4$ gauge theories and study their IR images in sigma-models to the Higgs and Coulomb branches.
In the presence of Omega deformations, a UV boundary condition defines a pair of modules for quantized algebras of chiral Higgs- and Coulomb-branch operators, respectively, whose structure we derive.
In the case of abelian theories, we use the formalism of hyperplane arrangements to make our constructions very explicit, and construct a half-BPS interface that implements the action of 3d mirror symmetry on gauge theories and boundary conditions.
Finally, by studying two-dimensional compactifications of 3d $\CN=4$ gauge theories and their boundary conditions, we propose a physical origin for \emph{symplectic duality} --- an equivalence of categories of modules associated to families of Higgs and Coulomb branches that has recently appeared in the mathematics literature, and generalizes classic results on Koszul duality in geometric representation theory.
We make several predictions about the structure of symplectic duality, and identify Koszul duality as a special case of wall crossing.}

\begin{document}

\maketitle

\section{Introduction}

In this paper, we introduce and study various families of half-BPS boundary conditions in three-dimensional $\CN=4$ gauge theories that preserve a two-dimensional $\CN=(2,2)$ super-Poincar\'e algebra. We then use these boundary conditions to try to understand a phenomenon known as ``symplectic duality'' in the mathematics literature, which, among other things, describes an equivalence of categories associated to the Higgs and Coulomb branches of 3d $\CN=4$ theories. Let us first say a bit about the physics.

A 3d $\CN=4$ gauge theory generically flows in the infrared to a sigma-model onto its Higgs $(\CM_H$) or Coulomb ($\CM_C$) branch of vacua, depending on the precise combinations of parameters that are turned on. Supersymmetry requires that these moduli spaces are hyperk\"ahler \cite{HKLR-HK}, which implies that in any fixed complex structure they become complex symplectic manifolds. Correspondingly, a UV boundary condition $\CB$ that preserves 2d $\CN = (2,2)$ supersymmetry must flow to holomorphic Lagrangian ``branes'' $\CB_H,\,\CB_C$ in the IR sigma-models, possibly enhanced by extra boundary degrees of freedom \cite{KRS, KR}. We use a combination of quantum and semi-classical methods to determine the form of these branes.

The holomorphic Lagrangians $\CB_H,\,\CB_C$ associated to a boundary condition $\CB$ also have an operator interpretation. The holomorphic functions on the Higgs and Coulomb branches are given by expectation values of scalar operators in two chiral rings $\C[\CM_H],$ $\C[\CM_C]$.
 From the perspective of 2d $\CN = (2,2)$ supersymmetry, the operators in $\C[\CM_H]$ are chiral, while the operators in $\C[\CM_C]$ are twisted-chiral.
As a holomorphic subvariety of the Higgs (Coulomb) branch, $\CB_H$ ($\CB_C$) simply encodes relations satisfied by the chiral-ring operators when they are brought to the boundary.

\begin{figure}[htb]
\centering
\includegraphics[width=2.4in]{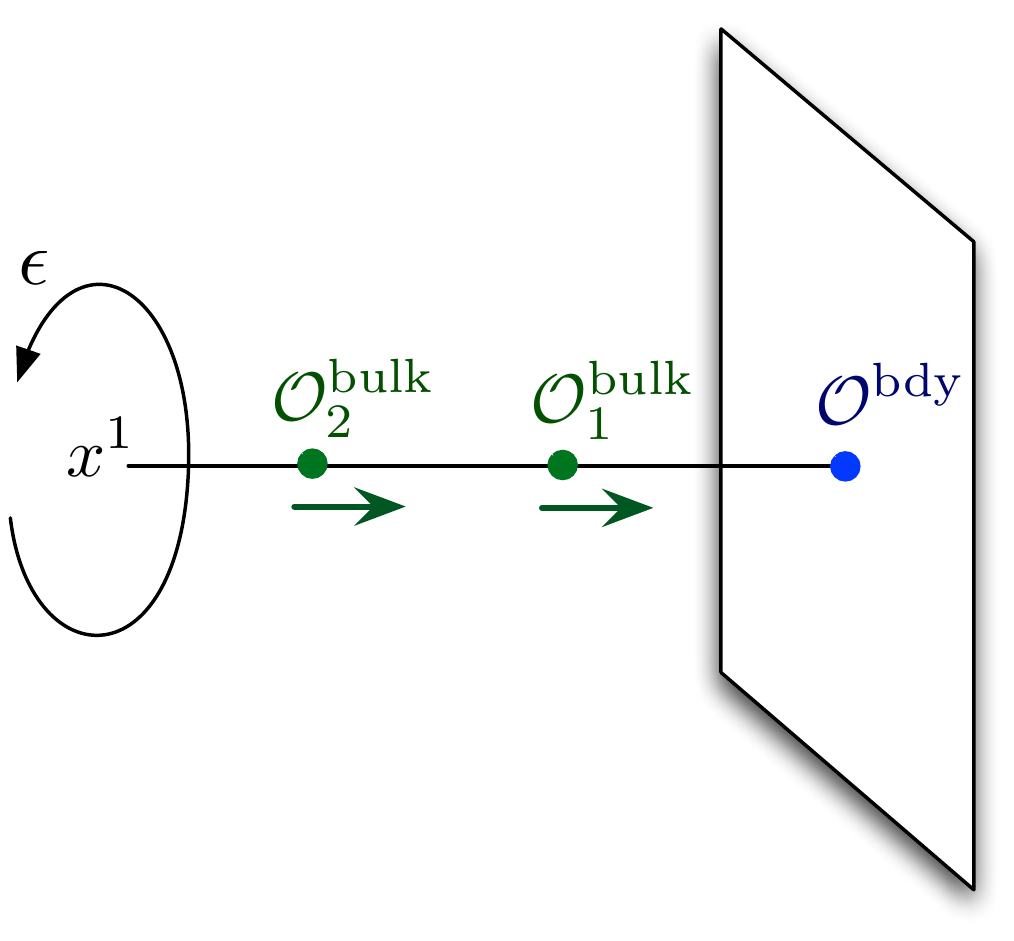}
\caption{Bulk operators in the Omega-background acting on boundary operators, which define a module for the bulk algebra $\hat \C[\CM_H]$. Here we have $\CO_2^{\rm bulk}\CO_1^{\rm bulk}|\CO^{\rm bdy}\rangle$.}
\label{fig:modules}
\end{figure}

There are two interesting deformations of 3d $\CN=4$ theories that turn the chiral rings into non-commutative algebras: standard and twisted Omega-backgrounds.
In the 3d $\CN=4$ context, this was studied in \cite{Yagi-quantization, BDG-Coulomb} (see below for other connections).
The Omega backgrounds mix supersymmetry transformations with rotations of some $\R^2_\epsilon\subset \R^3$, and effectively reduce the 3d theory to one-dimensional supersymmetric quantum mechanics supported on the fixed axis of rotations. In a standard (resp., twisted) Omega background, chiral Coulomb (Higgs) branch operators can be inserted at points on the fixed axis, in a particular order as in Figure \ref{fig:modules}. As one might expect in quantum mechanics, the product of operators becomes noncommutative, by an amount $\epsilon$. One therefore obtains a ``quantized,'' noncommutative operator algebra $\hat \C[\CM_C]$ ($\hat \C[\CM_H]$) that reduces to the ring $\C[\CM_C]$ ($\C[\CM_H]$) as $\epsilon\to 0$. Mathematically, these algebras are deformation quantizations.

\begin{figure}[htb]
\centering
\includegraphics[width=5in]{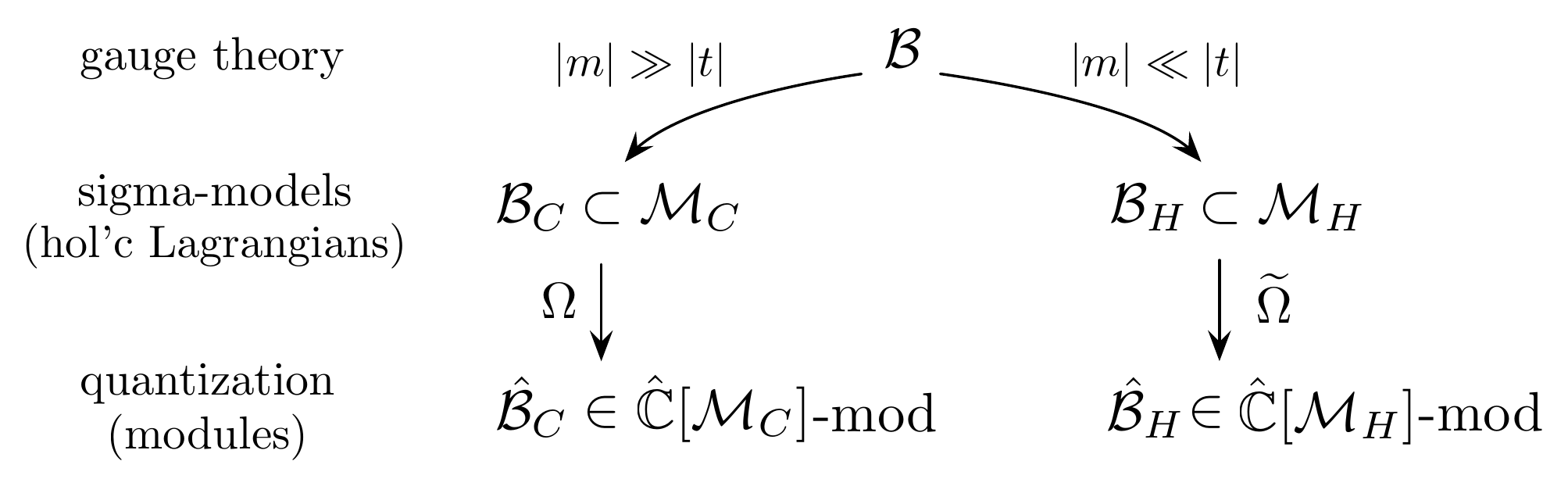}
\caption{The flow of a UV boundary condition to holomorphic Lagrangians in Coulomb- and Higgs-branch sigma-models, and its quantizations in the presence of Omega backgrounds.}
\label{fig:Bflow}
\end{figure}

A UV boundary condition $\CB$ will define a pair of \emph{modules} $\hat \CB_C,\,\hat \CB_H$ for the algebras $\hat\C[\CM_C]$ and $\hat\C[\CM_H]$. Heuristically, these modules are generated by some relations in $\hat\C[\CM_C],$ $\hat \C[\CM_H]$ that reduce to the classical equations defining holomorphic Lagrangians  $\CB_C,\,\CB_H$ when the deformation $\epsilon$ is turned off.
The situation is summarized in Figure~\ref{fig:Bflow}.

In the case of abelian gauge theories, the study of boundary conditions and their quantization is fully systematic, and leads to a rich geometric story that we will describe in some detail. The analysis is aided by tools from hypertoric geometry \cite{Bielawski-hypt, BielawskiDancer} (see also \cite{Proudfoot-survey} and references therein), which plays role analogous to that of toric geometry in abelian gauge theories (GLSM) with four supercharges. Three-dimensional mirror symmetry also acts in a systematic way on abelian theories \cite{IS, dBHOY}, and we find that it relates pairs of UV boundary conditions in mirror abelian theories. More so, using techniques from two-dimensional mirror symmetry, we will describe a 3d mirror symmetry interface that can be collided with any UV boundary condition to produce its mirror.

Many of the developments in this paper have close connections with previous work on boundary conditions, their RG flow, and the algebras of operators that act on them. As a small sampling:

\subsection*{Four Dimensions} 

Some of our constructions may be viewed as a dimensional reduction of half-BPS boundary conditions and interfaces for 4d $\CN=2$ theories studied in \cite{DG-Sdual, DGG, DGV-hybrid}, in turn inspired by Gaiotto and Witten's analysis of half-BPS boundary conditions in four-dimensional $\CN=4$ theory \cite{GW-boundary, GW-Sduality}.
In four dimensions, an Omega background quantizes the algebra of Coulomb-branch line operators, and boundary conditions produce modules for these algebras.%
\footnote{The idea that Omega backgrounds \cite{MNS, MNS-D, LNS-SW, Nek-SW} are related to quantization arose in \cite{NShatashvili} and many related works, including \cite{GW-surface, DG-RMQ, AGGTV, DGOT, NekWitten, Teschner-opers, GMNIII}; \cf\ the recent review \cite{Gukov-surface}.} %
 Some of our Coulomb-branch algebras and modules come from dimensional reductions of such 3d-4d systems. Our methods can be likely extended to compactifications on a finite-size circle. 

For example, a four-dimensional $\CN=2$ theory of class $\CS$ on a finite-size circle has a hyperk\"ahler Coulomb branch that is a Hitchin system \cite{BSV-Dtop, KapustinWitten, Gaiotto-dualities, GMN}. Half-BPS boundary conditions produce holomorphic Lagrangian submanifolds of the Hitchin system. As the radius of the circle is taken to zero size, the Hitchin system (partially) decompactifies to become a 3d Coulomb branch \cite{SW-3d}, supporting a holomorphic Lagrangian submanifold of the type we study here.

Alternatively, boundary conditions for 3d $\CN=4$ theories may be obtained from 4d $\CN=2$ theories with a surface operator, as in \cite{GW-surface, AGGTV, GMNIV, GGS}, by compactifying along the circle that links the surface operator.

\subsection*{Five Dimensions}
Some of our constructions can be dimensionally oxidized to half-BPS boundary conditions for 5d $\CN=1$ gauge theories. 
These gauge theories admit rather mysterious UV completions 
(see e.g. \cite{Seiberg:1996bd,Morrison:1996xf,Douglas:1996xp,Intriligator:1997pq,Aharony:1997ju}) and some boundary conditions may admit a UV completion as well \cite{GaiottoKim}.  
It would be interesting to explore the extension of our methods to five-dimensional gauge theories compactified on a two-torus of finite size. 

\subsection*{Three Dimensions}
Boundary conditions that preserve 2d $\CN=(2,2)$ supersymmetry are compatible with several topological twists, including a standard Rozansky-Witten twist \cite{RW} that effectively leads to a topological sigma-model with target $\CM_H$, and a ``twisted'' Rozansky-Witten twist that effectively leads to a topological sigma-model with target $\CM_C$.%
\footnote{These twists were first identified in the classification of Blau and Thompson \cite{BT-twists}.} %
In the topological sigma-models, boundary conditions generate a 2-category that was studied in \cite{KRS, KR}. Our present analysis of boundary conditions in gauge theory takes much inspiration from \cite{KRS, KR}. We will also make contact with the recent work of Teleman \cite{Teleman-MS} on some special boundary conditions in pure $\CN=4$ gauge theory.

If we break the bulk 3d $\CN=4$ symmetry to 3d $\CN=2$, say by adding a twisted mass for the R-symmetry, the supersymmetry preserved by our half-BPS boundary conditions is broken to 2d $\CN=(0,2)$. Such half-BPS boundary conditions for 3d $\CN=2$ theories were studied in \cite{OkazakiYamaguchi, GGP-walls}, and play a central role in the 4d-2d correspondence \cite{GGP-4d}, where they are labelled by four-manifolds with boundary.

We will occasionally combine boundary conditions and line operators in our constructions; the action of 3d mirror symmetry on line operators was studied in \cite{AG-loops}.

In upcoming work, Aganagic and Okounkov \cite{Okounkov-simonstalk} study holomorphic blocks (\cf~\cite{BDP-blocks}) of 3d $\CN=4$ theories. These are partition functions on $D^2\times S^1$, defined using a topological twist that treats Higgs and Coulomb branches symmetrically (in contrast to our Omega backgrounds). The theory on $D^2\times S^1$ has a boundary condition labelled by a vacuum, which can be constructed in the UV using our exceptional Dirichlet boundary conditions from Section \ref{sec:xD}. 3d mirror symmetry exchanges Higgs and Coulomb branches, and is found to produce interesting dualities of the holomorphic blocks, interpreted mathematically as elliptic stable envelopes.

The quantization of operator algebras $\hat\C[\CM_C]$, $\hat\C[\CM_H]$ in 3d super\emph{conformal} theories was recently studied in \cite{BPR-quantization}, using different methods than Omega backgrounds. It was found that superconformal symmetry puts additional interesting constraints on the structure constants of these algebras.

\subsection*{Two Dimensions}
 A dimensional reduction of our setup leads to boundary conditions for 2d $\CN=(4,4)$ theories. As we will explain in Section \ref{sec:SD}, the reduction is subtle, and depends on the relative scales of various parameters. One possible reduction produces 2d sigma models with target $\CM_H$ and boundary conditions of type (B,A,A), which played a prominent role in the gauge-theory approach to the geometric Langlands program \cite{KapustinWitten, GW-surface}. In the presence of an Omega-deformation $\R^2_\epsilon\times \R$, reduction along the circle linking the fixed axis leads to an A-twisted 2d theory, with the axis mapping to a ``canonical coisotropic brane'' $\CB_{cc}$ \cite{KapustinOrlov, NekWitten}, whose algebra of local operators $\text{Hom}(\CB_{cc},\CB_{cc})\simeq \hat \C[\CM_H]$ is known to be a deformation quantization of a chiral ring \cite{KapustinWitten}. A boundary condition in 3d leads to a second brane $\CB_H$ under this reduction, and the space of open string states $\text{Hom}(\CB_{cc},\CB_H)$ is exactly the module that we call $\hat \CB_H$ \cite{KapustinWitten, GW-branes}. This 2d setup was used by \cite{GW-branes} to construct representations of simple Lie algebras, connecting to much of the same mathematics that we study in this paper.
 
Two-dimensional $\CN=(4,4)$ sigma models with hyperk\"ahler targets (such as $\CM_H$) also appeared as effective theories of surface operators in \cite{GW-surface}. Therein, Gukov and Witten constructed noncommutative algebras of interfaces (line operators) in these sigma-models, generating an affine braid group action.
(Such affine braid group actions have played a central role in constructions of knot homology, both in mathematics and physics, \cf\ \cite{SeidelSmith, CautisKamnitzer-sl2, CautisKamnitzer-sln, Webster-quivercat, Webster-HRT}, \cite{GW-Jones}.) In the 2d reductions of 3d gauge theories that we study in Section \ref{sec:SD}, two commuting braid-group actions will appear. One of the two actions coincides with that of \cite{GW-surface}. We expect that the actions can be realized explicitly in terms of UV gauge-theory interfaces, along the lines of \cite{GaiottoKim}, but defer discussions of this to future work.

There are also many parallels between our constructions and boundary conditions for 2d $\CN=(2,2)$ theories. 
In the presence of mass and FI parameters, the boundary conditions in 3d $\CN=4$ theories share many properties with boundary conditions in A-twisted Landau-Ginzburg models \cite{HIV, GMW}, which generate Fukaya-Seidel categories \cite{Seidel-Fukaya}. We make extensive use of the tools of \cite{GMW} to describe the categories of boundary conditions in 2d reductions of massive 3d $\CN=4$ theories.

In a different direction, the maps that we construct between boundary conditions in 3d gauge theories and IR sigma-models are directly analogous to the recent analysis of \cite{HHP} for 2d $\CN=(2,2)$ gauge theories.

\subsection*{Partition Functions} 
It is possible to study many of our boundary conditions using partition functions on ``halves'' of symmetric spaces, such as half-spheres. These can be computed using localization, along the lines of \cite{HLP-wall, DGG-index} (4d) and \cite{SugishitaTerashima, HondaOkuda, HoriRomo, YoshidaSugiyama} (2d and 3d). We will investigate these partition functions in a future publication. Partition functions on a half-space are acted on by operators in the algebras $\hat\C[\CM_H]$ (or $\hat\C[\CM_C]$), and are annihilated by the operators that generate the modules $\hat \CB_H$ (or $\hat \CB_C$) -- \emph{i.e.} partition functions are \emph{solutions} for the difference/differential equations that we set up in the current paper.

\vspace{.5cm}

\subsection{Symplectic duality}
\label{sec:intro-SD}

This paper's underlying mathematical objective is to identify the precise physical underpinning of a beautiful subject known as \emph{symplectic duality}. As presented in the recent work of Braden, Licata, Proudfoot, and Webster \cite{BPW-I, BLPW-II}, symplectic duality is an equivalence between certain collections of structures attached to specific pairs $(\CM_H, \CM_C)$ of hyperk\"ahler cones. There is no general, systematic construction of such pairs. All known examples, however, arise in physics as the Higgs and Coulomb branches of three-dimensional $\CN=4$ gauge theories that
\begin{itemize}
\item[a)] have superconformal infrared fixed points; and
\item[b)] after deformation by mass and FI parameters, acquire isolated massive vacua.%
\footnote{There are several indications that this second property can be relaxed, but it is assumed in much of the current mathematics literature, and for simplicity we will assume it throughout this paper.}
\end{itemize}
It is thus generally expected that symplectic duality should encode mathematical aspects of three-dimensional mirror symmetry, which exchanges the Higgs and Coulomb branches of $\CN=4$ SCFT's.%
\footnote{There are several notions of ``3d mirror symmetry'' in the literature. The classic interpretation \cite{IS} involves a pair of UV gauge theories that flow to the same CFT, with Higgs and Coulomb branches interchanged. However, only a small subset of gauge theories have gauge-theory mirrors in this sense.
 More generally, one may regard 3d mirror symmetry as an involution of a 3d $\CN=4$ SCFT that exchanges the branches in its moduli space. This notion applies to any 3d $\CN=4$ SCFT, and is what we have in mind when we say that symplectic duality should be related to mirror symmetry.}

The most rudimentary aspects of symplectic duality can readily be given a direct physical interpretation. Consider a gauge theory that satisfies the two properties above. By tuning the relative magnitude of real mass and FI deformations, the massive vacua of the theory can either be identified with fixed points of isometries on a resolved $\CM_H$, or fixed points of isometries on a resolved $\CM_C$. This match between 
fixed points is a simple part of symplectic duality. 

Much less trivially, symplectic duality involves an equivalence of two categories $\CO_H$ and $\CO_C$ attached to the Higgs and Coulomb branches, whose spaces of morphisms have two distinct $\Z$ gradings. (The equivalence is a particular case of Koszul duality.)
The categories $\CO_H$ and $\CO_C$ have a somewhat intricate definition; but if one drops one of the gradings
they reduce to (derived) categories of lowest-weight modules for the quantized algebras $\hat \C[\CM_H]$ and $\hat \C[\CM_C]$. 
Symplectic duality gives large collections of pairs $(\hat \CB_H, \hat \CB_C)$ 
of modules for the two algebras that are mapped to each other under the equivalence.

Historically, symplectic duality has its origins in geometric representation theory. The prototypical example of categories $\CO_H$ and $\CO_C$ involves particular modules for a simple Lie algebra $\mathfrak g$ and its Langlands dual $\mathfrak g^\vee$. These categories first appeared in work of Bernstein-Gel'fand-Gel'fand (BGG) \cite{BGG}, were related to D-modules on flag manifolds in \cite{BeilinsonBernstein}, and were shown to be Koszul-dual by Beilinson, Ginzburg, and Soergel \cite{BGS}. (See \cite{Humphreys-book} for a review.) The physical theory related to this representation-theoretic example is the $\CN=4$ theory $T[G]$ introduced in \cite{GW-Sduality} in the context of four-dimensional S-duality. Its Higgs and Coulomb branches are cotangent bundles to the flag manifold for $G$ and its Langlands dual, respectively.

In order to give a physical underpinning to symplectic duality, we would like to find a class of physical objects in 3d gauge theories that could be mapped to 
$\hat \C[\CM_H]$ and $\hat \C[\CM_C]$ modules, in such a way that each physical object $\CB$ gives us a pair $(\hat \CB_H, \hat \CB_C)$ related by the duality. An obvious candidate is a half-BPS boundary condition of the type described above.

We compute the pairs of modules associated to a variety of simple boundary conditions in 3d gauge theories. When a comparison is possible, our results match 
the symplectic duality expectations. In other cases, the physical analysis makes some non-trivial predictions. In Section \ref{sec:SD} we push the 
comparison further and seek a physical origin for the doubly-graded categories at the heart of symplectic duality. This requires careful compactification to two dimensions. We summarize our major conceptual results on page \pageref{intro-SDresults}.

\subsection{A lightning review of $\CN=4$ 3d gauge theories} 

We now turn to a brief review of the structure of 3d $\CN=4$ gauge theories. For further detail, we refer the reader to the appendices or (\eg) our previous work \cite{BDG-Coulomb}.

We consider renormalizable 3d $\CN=4$ gauge theories. They are defined by the following data:
\begin{enumerate}
\item a compact gauge group $G$
\item a linear quaternionic representation $\CR \simeq \mathbb H^N$ of $G$. 
\end{enumerate}
A quaternionic representation means that $G$ acts as a subgroup of $USp(N)$, preserving the canonical hyperk\"ahler structure on quaternionic space $\mathbb H^N$.
We will restrict to the case where the representation decomposes as a sum of a complex representation and its conjugate: 
$\CR = R \oplus R^*$. This appears to be necessary for the theory to admit simple weakly coupled boundary conditions.

The gauge fields lie in vectormultiplets, whose bosonic components include an adjoint-valued triplet of real scalars $\vec\phi \in \mathfrak g^3$ in addition to the gauge connection $A_\mu$. The remaining matter fields are organized in $N$ hypermultiplets, whose bosonic components consist of $4N$ real scalars parametrizing $\mathbb H^N$. The theory has R-symmetry $SU(2)_C\times SU(2)_H$, with $\vec\phi$ transforming as a triplet of $SU(2)_C$ and the hypermultiplet scalars transforming as complex doublets of $SU(2)_H$.%
\footnote{There is a somewhat larger class of renormalizable $\CN=4$ gauge theories that can be defined by Lagrangians that involve both vectormultiplets and hypermultiplets and twisted vectormultiplets and twisted hypermultiplets \cite{Hosomichi:2008jd}. We will not consider them here.}

We will typically choose a splitting of the vectormultiplet scalars into real and complex parts $(\sigma, \varphi)\in \mathfrak g\oplus \mathfrak g_\C$, together with a splitting of the hypermultiplet scalars into pairs of complex fields $(X,Y) = (X^i,Y^i)_{i=1}^N \in R\oplus R^*$. The $SU(2)_C\times SU(2)_H$ R-symmetry rotates the complex splittings of vector and hypermultiplets; each particular splitting is left invariant by a maximal torus  $U(1)_C\times U(1)_H$.

The theory has flavor symmetry $G_C\times G_H$, where $G_C$ is the Pontryagin dual of the abelian part of $G$, essentially
\be G_C \simeq U(1)^{\text{$\#$ $U(1)$ factors in $G$}}\,; \ee
and $G_H$ is the normalizer of $G$ in $USp(N)$. The group $G_H$ is simply the residual symmetry acting on the hypermultiplets. The flavor symmetry $G_C$ is a topological symmetry that rotates the periodic dual photons $\gamma$, which are defined by $d\gamma = *dA_{U(1)}$ for each abelian factor in $G$. The group $G_C$ may enjoy a non-abelian enhancement in the infrared.

The Lagrangian is uniquely determined by the data $(G,\CR)$ together with three sets of dimensionful parameters: 
\begin{enumerate}
\item a gauge coupling $g^2$ for each factor in $G$,
\item a triplet of mass parameters $\vec m\in \mathfrak t_{G_H}^3$,
\item a triplet of FI parameters $\vec t \in \mathfrak t_{G_C}^3$. 
\end{enumerate} 
(Here $\mathfrak t_{G_H},\,\mathfrak t_{G_C}$ denote the Cartan subalgebras of $G_H$, $G_C$.) The masses and FI parameters are expectation values for scalars in background vectormultiplets (or twisted vectormultiplets) for the flavor symmetry group. The masses transform as a triplet of $SU(2)_C$ while the FI parameters transform as a triplet of $SU(2)_H$.
We split these parameters into real and complex parts $m_\R,m_\C$ and $t_\R,t_\C$.

The moduli space of vacua of the gauge theory is hyperk\"ahler. 
Classically, the moduli space is determined by the following equations:
\be
[ \vec \phi, \vec \phi] = 0\,, \qquad
(\vec \phi + \vec m) \cdot (X,Y) =0\,, \qquad
\vec \mu + \vec t=0\,.
\ee
Here the dot denotes the gauge and flavor action on the hypermultiplet scalars and $\vec \mu$ are the three hyperk\"ahler moment maps  
for the $G$ action on the hypermultiplets. 

We will decompose the moment maps into $\mu_\R$ and $\mu_\C$, the real and complex moment maps
computed with respect to the K\"ahler form $\omega = \sum_i \big(|dX^i|^2 + |dY^i|^2\big)$ and the holomorphic symplectic form $\Omega = \sum_i dX^i\wedge dY^i$, respectively. 
Concretely, if we denote by $T\in i\mathfrak g$ the \emph{Hermitian} symmetry generators we can write the moment maps as 
\be
\mu_\C = YTX\,, \qquad \mu_\R = X^\dagger T X - Y^\dagger T Y\, .
\ee
Likewise, we denote the real and complex moment maps for the $G_H$ flavor symmetry as $\mu_{H,\R}$ and $\mu_{H,\C}$.

When the mass parameters vanish, the moduli space contains a Higgs branch $\CM_H$ along which $(X,Y)$ get non-vanishing vacuum expectation values, $\varphi=\sigma=0$, and the gauge group is fully broken. The classical computation
\be \CM_H = \{\vec\mu +\vec t = 0\} /G \simeq \CR/\!/\!/G \ee
is exact, and identifies the Higgs branch as a hyperk\"ahler quotient. The chiral ring $\C[\CM_H]$ of holomorphic functions on the Higgs branch is generated by gauge-invariant polynomials in the $X$'s and $Y$'s, subject to the complex moment map constraint. It is a complex symplectic reduction of the free hypermultiplet ring $\C[X^i,Y^i]$,
\be \C[\CM_H] = \C[X^i,Y^i]^G/(\mu_\C+t_\C=0)\,.\ee

When the FI parameters vanish, the moduli space contains a Coulomb branch $\CM_C$ along which $X=Y=0$ and $\varphi$ and $\sigma$ get vacuum expectation values in the Cartan subalgebra of $\mathfrak g$. The gauge group is generically broken to its maximal torus $\mathbb T_G$, and upon dualizing the abelian gauge fields for $\mathbb T_G$ to periodic scalars, one arrives at the classical description
\be \CM_C^{\rm class} \simeq (\R^3\times S^1)^{\text{rk($G$)}}/\text{Weyl($G$)} \simeq (\C\times \C^*)^{\text{rk($G$)}}/\text{Weyl($G$)}\,.\ee
Perturbative and non-perturbative quantum corrections modify the geometry and topology of the Coulomb branch, in a way that was precisely described in \cite{BDG-Coulomb} (see also \cite{Nakajima-Coulomb, BFN}), and which we summarize later in Section \ref{sec:NC}. The chiral ring $\C[\CM_C]$ of holomorphic functions on the Coulomb branch is generated
by BPS monopole operators, dressed by polynomials in the $\varphi$ vectormultiplet scalars. 

Because of the second set of constraints $(\vec \phi + \vec m) \cdot (X,Y) =0$, the Higgs-branch and Coulomb-branch vevs obstruct each other. The full space of vacua is a direct sum of products of sub-manifolds of the Higgs and Coulomb branches. The FI parameters $t,t_\R$ resolve/deform the Higgs branch, either partially or fully. As they enforce non-zero hypermultiplet vevs, they restrict the possible vectormultiplet vevs and make some or all Coulomb branch directions massive.
The masses $m,m_\R$ resolve/deform the quantum Coulomb branch while making the Higgs branch massive, in the corresponding way.

We consider half-BPS boundary conditions that preserve a 2d $\CN = (2,2)$ sub-algebra of the 3d $\CN=4$ super-algebra.%
\footnote{Other boundary conditions exist which preserve other halves of the 
bulk supersymmetry, such as a 2d $\CN = (p,4-p)$ sub-algebra, but we will not study them here.} %
The choice of sub-algebra uniquely determines a choice of maximal torus $U(1)_C\times U(1)_H$ of the R-symmetry group that is left unbroken, becoming the standard R-symmetry of a 2d $\CN = (2,2)$ theory. Correspondingly, the choice of sub-algebra determines a complex splitting of the vectormultiplet and hypermultiplet scalars. The resulting complex fields become components of twisted-chiral and chiral multiplets (respectively) for the 2d $\CN = (2,2)$ supersymmetry. We refer to the appendices for further details.

\subsection{Structure and results}
\label{sec:structure}

In Sections \ref{sec:N}, \ref{sec:D}, and \ref{sec:xD}, we will introduce three families of $\CN=(2,2)$ boundary conditions for 3d $\CN=4$ gauge theories. We will require that boundary conditions admit a weakly-coupled Lagrangian description. The boundary conditions are classified by two basic pieces of data:
\begin{itemize}

\item A subgroup $H \subset G$ of the gauge symmetry that remains unbroken at the boundary. Two basic choices are $H=G$ and $H=\{id\}$, which correspond respectively to Neumann and Dirichlet boundary conditions for the gauge fields. Once $H$ is chosen, supersymmetry dictates the boundary conditions for the rest of the vectormultiplet scalars and fermions.

\item An $H$-invariant holomorphic Lagrangian splitting of the hypermultiplets $\CR = L\oplus L^*$, with hypermultiplet scalars $X_L\in L$ and $Y_L\in L^*$. The scalars in $L^*$ are given Dirichlet b.c., $Y_L\big|_\pd = c_L$, for some constants $c_L$ compatible with $H$ symmetry; then supersymmetry dictates the boundary conditions for the rest of the hypermultiplet scalars and fermions.

\end{itemize}

When $H=G$ and (necessarily) $c_L=0$, we obtain a minimal supersymmetric extension of Neumann boundary conditions for the gauge fields. These boundary conditions preserve $G_H$ but break $G_C$. We construct their IR images $(\CB_C,\CB_H)$ and the modules $(\hat\CB_C,\,\hat\CB_H)$ in Section \ref{sec:N}. While the Higgs-branch images are fairly straightforward to analyze, the Coulomb-branch images require a one-loop quantum correction, reminiscent of a classic calculation in 2d mirror symmetry \cite{Witten-phases, HoriVafa}.

When $H=\{id\}$ and $c_L$ is generic, both $G$ and $G_H$ are broken at the boundary, while $G_C$ is preserved. We call this a ``generic" Dirichlet boundary condition, and construct their IR images and modules in Section \ref{sec:D}. This time, the Coulomb-branch image can be found by analyzing the semi-classical $\CN=(2,2)$ BPS equations in the bulk (which play a role analogous to those of Nahm's equations in \cite{GW-boundary}). Understanding the modules for the quantized Coulomb-branch algebra requires the introduction of boundary monopole operators.

When $H=\{id\}$ but $c_L$ is chosen so that the flavor symmetry $G_H$ is preserved at the boundary, we obtain ``exceptional'' Dirichlet boundary conditions (Section \ref{sec:xD}). They preserve both $G_H$ and $G_C$, and (for appropriate choices of $L$) their IR images take the form of Lefschetz thimbles on both the Higgs and Coulomb branches. They are direct analogues of the thimble branes that generate the category of boundary conditions in a massive 2d A-model \cite{HIV, Seidel-Fukaya, GMW}. The modules corresponding to thimble branes are either Verma modules or their duals.

These basic boundary conditions may be further enhanced with boundary degrees of freedom, coupled to the bulk hypermultiplet and vectormultiplet fields in a supersymmetric way. We describe such enhancements and their effect on modules $(\hat\CB_C,\,\hat\CB_H)$ in Section \ref{sec:en}. We also present there a particularly interesting class of enhanced boundary conditions for pure $U(N)$ gauge theory related to the Toda integrable system and to recent work of Teleman \cite{Teleman-MS}.

Section \ref{sec:abel} is devoted to boundary conditions in abelian gauge theories. Both mirror symmetry and symplectic duality are very well understood in 
abelian examples and thus we are able to push the comparison between the two quite far. We review the technology of hyperplane arrangements and use it to characterize 
in detail the IR images and modules for all the basic boundary conditions.
We find explicitly that 3d mirror symmetry acts by swapping Neumann and generic Dirichlet boundary conditions, while preserving exceptional Dirichlet,
\be \begin{array}{c}\includegraphics[width=3in]{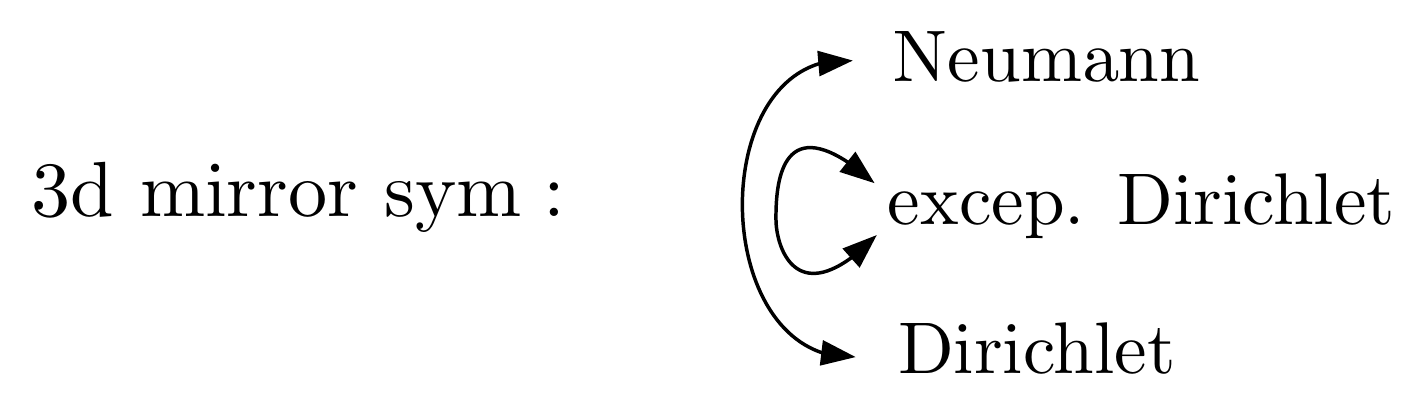}\end{array} \label{MS-intro}  \ee
and we construct half-BPS interfaces implementing mirror symmetry.

In Section \ref{sec:SD} we connect the physics of boundary conditions to symplectic duality. In the case of (massive) abelian theories, each of the three basic classes of boundary conditions produces a well-known set of modules in the categories $\CO_C$, $\CO_H$\,:
\be \label{mod-intro}
\begin{array}{c@{\;}|@{\;}c@{\;}|@{\;}c}
\CB & \CO_C & \CO_H \\\hline
\text{Neumann} & \text{tilting ($T$)} & \text{simple ($S$)} \\
\text{Dirichlet} & \text{simple ($S$)} & \text{tilting ($T$)} \\
\text{excep. Dirichlet} & \text{costandard ($\Lambda$)} &  \text{costandard ($\Lambda$)}
\end{array} \ee
Here ``simple'' modules are irreducible; ``costandard'' modules are an exceptional collection formed by successively extending simple modules, and are dual to ``standard'' or ``Verma'' modules; and ``tilting'' modules are formed by successively extending costandard modules, or (equivalently) by extending Verma modules in the reverse order. By varying the choice of Lagrangian splitting for hypermultiplets, we obtain all possible modules of the various types. Symplectic duality is meant to swap simple and tilting objects in $(\CO_C,\CO_H)$ while preserving costandard objects, and we see immediately that this corresponds to swapping Coulomb and Higgs branches.

In the correspondence \eqref{mod-intro}, there is actually a slight mismatch between the physics and mathematics, which embodies an interesting prediction. Namely, the Coulomb-branch images of Neumann b.c. and the Higgs-branch images of Dirichlet b.c. do not manifestly take the form of tilting modules. These images are not even lowest-weight modules, and do not (naively) belong in categories $\CO_C$, $\CO_H$.
 Rather, as we describe in Sections \ref{sec:NC}, \ref{sec:NC-eg}, \ref{sec:abel-HD}, \ref{sec:abel-CN},
 the images are generalizations of Whittaker modules --- generated by a vector that (roughly) is an eigenvector of the lowering operators. It turns out that the Whittaker modules have a natural deformation to extensions of lowest-weight Verma modules. Mathematically, the deformation is obtained by applying a Jacquet functor (Section \ref{sec:NC-tinf}). We conjecture that
\begin{itemize}
\item All tilting modules (and also all projective modules) in categories $\CO_C$ and $\CO_H$ can be obtained as deformations of generalized Whittaker modules.
\end{itemize}
This generalizes some known relations between Whittaker and tilting/projective modules in the classic BGG category $\CO$ \cite{Frenkel-Gaitsgory, Nadler-mtts, Campbell}.
For abelian theories, the conjecture is proven in \cite{Hilburn}.

As we mentioned before, symplectic duality is much more than a correspondence of some modules; in particular, it predicts a Koszul duality of derived categories $\CO_C,\CO_H$.
Obtaining this equivalence from physics requires a subtle reduction of three-dimensional theories to two dimensions, which we sketch in the remainder of Section \ref{sec:SD}.
\label{intro-SDresults}

The most important object in our construction is a two-dimensional theory $\CT_{2d}$, obtained by placing a 3d $\CN=4$ theory $\CT$ on a circle of radius $R$, turning on real mass and FI parameters $m_\R$, $t_\R$, and sending $R\to 0$, $m_\R\to\infty$, $t_\R\to\infty$ while holding $R\,m_\R t_\R$ fixed. For example, we may take
\be R\to 0\,;\qquad m= R^{\frac12}m_\R\,,\quad t=R^{\frac12} t_\R\quad\text{fixed}\,. \ee
In this limit, the BPS particles remaining in $\CT_{2d}$ originate from domain walls (rather than particles) in $\CT$.

The $\CN=(4,4)$ theory $\CT_{2d}$ admits a large family of topological supercharges $Q_{\zeta,\zeta'}$ (for $\zeta,\zeta'\in \cp^1$) and corresponding topological twists that are compatible with our boundary conditions. Among them is a distinguished supercharge $Q_{0,0}$ that preserves the entire torus $U(1)_C\times U(1)_H$ of the 3d R-symmetry. This turns out to be a B-type supercharge from the perspective of both Higgs and Coulomb-branch sigma models. On the other hand, the derived category $\CO_C$ (resp. $\CO_H$) most naturally arises as the category of boundary conditions in the $Q_{0,1}$ (resp. $Q_{1,0}$) topological twists, which are A-type twists from the perspective of the Coulomb (resp. Higgs) branches.
We propose that we can \emph{deform} the $Q_{0,0}$ twist of $\CT_{2d}$ to either $Q_{1,0}$ or $Q_{0,1}$ without changing the category of boundary conditions, thus obtaining an equivalence between $\CO_C$ and $\CO_H$,
\be \CO_C \;\overset{\sim}\longleftarrow\; \text{B-type $Q_{0,0}$ twist of $\CT_{2d}$} \;\overset{\sim}\longrightarrow\; \CO_H\,.\ee

There are several major advantages to working with the B-type $Q_{0,0}$ twist of $\CT_{2d}$. First, as mentioned above, this twist preserves a full $U(1)_C\times U(1)_H$ R-symmetry, leading to two $\Z$ gradings in the category of boundary conditions, one homological (meaning it is shifted by $Q_{0,0}$) and one internal (meaning it commutes with $Q_{0,0}$). We may then transport these two gradings to both categories $\CO_C$ and $\CO_H$. In the mathematics of categories $\CO_C,\CO_H$, the second, internal, grading is both essential in defining Koszul duality and famously mysterious. The physics here suggests a way to define it.

\begin{figure}[htb]
\centering
\includegraphics[width=5.4in]{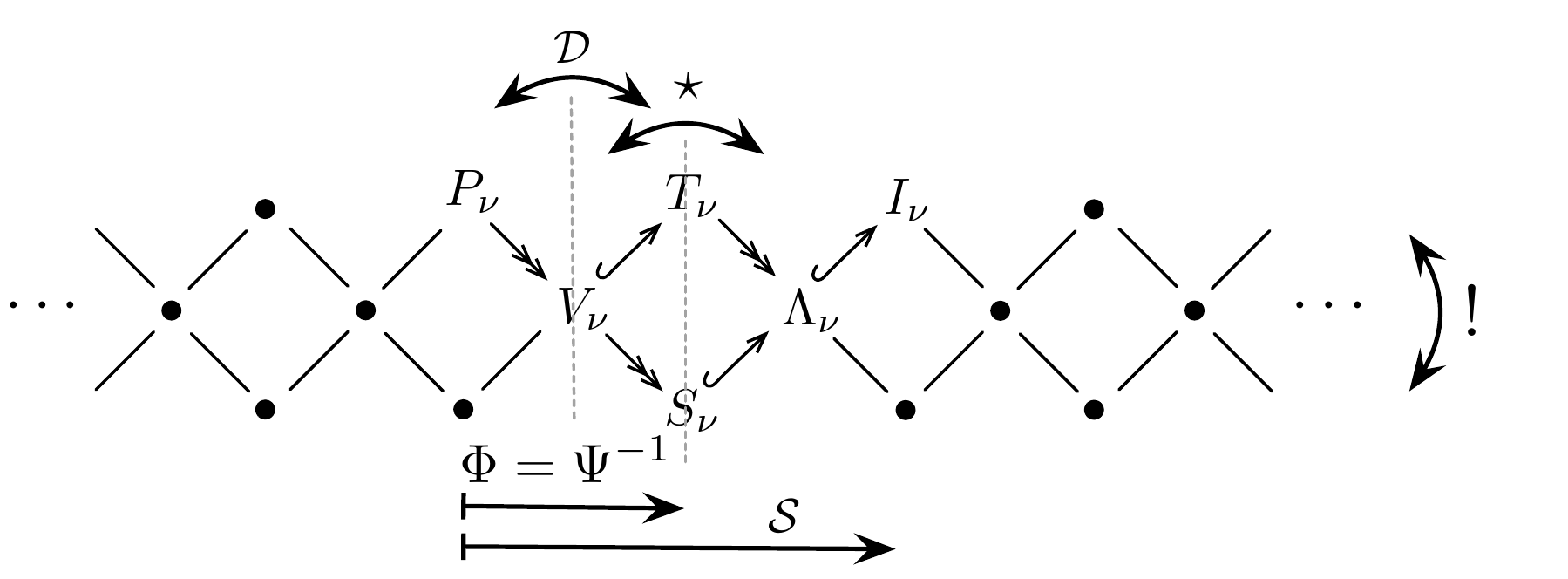}
\includegraphics[width=5in]{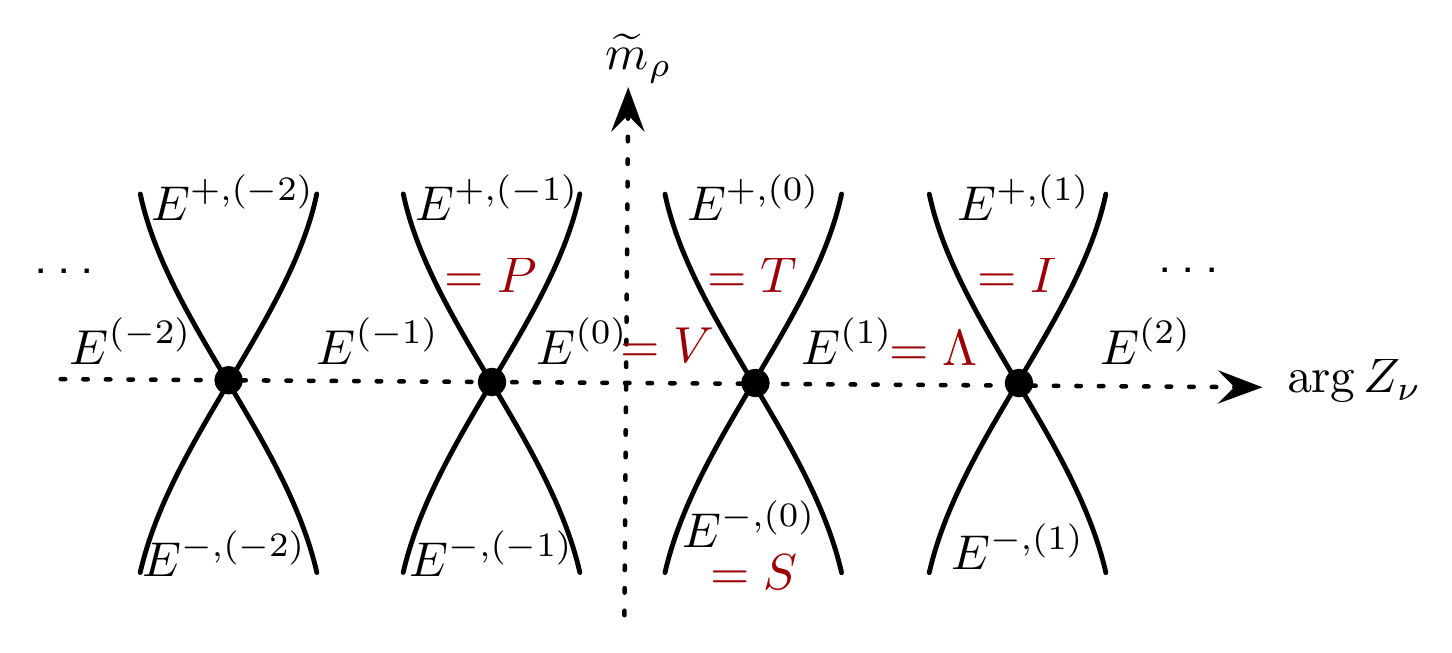}\hspace{.5in}
\caption{Top: A picture of derived category $\CO$, with six distinguished collections of modules, and various functors represented as isometries. Koszul duality is the vertical reflection~`$!$'. Bottom: chambers and generalized exceptional collections in the category of boundary conditions for the B-type twist of $\CT_{2d}$ for real $Z_\nu(m,t)$. Koszul duality is wall crossing from $\mathrm{Im}\, \wt m_\rho < 0$ to $\mathrm{Im}\, \wt m_\rho > 0$.}
\label{fig:O-intro}
\end{figure}

Second, a large set of functors that act on categories $\CO_C,\CO_H$ --- including Koszul duality and braiding actions --- all receive a common interpretation as \emph{wall-crossing} transformations in the category of boundary conditions for the $Q_{0,0}$ twist of $\CT_{2d}$. To get a flavor of this relation, consider the ``picture'' of derived category $\CO=\CO_H$ (say) at the top of Figure \ref{fig:O-intro} (explained in much greater detail in Sections \ref{sec:cast}--\ref{sec:tale}).%
\footnote{This picture is assembled by combining many mathematical results and conjectures on category $\CO$, including those of \cite{Maz-lectures, MOS, BLPW-gale, BLPW-hyp, BPW-I, BLPW-II, Losev-quantizations, Losev-CatO}.} %
There are six distinguished collections of modules in category $\CO$: simples (irreducibles) $S_\nu$, standards (Vermas) $V_\nu$, costandards $\Lambda_\nu$, projectives $P_\nu$, tiltings $T_\nu$, and injectives~$I_\nu$. The objects in each collection are labelled by vacua $\nu$ of our theory, and each collection generates the entire category. Every symmetry of the figure corresponds to an invertible functor from derived $\CO$ to itself or to the opposite category $\CO^{\rm op}$.

Similarly, the category of boundary conditions for the B-type twist of $\CT_{2d}$ has many generalized exceptional collections of objects labelled by the massive vacua of the theory.
Each generalized exceptional collection is associated to a \emph{chamber} in the space of parameters of the theory, which include $m,t$ and a twisted mass $\wt m_\rho$ for the anti-diagonal $U(1)_\rho$ subgroup of $U(1)_C\times U(1)_H$ (\ie\ for the symmetry that provides an internal grading). The chamber structure is controlled by $\wt m_\rho$ in addition to standard complex central charge functions $Z_\nu=Z_\nu(m,t)$, which depending bilinearly on complexified mass and FI parameters. A particular slice in parameter space is depicted on the bottom of Figure \ref{fig:O-intro}. It corresponds to real $Z_\nu(m,t)$ and infinitesimal imaginary $\wt m_\rho$. The generalized exceptional collections $E^{(n)}$, $E^{\pm,(n)}$ are in 1-1 correspondence with distinguished collections of objects in category $\CO$ at the top of the figure, and we propose to identify them. We also propose that Koszul duality can be interpreted as the wall-crossing transformation from negative imaginary $\wt m_\rho$ to positive imaginary $\wt m_\rho$. We expand on these ideas in Section \ref{sec:WC}.

The braiding of mass and FI parameters at $\wt m_\rho=0$ has been well studied in the mathematics literature and is known to be a manifestation of wall crossing. (A physical construction of this braiding was realized in \cite{GW-surface}.) In contrast, the wall crossing obtained by varying $\wt m_\rho$ seems to be new.

A third advantage of studying the B-type twist of $\CT_{2d}$ is that, via 2d mirror symmetry, this theory can be related to an A-twisted Landau-Ginzburg model with a very concrete superpotential (Section \ref{sec:LG}). 
When the underlying 3d $\CN=4$ theory is an A-type quiver gauge theory, the resulting superpotential coincides with the Yang-Yang functional for a rational Gaudin model \cite{NS-I, GK-3d}. 
In this case, the very A-twisted Landau-Ginzburg model appeared in recent work on knot homology \cite{Wfiveknots, GW-Jones}. 
More generally, the superpotential appears to govern the physics of an M2-M5 brane system that has appeared in many physical constructions of knot homology, related to the classic M5-M5' construction of \cite{OV, GSV}.
(Other B-twisted Landau-Ginzburg models have also been proposed to describe the same system, \eg\ \cite{GukovWalcher, GNSSS}, \cite{KR-MFI, KR-MFII}; their relation with $\CT_{2d}$ is still unclear.)

We will give a direct argument that the scaling limit that defined the theory $\CT_{2d}$ for an A-type quiver gauge theory should capture the low energy physics of M2 branes stretched between two orthogonal stacks of M5 branes. 
We hope to elaborate on the connection with knot homology in future work.

\section{Pure Neumann Boundary Conditions}
\label{sec:N}

In this section we focus on half-BPS Neumann boundary conditions that preserve 2d $\CN = (2,2)$ supersymmetry. We work out their infrared images and the modules they produce in Omega backgrounds. We devote special attention to the effect of real mass and FI deformations, which can cause some boundary conditions to break supersymmetry in the IR.

\subsection{Definition And Symmetries}
\label{sec:defN}

Our boundary conditions can obtained as the dimensional reduction of half-BPS 
Neumann boundary conditions for 5d $\CN=1$ gauge theories, 
which preserve a 4d $\CN=1$ super-Poincar\'e subalgebra 
of the full supersymmetry algebra. They are defined by a combination of standard Neumann b.c. for the gauge fields,
accompanied Dirichlet b.c. for the adjoint real scalar field $\sigma$ in the gauge multiplet.
The boundary conditions also set to zero an appropriate half of the gauginos. 

A concise justification for these boundary conditions can be given along the lines of \cite{GW-boundary}: the 5d gauge 
theory with gauge group $G$ can be re-cast as a 4d gauge theory with gauge group $\CG$, the group of 
maps from the half line into $G$. The complexified covariant derivative 
\begin{equation}
\CD_1 :=  D_1 + \sigma
\end{equation}
in the direction $x^1$ normal to the boundary behaves as a chiral multiplet and thus Dirichlet boundary conditions for 
$\sigma$ are compatible with the $F_{1\mu}=0$ Neumann boundary conditions for the gauge field.  

Upon dimensional reduction to three dimensions we recover the desired Neumann boundary conditions for three-dimensional 
$\CN=4$ gauge theories: 
\begin{equation}
F_{1\mu}\big|_\partial=0\,, \qquad\qquad \sigma\big|_\partial = 0\,, \qquad\qquad D_1 \varphi\big|_\partial =0\,, 
\end{equation}
where $\varphi$ is the complex adjoint scalar superpartner of the gauge field, which arises from the dimensional reduction of 
$A_4 + i A_5$. These boundary conditions preserve a 2d $\CN = (2,2)$ supersymmetry. 
They also classically preserve a $U(1)_H \times U(1)_C$ subgroup of the $SU(2)_H \times SU(2)_C$ R-symmetry of the bulk theory, which can be identified with the usual vector and axial R-symmetries on the boundary:
\be U(1)_H = U(1)_V\,,\qquad U(1)_C = U(1)_A\,. \label{HVCA} \ee

A more intrinsic three-dimensional definition of these boundary conditions can be obtained by writing 3d $\CN=4$ gauge theory as a two-dimensional $\CN=(2,2)$ theory with gauge group $\CG$, as outlined in Appendix \ref{app:2d}, and consistently imposing Neumann or Dirichlet boundary conditions for entire $\CN=(2,2)$ supermultiplets.

If the gauge group has an abelian factor, the boundary condition can be deformed by a boundary FI term and a boundary $\theta$ angle, which as usual are grouped into a complex parameter $t_{2d}$.
The boundary FI term shifts the boundary value of the abelian part 
$\sigma_{U(1)}$ of $\sigma$.
If we dualize the corresponding abelian gauge field $A_{U(1)}$ to a periodic scalar field $\gamma_{U(1)}$ (the ``dual photon"), 
which receives Dirichlet boundary conditions, 
the boundary $\theta$ angle will shift the boundary value of $\gamma_{U(1)}$ so that altogether
\begin{equation} 
(\sigma_{U(1)}+ i \gamma_{U(1)} )\big|_\partial = t_{2d}\,.
\end{equation}
Each abelian factor of the gauge group is associated to a ``topological" symmetry $U(1)_t\subset G_C$, whose current is $*\,F_{U(1)}$, and which rotates the dual photon. 
This symmetry is broken explicitly by Neumann boundary conditions, since $U(1)_t$ rotations will shift the boundary $\theta$ angle. 

We must also describe boundary conditions for the matter hypermultiplets. We first consider a single $\CN=4$ hypermultiplet with complex scalars $(X,Y)$. 
Two basic supersymmetric boundary conditions for the hypermultiplet are  \cite{DG-E7}
\begin{align} \label{basicH}  \CB_X: &\quad Y \big|_\partial=0  \qquad D_1 X\big|_\partial=0 \cr
 \CB_Y: &\quad  X\big|_\partial=0 \qquad  D_1 Y\big|_\partial=0\,.
\end{align}
The boundary conditions also set to zero an appropriate half of the fermions.
(In terms of $(2,2)$ supersymmetry, the bulk scalars $X$ and $Y$ are the leading components of chiral superfields, whose F-terms contain $D_1Y$ and $D_1X$, respectively, \cf\ Appendix \ref{app:hyper}.  The boundary conditions here follow from setting an entire chiral superfield to zero at the boundary.) The boundary values $X|_\pd$ or $Y|_\pd$ that survive behave as chiral operators under the boundary supersymmetry algebra.

These basic boundary conditions each preserve a $U(1)_f$ flavor symmetry that rotates $X$ with charge $1$ and $Y$ with charge $-1$.
The two boundary conditions $\CB_X$ and $\CB_Y$ can be related by a simple transformation involving an extra chiral multiplet $\Phi$ supported on the boundary.%
\footnote{The transformation was discussed in the context of 4d $\CN=2$ theories in \cite{DGG, DGV-hybrid}, and is closely related to the action of S-duality on boundary conditions of abelian 4d $\CN=4$ theory \cite{KS-mirror, Witten-sl2}.} %
For example, we can start from $\CB_X$ and add a boundary superpotential
\be W_{\rm bdy} = X\big|_\partial\,\Phi\,. \label{WXYPhi} \ee
The chiral field acts as a Lagrange multiplier setting $X\big|_\partial=0$, while the boundary superpotential relaxes the $Y\big|_\partial=0$ boundary condition to 
$Y\big|_\partial=\Phi$. Thus we recover $\CB_Y$. This relation implies the existence of a boundary mixed 't Hooft anomaly for $U(1)_f$ and $U(1)_A$.
If we normalize to 1 the coefficient of the mixed anomaly due to a chiral multiplet of $U(1)_f$ charge $1$, $\CB_X$ ($\CB_Y$)
has an anomaly coefficient of $1/2$ ($-1/2$).

When there are multiple hypermultiplets $\{X^i,Y^i\}_{i=1}^N$, one can again choose a basic boundary condition $\CB_X$ or $\CB_Y$ for each $i$, 
or more generally some $\CB_{L}$ associated to a Lagrangian splitting $L$ of the hypermultiplet scalars into two sets:
we use a $USp(N)$ rotation to re-organize the scalar fields into some new sets $(X_L, Y_L)$ and 
pick Neumann boundary conditions for $X_L$ and Dirichlet for $Y_L$.

In order to combine Neumann boundary conditions for the gauge fields and simple boundary conditions for the matter fields, we need the splitting $L$ 
to be gauge invariant. This is only possible if the hypermultiplets transform as a direct sum of a unitary representation of $G$ and its conjugate $R\oplus R^*$, or equivalently if $G$ acts as a subgroup of $U(N)\subset USp(N)$.%
\footnote{For example, this will not be possible if the matter fields include an odd number of ``half-hypermultiplets''.} %
We denote the corresponding boundary condition as $\CN_L$. 

If the gauge group has an abelian factor, the $\CN_L$ boundary condition generically breaks $U(1)_A$ via an anomaly. 
However, an appropriate linear combination $U(1)_A'$ of $U(1)_A$ and $U(1)_t$ is preserved, since both $U(1)_t$ and $U(1)_A$ are broken at the 
boundary by an amount proportional to $F_{23}$. If the boundary mixed anomaly coefficient is $n$, the unbroken 
symmetry current is $J_A - n J_t$.

\subsection{General structure of images}
\label{sec:genstruc}

In the presence of a boundary condition $\CB$, one may consider the moduli space of 
vacua of the full bulk-boundary system that preserve 2d $\CN = (2,2)$ supersymmetry. We refer to this as the IR ``image'' $\CB_{IR}$ of $\CB$. There is a natural map from the space of vacua $\CB_{IR}$ of the full system to the moduli space of vacua $\CM=\CM_C\cup\CM_H\cup...$ of the bulk theory. Denoting the image of this map as $\CL_{IR}$, we may give $\CB_{IR}$ the structure of a fibration
\be \hspace{.5in} \begin{array}{l}{\CB_{IR}}\\ \hspace{.2cm}  \downarrow \\  \CL_{IR} \subset \CM_C\cup \CM_H \cup... \end{array} \label{Bvac} \ee
We may further decompose $\CB_{IR}$ into components that project to particular branches of the bulk moduli space,
\be \CB_{IR} = \begin{array}{l}\CB_{C} \\\hspace{.2cm}\downarrow \\ \CL_C\subset  \CM_C \end{array} \cup \begin{array}{l}\CB_H \\\hspace{.2cm}\downarrow \\ \CL_H \subset \CM_H \end{array} \cup ...\,, \label{BCHvac} \ee
leading to the notion of Coulomb and Higgs-branch images $\CB_C,\,\CB_H$.

Just as 3d $\CN=4$ supersymmetry ensures that all components of the bulk moduli space are hyperk\"ahler \cite{HKLR-HK}, 2d $\CN=(2,2)$ supersymmetry ensures that the IR images of boundary conditions are supported on holomorphic Lagrangian submanifolds $\CL_C\subset \CM_C$ and $\CL_H\subset \CM_H$. More precisely, $\CL_C$ and $\CL_H$ should be holomorphic Lagrangian at smooth points, away from potential singularities.

A quick but indirect proof of this claim is to note that topological boundary conditions in Rozansky-Witten theory are supported on holomorphic Lagrangian submanifolds of the target space \cite{KRS}.
At low energies, away from singularities, our bulk gauge theory has an effective description as an $\CN=4$ sigma-model with target space $\CM_H$ or $\CM_C$, each of which admits a topological twist that leads to a Rozansky-Witten theory. $\CN = (2,2)$ boundary conditions preserve the topological supercharges, so they become topological boundary conditions of the type studied by \cite{KRS}. A more direct argument is given in Appendix \ref{app:Lag}.

For most of the boundary conditions we study in this paper, the full moduli spaces $\CB_{C},\CB_H$ and their projections to the bulk vacua $\CL_C,\CL_H$ will be identical, \ie\ the projections in \eqref{BCHvac} are one-to-one. In physical terms, this means that for every bulk vacuum consistent with the boundary condition $\CB$, there is a unique vacuum of the full bulk-boundary system. Of course, this need not be true in general, and it is always possible to enhance a boundary condition with additional boundary degrees of freedom so that the projections in \eqref{BCHvac} are highly non-trivial.

\subsection{Higgs-branch image}
\label{sec:NHiggs}

Now, let us return to Neumann boundary conditions.
In this section, we are interested in vacua which project to Higgs branch vacua. Classically, such vacua are described by field configurations that satisfy the boundary conditions at $x^1=0$ and possibly evolve as a function of $x^1$ 
according to the BPS equations of 2d $\CN = (2,2)$ supersymmetry.
We refer to Appendix \ref{app:2d} for the full set of BPS 
equations. 

To begin with, we set real mass and FI parameters to zero and consider the Higgs branch as a complex manifold. In this case, we only need the simple holomorphic BPS equations 
\begin{equation}
\mu_\C(X,Y) + t_\C= 0\,, \qquad \qquad \CD_1 X = 0\,, \qquad \CD_1 Y =0\,,
\end{equation}
where the complex moment map $\mu_\C(X,Y) \in \mathfrak g^*$ is defined as 
\begin{equation}
\mu_\C(X,Y) := YTX\,,
\end{equation}
with $T$ a generator of the gauge group action on the hypermultiplet fields. We denote the set of complex FI parameters as $t_\C$, implicitly identifying them with an element in the abelian factor of $\mathfrak g^*$. 

As the hypermultiplet vevs are covariantly constant, gauge-invariant polynomials in $X$ and $Y$ 
must have the same value at $x^1=0$ and $x^1 = \infty$. 
Thus the Higgs branch image $\CN^{(H)}_L$ of the space of vacua of a simple Neumann boundary conditions $\CN_L$ consists classically of 
the complex submanifold of the full Higgs branch $\CM_H$ defined by the $\CB_L$ boundary conditions on the elementary fields. 
Mathematically, this is the image of $L$ under the hyper-K\"ahler quotient that defines the Higgs branch; it is automatically a holomorphic Lagrangian submanifold of $\CM_H$.

The Higgs branch of a 3d $\CN=4$ gauge theory is not subject to quantum corrections. We similarly expect $\CN^{(H)}_L$
to be uncorrected. Quantum corrections to the complex
geometry of $\CN^{(H)}_L$ would take the form of boundary superpotential terms, which would be incompatible with 
the $U(1)_V$ R-symmetry preserved by the $\CN_L$ boundary conditions.%
\footnote{It should be also possible to formulate the problem 
in a B-twisted version of the system. The B-twist of the 2d $(2,2)$ supersymmetry algebra preserved by the boundary 
corresponds to the Rozansky-Witten twist of the bulk gauge theory.}

The geometry of $\CN^{(H)}_L$ is also encoded in the chiral ring $\C[\CN^{(H)}_L]$ of boundary local operators.
In the bulk, there is a chiral ring $\C[\CM_H]$ of protected operators whose vevs give holomorphic functions on the Higgs branch. By bringing bulk operators to the boundary, one obtains a map
\be \C[\CM_H] \to \C[\CN^{(H)}_L]\,. \label{bulkboundaryring} \ee
For $\CN_L$ boundary conditions, this map is a surjection, and $\C[\CN^{(H)}_L]$ simply consists of gauge-invariant polynomials in the 
$X_L$ scalar fields that survive at the boundary. (The normal derivatives $\CD_1 Y_L$ also survive at the boundary are chiral, but they are exact in the chiral ring.) 
Alternatively, the kernel of \eqref{bulkboundaryring} contains the bulk operators that vanish when brought to the boundary. Formally, these form an ideal $\CI$ in the bulk ring, and we have $\C[\CN^{(H)}_L] =  \C[\CM_H]/\CI$.

\subsubsection{Quantum Higgs-branch image}
\label{sec:qNHiggs}

As discussed in the introduction, there is a variant of the notion of boundary chiral ring that will play a crucial role in this paper. Boundary conditions that preserve $U(1)_A$ R-symmetry are compatible with a twisted $\wt\Omega$-deformation in the plane parallel to the boundary. This is a mirror of the standard $\Omega$-deformation. The $\wt\Omega$-deformation is known to localize a non-linear sigma model with hyperk\"ahler target space $\CM$ to a supersymmetric quantum mechanics whose operator algebra $\hat\C[\CM]$ quantizes the Poisson algebra $\C[\CM]$ of holomorphic functions on $\CM$ \cite{Yagi-quantization}.
We similarly expect the $\wt\Omega$-deformation to localize a gauge theory to a gauged supersymmetric quantum mechanics, in which a quantization of the chiral ring $\hat \C[\CM_H]$ appears as the gauge-invariant part of the operator algebra associated to a quantization of the matter fields \cite{BDG-Coulomb}.

Concretely, our starting point is $N$ copies of the Heisenberg algebra
\be [\hat Y_i,\hat X_j] = \epsilon\,\delta_{ij}\,, \ee
which quantizes the ring $\C[T^*\C^N]$ of hypermultiplet scalars. Call this algebra $H$.
Gauge transformations are generated by the complex moment map operator 
\begin{equation}
\hat \mu_\C(\hat X,\hat Y) = \; :\!\! \hat Y T\hat X\!\!: \; = \; :\!\! \hat Y_L T_L\hat X_L\!\!:  \,.
\end{equation}
(We emphasize that this in independent of the Lagrangian splitting, as long as the generators $T$ are appropriately redefined.)
As the classical moment map is quadratic in the fields, the quantum moment map is well defined up to a constant, which we fix by normal ordering. The ambiguity only affects the abelian factors of the gauge group, and can be absorbed in the choice of complex FI parameters $t_\C$. 

In order to obtain $\hat \C[\CM_H]$, we quotient the Heisenberg algebra by either the left or right ideal generated by the complex moment map constraint $\hat\mu_\C + t_\C$, and then restrict to gauge-invariant operators. Formally, 
\be  \label{qCMH}
\hat \C[\CM_H] = \big( (\hat\mu_\C + t_\C)H\backslash H \big)^G = \big( H/ H(\hat\mu_\C + t_\C) \big)^G\,. \ee
Equivalently, we can restrict first to the gauge-invariant part of the Heisenberg algebra, $H^G$. Inside $H^G$, the complex moment map constraint forms an ordinary two-sided ideal, which can be expressed as $\big( (\hat\mu_\C + t_\C)H\big)^G$ or $\big( H (\hat\mu_\C + t_\C)\big)^G$, or in abelian theories simply as $H^G(\hat\mu_\C+t_\C)$. Thus,
\be \hat\C[\CM_H] = H^G\big/\big( (\hat\mu_\C + t_\C)H\big)^G =  H^G\big/ \big( H(\hat\mu_\C + t_\C)\big)^G\,. \label{qCMH2} \ee
The equivalence of all these descriptions follows from basic results in representation theory, which are collected (\eg) in \cite{McGN-der}.%
\footnote{For example, to see that $\big( H/H(\hat\mu_\C + t_\C) \big)^G$ is equivalent to $H^G\big/\big( H(\hat\mu_\C + t_\C)\big)^G$, one may start with the exact sequence of $G$-modules $0\to H(\hat\mu_\C+t_\C) \to H \to H/H(\hat\mu_\C+t_\C) \to 0$. Since $G$ is compact, the functor of taking $G$-invariants is exact, whence $0\to \big(H(\hat\mu_\C+t_\C)\big)^G \to H^G \to \big(H/H(\hat\mu_\C+t_\C)\big)^G \to 0$ is again an exact sequence that provides the desired isomorphism.}

In the presence of a boundary condition $\CB$, the boundary chiral operators are restricted to lie at the 
origin of the of the $\wt\Omega$-deformation plane as well. Thus the $\tilde \Omega$-deformation kills the conventional 
notion of boundary chiral ring. It is still possible, though, to consider the action of protected bulk operators on the space
of boundary chiral operators. We thus obtain a module $\hat\CB^{(H)}$ for the quantum algebra $\hat \C[\CM_H]$.
We will use a convention such that right boundary conditions correspond to 
left modules for the bulk quantum algebra, so that bulk operators act from the left both 
in space-time and in equations (as in Figure \ref{fig:modules}).
Similarly, left boundary conditions correspond to 
right modules and interfaces would correspond to bimodules. 

If we specialize to Neumann boundary conditions, the module $\hat \CN_L^{(H)}$ can be identified with 
the space of gauge-invariant polynomials in $X_L$, with the operators $\hat X_L$ and $\hat Y_L$ 
acting as
\begin{equation} \label{MNHeis}
\hat X_L \cdot p(X_L) = X_L p(X_L)\,, \qquad \qquad \hat Y_L \cdot p(X_L) = \epsilon \partial_{X_L} p(X_L)\,.
\end{equation} 
If we denote by $|\CN_L\rangle$ the state in the quantum mechanics created by the boundary condition at $x^1=0$ with 
\begin{equation}
\hat Y_L |\CN_L\rangle =0\,,
\end{equation}
the elements of the module are $p(\hat X_L)|\CN_L\rangle$\footnote{We abuse notation by using a `ket' to denote elements of a module even in the absence of an inner product.}.
We will often shorten this to $p(\hat X_L)\big|$.

If the gauge group includes an abelian factor, we need to take into account the effect of the breaking of 
$U(1)_t$ and the possible anomaly in $U(1)_A$. The latter is of course worrisome, as it threatens to 
make the $\wt\Omega$-deformation inconsistent. Happily, the existence of an unbroken combination of
$U(1)_A$ and $U(1)_t$ saves the day. In the absence of the anomaly, the breaking of $U(1)_t$
would require one to set $t_\C$ to zero, as it is (the mirror of) a twisted mass for $U(1)_t$. 
In the presence of an anomaly with coefficient~$n$, one expects to set $t_\C = -n \epsilon$, as the 
$U(1)_t$ generator has to be added to the $U(1)_A$ generator employed in the $\wt\Omega$-deformation.

This expectation agrees well with our construction. In the absence of an anomaly, we would expect that the gauge-invariant elements of our module are precisely
\be   p(\hat X_L)\big| \qquad\text{s.t.}\qquad \hat \mu_\C \cdot p(\hat X_L)\big| = 0\,, \ee
since $\hat \mu_\C$ is the generator of gauge transformations. In particular, the identity operator $1\big|$ should be annihilated by $\hat\mu_\C$. In the presence of an anomaly, we instead find that the identity and other gauge-invariant operators are annihilated by
\be \hat X_L T_L \hat Y_L  = \;:\!\!\hat Y_L T_L X_L\!\!:  - \frac{\epsilon}{2} \mathrm{Tr}(T_L) = \hat\mu_\C + t_\C \ee
where the anomaly coefficient is precisely $n = \frac{1}{2} \mathrm{Tr}(T_L)$. We thus obtain a module for \eqref{qCMH} with $t_\C = - n\epsilon$ as desired.

\subsubsection{Twisting with line operators}
\label{sec:lineH}

The above restriction on the values of $t_\C$ can be relaxed to a more general value
\be t_\C = (k-n)\epsilon\,\qquad k\in \Z  \label{tCk} \ee
by adding a supersymmetric abelian Wilson line of charge $k$ along the axis of the $\wt\Omega$-background geometry, perpendicular to the boundary. In the presence of the Wilson line, local operators at the boundary must have gauge charge $-k$. Correspondingly, the elements of the module $\hat \CN_L^{(H)}$ are polynomials $p(\hat X_L)\big|$ that satisfy
\be (\hat \mu + t_\C)\cdot p(\hat X_L)\big| = (\hat X_L T_L \hat Y_L + k\epsilon)\cdot p(\hat X_L)\big| = 0\,.\ee

It is also possible to include non-abelian line operators, allowing for a rich generalization of our story and connections to \cite{AG-loops}, which we leave for a future publication. 

\subsubsection{Effect of real FI and real masses}
\label{sec:NH-mt}

Boundary conditions preserving 2d $\CN = (2,2)$ supersymmetry are compatible with both real mass and real FI 
deformations of the bulk gauge theory. This should be contrasted with the 
complex mass and FI deformations, which behave as twisted masses from the point of view of 
2d $\CN = (2,2)$ supersymmetry and thus are only available 
if the boundary conditions preserve the corresponding bulk global symmetries. 

Real FI parameters $t_\R$, when available, (partially) resolve the Higgs branch of vacua. 
Some of the Neumann $\CN_L$ boundary conditions may not be compatible with the resolution:
it may be impossible to satisfy the real moment map constraint on the 
locus $Y_L=0$, so that no supersymmetric vacuum exists for the system. 
The list of $t_\R$-feasible $\CN_L$ boundary conditions will depend on a choice of 
``chamber'' in the real FI parameter space. 

Each real mass deformation $m_\R$ is associated to an infinitesimal global symmetry transformation on the Higgs branch, and thus to a $\mathfrak{u}(1)_m$ subalgebra of the flavor symmetry~$\mathfrak g_H$. The mass $m_\R$ itself may be thought of as the generator of this subalgebra. Turning on a real mass deformation restricts the bulk Higgs branch to a submanifold $\CM_H^{0}[m_\R]$
of fixed points under $m_\R$. The fixed-point manifold is union of components
\be \CM_H^{0}[m_\R] = \bigcup_\nu \CM_H^{0}[m_\R^\nu] \ee
labelled by the specific inequivalent lifts $m_\R^{\nu}\in \text{Cartan}(\mathfrak u(N))$ 
of $m_\R\in \mathfrak t_H\subset \mathfrak g_H$ to a combination of global \emph{and} gauge symmetry Cartan generators 
that fix the expectation values of the matter hypermultiplets. The different components $\CM_H^{0}[m^\nu_\R]$
may intersect in the Higgs branch, but are actually separated along the Coulomb branch by different 
vevs for the Coulomb branch scalar $\sigma$, encoded in $m_\R^{\nu}$.

Interestingly, the moduli space of 2d vacua in the presence of a boundary condition is {\it not} 
restricted to the fixed points of $m_\R$. In order to understand this observation, it useful to remember that 
$m_\R$ is the expectation value of the real scalar for a background vector multiplet, and thus 
in the presence of $m_\R$ the complexified covariant derivative normal to the boundary becomes 
\begin{equation} \label{CD1}
\CD_1 := D_1 + \sigma + m_\R 
\end{equation}
(with $\sigma$ and $m_\R$ acting in the appropriate representation of $G$ and $G_H$).
The gauge invariant combinations of $X$ and $Y$ will now grow or decay 
exponentially along the $x^1$ direction depending on their flavor charges. 
On the Higgs branch, this flow can be identified with inverse gradient flow for the real moment
map%
\footnote{It is well known that BPS equations in $\CN=1$ supersymmetric quantum mechanics produce gradient flow with respect to a real superpotential (``Morse function'') \cite{Witten-Morse}. The structure we find for 3d $\CN=4$ theory can be understood by reducing it to supersymmetric quantum mechanics with a real superpotential equal (modulo F-terms) to the real moment map $m_\R\cdot \mu_{H,\R} + \sigma\cdot \mu_\R$. On the Higgs branch, this leads to gradient flows of \eqref{hm}. In the full gauge theory, one must also vary $\sigma$, leading to the additional equation \eqref{D1s} below. A similar structure appeared in 2d $\CN=(2,2)$ gauged sigma models studied in \cite{Witten-path}.}
\be h_m = m_\R \cdot \mu_{H,\R}  \label{hm} \ee
for the $\mathfrak{u}(1)_m$ symmetry generated by $m_\R$.
Thus a necessary condition for a point in $\CN^{(H)}_L$ to define (classically) a 2d vacuum 
is that it will flow to the fixed locus under this vector field. 

Geometrically, one may define submanifolds $\CM_H^{<}[m_\R]$ ($\CM_H^{>}[m_\R]$) containing the points that flow to $\CM_H^{0}[m_\R]$ under gradient flow (inverse gradient flow); then the potential 2d vacua exist on intersections of $\CN^{(H)}_L$ with these submanifolds,
\be \label{Nflows}
\text{2d vacua}:\qquad \begin{array}{c@{\qquad}c} \CN_L^{(H)}\cap \CM_H^{>}[m_\R] & \text{(left b.c.)} \\[.1cm]
\CN_L^{(H)}\cap \CM_H^{<}[m_\R] & \text{(right b.c.)}  \end{array}\,.
\ee
If the intersections in \eqref{Nflows} are empty, then the boundary condition under consideration breaks supersymmetry. This never happens for $\CN_L$ boundary conditions, but may occur in more general examples.

An elementary example is provided by the theory of a free hypermultiplet $(X,Y)$. The real moment map for the $U(1)$ flavor symmetry that rotates $X,Y$ with opposite charges is $\mu_{H,\R} = |X|^2-|Y|^2$, and $h_m = m(|X|^2-|Y|^2)$. The Higgs branch is $\CM_H=\C^2$. For positive $m$, the bulk vacuum lies at $\CM_H^0[m_\R] = \{X=Y=0\}$ and the gradient-flow manifolds are $\CM_H^{>}[m_\R] = \{Y=0\}$ and $\CM_H^{<}[m_\R]=\{X=0\}$. Correspondingly, the left boundary condition $\CB_X$ has a full $\C$ worth of classical 2d vacua, while the left boundary condition $\CB_Y$ has the single vacuum $X=Y=0$.

We conjecture that condition \eqref{Nflows} is also sufficient for the existence of 2d vacua, at least for appropriate values of the 2d FI parameters. 
If we could replace the gauge theory with a sigma model 
with target $\CM_H$ this would automatically be true. 
Proving it in the gauge theory requires looking at the (2,2) D-term equation (\cf\ \eqref{22D})
\begin{equation} \label{D1s}
D_1 \sigma + g_{YM}^2 \mu_\R = 0\,.
\end{equation}
If the 2d FI parameters set the value of $\sigma$ at the boundary to the same value they assume at infinity, determined by the requirement that 
the vevs of $X$ and $Y$ at infinity are annihilated by $\sigma + m_\R = m^\nu_\R$, we can take $\sigma$ to be constant. The 
gradient flow of $X$ and $Y$ is then solved by simple exponentials. 
For general 2d FI parameters the statement is likely to remain true, but a proof would require some analysis. 

A full description of the moduli space of vacua of the system should specify 
the projection onto the space of the bulk vacua, \ie\ the projection of  $\CN^{(H)}_L \cap \CM_H^{>}[m_\R]$
onto the fixed locus $\CM_H^{0}[m_\R]$. Clearly, the projection associates to each point of $\CM_H^{>}[m_\R]$ the endpoint of 
the gradient flow into the fixed locus. 

It is also easy to describe the behavior of chiral ring operators when restricted to gradient-flow manifolds.
If we decompose the Higgs-branch chiral ring $\C[\CM_H]$ into subspaces with positive, zero, and negative charges under $m_\R$ as
\be
\C[\CM_H] = \C[\CM_H]_> \oplus \C[\CM_H]_0 \oplus \C[\CM_H]_<\,,
\label{eq:mdecomp}
\ee
then every element in $\C[\CM_H]_>$ will vanish on $\CM_H^{<}[m_\R]$, every element in $\C[\CM_H]_<$ will vanish on $\CM_H^{>}[m_\R]$, 
and every element in $\C[\CM_H]_>$ and $\C[\CM_H]_<$ will vanish on $\CM^{0}[m_\R]$.

We can further lift this to a gauge-theory statement. For every choice of $m_\R^\nu$ labeling a component $\CM_H^{0}[m^\nu_\R]$ of the $m_\R$-fixed locus, we decompose the hypermultiplet scalar fields into subspaces of positive, zero, or negative $m_\R^\nu$ charge. Then if we compute the gradient flows at constant $\sigma$, we have
\begin{itemize} \label{Xpm}
\item $\CM_H^{<}[m^\nu_\R]$
is defined by setting to zero $X^+_{m^\nu_\R}$ and $Y^+_{m^\nu_\R}$ of positive charge, 
\item $\CM_H^{>}[m^\nu_\R]$
is defined by setting to zero $X^-_{m^\nu_\R}$ and $Y^-_{m^\nu_\R}$ of negative charge, 
\item $\CM_H^{0}[m^\nu_\R]$
is defined by setting to zero $X^\pm_{m^\nu_\R}$ and $Y^\pm_{m^\nu_\R}$ of non-zero charge. 
\end{itemize}

Altogether, the inclusion of real masses has two effects on our boundary conditions: it restricts the full moduli space of 2d vacua as in \eqref{Nflows}, 
but it may effectively \emph{enlarge} the space of (classical) 2d vacua compatible with a single bulk vacuum $\nu$.
It is important to remember that we are giving here a classical description of the two-dimensional space of vacua. 
If there is a continuous moduli space of classical 2d vacua that are associated to a single bulk vacuum, 
the system may become gapless, strongly coupled, or unstable at low energy. If the moduli space is non-compact, the situation is especially bad; the study of two-dimensional theories with non-compact moduli, such as cigar sigma-models (\cf\ \cite{HoriVafa, HoriKapustin}), suggests that supersymmetry will be broken. 

This complication will occur often for $\CN_L$ boundary conditions as one adds real mass deformations. If a left
$\CN_L$ imposes Neumann boundary conditions on matter fields with negative charge under $m^\nu_\R$, 
the system will typically have a branch of classical 2d vacua parameterized by expectation values of these fields, which projects down to a fixed 
bulk vacuum\footnote{The problem could be ameliorated by turning on complex mass deformations $m_\C$ in the same direction as $m_\R$: these suppress expectation values of charged fields and force the system back to $\CM_H^{0}[m^\nu_\R]$ (see Section \ref{sec:NC-SQED} for an example).}.

Altogether, it is tempting to refer to boundary conditions for which the intersections \eqref{Nflows} are unbounded as ``$m_\R$-infeasible.'' We expect that they break supersymmetry in the IR for given values of $m_\R$.
In general, for any UV boundary condition $\CB$ with a Higgs-branch image $\CB^{(H)}$, we say
\be \label{m-feasible}
\text{$\CB$ is $m_\R$-feasible} \quad \Leftrightarrow \quad \begin{cases} \text{$\CB^{(H)}\cap \CM_H^>[m_\R]$ is nonempty and bounded} & \text{(left b.c.)} \\
\text{$\CB^{(H)}\cap \CM_H^<[m_\R]$ is nonempty and bounded} & \text{(right b.c.)} \end{cases}\,.
\ee

If we turn on both real FI parameters and real masses, the theory will generically admit dynamical BPS domain walls 
that interpolate between vacua of the theory, associated to gradient flow solutions interpolating between the corresponding 
fixed points. The tension of these domain walls is controlled by a central charge 
equal to the difference in the value of $m_\R \cdot \mu_{H,\R}$ at the fixed points (see Appendix \ref{app:central} for details).
These domain walls preserve the same supersymmetry as the boundary conditions. The existence of these domain walls, which can lie at arbitrary distance from a boundary, may result in non-compact directions in the moduli spaces of 2d vacua. 

\subsection{Examples}
\label{sec:NH-eg}

\subsubsection{SQED}
\label{sec:NH-SQED}

We consider a $U(1)$ gauge theory with $N$ hypermultiplets $(X_i,Y_i)$ of charge $(+1,-1)$ under the gauge symmetry. The theory has a topological $U(1)_t$ symmetry and a $G_H=PSU(N)$ flavor symmetry acting on the hypermultiplets. The real and complex moment maps for the $U(1)$ gauge symmetry are
\be \mu_\R = \sum_{i=1}^N(|X_i|^2 - |Y_i|^2)\,,\qquad \mu_\C = \sum_{i=1}^N X_iY_i\,. \label{SQED_mom} \ee
The Higgs branch $\CM_H$ is the hyperk\"ahler quotient by the $U(1)$ symmetry with moment map constraints $\mu_\R +t_\R =0$ and $\mu_\C +t_\C=0$. 

In order to study Neumann boundary conditions, we must set the complex FI parameter $t_\C$ to zero. Then for $t_\R>0$ ($t_\R<0$) the Higgs branch is identified as the cotangent bundle $T^*\cp^{N-1}$, with the $Y$'s (the $X$'s) providing homogeneous coordinates for the base. At $t_\R = 0$, the Higgs branch becomes singular, and can be identified as the minimal nilpotent orbit inside $\mathfrak{sl}_{N,\C}$.
The chiral ring $\C[\CM_H]$ is generated by the gauge invariant bilinears $X_iY_j$ subject to the vanishing of the complex moment map. 

A general Neumann boundary condition is labelled by a sign vector $\varepsilon = (\varepsilon_1,\ldots,\varepsilon_{N} )$,
\be 
\CN_\varepsilon\,:\quad \text{Neumann for gauge multiplet and} \qquad \begin{cases} \CB_{X_i} & \varepsilon_i = + \\  \CB_{Y_i} & \varepsilon_i = - \end{cases}\,. \label{SQED_Neumann} 
\ee
These are clearly compatible with the vanishing of the complex moment map when $t_\C=0$ and define holomorphic Lagrangian submanifolds of the Higgs branch. The boundary conditions with all $\varepsilon_i= +$ or all $\varepsilon_i = -$ preserve the full $PSU(N)$ flavor symmetry. In the other cases, the flavor symmetry is broken to a Levi subgroup. The naive axial anomaly in the presence of an $\CN_\varepsilon$ boundary condition is
\be
n = \frac12 \sum\varepsilon_i\,, \label{SQED_n} \ee
which must be compensated be redefining the axial current by a multiple of $U(1)_t$. It is easy to find the images of these boundary conditions on the Higgs branch, for (say) positive $t_\R>0$:
\begin{itemize}
\item $\CN_{-\cdots-}$ is the vanishing cycle $\mathbb{CP}^{N-1}$.
\item $\CN_{+-\ldots-}$ and its permutations are the conormal bundles to the $N$ coordinate hyperplanes in $\mathbb{CP}^{N-1}$.
\item A general $\CN_\varepsilon$ is the conormal bundle to the space of complex lines in $\C^{N}$ that lie inside the subspace $\{Y_i = 0 \, | \, \varepsilon_i = +\}$.
\item $\CN_{+\cdots+}$ is $t_\R$-infeasible: it has no supersymmetric vacua when $t_\R>0$.
\end{itemize}

\begin{figure}[htb]
\centering
\includegraphics[width=4.3in]{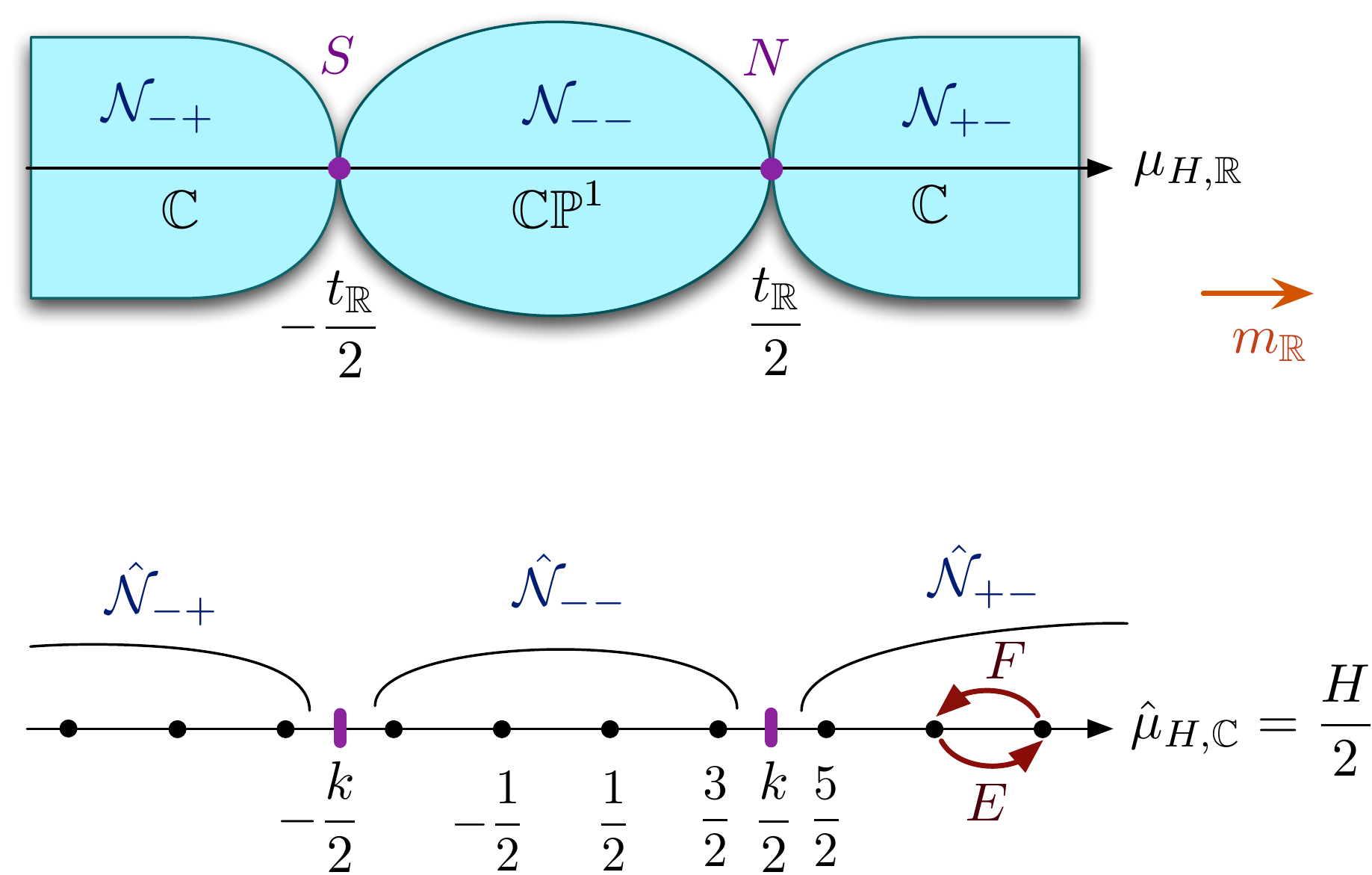}
\caption{Top: the Higgs-branch images of Neumann boundary conditions $\CN_\varepsilon$ for SQED with $N=2$ hypermultiplets with $t_\R>0$. Bottom: the corresponding $\mathfrak{sl}_2$ modules that these boundary conditions define in the $\wt\Omega$-background, for $t_\C/\epsilon = k$ a positive integer (here $k=4$).}
\label{fig:cp1NH}
\end{figure}

In the case of $N=2$ hypermultiplets, where $\CM_H=T^*\cp^1$, we can depict the images of $\CN_\varepsilon$ boundary conditions as in the top of Figure \ref{fig:cp1NH}. All the images lie on the holomorphic Lagrangian slice of the Higgs branch with $X_iY_i=0$ $\forall i$, which contains $\cp^1$ together with the fibers at its north and south poles. This slice is an $S^1$ fibration over the real line parameterized by the real moment map for the Cartan subalgebra of the $PSU(2)$ flavor symmetry 
\be
\mu_{H,\R} = \frac12(|X_1|^2-|Y_1|^2)-\frac12(|X_2|^2-|Y_2|^2)\, .
\ee
The fibers degenerate at the points $\mu_{H,\R} =\pm t_\R/2$, cutting out the $\cp^1$ and its fibers at the north and south poles.

Now consider turning on a real mass $m_\R$, associated to the Cartan subalgebra of the flavor symmetry group $PSU(2)$, which rotates the Higgs branch around the axis in Figure \ref{fig:cp1NH}. (We continue to specialize to the case $N=2$.) There are two bulk vacua, or fixed points of the rotation: the North pole of $\cp^1$, where $\mu_{H,\R}=\frac{t_\R}{2}$ and $\sigma=-\frac{m_\R}{2}$; and the South pole, where $\mu_{H,\R}=-\frac{t_\R}{2}$ and $\sigma=\frac{m_\R}{2}$. Gradient flows for the real moment map $h_m = m_\R \, \mu_{H,\R}$ preserve the slice $X_iY_i=0$ depicted in Figure \eqref{fig:cp1NH}.
Depending on the sign of $m_\R$, one may have either gradient flows from the North to the South pole, or vice versa, corresponding to the existence of a single dynamical domain wall 
between the two vacua. 

Without loss of generality, we can analyze in detail the case $m_\R>0$ and focus on right boundary conditions. In the notation of Section \ref{sec:NH-mt}, the locus $\CM_H^{<}[m_\R]$ contains the fiber at the South pole (which flows to the South pole) and the $\cp^1$ itself (which flows to the North pole).
Thus the boundary conditions have the following 2d moduli spaces:
\begin{itemize}
\item  $\CN_{--}$: In the South bulk vacuum, we have a single 2d vacuum. In the North bulk vacuum, there is a $\mathbb{CP}^{1}$ space of classical vacua,
although the region near the South pole of $\mathbb{CP}^{1}$ corresponds to a dynamical domain wall detached from the boundary and thus 
may lie at infinite distance in field space. The quantum dynamics of the system may be subtle.  
\item $\CN_{-+}$: In the South bulk vacuum, the classical moduli space of 2d vacua coincides with the noncompact North pole fiber. The quantum dynamics 
of the system will be non-trivial. Analogy with a 2d cigar sigma-model suggests that SUSY will be broken, so that the boundary condition is ``$m_\R$-infeasible.''
In the North bulk vacuum, we have no supersymmetric 2d vacua, unless we allow for a dynamical domain wall at infinite distance.
\item $\CN_{+-}$: In the South bulk vacuum, we have no supersymmetric 2d vacua. In the North bulk vacuum, we have a single 2d vacuum. 
\item $\CN_{++}$: Supersymmetry is broken ($t_\R$-infeasible). 
\end{itemize}
For general $N\geq 2$, the situation is similar. Geometrically, a choice of mass parameters $m_\R$ defines a standard flag inside $\C^N$, and the boundary conditions $\CN_\varepsilon$ that have continuous 2d moduli spaces are precisely those for which the subspace $\{Y_i\,|\,\varepsilon_i=-\}$ is compatible with the flag. The associated moduli spaces are conormal bundles to Schubert cells.

\subsubsection{SQED, quantized}
\label{sec:NH-qSQED}

In the presence of an $\wt\Omega$-background with equivariant parameter $\epsilon$, the Higgs-branch chiral ring becomes a non-commutative algebra, which isomorphic to a central quotient of the enveloping algebra of $\mathfrak{sl}_N$, \cf\ \cite{BDG-Coulomb}. Explicitly, the quantized chiral ring $\hat \C[\CM_H]$ is obtained by starting with $N$ copies of the Heisenberg algebra generated by $\hat X_i,\hat Y_i$ with $[\hat Y_i,\hat X_j]=\epsilon\,\delta_{ij}$, restricting to gauge-invariant operators --- which form a subalgebra generated by the binomials $\hat X_i\hat Y_j$ --- and imposing the complex moment-map constraint
\be
\hat \mu_\C + t_\C = \sum_{i=1}^{N} :\! \hat X_i \, \hat Y_i \! : + \,  t_\C  = 0 \, .
\label{cmm-SQED}
\ee
The generators of $\mathfrak{sl}_N$ are identified as follows:
\begin{itemize}
\item $\hat X_i\hat Y_j$ with $i<j$ are raising operators,
\item $\hat X_i\hat Y_j$ with $i>j$ are lowering operators, 
\item Differences of $\hat X_i\hat Y_i$ are the Cartan generators.
\end{itemize}
The complex FI parameter $t_\C$ determines the values of all the Casimir operators through the complex moment map constraint~\eqref{cmm-SQED}.

As noted above, the Neumann boundary condition $\CN_\varepsilon$ naively has an axial anomaly with coefficient $n= \frac{1}{2}\sum_i \varepsilon_i$. Following Section \ref{sec:qNHiggs}, a consequence is that we must choose $t_\C = - \epsilon n$ in order for the moment map to annihilate the identity operator on the boundary. Indeed, we find
\be
\begin{aligned}
\mu_\C - \frac{\epsilon}{2}\sum_i\varepsilon_i  = \sum_{\varepsilon_i = +} \hat X_i \, \hat Y_i  + \sum_{\varepsilon_i = -} \hat Y_i \, \hat X_i \,,
\end{aligned}
\ee
which annihilates the identity operator since $Y_i |\CN_\varepsilon\rangle= 0$ for $\varepsilon_i = +$ and $X_i |\CN_\varepsilon\rangle = 0$ for $\varepsilon_i = -$.
With this $\varepsilon$-dependent choice of $t_\C$, we find that
\begin{itemize}
\item $\hat \CN_{-...-}$ and $\hat\CN_{+...+}$ are trivial modules containing only the identity operator;
\item $\hat \CN_{+...+-...-}$ are infinite-dimensional modules containing gauge-invariant boundary operators of the form 
\[ \prod_{\varepsilon_i=+}X_i^{a_i}\prod_{\varepsilon_i=-} Y_i^{b_i}|\CN_\varepsilon\rangle\] with $\sum a_i-\sum b_i = 0$.
\end{itemize}
All of these representations are irreducible.

If we include Wilson lines that set $t_\C = (k-n)\epsilon$ for $k\in \Z$ and allow charged operators on the boundary, then we find for $k\geq 0$ we find that $\hat \CN_{-...-}$ produces the $k$-th symmetric power of the anti-fundamental representation of $\mathfrak{sl}_N$ (while $\hat \CN_{+...+}$ admits no boundary operators); and for $k\leq 0$, $\hat \CN_{+...+}$ produces the $|k|$-th symmetric power of the fundamental (while $\hat \CN_{-...-}$ admits no boundary operators). The other infinite-dimensional representations are irreducible quotients of Verma modules.

We can illustrate this in more detail for $N=2$. (For $N=3$, see Section \ref{sec:abel}.) Let us introduce the notation
\be \label{EFH-XY}
H = 2\hat\mu_{H,\C} = \hat X_1\, \hat Y_1 - \hat X_2 \, \hat Y_2 \,, \qquad E = \hat X_1 \, \hat Y_2\,, \qquad F = \hat X_2 \, \hat Y_1\,,
\ee
for the bulk gauge-invariant operators. These are simply the components of the complex moment map for the $PSU(2)$ flavor symmetry.
Note also that $\hat X_1\,\hat Y_1+\hat X_2\hat Y_2 = -t_\C -1$. It is a straightforward computation to check that
\be
[ H, E] = 2 \epsilon E \qquad [ H,  F] = -2 \epsilon F \qquad [E,F]=\epsilon H 
\ee
and the quadratic Casimir is
\be
C_2 = \frac{1}{2}H^2+EF+FE = \frac{1}{2}(t_\C^2-\epsilon^2) \, .
\ee

To visualize modules for this algebra, we draw the weight spaces of $\hat\mu_{H,\C}$ at the bottom of Figure \ref{fig:cp1NH}. The operators $E$ and $F$ raise and lower the weights. We suppose that a combination of Wilson lines and anomaly shifts sets $t_\C = k\epsilon$ with $k\geq 1$. Then there are two distinguished weight spaces at $H = \pm\, k$ where the operators $:\!\hat X_1\hat Y_1\!:$ and $:\!\hat X_2\hat Y_2\!:$ (respectively) have eigenvalue zero. (These weight spaces are never realized in modules.) The modules $\hat \CN_\varepsilon$ contain weight spaces lying on one side or the other of the distinguished ones, as shown in Figure~\ref{fig:cp1NH}.
Namely,
\begin{itemize}
\item $\hat \CN_{--}$ is the $k$-dimensional irreducible representation of $\mathfrak{sl}_2$~\footnote{The boundary operators in this case are $Y_1^{k_1}Y_2^{k_2}| \CN_{--} \rangle$ with $k_1+k_2 = k-1$, reproducing the Borel-Weil construction of the finite dimensional representations of $\mathfrak{sl}_2$.}.
\item $\hat \CN_{-+}$ is an irreducible highest-weight Verma module, generated from the highest-weight vector $Y_1^k|\CN_{-+}\rangle$ by acting with $F^a$.
\item $\hat \CN_{+-}$ is (similarly) an irreducible lowest-weight Verma module.
\item $\hat \CN_{++}$ admits no boundary operators.
\end{itemize}

\subsubsection{SQCD}
\label{sec:NH-SQCD}

Now consider a $G=U(K)$ gauge theory with $N$ hypermultiplets $(X^i,Y_i)=(X^i_a,Y_i^a)$ in the fundamental representation of the gauge group. There is a topological $G_C=U(1)_t$ symmetry due to the $U(1) \subset U(N)$ factor of the gauge group, and a Higgs-branch flavor symmetry $G_H=PSU(N)$. The Higgs-branch chiral ring consists of polynomials in the gauge-invariant bilinears $\sum_a Y^a_i X_a^j$ (\ie\ the components of the moment map $\mu_{H,\C}$ for $G_H$) subject to the vanishing of the complex moment map for $G$,
\be
{\left( \mu_{\C} +t_\C  \right)_a}^b = \sum_{i=1}^{N} X_a^i \, Y^b_i  + t_\C\, \delta_a{}^b = 0\,.
\label{SQCD_cmm}
\ee

As we consider Neumann boundary conditions, the choice of boundary condition $\CB_L$ for the matter fields must preserve the full $U(K)$ gauge symmetry. As before, we must set $t_\C=0$ and we will first assume that $t_\R = 0$. The Higgs branch is then identified with the closure of the nilpotent orbit $\overline\CO_{\rho}\subset \mathfrak{sl}_N$ whose dual partition is $\rho^T = [N-K,K]$~\cite{GW-Sduality}. (In other words, it is the nilpotent orbit with $K$ Jordan blocks of size 2 and $N-2K$ trivial Jordan blocks of size 1.) A Neumann boundary condition is again labelled by a sign vector,
\be 
\CN_\varepsilon\,:\quad \text{Neumann for gauge multiplet and} \qquad \begin{cases} \CB_{X^i} & \varepsilon_i = + \\  \CB_{Y_i} & \varepsilon_i = - \end{cases}\,, \label{SQCD_Neumann} 
\ee
where now, for example, $ \CB_{X}$ means $Y|=0$ for all gauge components of $Y$.

The quantum Higgs-branch algebra $\hat C[\CM_H]$ is generated by the traceless part of the meson matrix $\hat M_i^j = \sum_a \hat Y^a_i \hat X_a^j$, which is the quantum moment map for the $G_F = PSU(N)$ flavor symmetry group. Thus the algebra may again be described as a central quotient 
of the universal enveloping algebra of $\mathfrak{sl}_N$. Similarly, the modules may be described as representations of $\mathfrak{sl}_N$.

A real FI parameter resolves the singularity of the Higgs branch, which becomes the cotangent bundle of a Grassmannian: $T^*Gr(K,N)$. We must now take into account the real moment map constraint
\be
\sum_{i=1}^{N} \left( X_i X_i^\dagger - Y_i^\dagger Y_i \right)  + t_\R = 0 \, .
\label{SQCD_rmm}
\ee
Assuming that $t_\R>0$, the base $Gr(K,N)$ is parameterized by the $Y$'s: the $K\times N$ matrix of the $Y$'s specifies the embedding of a $K$-plane in $N$-space. The Neumann boundary condition $\CN_\varepsilon$ is feasible provided the number of fundamental hypermultiplets with $\CB_X$ type boundary conditions, or equivalently the number of $+$ signs in $\varepsilon$, is less than $K$. Otherwise, there are no supersymmetric vacua. The image of a feasible boundary condition $\CN_\varepsilon^{(H)}$ then becomes the conormal bundle to the space of $K$-planes inside the subspace $\{ Y_i = 0 \, | \, \varepsilon_i = +\} \subset \mathbb{C}^N$. In particular, the image of the boundary condition $\CN_{-\cdots-}$ is simply the base $Gr(K,N)$.

If generic real masses $m_\R$ are turned on, the bulk theory has ${N \choose K}$ massive vacua $\nu$, labelled by 
subsets of $K$ $Y_i$'s. In each vacuum, the corresponding $K\times K$ submatrix of the $Y_i$ gets a vev proportional to the identity.
Correspondingly, the lift $m_\R^\nu = \sigma^\nu + m_\R$ is the unique lift of $m_\R$ to a generator of gauge and flavor symmetries that preserves the vev of the $Y_i$. Then the component $\CM_H^{>}[m_\R^\nu]$ of 
$\CM_H^{>}[m_\R]$ that flows to a given vacuum $\nu$ is given by a collection of equations of the general form 
\be
X_< =0 \qquad\text{or} \qquad Y_< =0
\ee
setting to zero the fields of negative charge under $\sigma^\nu+m_\R$. For example, if $(K,N)=(2,3)$ and $t_\R>0$, the first vacuum takes the form
\be 
X = \left( \begin{array}{cc|c} 0 & 0 & 0 \\ 0 & 0 & 0 \end{array}\right)\,,\qquad
Y^T =  \left( \begin{array}{cc|c} c & 0 & 0 \\ 0 & c & 0 \end{array}\right)\,\ee
with $|c|^2 = t_\R$, and the corresponding lift has
\be
\sigma^\nu=\left(\begin{smallmatrix} -m_{1,\R} & 0 \\ 0 & -m_{2,\R} \end{smallmatrix}\right) \, .
\ee 
For the ordering $m_{1,\R}<m_{2,\R}<m_{3,\R}$, the thimble $\CM_H^{>}[m_\R^\nu]$ is the image of
\be 
X = \left( \begin{array}{cc|c} * & * & * \\ 0 & * & * \end{array}\right)\,,\qquad
Y^T =  \left( \begin{array}{cc|c} * & 0 & 0 \\ * & * & 0 \end{array}\right)\,\ee
under hyperk\"ahler reduction.

More geometrically, a generic choice of real masses $m_{i,\R}$, puts an ordering on the $N$ fields $Y_i$, and thus defines a standard flag in $\C^N$. The submanifolds $\CM_H^{>}[m_\R^\nu]$ are conormal bundles to the ${N\choose K}$ Schubert cells in $Gr(K,N)$ with respect to this flag. The moduli space of (classical) 2d vacua associated to a boundary condition $\CN_\varepsilon$ is obtained by intersecting the images $\CN_\epsilon^{(H)}$ with Schubert cells.

\subsection{Coulomb-branch image}
\label{sec:NC}

We assume here that our gauge theory admits a Coulomb branch in which all matter fields are massive. 
Classically, the Coulomb branch of a theory with gauge group $G$
is parameterized by generic Cartan-valued vevs of the 
adjoint real $\sigma$ and complex $\varphi$ scalars, together with the dual photons for the unbroken Cartan subalgebra.
The $\varphi$ expectation values prevent the matter fields from getting expectation values even in the 2d sense.
The classical moduli space $\CN^{(C)}_L$ of 2d vacua in the presence of $\CN_L$ boundary conditions is thus 
parameterized by generic values of $\varphi$ and fixed values of $\sigma$ determined by the boundary FI parameters $t_{2d}$. 

The Coulomb branch of $\CN=4$ gauge theories is subject to important quantum corrections. 
These include one-loop effects and instanton corrections. Our purpose here is to determine the 
corresponding corrections to $\CN^{(C)}_L$. 

In abelian gauge theories, the Coulomb branch only receives one-loop corrections \cite{SW-3d, IS, dBHOY}.
As a complex manifold, it is described by 
the expectation values of the complex scalars $\varphi$ valued in the Lie algebra of $G$ 
and of BPS 't Hooft operators (monopole operators) $v_A$ labelled by a magnetic charge $A$, \ie\ a cocharacter $A \in \mathrm{Hom}(U(1),G)$. 
The quantum-corrected chiral-ring relations take the form \cite{BKW-monopoles, CHZ-Hilbert, BDG-Coulomb}
\begin{equation}
v_A v_B = v_{A+B} P_{A,B}(\varphi, m_\C)\,,
\end{equation}
where $m_\C$ are complex mass deformation parameters and $P_{A,B}(\varphi, m_\C)$ is a product of contributions from all hypermultiplets
\begin{equation} \label{PAB}
P_{A,B}(\varphi,m) =  \prod_{\text{$i$ s.t. $Q^i_A Q^i_B<0$}} M_i^{\text{min}(|Q^i_A|,\, |Q^i_B|)}  \;=\; \prod_{i=1}^N M_i^{(Q_{A}^i)_+ + (Q_{B}^i)_+ - (Q_{A+B}^i)_+}\,.
\end{equation}
Here $Q_A^i$ is the charge of $X_i$ under the gauge symmetry generator $A$, $(x)_+ = \mathrm{max}(x,0)$ 
and $M_i$ is the effective complex mass of the $i$-th hypermultiplet, a linear combination of $\varphi$ and $m_\C$. (In parallel with the effective real mass in \eqref{CD1}, we could write $M_i = (\varphi T+m_\C T^H)_i$\,.)

Notice that the middle expression in \eqref{PAB} makes it clear that $P_{A,B}(\varphi,m_\C)$ is independent of the choice of Lagrangian splitting $L$ for the hypermultiplets: changing the splitting sends $(Q_A^i,Q_B^i,M_i)\to (-Q_A^i,-Q_B^i,-M_i)$ for some $i$'s, leaving the product invariant up to a sign that can be absorbed in the definition of the $v_A$'s.
Thus we could equivalently write
\begin{equation}
P_{A,B}(\varphi,m) = \prod_{i=1}^N M_{L,i}^{(Q_{A,L}^i)_+ + (Q_{B,L}^i)_+ - (Q_{A+B,L}^i)_+}   \qquad \qquad 
\end{equation}
where $Q_{A,L}^i$ is the charge of $X_{L,i}$ under the gauge symmetry generator $A$.

We claim that the quantum-corrected space of vacua $\CN^{(C)}_L$ is the submanifold of the Coulomb branch defined by 
the relations
\begin{equation} \label{NC-abel}
\CN_L^{(C)}\,:\qquad \begin{cases} \ds v_A  = \;  \xi_{-A} \hspace{-.3cm}\prod_{\text{$i$ s.t. $Q_{A,L}^i>0$}} \hspace{-.3cm} M_{L,i}^{|Q_{A,L}^i|} =  \xi_{-A} \prod_{i=1}^N M_{L,i}^{(Q_{A,L}^i)_+}  & \text{left b.c.} \\
\ds v_A  = \;  \xi_{A} \hspace{-.3cm}\prod_{\text{$i$ s.t. $Q_{A,L}^i<0$}} \hspace{-.3cm} M_{L,i}^{|Q_{A,L}^i|}  =  \xi_{A} \prod_{i=1}^N M_{L,i}^{(-Q_{A,L}^i)_+} & \text{right b.c.}
\end{cases}
\end{equation}
where $\xi_A = e^{-A\cdot t_{2d}}$.
The most basic check of our claim is that it has the correct symmetry. For (say) a left boundary condition, the left hand side of the equation has topological 
$U(1)_t$ charge $A$, while the right hand side has charge $0$. The left hand side has axial R-charge%
\footnote{Throughout the paper we denote the axial R-symmetry as $U(1)_A$, not to be confused with the cocharacter $A$ appearing here.} %
$\frac{1}{2} \sum_i |Q_A^i|$, while the right hand side has charge $\frac{1}{2} \sum_i (Q_{A,L}^i)_+$. The 
mismatch is $\frac{1}{2} \sum_i Q_A^i$. As we discussed in the previous section, the $\CN_L$ boundary conditions preserve 
the difference between the axial R-symmetry generator and a $U(1)_t$ generator proportional to the anomaly coefficient 
$\frac{1}{2} \sum_i Q^i$. 

We will subject our claim to several other checks throughout the draft. 
Here we can give an intuitive motivation for our claim. The field configuration of a monopole operator approaching a Neumann boundary condition 
is the same as the field configuration for a monopole approaching a second monopole of opposite charge. 
The right hand side of the relation \eqref{NC-abel} for a left boundary condition is similar to the right hand side of $v_A v_{-A}$ but only includes contributions from the 
half of the hypermultiplet fields that survive at the boundary. The slightly different behavior of left and right boundary conditions will be justified in Section \ref{sec:en}, by calculating effective twisted superpotentials at the boundary.

Let us now consider non-abelian gauge theories. The main result of \cite{BDG-Coulomb} is a description of the Coulomb branch of a general nonabelian gauge theory in terms of an ``abelianization map''. A complementary approach appeared in the mathematical literature in~\cite{Nakajima-Coulomb, BFN}.
Essentially, the expectation values of nonabelian Coulomb branch operators are written as certain rational functions of 
a set of variables $\varphi_a$, $v_A$ associated to the Cartan subalgebra of the gauge group $G$,
which satisfy the relations 
\begin{equation}
v_A v_B = v_{A+B} \frac{P_{A,B}(\varphi, m_\C)}{P^W_{A,B}(\varphi)}
\end{equation}
where the numerator is computed as before from the complex masses of hypermultiplets and the denominator is the analogous expression 
involving the complex masses of vectormultiplets.

We propose that the quantum corrected space of vacua $\CN^{(C)}_L$ is the submanifold of the Coulomb branch defined by the pullback under the abelianization map of 
the relations 
\begin{equation} \label{NC-nonabel}
\qquad v_A =  \xi_{-A} \frac{\prod_{i=1}^N M_{L,i}^{(Q_{A,L}^i)_+}}{\prod_{\text{roots $\alpha$}} (\alpha \cdot \varphi)^{(\alpha \cdot A)_+}}\qquad \text{(left b.c.)}
\end{equation}
where $\xi_A = e^{A\cdot t_{2d}}$. 
We will verify through concrete examples that this definition gives a well-defined locus in the Coulomb branch, 
setting the vevs of nonabelian monopole operators to appropriate polynomials in $\varphi$.

\subsubsection{Images and the integrable system}
\label{sec:intsys}

A useful perspective on Coulomb-branch images of various boundary conditions comes from viewing the Coulomb branch as a complex integrable system (\cf\ \cite{Nakajima-Coulomb, BFN}). Namely, there is a natural holomorphic projection
\be  \CM_C \overset{\pi}{\longrightarrow} \mathfrak t_\C/W \label{intsys} \ee
that comes from ``forgetting'' about monopole operators. Here $\mathfrak t_\C$ is the complexified Cartan subalgebra of the gauge group $G$, and $W$ the Weyl group, and the base $\mathfrak t_\C/W$ is parameterized by gauge-invariant polynomials in the $\varphi$ fields, \eg\ $\Tr(\varphi^n)$. This is an integrable system in the sense that the base is mid-dimensional and any functions $f(\varphi),g(\varphi)$ that are pulled back from the base Poisson-commute with respect to the holomorphic symplectic form $\Omega$. Moreover, each fiber of \eqref{intsys} is a holomorphic Lagrangian submanifold. The generic fiber is isomorphic to $T_\C^\vee \simeq (\C^*)^{\text{rank G}}$ (the dual of the maximal torus of $G_\C$) as a complex manifold, but interesting singular fibers may arise at complex codimension-one loci in the base.

This integrable system is analogous to the Seiberg-Witten integrable system that describes the Coulomb branch of a four-dimensional $\CN=2$ gauge theory on $\R^3\times S^1$ \cite{DonagiWitten}. In the four-dimensional case, the generic fibers are ``abelian varieties,'' \ie\ tori $(T^2)^{\text{rank}(G)}$ with an interesting complex structure. In contrast, for the purely three-dimensional theories considered here, the fibers are (partially) non-compact.

The Coulomb branch image of a Neumann boundary condition $\CN_L^{(C)}$ is a holomorphic section of this integrable system
\be \CN_L^{(C)}\,:\quad\text{section of $\CM_C \overset{\pi}{\longrightarrow} \mathfrak t_\C/W$} \ee 
that depends on the choice of Lagrangian splitting $L$ and the boundary FI parameter $t_{2d}$.

\subsubsection{Quantum Coulomb-branch image}
\label{sec:NC-q}

Just as a twisted $\wt \Omega$-deformation quantized the chiral ring of the Higgs branch, an ordinary $\Omega$-deformation with parameter $\epsilon$ quantizes the chiral ring of the Coulomb branch.
For an abelian theory, the algebra $\hat \C[\CM_C]$ is generated by 
operators $\hat \varphi$, $\hat v_A$. The $\hat \varphi$ commute with each other and generate $U(1)_t$ transformations of the $\hat v_A$,
\begin{subequations} \label{Cqring}
\begin{equation} \label{Cqring-a}
[\hat \varphi_a,\hat v_A] = \epsilon\, A_a \hat v_A\,
\end{equation}
where the index `$a$' labels generators of the Cartan subalgebra of the gauge group $G$. The ring relations are quantized to 
\begin{equation} \label{Cqring-b}
\hat v_A \hat v_B = P^\ell_{A,B}(\hat \varphi, m_\C) \hat v_{A+B} P^r_{A,B}(\hat \varphi, m_\C)
\end{equation}
\end{subequations}
with 
\begin{align}
P^\ell_{A,B}(\hat \varphi,m) &= \hspace{-.3cm}  \prod_{\substack{\text{$i$ s.t. $|Q_A^i|\leq |Q_B^i|$,}\\[.05cm] Q_A^i Q_B^i<0}} \hspace{-.3cm}  [\hat M_i]^{-Q_A^i}\,, \qquad\;
 P^r_{A,B}(\varphi,m) &=  \hspace{-.3cm}  \prod_{\substack{\text{$i$ s.t. $|Q_A^i|> |Q_B^i|$,}\\[.05cm] Q_A^i Q_B^i<0}}  \hspace{-.3cm}  [\hat M_i]^{Q_B^i}\,,
\end{align}
and the quantum exponentials
\be \label{q-exp}
[a]^b := \begin{cases} \prod_{i=1}^{b} (a+(i-\frac12)\epsilon) & b>0 \\
 \prod_{i=1}^{|b|} (a-(i-\frac12)\epsilon) & b <0 \\
 1 & b=0\,. \end{cases}
\ee
It follows from the property $[a]^{b}=(-1)^b[-a]^{-b}$ that \eqref{Cqring-b} is independent of a choice of Lagrangian splitting, up to a sign as in the classical case.

We claim that the left module $\hat \CN_L^{(C)}$ is generated from an identity vector $|\CN_L\rangle$, 
which satisfies 
\begin{equation}\label{NC-qabel}
\begin{array}{ll}\hat \CN_L^{(C)}\;:\qquad \hat v_A |\CN_L\rangle & \ds =\, \xi_A \prod_i [\hat M_{L,i}]^{(-Q_{A,L}^i)_+} |\CN_L\rangle\,. \\[.2cm]
&\ds =\, \xi_A \hspace{-.3cm} \prod_{\text{$i$ s.t. $Q_{A,L}^i<0$}} \hspace{-.3cm} [\hat M_{i}]^{-Q_{A}^i} |\CN_L\rangle \qquad (\text{up to sign})\,.
\end{array}
\end{equation}
This expression is consistent with the quantum chiral-ring relations above.%
\footnote{An easy way to see this is to use the mirror Higgs-branch formulas from Section \ref{sec:abel-HD}.} %
Abstractly, we may describe the module as a quotient $\hat \CN_L^{(C)}=\hat \C[\CM_C]\big/\CI$, where $\CI$ is the left ideal generated by the elements $(\hat v_A -  \xi_A \prod_i [\hat M_{L,i}]^{(-Q_{A,L}^i)_+})$.

The nonabelian version of these formulas is 
\begin{equation} \label{Cqring-nonabel}
\hat v_A \hat v_B = \frac{P^\ell_{A,B}(\hat \varphi^{\mathrm{ab}}, m_\C)}{P^{W,\ell}_{A,B}(\hat \varphi^{\mathrm{ab}})} \hat v_{A+B} \frac{P^r_{A,B}(\hat \varphi^{\mathrm{ab}}, m_\C)}{P^{W,r}_{A,B}(\hat \varphi^{\mathrm{ab}})}\,,
\end{equation}
where the numerator is computed as before from the complex masses of hypermultiplets and the denominator is the analogous expression 
involving the complex masses of vectormultiplets, up to a crucial shift of $- \frac{\epsilon}{2}$.
Thus we expect to be able to build a module starting from the relation
\begin{equation} \label{NC-qnonabel}
\hat v_A |\CN_L\rangle = \xi_A \frac{\prod_i [\hat M_{L,i}]^{(-Q_{A,L}^i)_+} }{\prod_{\text{roots $\alpha$}} [\alpha \cdot \hat \varphi^{\mathrm{ab}}- \frac{\epsilon}{2}]^{(-\alpha \cdot A)_+}}|\CN_L\rangle
\end{equation}
Notice that although the relation involves a non-trivial denominator, we expect it to reduce to a polynomial 
relation when inserted in the quantum non-abelianization map, so that quantum nonabelian monopole operators 
act on $|\CN_L\rangle$ as the multiplication by appropriate polynomials in $\varphi$.

\subsubsection{Twisting with vortex operators}
\label{sec:lineC}

At generic values of the complex masses $m_\C$, we will see in examples that the bulk algebra $\hat \C[\CM_C]$ has a collection of irreducible Verma modules with no interesting maps or extensions between them. Modules such as $\hat \CN_L^{(C)}$ are isomorphic to direct sums of Verma modules. Much more interesting structure arises when the classical complex masses are set to zero, and the quantum parameters entering the algebras \eqref{Cqring}, \eqref{Cqring-nonabel} are integer or half-integer multiples of $\epsilon$,
\be m_\C = k\epsilon\,. \label{mk} \ee

Such a specialization of equivariant parameters in an $\Omega$-background is quite familiar. We interpret integral shifts in 
$m_\C$ as coming from the insertion of line operators in the theory that are the mirrors of the abelian Wilson lines of Section \ref{sec:lineH}. These operators are a special case of a large class that can be defined by coupling the 3d theory to a one-dimensional quantum mechanics \cite{AG-loops}. (Operators in this class are mirror to more general Wilson lines.) Again, the inclusion of 
general line defects compatible with an $\Omega$-background is a very interesting generalization of our setup, which we 
leave for future work.  

\subsubsection{Monodromy}
\label{sec:NC-mon}

Since Neumann boundary conditions depend on parameters $\xi_A = e^{A\cdot t_{2d}}$, we may ask how their physics changes as these parameters are varied. In particular, the complex parameters $t_{2d}$ include boundary theta angles, and nontrivial monodromy can arise as we send $t_{2d}\to t_{2d}+2\pi i$.%
\footnote{Such monodromies play many fundamental roles in quantum field theory and string theory; they are analogous to Berry's phase in quantum mechanics \cite{Berry}.}

Both the Higgs-branch images of boundary conditions and their quantization in the $\wt\Omega$-background are insensitive to this effect: the boundary theta-angles do not enter into their definition. More concretely, the parameters $t_{2d}$ can be thought of as expectation values of twisted-chiral operators on the boundary, which do not enter the protected (chiral) sector of Higgs-branch physics that we have been exploring. On the other hand, twisted-chiral operators can and do enter the description of Coulomb-branch images and their quantization; and varying $t_{2d}$ turns out to affect the quantization.

In the presence of an $\Omega$-background, the 3d theory is reduced to a one-dimensional quantum mechanics, and we have seen that boundary operators generate a vector space $\hat \CN_L^{(C)}$. This vector space is fibered over the space of boundary parameters, and has a flat connection $\Theta$ given by 
\be \Theta = \epsilon\, \frac{\pd}{\pd t_{2d}} = \epsilon \,\xi\frac{\pd}{\pd\xi}.  \label{eq:NC-con} \ee
To find the action of $\Theta$ on the identity operator, we observe that the Neumann boundary condition contains a boundary twisted-superpotential coupling
\be \wt W = t_{2d}\cdot \varphi_{\rm ab}\,, \ee
where $\varphi_{\rm ab}$ are the abelian parts of $\varphi$ (equivalently, they are the complex moment maps for the topological symmetry on the Coulomb branch). Then we expect
\be \Theta |\CN_L\rangle =  \Big(\frac{\pd}{\pd t_{2d}}\wt W\Big)|\CN_L\rangle = \varphi_{\rm ab}|\CN_L\rangle\,.\ee
Exponentiating the action of $\Theta$ produces a $\hat \C[\CM_C]$-linear monodromy endomorphism on $\hat \CN_L^{(C)}$.

\subsubsection{Effect of real FI and real masses}
\label{sec:NC-mt}

Just like on the Higgs branch, turning on particular real FI parameters and real masses will affect the images of boundary conditions, possibly causing them to break supersymmetry -- rendering them ``infeasible.'' In fact, we expect that for given values of $(m_\R,t_\R)$, the Coulomb-branch image of a boundary condition breaks supersymmetry if and only if its Higgs-branch images does. In the case of Neumann b.c., the Higgs-branch analysis of $t_\R$-feasibility was straightforward, but the analysis of $m_\R$-feasibility was subtle; the same turns out to be true on the Coulomb branch.

When available, real mass parameters will (partially) resolve the singularities of the Coulomb branch of vacua. In terms of the integrable system \eqref{intsys}, singularities lie (at worst) over complex codimension-one loci of the base, while the image $\CN_L^{(C)}$ of a Neumann boundary condition is a section. Thus, generic points of $\CN_L^{(C)}$ are disjoint from the singularities of the Coulomb branch, and we naively expect that these boundary conditions survive any potential resolution. This is related to the observation on the Higgs branch that the intersections \eqref{Nflows} of $\CN_L^{(H)}$ and gradient-flow cycles $\CM_H^{\gtrless}[m_\R]$ are always non-empty (Section \ref{sec:NH-mt}).

If $\CN_L^{(C)}$ does intersect the singular locus, one should more carefully determine its intersection with cycles that resolve it. This is an interesting and possibly hard problem, since our description of $\CN_L^{(C)}$ in \eqref{NC-abel}--\eqref{NC-nonabel} was in terms of global holomorphic functions on the Coulomb branch, which cannot directly detect a resolution. We expect to encounter difficulties whenever the corresponding Higgs-branch intersections \eqref{Nflows} are unbounded, and we will see this in the examples. 

Next, we can look at real FI parameters. From the perspective of a sigma-model, any choice of FI parameters $t_\R$ generates a particular (infinitesimal) $U(1)_t$ isometry of the Coulomb branch --- playing a role analogous to that of real masses for the Higgs branch. The bulk vacua lie at fixed points of $U(1)_t$, which we denote as $\CM_C^{0}[t_\R]$.
The charge of an abelianized monopole operator $v_A$ under $t_\R$ is $t_\R\cdot A$. Thus, using the chiral-ring relation $v_Av_{-A} = P_{A,-A}(\varphi,m_\C)$, we see that on $\CM_C^{0}[t_\R]$
\be \label{defMC0t}
P_{A,-A}(\varphi,m_\C) = \prod_{1\leq i\leq N} M_i^{|Q_A^i|} = 0
\qquad \text{$\forall$ $A$ \;s.t.\; $t_\R\cdot A\neq 0$}\,,  
\ee
and so some combination of the effective complex masses $M_i$ must also vanish. This is natural: in the presence of nonzero $t_\R$, some combination of hypermultiplets must be able to get a vev. The fixed locus has a number of different components
\be \CM_C^{0}[t_\R] = \bigcup_\nu \CM_C^{0}[t_\R^\nu]\,, \ee
labelled by the different combinations of nonzero hypermultiplets. The choice can be encoded in the value of the moment maps $\mu_{H,\R}$ for the Higgs-branch flavor symmetry on a given component.

Just as in Section \ref{sec:NH-mt}, the fixed-point locus $\CM_C^0[t_\R]$ labels the bulk vacua, but we expect that the 2d moduli space in the presence of a boundary condition is the intersection
\be \label{NCflows}
\text{2d vacua}:\qquad \begin{array}{c@{\qquad}c} \CN_L^{(C)}\cap \CM_C^{>}[t_\R] & \text{(left b.c.)} \\[.1cm]
\CN_L^{(C)}\cap \CM_C^{<}[t_\R] & \text{(right b.c.)}  \end{array}\,,
\ee
where $\CM_C^{<}[t_\R]$ ($\CM_C^{>}[t_\R]$) is the submanifold containing points that flow to $\CM_C^0[t_\R]$ under gradient flow (inverse gradient flow) with respect to the real moment map $h_t$ for $U(1)_t$. Classically, this is just
\be h_t = t_\R \cdot \sigma_{\rm ab}\,, \label{defht} \ee
where $\sigma_{\rm ab}$ denotes the abelian part of $\sigma$. We can further decompose the submanifolds $\CM_C^{\gtrless}[t_\R]$ into components $\CM_C^{\gtrless}[t_\R^\nu]$ labelled by component of $\CM_C^{0}[t_\R]$ to which they flow.

To describe the gradient-flow manifolds more explicitly, we split the chiral ring as
\be \C[\CM_C] = \C[\CM_C]_< \oplus \C[\CM_C]_0 \oplus \C[\CM_C]_>\,, \label{eq:cbcrdecomp} \ee
where $\C[\CM_C]_0$ contains $\varphi$ and monopole operators with $t_\R\cdot A=0$, and $\C[\CM_C]_<$ ($\C[\CM_C]_>$) are generated over $\C[\CM_C]_0$ by monopole operators with $t_\R\cdot A< 0$ ($t_\R\cdot A>0$). Then all of $\C[\CM_C]_>$ vanishes on $\CM_C^<[t_\R]$ and all of $\C[\CM_C]_<$ vanishes on $\CM_C^>[t_\R]$. For a generic complex mass deformation, this property defines the gradient-flow manifolds. When complex masses vanish and real masses resolve the Coulomb branch, more care is needed; for abelian theories, we will provide a full description of $\CM_C^{\gtrless}[t_\R]$ in Section~\ref{sec:abel-Cmod}.

In terms of the integrable system $\CM_C \overset{\pi}{\longrightarrow} \mathfrak t_\C/W$ \eqref{intsys}, the fixed locus $\CM_C^0[t_\R]$ is supported on a proper complex submanifold of the base, defined by \eqref{defMC0t}. Moreover, the gradient flow of $h_t$ lies strictly along the fibers (\ie\ it commutes with the projection $\pi$).%
\footnote{This is intuitively clear from the semi-classical description of $h_t$ \eqref{defht}. Alternatively, we may observe that gradient flow of $h_t$ combines with the $U(1)_t$ action to produce a \emph{holomorphic} $\C^*_t$ action on $\CM_C$, as a complex symplectic manifold. (Gradient flow corresponds to dilations, in the noncompact directions of $\C^*_t$.)
The entire $\C^*_t$ action is generated by a holomorphic vector field that can be expressed as $\Omega^{-1}d(t_\R\cdot \varphi_{\rm ab})$, where $\varphi_{\rm ab}$ is the exact complex moment map for $U(1)_t$. Since $\varphi_{\rm ab}$ is a function on the base of the integrable system, the holomorphic vector field must be tangent to the fibers. Therefore, $\C^*_t$ acts only along the fibers.} %
Therefore, the gradient-flow cycles $\CM_C^>[t_\R]$ and $\CM_C^<[t_\R]$ extend in the fiber directions. In the special case that the $U(1)_t$ action has isolated fixed points $\nu$, each component $\CM_C^>[t_\R^\nu]$ and $\CM_C^<[t_\R^\nu]$ must be supported on a \emph{single} singular fiber of the integrable system, containing the fixed point $\nu$.

It follows from this picture, together with the fact that $\CN_L^{(C)}$ is a section, that if $U(t)_t$ has isolated fixed points the intersections \eqref{NCflows} are discrete. Thus, $\CN_L$ is $t_\R$-feasible if and only if the corresponding intersection is non-empty --- with no additional subtleties arising from noncompact 2d moduli spaces. This matches the simple analysis of $t_\R$-feasibility on the Higgs branch. We expect that the intersections \eqref{NCflows} are non-empty precisely when $\CN_L^{(H)}$ is compatible with the $t_\R$-resolution of the Higgs branch.

It is a bit tricky to characterize the gradient-flow loci in the full gauge theory rather than a sigma model. The classical D-term BPS equations 
are likely inadequate to describe the flow on the quantum-corrected Coulomb branch. They are also rather complicated, as they 
involve a non-trivial evolution of both $\sigma$ and of the scalar fields which receive vevs at the fixed point.
We will not attempt to analyze further the 2d dynamics induced by real FI and mass parameters from the Coulomb branch perspective. 

\subsubsection{The $t_{2d}\to \infty$ limit}
\label{sec:NC-tinf}

In the standard framework of symplectic duality, the categories $\CO_H$ and $\CO_C$ (Section \ref{sec:intro-SD}) depend on choices of parameters $m_\R,t_\R$, respectively. These categories are defined as categories of \emph{lowest-weight}%
\footnote{In the literature, one often encounters ``highest-weight'' modules instead; this is purely a matter of convention.} %
modules with respect to the corresponding actions of $U(1)_m$ and $U(1)_t$. This means that 1) the modules admit an action of these isometries; 1$'$) they decompose into finite-dimensional generalized weight spaces; and 2) all operators in the quantum algebras $\hat \C[\CM_H]_<$ and $\hat \C[\CM_C]_<$ with negative $U(1)_m$, $U(1)_t$ charge act nilpotently on the modules. The modules can be understood as a quantization of holomorphic-Lagrangian boundary conditions in $\CM_H$, $\CM_C$ sigma-models that 1) preserve $U(1)_m$, $U(1)_t$, and moreover 2) are supported entirely on gradient-flow cycles  $\CM_H^>[t_\C]$, $\CM_C^>[t_\R]$ (if we think of them as right boundary conditions).

The Higgs-branch images of Neumann b.c. $\CN_L^{(H)}$ all preserved $U(1)_m$. Moreover, the ones that were $m_\R$-feasible were actually supported on $\CM_H^>[m_\R]$. Correspondingly, their quantizations become good objects in the standard category $\CO_H$.

In contrast, the Coulomb-branch images of $\CN_L^{(C)}$ all break $U(1)_t$ and do not lead to the standard sort of objects one encounters in $\CO_C$.

\begin{figure}[htb]
\centering
\includegraphics[width=5.8in]{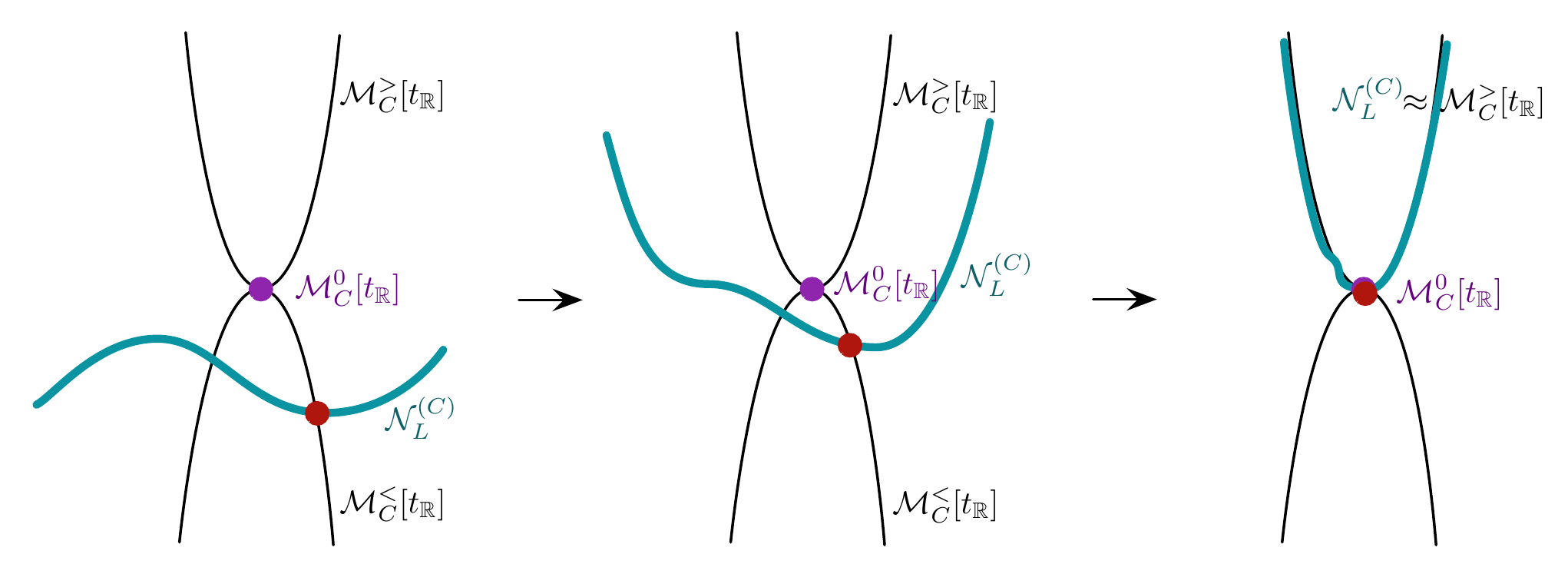}
\caption{Applying gradient flow to deform $\CN_L^{(C)}$ into a holomorphic Lagrangian that is invariant under $U(1)_t$ and supported on the upward-flow cycles $\CM_C^>[t_\R]$. The intersection with downward-flow cycles $\CM_C^<[t_\R]$ is preserved, and slides toward the vacuum locus $\CM_C^0[t_\R]$.}
\label{fig:flow}
\end{figure}

One way to ameliorate this problem is to deform the images $\CN_L^{(C)}$ and the corresponding modules so that they become $U(1)_t$ invariant, without changing the essential properties of the intersections $\CN_L^{(C)}\cap \CM_C^<[t_\R]$ (for, say, right boundary conditions) that define vacua of the bulk-boundary system.
The appropriate deformation is suggested by working in a massive sigma-model (with ``mass'' $t_\R$) and using the $(2,2)$ BPS equations.
While at the boundary itself the operators $\varphi$, $v_A$ obey \eqref{NC-abel}, \eqref{NC-nonabel} (or the corresponding quantized versions), the expectation values of these operators away from the boundary are governed by gradient flow with respect to the real moment map of the $U(1)_t$ isometry.
As we move very far away from the boundary, the image $\CN_L^{(C)}$ becomes deformed by an infinite gradient flow, and its support converges to components of $\CM_C^>[t_\R]$, precisely as desired for symplectic-duality applications. Moreover, the intersections of $\CN_L^{(C)}$ and the downward-flow cycles $\CM_C^<[t_\R]$ are (necessarily) preserved; intersection points just ``slide'' along $\CM_C^<[t_\R]$ toward the fixed-points $\CM_C^0[t_\R]$, according to gradient flow. We depict this process in Figure \ref{fig:flow}.

The effect of this deformation on the operator equations \eqref{NC-abel}, \eqref{NC-nonabel} and corresponding modules is easy to describe. For any chiral operator $\CO$, the gradient-flow is given by
\be \CD_1 \CO  = (\pd_1 +q_t^\CO)\CO= 0\,,\ee
where $q_t^\CO$ is the charge of $\CO$ under $U(1)_t$.
Thus, the LHS and RHS of equations \eqref{NC-abel}, \eqref{NC-nonabel} simply get rescaled by their $U(1)_t$ charges. Since $U(1)_t$ invariance would be restored by making the 2d FI parameters $t_{2d}$ dynamical, the deformation may be encoded by replacing
\be t_{2d} \to t_{2d} - \lambda t_\R\,,\qquad \xi_A \to e^{\lambda( t_\R\cdot A)}\xi_A\,,  \label{t-limit} \ee
and sending $\lambda\to\infty$. (In the physical setup above, $\lambda$ is the distance away from the actual boundary.)
In essence, this limit just sends $t_{2d}\to\infty$ in a particular direction.

In the case when the action of $U(1)_t$ on the Coulomb branch has isolated fixed points $\nu$, the deformation of $\CN_L^{(C)}$ converges to a union of $\CM_C^>[t_\R^\nu]$ cycles, one for every intersection between $\CN_L^{(C)}$ and the dual $\CM_C^<[t_\R^\nu]$ cycles. The limit of the module $\hat \CN_L^{(C)}$  turns out to be much more interesting and subtle: when $m_\C$ is generic it converges to a direct sum of the lowest-weight modules obtained by quantizing $\CM_C^>[t_\R^\nu]$ cycles --- \ie\ to a sum of Verma modules --- but for quantized values of $m_\C$ as in \eqref{mk} it converges to a nontrivial extension of the same Verma modules.

Moreover, since some information about the phase of $\xi_A$ (the imaginary part of $t_{2d}$) is preserved in the limit \eqref{t-limit}, the limiting modules retain an action of the monodromy from Section \ref{sec:NC-mon}. For generic $m_\C$ the monodromy will act by a scalar on each irreducible Verma module but for quantized values of $m_\C$ the action will be quite interesting.

The procedure of taking the $t_{2d} \to \infty$ limit of the module $\hat \CN_L^{(C)}$ translates to a very precise mathematical prescription. As discussed in Section \ref{sec:NC-mon}, the connection defined by \eqref{eq:NC-con} makes $\hat \CN_L^{(C)}$ into a local system of modules fibered over the $\C^*$ of exponentiated boundary parameters $e^{t_{2d}}$. The $t_{2d} \to \infty$ limit of $\hat \CN_L^{(C)}$ is obtained as the nearby cycles of this local system. A variant of a theorem of Emerton, Nadler, and Vilonen \cite{ENV-gjf}, to appear in \cite{Hilburn}, shows that we can compute these nearby cycles by applying a variant of the Jacquet functor to the fiber of $\hat \CN_L^{(C)}$ over $1 \in \C^*$.

We will content ourselves with a brief description of the Jacquet functor $\mathfrak{J}$. Let $\hat \C[\CM_C]_{>0}$ be as in the decomposition \eqref{eq:cbcrdecomp}. Define $\mathfrak{J}(\hat \CN_L^{(C)})$ to be the direct sum of the generalized weight spaces for the infinitesimal $U(1)_t$ symmetry in the completed module
\[
\varprojlim_{k} \; \hat \CN_L^{(C)} / (\hat \C[\CM_C]_{>0})^k \hat \CN_L^{(C)}.
\]
Intuitively, ``completion'' means that we allow ourselves to work with formal power series in $\hat \C[\CM_C]_{>0}$. The modules $\hat \CN_L^{(C)}$ are finitely generated as $\hat \C[\CM_C]_{>0}$-modules so the discussion in \cite[Section~5]{Ginzburg-pi} shows that $\mathfrak{J}(\hat \CN_L^{(C)})$ is in $\CO_C$.

In terms of representation theory, the modules $\hat \CN_L^{(C)}$ for fixed $\xi$ are generalizations of \emph{Whittaker modules}. An ordinary Whittaker module would set lowering operators (operators with negative charge under some $U(1)$ isometry) equal to constants, while a Neumann boundary condition more generally sets lowering operators equal to a function of the neutral $\varphi$'s. The main result of \cite{Hilburn} is that, for abelian theories, the $t_{2d} \to \infty$ limits of Neumann b.c. are exactly the twisted projective modules in $\CO_C$. In particular, all projective and tilting modules in  $\CO_C$ arise this way. In geometric representation theory, it is known that a non-degenerate Whittaker module over a semisimple Lie algebra can be averaged or degenerated to give the "big" projective module in the BGG category $\CO$ \cite{Frenkel-Gaitsgory, Nadler-mtts, Campbell}. Our analysis of Neumann b.c. suggests that this construction of a particular projective/tilting module admits a vast generalization.

\subsection{Examples}
\label{sec:NC-eg}

\subsubsection{SQED}
\label{sec:NC-SQED}

For a $G=U(1)$ gauge theory, the cocharacters $A\in \text{Hom}(U(1),G)\simeq \Z$ are just integers.
The chiral ring $\C[\CM_C]$ is generated by $\varphi$ and by fundamental monopole operators $v_\pm$, with
\be
v_A = \begin{cases}
(v_+)^A  & \mathrm{if} \quad A \geq 0\\
(v_-)^{|A|} & \mathrm{if} \quad A < 0\, .
\end{cases}
\ee
The operator $v_A$ has charge $A$ under the Coulomb-branch isometry $G_C\simeq U(1)$.
In a theory with $N$ fundamental hypermultiplets, the fundamental monopoles obey the chiral-ring relation
\be
v_+v_- = \prod_{i=1}^N ( \varphi + m_{\C,i})\, ,
\label{eq:SQEDrel}
\ee
where we have introduced complex masses $m_{\C,i}$ for the $PSU(N)$ flavor symmetry, normalized so that
 $\sum_i m_{\C,i} = 0$. In absence of complex mass parameters, the Coulomb branch is $\mathbb{C}^2 / \mathbb{Z}_N$.%
\footnote{This description is exact in the far IR, at infinite gauge coupling. Otherwise the metric on the Coulomb branch is that of a singular or resolved/deformed Taub-NUT space \cite{SW-3d, IS, dBHOY}.} %
Turning on complex masses gives a smooth deformation thereof.

The infrared image of the right boundary condition $\CN_\varepsilon$ is 
\be \label{NC-SQED}
v_+ = \xi \prod_{\text{$i$ s.t. $\varepsilon_i = -$}} (\varphi + m_{\C,i})\,, \qquad  v_- = \xi^{-1} \prod_{\text{$i$ s.t. $\varepsilon_i = +$}}(\varphi + m_{\C,i})\,,
\ee
which is clearly compatible with the relation~\eqref{eq:SQEDrel}. For example, with $N=2$ hypers,
\be \label{NC-SQED2}
\begin{array}{r@{\quad}l@{\quad}l}
\CN_{--}^{(C)}: & v_+ = \xi(\varphi+m_\C/2)(\varphi-m_\C/2)\,, & v_- = \xi^{-1}\,, \\
\CN_{-+}^{(C)}: & v_+ = \xi(\varphi+m_\C/2)\,, & v_- = \xi^{-1}(\varphi-m_\C/2)\,, \\
\CN_{+-}^{(C)}: & v_+ = \xi(\varphi-m_\C/2)\,, & v_- = \xi^{-1}(\varphi+m_\C/2)\,, \\
\CN_{++}^{(C)}: & v_+ = \xi\,, & v_- = \xi^{-1}(\varphi+m_\C/2)(\varphi-m_\C/2)\,, \\
\end{array}
\ee
which are all compatible with the ring relation $v_+v_- = (\varphi+\frac{m_\C}{2})(\varphi-\frac{m_\C}{2})$. 

Turning on real masses resolves the Coulomb branch into an ALE space, with a familiar string of $N-1$ $\cp^1$ exceptional divisors. 
We investigate how this affects Neumann boundary conditions for $N=2$. In this case, the resolved Coulomb branch at $m_\C=0$ is $T^*\cp^1$.
The compact $\cp^1$ can be parameterized by a choice of a null eigenline for the matrix 
\begin{equation}
\begin{pmatrix} \varphi & v_+ \cr -v_- & - \varphi \end{pmatrix}\,.
\end{equation}
The sign of $m_\R$ dictates whether to take left or right eigenlines. If we additionally turn on the deformation $m_\C$, 
we should look instead for null eigenlines of 
\begin{equation}
\begin{pmatrix} \frac{m_\C}{2}+ \varphi & v_+ \cr -v_- & \frac{m_\C}{2}- \varphi \end{pmatrix}\,,
\end{equation}
which are unique even at $\varphi=v_\pm=0$ (hence the $\cp^1$ is eliminated).

The image of $\CN_{++}$ in the resolved Coulomb branch is uncontroversial: 
\begin{equation}
\begin{pmatrix} \varphi & \xi \cr -\xi^{-1} \varphi^2 & - \varphi \end{pmatrix}
\end{equation}
has left and right eigenlines generated by $(\varphi \quad \xi)$ and ${\xi \choose - \varphi}$
which have an obvious $\varphi \to 0$ limit. Similar considerations apply to $\CN_{--}$.

On the other hand, the behavior of $\CN_{+-}$ and $\CN_{-+}$ is more subtle. If we set $m_\C=0$ the images of the two boundary conditions appear to be 
identical. The matrix 
\begin{equation}
\begin{pmatrix} \varphi & \xi \varphi \cr -\xi^{-1} \varphi & - \varphi \end{pmatrix}
\end{equation}
naively admits the whole $\mathbb{CP}^1$ worth of left and right eigenlines. 
If we turn on $m_\C$, however, we see a different story: the matrix for $\CN_{+-}$
\begin{equation}
\begin{pmatrix} \frac{m_\C}{2}+ \varphi & \xi (\frac{m_\C}{2} +\varphi) \cr \xi^{-1} (\frac{m_\C}{2}-\varphi) & \frac{m_\C}{2}- \varphi \end{pmatrix}
\end{equation}
has a simple right eigenline generated by ${\xi \choose - 1}$ but a left eigenline $(\varphi-\frac{m_\C}{2} \quad \xi (\varphi+ \frac{m_\C}{2}))$
for which the $m_\C \to 0$ and $\varphi \to 0$ limits do not commute. The opposite is true for $\CN_{-+}$
This erratic behavior seems likely related to the unbounded moduli space of classical 2d vacua we encountered in the Higgs branch analysis.

Finally, let us turn on a real FI parameter $t_\R$, corresponding to a choice of infinitesimal generator for the flavor symmetry $G_C\simeq U(1)_t$. Under $U(1)_t$, each monopole operator $v_A$ has charge $A\,t_\R$. In the presence of a generic $m_\C$ deformation, there are $N$ massive vacua $\nu_i$ at $v_+=v_-=0$ and $\varphi=-m_{\C,i}$. The corresponding gradient-flow manifolds are
\be \CM_C^<[t_\R^{\nu_i}] =\{v_+=\varphi+m_{\C,i}=0\}\,,\qquad \CM_C^>[t_\R^{\nu_i}] =\{v_-=\varphi+m_{\C,i}=0\} \ee
for positive $t_\R$; the roles of $\CM_C^<$ and $\CM_C^>$ are swapped for negative $t_\R$. As usual, we denote by $\CM_C^<[t_\R]$ and $\CM_C^>[t_\R]$ the sum of gradient-flow manifolds attached to all vacua.

Suppose that $t_\R>0$ and that $\CN_\varepsilon$ is a right boundary condition. Then it is easy to see from \eqref{NC-SQED} that $\CN_\varepsilon^{(C)}$ intersects $\CM_C^<[t_\R^{\nu_i}]$ (and the intersection is transverse) if and only if $\varepsilon_i=-$. Thus  $\CN_\varepsilon^{(C)}\cap \CM_C^<[t_\R]$ is discrete, and nonempty so long as $\varepsilon\neq (++...+)$. We conclude that all the $\CN_\varepsilon$ boundary conditions are $t_\R$-feasible except for $\CN_{++...+}$, which breaks SUSY. This agrees with the Higgs-branch analysis based on resolutions.

Following Section \ref{sec:NC-tinf}, we may also deform $\CN_\varepsilon^{(C)}$ by an infinite (positive) gradient flow while preserving the intersections $\CN_\varepsilon^{(C)}\cap \CM_C^<[t_\R]$. For $t_\R>0$ ($t_\R<0$), this amounts to sending $\xi\to\infty$ ($\xi \to 0$). For example, for $N=2$ flavors and $\xi\to\infty$ we obtain limits
\be \label{NC-SQED3}
\begin{array}{r@{\quad}l@{\quad}l}
\CN_{--}^{(C)}: & 0 = (\varphi+m_\C/2)(\varphi-m_\C/2)\,, & v_- = 0\,, \\
\CN_{-+}^{(C)}: & 0 = (\varphi+m_\C/2)\,, & v_- = 0\,, \\
\CN_{+-}^{(C)}: & 0 = (\varphi-m_\C/2)\,, & v_- = 0\,, \\
\CN_{++}^{(C)}: & v_+ = \infty\,, & v_- = 0\,. \\
\end{array}
\ee
The first three are supported on $\CM_C^>[t_\R^{\nu_i}]$ cycles, while the image of $\CN_{++}^{(C)}$ slides off to infinity in the Coulomb branch, indicating that it does not support a supersymmetric vacuum.

\subsubsection{SQED, quantized}
\label{sec:NC-qSQED}

In the $\Omega$-background with equivariant parameter $\epsilon$, the Coulomb-branch chiral ring is deformed to the non-commutative algebra $\hat \C[\CM_C]$,
\be
\left[ \hat\varphi , \hat v_\pm \right] = \pm \epsilon \hat v_\pm\,, \qquad \hat v_+ \hat v_- = \prod_{i=1}^N \Big( \, \varphi+m_i-\frac{\epsilon}{2} \, \Big)\,, \qquad \hat v_- \hat v_+ = \prod_{i=1}^N \Big( \, \varphi+m_i+\frac{\epsilon}{2} \,  \Big)\, .
\ee
The deformation quantization of the singularity $\mathbb{C}^2 / \mathbb{Z}_N$ is a member of many interesting families of algebras that appear in the mathematical literature such as finite $W$-algebras \cite{Premet}, symplectic reflection algebras \cite{Gordon-Cherednik, EGGO}, and hypertoric enveloping algebras \cite{BLPW-hyp}. The right boundary condition $\CN_\varepsilon$ produces a left module for the algebra $\hat \C[\CM_C]$ that is generated from an identity state $|\CN_\varepsilon\rangle$, which satisfies
\be
\begin{aligned}
\hat v_+ |\, \CN_\varepsilon \, \rangle & = \xi \prod_{\text{$i$ s.t. $\varepsilon_i = -$}} \Big( \, \hat \varphi + m_i - \frac{\epsilon}{2} \, \Big)|\, \CN_\varepsilon \, \rangle\,,  \\
  \hat v_- |\, \CN_\varepsilon \, \rangle & = \xi^{-1} \prod_{\text{$i$ s.t. $\varepsilon_i = +$}} \Big( \, \hat \varphi + m_i+\frac{\epsilon}{2} \, \Big)|\, \CN_\varepsilon \rangle\,.
\end{aligned}
\ee
The states of the module can be uniquely represented as $p(\varphi)|\, \CN_\varepsilon \, \rangle$ (or in shorthand $p(\varphi)\big|$), where $p$ is a polynomial in the boundary operator $\hat\varphi$.

Let us now focus on the special case $N=2$. We first define the operators
\be
H = 2 \hat \varphi \qquad E = \hat v_+ \qquad F = - \hat v_- 
\ee
and parameterize the complex masses as $m_{1} = - m_{2} = m_\C / 2$. It is then straightforward to check that we generate a central quotient of the universal enveloping algebra $U(\mathfrak{sl}_2)$ with the quadratic Casimir element fixed to
\be \label{C-SQED-Cas}
C_2 = EF+FE+\frac12 H^2 =  \frac{1}{2}(m_\C^2 - \epsilon^2)\, .
\ee
The modules $\hat\CN_\varepsilon^{(C)}$ produced by Neumann boundary conditions are generated from identity states (in shorthand, `$|$') that satisfy
\be \label{NC-qSQED2}
\begin{array}{r@{\quad}l@{\quad\;\;}l}
\hat\CN_{--}: & E\big| = \frac14\xi(H+m_\C-\epsilon)(H-m_\C-\epsilon)\big|\,, & F\big| = - \xi^{-1}\big|\,, \\[.1cm]
\hat\CN_{-+}: & E\big| = \frac12\xi(H+m_\C-\epsilon)\big|\,, & F\big| = -\frac12\xi^{-1}(H-m_\C+\epsilon)\big|\,, \\[.1cm]
\hat\CN_{+-}: & E\big| = \frac12\xi(H-m_\C-\epsilon)\big|\,, & F\big| = -\frac12\xi^{-1}(H+m_\C+\epsilon)\big|\,, \\[.1cm]
\hat\CN_{++}: & E\big| = \xi\big|\,, & \hspace{-.97in}F\big| = -\frac14\xi^{-1}(H+m_\C+\epsilon)(H-m_\C+\epsilon)\big|\,.
\end{array}
\ee
Note that in each case only the relation for $E$ or for $F$ is required to define the module; the other relation follows automatically from the Casimir identity \eqref{C-SQED-Cas}.

The modules $\hat\CN_{++}$ and $\hat\CN_{--}$ are known as Whittaker modules for the raising and lowering operators, respectively. The modules $\hat\CN_{+-}$ and $\hat\CN_{-+}$ are less conventional. In contrast to the Higgs-branch analysis of Section \ref{sec:NH-qSQED}, none of the modules in \eqref{NC-qSQED2} are highest-weight or lowest-weight. This is a direct consequence of the fact that Neumann boundary conditions break the topological $U(1)_t$ symmetry, preventing these modules from being graded.

As discussed in Section \ref{sec:NC-tinf}, we can obtain weight modules by sending $t_{2d}\to\infty$ in a particular direction, depending on a choice of real FI parameter $t_\R$. Let us choose $t_\R<0$, which corresponds to $t_{2d}\to-\infty$ or equivalently $\xi \to 0$. From our previous discussion we expect that when $m_\C$ is generic the $\xi\to 0$ limit of $\hat \CN_\varepsilon^{(C)}$ is a direct sum of lowest-weight Verma modules (corresponding to a quantization of the classical cycles $\CM_C^>[t_\R^{\nu_i}]$) but when $m_\C = k\epsilon$ for integer $k$ the limit will be a possibly non-trivial extension of Verma modules. Let us illustrate these facts in our example. Assume we have turned off the classical complex mass and introduced a line operator as in Section \ref{sec:lineC}, so that $m_\C = k\epsilon$ for integer $k$.

\begin{figure}[htb]
\centering
\includegraphics[width=4.5in]{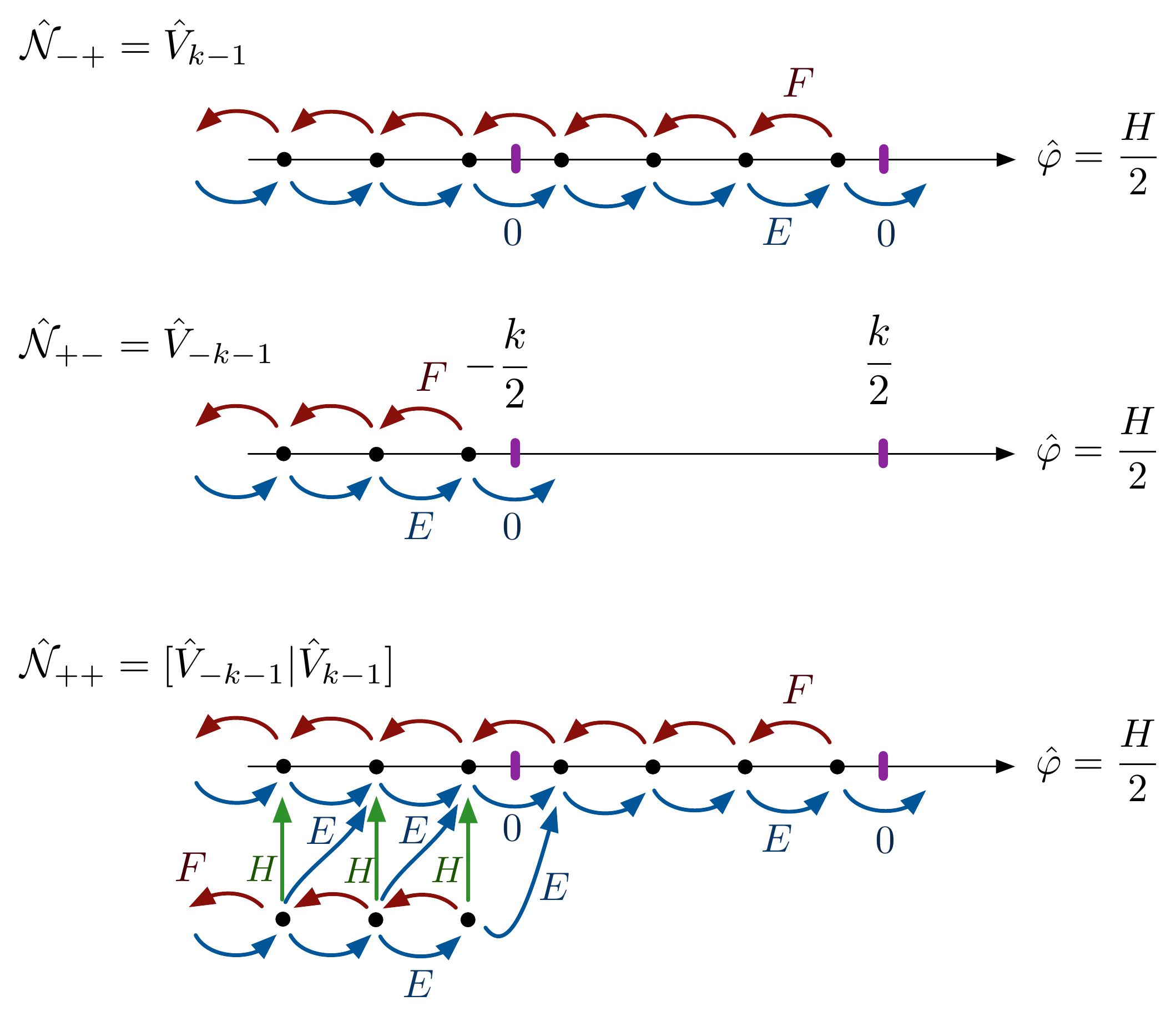}
\caption{The highest-weight modules isomorphic to $\hat \CN_\varepsilon$ modules for $U(1)$ theory with $N=2$ hypermultiplets. The complex mass is $m_\C = k\epsilon$ with $k=4$ (compare Figure \ref{fig:cp1NH}). The modules $\hat \CN_{-+}$ and $\hat \CN_{+-}$ are Verma modules $\hat V_{k-1}$, $\hat V_{-k-1}$ of lowest weights $k-1$ and $-k-1$, while $\hat \CN_{++}$ is an extension of $\hat V_{-k-1}$ by $\hat V_{k-1}$. }
\label{fig:proj}
\end{figure}

Consider the module $\hat \CN_{-+}$. It has a basis $|n\rangle := F^n\big|$ with $n\geq 0$, on which the algebra generators act as
\be 
\label{SQED-N+-}
\begin{aligned}
F|n\rangle & = |n+1\rangle\,,\\[.2cm]
H|n\rangle & = (k-2n-1)\epsilon|n\rangle -2\xi|n+1\rangle\,,\\[.2cm]
E|n\rangle & = n(k-n)\epsilon^2|n-1\rangle +\xi(k-2n-1)\epsilon|n\rangle -\xi^2|n+1\rangle\,.
\end{aligned}
\ee
To compute the Jacquet module of $\hat \CN_{-+}$ we allow formal power series in $F$ and look for generalized eigenvectors of $H$. It is easy to see that the vector 
\be \widetilde{ | 0 \rangle } = \sum_{n=0}^\infty \frac{(-\xi)^n}{\epsilon^n n!}|n\rangle = e^{-\xi F/\epsilon}\big|\,.\ee
is an eigenvector of $H$ and a null vector of $E$. The remaining $H$ eigenvectors are are $\widetilde{|n \rangle} :=F^n \widetilde{|0\rangle} = e^{-\xi F/\epsilon}|n\rangle$ which satisfy the relations
\be F \widetilde{|n\rangle} = \widetilde{ |n+1\rangle} \,,\quad 
H \widetilde{|n\rangle}=(k-2n-1)\epsilon \widetilde{|n\rangle}\,,\quad
E \widetilde{|n\rangle} = n(k-n)\epsilon^2 \widetilde{|n-1\rangle} \,, \label{V-SQED-1} \ee
and hence span a Verma module $\hat V_{k-1}$ of lowest weight $k-1$, as illustrated at the top of Figure \ref{fig:proj}.

A similar computation shows that the Jacquet module of $\hat \CN_{+-}$ is isomorphic to a Verma module $\hat V_{-k-1}$ with lowest weight $-k-1$, illustrated in the middle of Figure \ref{fig:proj}.

Now consider $\hat \CN_{--}$. The equation $F\big| = - \xi^{-1}\big|$ suggests that working with power series in $\xi$ will not help us find eigenvectors for $H$. In fact, the Jacquet module of $\hat \CN_{--}$ is $0$. This is consistent with the fact that the boundary condition $\CN_{--}$ breaks supersymmetry when $t_\R<0$.

The most interesting module is $\hat \CN_{++}$. For simplicity we will assume $k=1$ so we are looking at the regular block of $\CO$. Then $\hat \CN_{++}$ has a basis
\be
\begin{aligned}
|n\rangle_+ &=  F^n(H+2\epsilon) \big| \\
|n+1\rangle_- &=  \xi^{-1} F^n H \big|
\end{aligned}
\ee
for $n\geq0$, on which the algebra generators act by
\begin{subequations} \label{proj-action}
\be 
\begin{aligned}
 F|n\rangle_+ & = |n+1\rangle_+  \\[.1cm]
 H|n\rangle_+ & =  -2n\epsilon|n\rangle_+  - \frac{2\xi}{\epsilon}|n+1\rangle_+ + \frac{2\xi^2}{\epsilon} |n+2\rangle_-  \\[.1cm]
 E|n\rangle_+ & = -n(n-1)\epsilon^2|n-1\rangle_+ - 2n\xi|n\rangle_+ +  (2n+1)\xi^2|n+1\rangle_-
\end{aligned}
\ee
and
\be 
\begin{aligned}
F|n\rangle_- & = |n+1\rangle_- \\[.1cm]
H|n\rangle_- & =  -2n\epsilon|n\rangle_- - \frac{2}{\epsilon} | n\rangle_+  + \frac{2\xi}{\epsilon}|n+1\rangle_-  \\[.1cm]
E|n\rangle_- & = -n(n-1)\epsilon^2|n-1\rangle_- - (2n-1)| n-1\rangle_+ + 2n\xi|n\rangle_- \, .
\end{aligned}
\ee
\end{subequations}

Working with formal power series, we can modify the basis $|n\rangle_\pm$ order by order, so that all $O(\xi)$ terms in the action \eqref{proj-action} are eliminated. Explicitly, the modified basis is given by
\be
\begin{array}{rl}
\widetilde{|n \rangle}_+ &= \ds \sum_{\ell=0}^{\infty}\frac{1}{(\ell!)^2}\left(-\frac{\xi F}{\epsilon}\right)^\ell |n\rangle_+ 
 + \frac{\xi^2}{\epsilon}\sum_{\ell=0}^\infty \frac{1}{\ell!(\ell+2)!}\left(-\frac{\xi F}{\epsilon}\right)^\ell|n+2\rangle_-\,, \\[.5cm]
\widetilde{|n\rangle}_- &= \ds \frac{1}{\epsilon}\sum_{\ell=1}^\infty\frac{2H_\ell}{(\ell!)^2}\left(-\frac{\xi F}{\epsilon}\right)^\ell|n\rangle_+ + \left[1 + \sum_{\ell = 1}^{\infty} \frac{2(\ell-1)\ell H_{\ell-1}-1}{(\ell!)^2} \left(-\frac{\xi F}{\epsilon}\right)^\ell \right] |n\rangle_-\,,
\end{array}
\ee
where $H_\ell=\sum_{m=1}^\ell \frac{1}{m}$ are the harmonic numbers. Equivalently,
\be
\begin{array}{rl}
\widetilde{|n \rangle}_+ &= J_0\Big( 2\sqrt{\frac{ \xi F}{\epsilon}} \Big) |n\rangle_+ + \xi J_2\Big( 2\sqrt{\frac{ \xi F}{\epsilon}} \Big) |n+1\rangle_-\,, \\[.2cm]
 \widetilde{|n\rangle}_- &= 
  -\frac{\pi}{\epsilon} Y_0\Big( 2\sqrt{\frac{ \xi F}{\epsilon}} \Big)|n\rangle_+ - \frac{\pi\xi F}{\epsilon} Y_2\Big( 2\sqrt{\frac{ \xi F}{\epsilon}} \Big)|n\rangle_-  + \frac1\epsilon(\log x+2\gamma)\wt{|n\rangle}_+\,, \\
\end{array}
\ee
where $J_m$ and $Y_m$ are Bessel functions of the first and second kind, respectively.
The modified basis vectors are generalized eigenvectors for $H$, with
\be \begin{array}{l@{\quad}l}
F\wt{|n\rangle}_+ = \wt{|n+1\rangle}_+\,, & F\wt{|n\rangle}_- = \wt{|n+1}\rangle_-\,, \\[.1cm]
H\wt{|n\rangle}_+ = -2n\epsilon\wt{|n\rangle}_+ \,, & H\wt{|n\rangle}_- = -2n\epsilon\wt{|n\rangle}_-- \frac2\epsilon\wt{|n\rangle}_+\,, \\[.1cm]
E\wt{|n\rangle}_+ = -n(n-1)\epsilon^2\wt{|n-1\rangle}_+\,, &
E\wt{|n\rangle}_- = -n(n-1)\epsilon^2\wt{|n-1\rangle}_- - (2n-1)\wt{|n-1\rangle}_+\,.
\end{array}\ee
Notice that $H$ cannot be diagonalized, but rather has 2-dimensional Jordan blocks spanned by each pair $\wt{|n\rangle}_\pm$. 
The Jacquet module of $\hat \CN_{++}$ then takes the form of a nontrivial extension
\be 0 \to \hat V_{0}\to \hat \CN_{++} \to \hat V_{-2}\to 0\,; \label{proj-21} \ee
it has a submodule $\hat V_0$ spanned by the $\wt{|n\rangle}_+$ and a quotient $\hat V_{-2}$ spanned by the $\wt{|n\rangle}_-$ (modulo the $\wt{|n\rangle}_+$).

For general $k$, the Jacquet module of $\hat \CN_{++}$ turns out to be an extension
\be 0 \to \hat V_{|k|-1}\to \hat \CN_{++} \to \hat V_{-|k|-1}\to 0\,. \label{proj-2} \ee
This is known as the big projective module in category $\CO$, illustrated in Figure~\ref{fig:proj}.

In summary, for $m_\C = k \epsilon$ with $k\geq 1$ we have
\begin{itemize}
\item $\hat \CN_{-+}$ is isomorphic to the (reducible) lowest-weight Verma module $\hat V_{k-1}$\, ,
\item $\hat \CN_{+-}$ is isomorphic to the (irreducible) lowest-weight Verma module $\hat V_{-k-1}$\, ,
\item $\hat \CN_{++}$ is an extension $ 0\to \hat V_{k-1} \to \hat \CN_{++} \to \hat V_{-k-1} \to 0$\, ,
\item $\hat \CN_{--}$ is not isomorphic to a lowest-weight module.
\end{itemize}
Had we instead chosen $t_\R>0$ and sent $\xi\to \infty$, we would have found that $\hat \CN_{+-}$ and $\hat \CN_{-+}$ were still Verma modules, that $\hat \CN_{--}$ is an extension, and that $\hat \CN_{++}$ has no regular limit (or isomorphism with a lowest-weight module).

The module $\hat \CN_{++}$ is our first example that undergoes interesting monodromy as the 2d theta-angle is varied $t_{2d}\to t_{2d}+2\pi i$. Recall from Section \ref{sec:NC-mon} that the infinitesimal monodromy is generated by $\Theta = \pd/\pd t_{2d} = \xi \pd/\pd \xi$, which acts on the identity as
\be \Theta \big| = - \hat \varphi\big| = - \tfrac12 H\big|\,.\ee
The monodromy survives the $\xi\to 0$ limit. In terms of the extension \eqref{proj-2}, we find that $\Theta$ acts by mapping $\hat V_{-|k|-1}$ into $\hat V_{|k|-1}$ (as a submodule), and sending all other vectors to zero.  (In Figure \ref{fig:proj}, $\Theta$ maps all weight spaces of $\hat \CN_{++}$ upward.)

\subsubsection{SQCD}
\label{sec:NC-SQCD}

We now consider $U(K)$ SQCD with $N$ fundamental hypermultiplets. The abelianized chiral ring is generated by the vectormultiplet eigenvalues and abelian monopole operators $\{ \varphi_a, v^\pm_a \}$ labelled by the weights of the fundamental representation of $U(K)$, with $a=1,\ldots,K$. It is convenient to define $u_a^+ = v_a^+$ and $u_a^- = (-1)^Kv_a^-$ and we follow this convention in what follows. 

The abelian coordinates are subject to the relations
\be
u_a^+ u_a^- = - \frac{P(\varphi_a) }{\prod_{b \neq a} (\varphi_a - \varphi_b)^2}
\label{eq:nonabring}
\ee
where $P(z) = \prod_{i=1}^{N}(z-m_i)$ is the matter polynomial whose roots are the complex masses $m_i$ (obeying $\sum_{\alpha=1}^{N} m_i = 0$) for the $PSU(N)$ global symmetry acting on the hypermultiplets.

The genuine chiral ring is generated by the gauge invariant polynomials in $\varphi$ and non-abelian monopole operators. In order to describe this ring, we first introduce the monic degree $K$ polynomial
\be
Q(z) = \prod_{a=1}^{K}(z-\varphi_a) = z^{K} - \Tr(\varphi) z^{K-1} + \cdots + (-1)^{K} \det ( \varphi ) \, .
\ee
whose components form a basis in the vector space of gauge invariant polynomials of $\varphi$. Second, we introduce the degree $K-1$ polynomials
\be
U^\pm(z) =  \sum_{a=1}^{K} u^\pm_a \prod_{b\neq a}(\varphi_a-\varphi_b) 
\ee
whose components are the non-abelian monopole operators labelled by the cocharacters $A = (\pm1, 0,\ldots,0)$ (i.e. the fundamental and anti-fundamental representations of $U(K)$) and dressed by the invariant polynomials in the unbroken $U(K-1)$ factor. Finally, the chiral ring relations are 
\be
U^+(z) U^-(z) = - P(z) \;\; \mathrm{mod}\;\; Q(z)\, ,
\ee
or equivalently
\be
Q(z) \, \tilde Q(z) - U^+(z) \, U^-(z) = P(z)\, ,
\label{eq:SQCD}
\ee 
where
\be
\deg\, \tilde Q(z) = 
\begin{cases}
K - 2 & \mathrm{if} \quad N \leq 2K-2 \\
N-K & \mathrm{otherwise}
\end{cases}\, ,
\ee
and $P(z) = \prod_{i=1}^{N}(z-m_i)$ is the characteristic matter polynomial. The components of $\tilde Q(z)$ are dressed monopole operators labelled by the cocharacter $A = (1,0,\ldots,0,-1)$ (i.e. the adjoint representation of $U(K)$). 

The Coulomb branch is identified with the moduli space of $PSU(2)$ monopoles with magnetic charge $N - 2 K$ at infinity and $N$ fundamental Dirac singularities. The moduli space is parametrized by the $PGL(2,\mathbb{C})$-valued scattering matrix 
\be
S (z) = \begin{pmatrix}
Q(z) & U^+(z) \\
U^-(z) & \tilde Q(z)
\end{pmatrix}
\ee
whose determinant is equal to $P(z)$. Via the Nahm transform, it is simultaneously the moduli space of solutions to the Nahm equations on an interval with appropriate boundary conditions. For 
$N \geq 2 K$ this identifies it (in the absence of real and complex mass parameters) with the intersection $\CN \cap \CS_\rho$ where $\CS_\rho$ is the Slodowy slice transverse to the nilpotent orbit $\CO_\rho$ with partition $\rho=(N-K,K)$ and $\CN $ is the nilpotent cone.

Let us now consider the image of (right, say) Neumann boundary conditions $\CN_\varepsilon$. The abelian monopole operators at the boundary obey
\be
u^+_a  =  \frac{ \xi \, P_-(\varphi_a)}{\prod_{b\neq a}(\varphi_a-\varphi_b) } \qquad u^-_a  = - \frac{ \xi^{-1} P_+(\varphi_a)}{\prod_{b\neq a}(\varphi_a-\varphi_b)  } \, .
\label{eq:abelneu}
\ee
where
\be
P_-(z) = \prod_{i, \varepsilon_i = -}(z - m_i) \qquad P_+(z) = \prod_{i, \varepsilon_i = +}(z - m_i)
\ee
are the matter polynomials for the hypermultiplets with $\CB_X$ and $\CB_Y$ boundary conditions respectively. To express the Neumann boundary condition as a module for the full Coulomb branch chiral ring, we must express it in terms the of non-abelian monopole operators. We find
\be
U^-(z) = - \xi^{-1}  P_+(z)  \;\; \mathrm{mod} \;\; Q(z)
\qquad\qquad U^+(z) = \xi P_-(z) \;\; \mathrm{mod} \;\; Q(z) \, ,
\ee
which are compatible with the chiral ring relations since $P_+(z)P_-(z) = P(z)$.

The quantized version of these boundary conditions is readily described.
The abelianized algebra has generators $\hat u_a^\pm$, $\hat\varphi_a$, and the inverses of W-boson masses $(\hat\varphi_a-\hat\varphi_b)^{-1}$. See \cite{BDG-Coulomb} for details. 
The nonabelian quantized algebra is expected to be generated by quantized versions of the classical generators \ie\ quantized versions of the coefficients of $Q(z)$ and $U^\pm(z)$, namely
\be \HAT Q(z) = \prod_{a=1}^{K} (z-\HAT\varphi_a)\,,\qquad \HAT U^\pm(z) = \sum_{a=1}^{K} \HAT u_a^\pm \prod_{b\neq a}(z-\HAT\varphi_b)\,, \ee
Starting from the abelianized relations
\be
\hat u^+_a |\, \CN_\varepsilon \, \rangle = \xi \frac{P_-(\hat \varphi_a - \frac{\epsilon}{2} )}{\prod_{b\neq a}(\hat \varphi_a-\hat \varphi_b) }|\, \CN_\varepsilon \, \rangle  \qquad  \hat u^-_a |\, \CN_\varepsilon \, \rangle = - \frac{ \xi^{-1} P_+(\hat \varphi_a+ \frac{\epsilon}{2})}{\prod_{b\neq a}(\hat \varphi_a-\hat \varphi_b)  }|\, \CN_\varepsilon \, \rangle 
\ee
we obtain the module relations 
\begin{align}
\hat U^+(z) |\, \CN_\varepsilon \, \rangle &= \left[\xi P_-(z - \frac{\epsilon}{2}) \;\; \mathrm{mod} \;\; Q(z)  \right]|\, \CN_\varepsilon \, \rangle  \cr  
 \hat U^-(z) |\, \CN_\varepsilon \, \rangle &= - \left[ \xi^{-1} P_+(z+ \frac{\epsilon}{2}) \;\; \mathrm{mod} \;\; Q(z) \right]|\, \CN_\varepsilon \, \rangle 
\end{align}
The module consists of elements of the form $p(\varphi)|\, \CN_\varepsilon \, \rangle$ for Weyl-invariant polynomials $p$,
corresponding to polynomials of the $\varphi|_\partial$ boundary operator. 

We will consider $t_{2d}\to\infty$ limits of such modules and relate them to projective/tilting representations in a separate paper.

\section{Generic Dirichlet Boundary Conditions} 
\label{sec:D}

\subsection{Definition And Symmetries}
\label{sec:defD}

Just as in the case of Neumann boundary conditions, the basic Dirichlet boundary for a 3d $\CN=4$ gauge multiplet that preserves 2d $\CN=(2,2)$ supersymmetry can be obtained by dimensional 
reduction from a Dirichlet boundary condition in 5d $\CN=1$ gauge theory preserving 4d $\CN=1$ supersymmetry. In five dimensions, the boundary condition simply sets to zero 
(or to a constant background flat connection) the components of the gauge field parallel to the boundary
\be A_\parallel\big|_\pd = 0\,. \label{DA} \ee
Preserving 4d $\CN=1$ supersymmetry at the boundary then requires that an appropriate half of the gauginos vanish, and that the real scalar $\sigma$ has a Neumann-like boundary condition
\be \frac{1}{g^2}\pd_1\sigma + \mu_\R + t_\R\,\Big|_\pd = 0\,, \label{Dsigma} \ee
where $\mu_\R$ is the real moment map for the gauge group action on the matter fields and $t_\R$ is a real FI parameter. (Here we use `$g^{-2}$' schematically to denote the metric for the gauge kinetic terms.)
Reducing to three dimensions, we find that \eqref{DA} and \eqref{Dsigma} still hold, and moreover the complex scalar is set to zero at the boundary
\be \varphi\big|_\pd = 0\,. \label{phizero} \ee

A Dirichlet boundary condition breaks the gauge symmetry $G$ at the boundary. In the absence of matter, global gauge transformations at the boundary generate a boundary flavor symmetry $G_\pd$. 
In addition, the topological flavor symmetry $U(1)_t$ associated to abelian factors in $G$ is preserved, 
as are the R-symmetries $U(1)_A$ and $U(1)_V$.

It is possible to deform the boundary condition \eqref{phizero} to
\be \varphi\big|_\pd = \varphi_0 \label{phi0} \ee
for some nonzero constant $\varphi_0$. This deformation breaks $U(1)_A$, and breaks $G_\pd$ to the stabilizer of $\varphi_0$ (the subgroup that acts trivially on $\varphi_0$). For example, if $\varphi_0$ is generic, then $G_\pd$ will be broken to a maximal torus. The boundary value $\varphi_0$ can be interpreted as a two-dimensional twisted mass for $G_\pd$.

If there are matter hypermultiplets, additional boundary conditions need to be specified.
The simplest choices are labelled by a Lagrangian splitting $(X_L,Y_L)$ of the complex hypermultiplet scalars, as in Section \ref{sec:N}. Given such a splitting, we define the boundary condition
\be \CD_{L}\,:\qquad \text{Dirichlet for gauge multiplet},\qquad Y_L\big|_\pd = c\,,  \label{DL}\ee
for a constant vector $c$. In addition, as described in Section \ref{sec:defN}, preserving 2d $\CN=(2,2)$ supersymmetry dictates that the $X_L$ have a Neumann-like boundary condition $\CD_1X_L\big|_\pd=0$, and that half the fermions are set to zero.

The physical properties of these boundary conditions depend strongly on the chosen value for $c$. The most obvious choice would be to set $c=0$. However, this choice comes with a significant complication: it leads to unbounded moduli spaces of classical 2d vacua fibered above every bulk vacuum (much as in Section \ref{sec:NH-mt}). Roughly, one cannot quotient by the gauge group at the boundary (since it is broken), nor impose D-term constraints (they are absorbed into $\pd_1\sigma$ via \eqref{Dsigma}). The only way to cut down hypermultiplet degrees of freedom at the boundary is with F-term constraints, which become trivial when $c=0$. Thus, above every bulk vacuum one finds a complexified gauge orbit of 2d vacua.

In order to ameliorate the problem, we could introduce just enough non-zero components in $c$ to break $G_\pd$ completely. 
The resulting boundary conditions are interesting, and we will return to them later in Section \ref{sec:xD}. 

In the remainder of the current section, we focus on the case that the vector $c$ specified in \eqref{DL} is nonzero and as generic as possible, subject to the following constraint.
Notice that any nonzero $c$ explicitly breaks the vector R-symmetry $U(1)_V=U(1)_H$ (\cf\ \eqref{HVCA}), which rotates all $X_L,Y_L$ with charge $+\frac12$.
We nevertheless want $c$ to preserve a combination $U(1)_V'$ of $U(1)_V$ and a $U(1)$ subgroup of $G_\partial$ and the flavor symmetry $G_H$.
This property is mirror to the anomaly inflow that evaded the axial anomaly and allowed $U(1)_A$ to be preserved in the case of Neumann boundary conditions.
Occasionally there are multiple ways to preserve a $U(1)_V'$ symmetry, in which case we denote the boundary condition as $\CD_{L,c}$ to emphasize its dependence on $c$.

When the flavor symmetry $G_H$ is abelian, our genericity assumption on $c$ usually constrains all the complex masses $m_\C$ and vevs of the complex vectormultiplet scalars $\varphi_0$ to vanish at the boundary.
In contrast, if the flavor symmetry is nonabelian, even a generic $c$ may not be sufficient to break $G_H$ completely and some complex mass deformation parameters may survive. 
In either case, complex FI parameters $t_\C$ can freely be turned on.

\subsection{Higgs-branch image}
\label{sec:DH}

We assume for simplicity that the gauge group $G$ acts faithfully on the hypermultiplets, as a subgroup of $U(N)\subset USp(N)$. In other words, at a generic point on the Higgs branch, 
all Coulomb-branch degrees of freedom are massive. 

We analyze the Higgs-branch image of $\CD_{L}$ by relating boundary degrees of freedom to the (bulk) vevs of gauge-invariant chiral operators $\CO$. 
As always, gauge-invariant chiral operators obey $\pd_1\CO = 0$ in the absence of real masses, so they are constant throughout the bulk.

The values of the hypermultiplet fields $X_L$ at the boundary are constrained by the complex moment-map conditions%
\footnote{In contrast, the $X_L$ are \emph{not} constrained by real moment map conditions, since the gauge symmetry is broken at the boundary. Indeed, the real moment map is absorbed into $\pd_1\sigma$ via~\eqref{Dsigma}.}
\be \mu_\C +t_\C\big|_\pd = Y_LT_LX_L +t_\C\big|_\pd = c\, T_L X_L\big|_\pd + t_\C = 0\,. \label{DLmuC} \ee
These are $\dim (G)$ independent constraints, which leave behind $N-\dim(G)$ complex degrees of freedom. The gauge-invariant operators in the bulk are polynomials $\CO(X_L,Y_L)$ that obey the bulk moment-map constraints $\mu_\C+t_\C=0$ and obey 
\be
\CO(X_L,Y_L) = \CO(X_L,Y_L)|_\pd = \CO(X_L|_\pd,c)
\ee
when brought to the boundary. The image of the boundary condition $\CD_L$ on the Higgs branch is simply the submanifold $\CD_{L,c}^{(H)}\subset \CM_H$ on which the relations $\CO(X_L,Y_L)=\CO(X_L,c)$ can be satisfied. Abstractly, if we interpret the Higgs branch as a complex symplectic quotient $\CM_H = \C^{2N}/\!/G_\C$, then
\be \begin{array}{ll} \CD_{L}^{(H)}\quad &= \quad \{\text{image of $Y_L = c$ under complex symplectic quotient}\} \\
&\simeq \quad (Y_L=c)\cap (\mu_\C+t_\C=0)\,. \end{array} \label{DLH} \ee
Note that the orbits of $G_\C$ in $\C^{2N}$ are transverse to $Y_L=c$ (this is what it means for the boundary gauge symmetry to be fully broken), so $\CD_{L}^{(H)}$ is simply isomorphic to $(Y_L=c)\cap (\mu_\C+t_\C=0)$, and becomes a holomorphic Lagrangian submanifold of  $\CM_H$.

In practice, it is useful to rewrite the relations $\CO(X_L,Y_L)=\CO(X_L,c)$ as relations $f(\CO_i;c)=0$ among the gauge-invariant operators themselves. The latter equations can only depend on combinations of the $c$'s that are themselves invariant under the (broken) symmetry $G_\pd$ at the boundary. These invariant combinations, which we denote as $\wt\xi$, will play a role that mirrors the role of the exponentiated 2d FI parameters $\xi = e^{-t_{2d}}$ that entered Neumann boundary conditions. 

\subsubsection{Effect of real FI and real masses}
\label{sec:DH-mt}

The effect of real FI parameters on the Dirichlet boundary conditions $\CD_{L,c}$ is straightforward: the image of $Y_L = c$ carves out a particular locus in the resolved Higgs branch.

The effect of real mass deformations is similarly straightforward. Just as in Section \ref{sec:NH-mt}, the bulk Higgs branch is restricted to fixed loci of the symmetry generated by $m_\R$, labelled by lifts $m_\R^\nu$. The moduli space of classical 2d vacua compatible with a bulk vacuum is determined by the intersection of $\CD_L^{(H)}$ with $\CM_H^>[m_\R^\nu]$ or $\CM_H^<[m_\R^\nu]$, depending on whether one has a left or right boundary condition. Alternatively, we can describe this as the intersection of the locus $Y_L = c$ with $X^+_{m^\nu_\R}=0$ and $Y^+_{m^\nu_\R}=0$ or 
with $X^-_{m^\nu_\R}=0$ and $Y^-_{m^\nu_\R}=0$, as on page \pageref{Xpm}. Notice that $\CD_{L,c}$ boundary conditions will be $m_\R$-infeasible if some non-zero $c$ has non-zero charge under all possible $m^\nu_\R$. 

\subsubsection{Quantum Higgs-branch  image}
\label{sec:qDH}

Upon turning on a twisted $\wt \Omega$-background, Dirichlet boundary conditions $\CD_{L}$ with generic $c$ should become modules $\hat\CD^{(H)}_{L}$ for the quantized algebra $\hat\C[\CM_H]$ of holomorphic functions on the Higgs branch.

The quantized algebra $\hat \C[\CM_H]$ can be constructed in several equivalent ways \eqref{qCMH}--\eqref{qCMH2}. 
For the purpose of studying left (right) Dirichlet boundary conditions, it is most convenient to start with the $N$-dimensional Heisenberg algebra generated by $\hat X,\hat Y$, quotient by the left (right) ideal $H(\hat\mu_\C+t_\C)$ (resp. $(\hat\mu_\C+t_\C)H$) and then restrict to $G$-invariant operators. Thus, for right boundary conditions, we have
\be \hat \C[\CM_H] = \big( (\hat \mu_\C +t_\C)H\backslash H\big)^G\,.
 \label{CMH-D} \ee

 The boundary condition $Y_L\big|_\pd = c$ for the hypermultiplets generates a left ideal $H(\hat Y_L - c)$ in the Heisenberg algebra $H$, and a corresponding module $M_{L} = H/H(\hat Y_L - c)$ for $H$. Explicitly, $M_L$ is the module whose vectors are polynomials $p(X_L)$ in the chiral operators that survive at the boundary, with bulk operators acting as follows:
\begin{enumerate}
\item $\hat X_L$ is multiplication by $X_L$\, ,
\item $\hat Y_L = \epsilon\,\pd_{X_L} + c$\, .
\end{enumerate} 
Then, to obtain a module for the bulk chiral ring \eqref{CMH-D}, we impose the complex moment map constraint, \ie\ we quotient $M_L$ by the subspace of vectors of the form $(\hat \mu_\C + t_\C)(...)$,
\be \hat \CD_L^{(H)} \;=\; M_L / (\hat\mu+t_\C)M_L\,. \label{qDH} \ee
This is a module for $(\hat \mu_\C +t_\C)H\backslash H$, and therefore for the gauge-invariant subalgebra $\big( (\hat \mu_\C +t_\C)H\backslash H\big)^G$.
We take this as a tentative definition for $\hat \CD_L^{(H)}$.

Notice that we do \emph{not} take gauge-invariant (or covariant) vectors of $M_L$ to form the module $\hat \CD_L^{(H)}$. This would be inappropriate, since Dirichlet boundary conditions break gauge symmetry; and indeed for general $t_\C$ there are no vectors $m\in M_L$ that satisfy $(\hat\mu+t_\C)m=0$. In contrast, in the case of Neumann boundary conditions with zero $t_\C$ (quantized $t_\C$), the quotient \eqref{qDH} was completely equivalent to taking gauge-invariant (gauge-covariant) vectors.

For a generic infinite-dimensional module, the quotient operation may be worrisome. In the current setup, though, we can obtain a concrete, finite description of $\hat \CD_{L}^{(H)}$.
The basic idea is to use the $U(1)_V'$ symmetry to put a filtration on the module $M_L$ (compatible with the action of $\hat \C[\CM_H]$) whose filtered subspaces are finite dimensional. The equivalence $(\hat\mu_\C+t_\C)m\sim 0$ then relates elements within these finite subspaces.

To illustrate this point, suppose for simplicity that we can set $Y_L\big|_\pd = c$ with all $c$ nonzero while preserving a modified R-symmetry $U(1)_V'$.
The charges of $(X_L,Y_L)$ under $U(1)_V'$ must be $(1,0)$.%
\footnote{Under $U(1)_V$ the charges are $(1/2,1/2)$, and under any abelian flavor symmetry the charges are of the form $(q,-q)$; since $U(1)_V'$ is a combination of $U(1)_V$ and a flavor symmetry, the new charges must be $(1,0)$.} %
The classical polynomial algebra in the $X_L$ and $Y_L$ is graded by $U(1)_V'$, whereas the quantized Heisenberg algebra is filtered. Explicitly,
\be H = \bigcup_{n\geq 0}\CF_n H\,,\qquad  \CF_0H \subset \CF_1 H \subset \CF_2 H \subset \cdots\,,\ee
with
\be \CF_n H := \{\text{polynomials in $\hat X_L,\hat Y_L$ with degree $\leq n$ in $\hat X_L$}\}\ee
and
\be \CF_n H  \cdot \CF_m H \subset \CF_{n+m} H\,. \ee
The reason we find a filtration rather than a true grading is that the $\wt\Omega$-background (slightly) breaks $U(1)_V'$:
 the basic commutator $[ \hat X , \hat Y ]=\epsilon$ relates elements of charge $1$ to an element $\epsilon$ of charge 0.%
\footnote{Given any filtration, one can canonically construct a graded algebra
${\rm gr}\, \CF_\bullet H := \oplus_{n\geq 0} \CF_nH/\CF_{n-1}H\,.$
In the present case it trivializes the commutator and reproduces the classical polynomial algebra in $X_L,Y_L$. In the mathematical theory of deformation quantization, one typically requires that a quantization of a classical algebra be filtered, in such a way that the associated graded algebra reproduces the original classical algebra. This is so both for the Heisenberg algebra and for all our chiral rings $\hat\C[\CM_H]$, $\hat \C[\CM_C]$, with the filtration induced by the appropriate R-symmetry.\label{foot:gr}}
The filtration on $H$ induces a filtration on the quotient $\hat \C[\CM_H]$, which is a generalization of the $U(1)_V'$ grading on the chiral ring $\C[\CM_H]$.

Now, the reason for using $U(1)_V'$ rather than $U(1)_V$ as a symmetry is that the former is preserved by the boundary conditions. This translates to the fact that the module $M_L$ and its quotient $\hat\CD_L^{(H)}$ are filtered by $U(1)_V'$ in a manner compatible with the actions of $H$ and $\hat \C[\CM_H]$, respectively.%
\footnote{Compatibility means that $(\CF_n H)\cdot \CF_m M_L \subset \CF_{m+n}M_L$, and similarly for $\hat \CD_L^{(H)}$.} %
Recall that a basis for $M_L$ is given by polynomials $p(X_L)|$. We set
\be \CF_n M_L = \{\text{polynomials $p(X_L)$ of degree $\leq n$}\}\,.\ee
Each $\CF_nM_L$ is finite-dimensional. Moreover, since the complex moment map $\hat\mu_\C$ lies in $\CF_1H$, the relations $(\hat \mu_\C+t_\C)m\sim 0$ relate elements in $\CF_{n+1}M_L$ to elements in $\CF_{n}M_L$. Thus the relations can consistently be restricted to finite-dimensional subspaces. The quotient $\hat \CD_L^{(H)} = M_L/(\hat \mu_\C+t_\C)M_L$ is unambiguously defined and acquires an induced filtration.

To be even more explicit, the moment map acts as
\be
\hat \mu_\C +t_\C \;=  X_L T_Lc + t_\C' + \epsilon  X_L T_L \pd_{X_L}
\ee
(where $t_\C'$ has absorbed a factor of $\frac12\epsilon\,{\rm Tr}(T_L)$ from undoing the normal-ordering in $\hat\mu_\C$), and can be thought of as a quantum deformation of a simple linear multiplication operator.
To give a concrete definition of $\hat \CD_{L}^{(H)}$ we choose a maximal subspace $N_{L} \subset M_{L}$ transverse to the space of polynomials 
of the form $(X_L T_L c) p(X_L)$. This choice can be made separately in each finite piece of the $U(1)_V'$ filtration of $M_L$.
After acting with some element of $\hat\C[\CM_H]$ on a vector in $N_{L}$, we can bring it back to $N_{L}$ by recursively replacing 
$(\hat X_L T_L c) p(X_L)$ with $-(t_\C' + \epsilon \hat X_L \,\pd_{X_L}) p(X_L)$, starting from the terms with highest $U(1)_V'$ charge and progressing to lower 
$U(1)_V'$ charge. The process will stop after finitely many steps.
We identify the module $\hat \CD_{L}^{(H)}$ with the space $N_{L}$, equipped with the $\hat \C[\CM_H]$ action we just described.

Different choices of subspace $N_{L}$ will give equivalent presentations of the module $\hat \CD_{L}^{(H)}$. Once we have demonstrated the existence of $\hat \CD_{L}^{(H)}$,
though, we can give a more intrinsic definition as follows. Before quantization, the Higgs-branch image $\CD_{L}^{(H)}$ was a Lagrangian submanifold 
of the Higgs branch defined by $\CO(X_L,Y_L)=\CO(X_L,c)$, and the boundary chiral ring coincided with the boundary image of bulk gauge-invariant operators. It is natural to assume that the same statement remains true after quantization, so that $\hat \CD_{L}^{(H)}$ can simply be generated from the identity vector $|\CD_L\rangle$, usually denoted `$|$', by acting with all of $\hat \C[\CM_H]$. After all, $\epsilon$-corrections to the classical calculation involve terms that are subleading in $U(1)_V'$ charge. (To formalize this statement, one again uses the $U(1)_V'$ filtration.)
Therefore, $\hat \CD_{L}^{(H)}$ can be described intrinsically as the quotient of $\hat \C[\CM_H]$ by the left ideal containing all bulk operators that annihilate the identity --- relations stemming from $\hat Y_L| = c|$.

There is an alternative way to study Dirichlet boundary conditions: one may represent them as Neumann boundary conditions enriched by an auxiliary 
compensator field which lives at the boundary and can be used to Higgs away the boundary gauge symmetry. 
In Appendix \ref{app:Dir} we show how to use such a description to compute the quantum Higgs branch image of Dirichlet boundary conditions in abelian theories.
We obtain the same answer as we found in this section.  

\subsubsection{Monodromy}
\label{sec:DH-mon}

Just as the quantization of the images of Neumann boundary conditions on the Coulomb branch experienced a nontrivial monodromy as the boundary theta-angles were varied $t_{2d}\to t_{2d}+2\pi i$, the quantized Higgs-branch images of Dirichlet boundary conditions may experience a monodromy as the phases of the boundary parameters $c_i$ are varied. We expect that the physics of a Dirichlet boundary condition is largely independent of the $c_i$ as long as these parameters are kept generic. However, when some of the $c_i$ are tuned to special values (such as zero), extra flavor symmetry emerges and the boundary condition undergoes a phase transition. These special values occur at complex codimension-one loci; winding around these loci, for example sending
\be c_i \to e^{2\pi i}\,c_i\,,\ee
may generate monodromy.

In terms of modules, the generator of an infinitesimal phase rotation of $c_i$ is $\Theta_i = \epsilon \, c_i \pd_{c_i}$. Acting on the identity, we expect $\epsilon \pd_{c_i}\big| = (X_L)_i\big|$. Combining this with $c_i\big| = (\hat Y_L)_i\big|$ we therefore expect
\be \Theta_i\big| = \; ( \hat  X_L)_i ( \hat Y_L)_i\big|\; = \; \hat X_i\hat Y_i\big|\,. \label{D-mon}\ee

\subsubsection{The $c\to\infty$ limit}
\label{sec:DH-cinf}

As discussed in Section \ref{sec:NC-tinf}, the mathematical definition of the category $\CO_H$ of modules associated to the Higgs branch involves lowest-weight modules with respect to some choice of mass parameters $m_\R$, and a corresponding $U(1)_m\subset G_H$ action on $\hat \C[\CM_H]$.
As left modules, these are quantizations of holomorphic Lagrangians supported on gradient-flow cycles $\CM_H^>[t_\R]$.
In contrast, generic Dirichlet boundary conditions break (all or part of) the flavor symmetry $G_H$; their Higgs-branch images are not supported on $\CM_H^>[m_\R]$, and the corresponding modules cannot be lowest-weight modules.

Nevertheless, we can deform $\CD_L^{(H)}$ and $\hat \CD_L^{(H)}$ into $\CM_H^>[m_\R]$ cycles and lowest-weight modules by following the same logic as in Section \ref{sec:NC-tinf}. Namely, in the presence of nonzero $m_\R$, we apply an infinite (positive) gradient flow for the real moment map $h_m=m_\R\cdot \mu_{H,\R}$. This rescales chiral gauge-invariant operators by an amount proportional to their charge under $U(1)_m$. Equivalently, letting $q_L$ denote the charge of $Y_L$ under $U(1)_m$, we may describe the limit as rescaling
\be c \to e^{\lambda q_L} c\,, \label{cinf} \ee
and sending $\lambda\to \infty$.

\subsection{Examples}
\label{sec:DH-eg}

\subsubsection{SQED}
\label{sec:DH-SQED}

The basic Dirichlet boundary conditions for SQED with $N$ hypermultiplets are labelled by a sign vector $\varepsilon\in\{\pm\}^N$, much as in Section \ref{sec:NH-eg}. Namely,
\be \CD_\varepsilon\,:\qquad A_\parallel\big|_\pd = 0\,,\qquad \varphi\big|_\pd = 0\,,\qquad \begin{cases} X_i\big|_\pd = c_i & \varepsilon_i = - \\ Y_i\big|_\pd = c_i & \varepsilon_i = + \end{cases}\,. \label{SQED-De}\ee
For generic $c_i$, these boundary conditions appear to break both $G_\pd = U(1)_\pd$ and the Higgs-branch flavor symmetry $G_H=PSU(N)$. However, it turns out that a hidden subgroup $U(N_+-1)\times U(N_--1)\subset G_H$  remains unbroken, where $N_+$ ($N_-$) are the number of $\varepsilon_i=+$ $(\varepsilon_-=-)$. For example, if $X_i\big|_\pd=c_i$ for $i\leq N_-$ and $Y_i\big|_\pd = c_i$ for $i> N_-$, we can use the ostensibly broken $PSU(N)$ to rotate the $X$'s and $Y$'s (and correspondingly the boundary condition) to the form
\be (X_1,...,X_{N_-})\big|_\pd = (c',0,...,0)\,,\qquad (Y_{N_-+1},...,Y_N)\big|_\pd = (c'',0,...,0)\,,
\label{SQED-sym} \ee
making the unbroken flavor symmetry manifest. (If $\varepsilon_i\equiv +$ or $\varepsilon_i\equiv +$ for all $i$, then only one $c_i$ survives and the unbroken symmetry is $SU(N-1)$.)
The extra symmetry will be relevant when considering Coulomb-branch images because it allows complex masses to be turned on. Here, for the most part, we work with the generic boundary condition \eqref{SQED-De}.

Consider the boundary condition $\CD_{++...+}$, where $Y_i\big|_\pd = c_i$ for all $i$. This implies that the gauge-invariant operators $X_iY_j$ satisfy
\be c_{j'} (X_iY_j) = c_j(X_i Y_{j'})\qquad \forall\;i,j,j'\,. \label{SQED-D+} \ee
These equations define a holomorphic Lagrangian submanifold of the Higgs branch $\CM_H$, which depends only on the $N-1$ ratios $\wt \xi_i = c_i/c_{i+1}$. 
The conserved R-symmetry $U(1)_V'$ is a combination of the bulk $U(1)_V$ and boundary $U(1)_\pd$. 
Indeed, the apparent violation of $U(1)_V$ by the boundary condition $Y_i=c_i$ can be entirely compensated by a $U(1)_\pd$ rotation at the boundary. The $U(1)_V'$ charges of the hypermultiplet fields $(X_i,Y_i)$ are $(1,0)$ for all $i$.

The other boundary conditions $\CD_\varepsilon$ all have Higgs-branch images similar to \eqref{SQED-D+}. Indeed, \eqref{SQED-De} allows us to relate any gauge-invariant operator (meson) $X_i Y_j$ with $i\neq j$ to a polynomial in $X_iY_i$ and $X_jY_j$\,:
\be X_i Y_j = \begin{cases} c_i c_j  & \varepsilon_i = -,\; \varepsilon_j = + \\
 \frac{c_i}{c_j} X_j Y_j & \varepsilon_i = -,\; \varepsilon_j = - \\
 \frac{c_j}{c_i} X_i Y_i & \varepsilon_i = +,\;\varepsilon_j = + \\
 \frac{1}{c_ic_j} (X_iY_i)(X_jY_j) & \varepsilon_i = +,\;\varepsilon_j = - \end{cases}\,.
 \label{SQED-DeH}
 \ee
Together with the moment-map constraint $\sum_i X_iY_i + t_\C=0$, these define the holomorphic Lagrangian $\CD_\varepsilon^{(H)}$. It depends only on $N-1$ products or quotients of the $c_i$.
Notice that for $N=2$ hypermultiplets, these relations are identical in form to the image of Neumann boundary conditions \eqref{NC-SQED2} on the Coulomb branch. In Section \ref{sec:abel}, we will argue more generally that Dirichlet boundary conditions are the mirrors of Neumann boundary conditions. (In the case of $N=2$ hypermultiplets, the theory is self-mirror, explaining the observed similarity.)

The quantization of the boundary conditions \eqref{SQED-DeH} is straightforward. Let us start with the boundary condition $\CD_{++...+}$, and follow the direct approach of Section \ref{sec:qDH}. The module $M_\varepsilon$ for the Heisenberg algebra has a basis of polynomials $f(X_1,...,X_N)|$, with $\hat X_i$ acting as multiplication and $\hat Y_i=\epsilon\pd_i+c_i$.
Writing the moment map as 
\be
\hat\mu_\C = \, :\!\hat X\cdot\hat Y \!: \, = \hat X\cdot \hat Y + \frac{N}{2}\epsilon \, ,
\ee
we find that the moment-map constraint imposes an equivalence
\be \textstyle (\hat \mu_\C+t_\C)f(X_1,...,X_N) = \big(  \sum_i X_i ( \epsilon \pd_i + c_i) +\tfrac{N}{2}\epsilon+t_\C\big)f(X_1,...,X_N) \simeq 0\,. \label{SQED-Dequiv} \ee
We can use this to eliminate (say) $X_N$, producing a basis for the quotient $\hat \CD_\varepsilon^{(H)} = M_\varepsilon/(\hat\mu_\C+t_\C)M_\varepsilon$ that consists of polynomials $p(X_1,...,X_{N-1})$. In the formalism of Section \ref{sec:qDH}, these polynomials generate a subspace $N_\varepsilon\subset M_\varepsilon$ that is transverse to the leading term $\sum_i c_iX_i$ of the moment map. The action of the mesons  $\hat X_i \hat Y_j$ for $i<N$, $j<N$ does not leave the subspace $N_\varepsilon$:
\be
(\hat X_i \hat Y_j) p(X) = X_i (c_j + \epsilon \pd_j) p(X) \,, \qquad  i,j<N \,,
\ee
whereas the action of the remaining mesons needs to be brought back to $N_\varepsilon$ by using the moment map relation \eqref{SQED-Dequiv}:
\begin{align}
(\hat X_i \hat Y_N) p(X) &= c_N X_i \,p(X)\,,  \notag\\
(\hat X_N \hat Y_j) p(X) &= \textstyle  - \tfrac{1}{c_N} \left[ \, \sum\limits_{i<N} X_i \,(c_i +\epsilon \pd_i) + \tfrac{N}{2} \epsilon + t_\C \, \right] (c_j + \epsilon \pd_j) p(X)\,,  \label{mesons-mu} \\
(\hat X_N \hat Y_N) p(X) &= \textstyle - \left[ \, \sum\limits_{i<N} X_i \,(c_i + \epsilon\pd_i) + \tfrac{N}{2} \epsilon + t_\C \, \right] p(X)\,.  \notag
\end{align}

Alternatively, following the intrinsic approach at the end of Section \ref{sec:qDH}, we are lead to identify the module $\hat\CD_\varepsilon^{(H)}$ (for $\varepsilon=(+,...,+)$) with the quotient of the bulk algebra $\hat\C[\CM_H]$ by the relations that the gauge-invariant operators satisfy when acting on the identity, namely
\be \hat X_i\hat Y_j  = \frac{c_j}{c_{i}} \hat X_i \hat Y_i  \qquad \forall\,i,j\,. \ee
A natural basis for this module is given by polynomials in any $N-1$ of the $N$ Cartan generators $H_i := \hat X_i\hat Y_i$, which satisfy $\sum_i H_i = -t_\C$ due to the complex moment-map constraint. It is easy to see that this basis is equivalent to one above, since 
\be
p(H_1,...,H_{N-1}) = p(c_1X_1,...,c_{N-1}X_{N-1})+\ldots
\ee
up to terms of lower degree. The modules $\hat\CD_\varepsilon^{(H)}$ for other choices of sign vector $\varepsilon$ can be treated in a similar way. 

For $N=2$, we find that the modules for $\CD_\varepsilon$, written in intrinsic form, take precisely the same form as the Coulomb-branch modules for Neumann boundary conditions 
that we encountered in Section \ref{sec:NC-qSQED}. Again, this is a special case of a more general mirror-symmetry relation.

To be explicit, let us recall that the bulk quantized algebra is a central quotient of the universal enveloping algebra $U(\mathfrak{sl}_2)$ with the quadratic Casimir element $C_2$ determined by the complex FI parameter. Let us use generators $E=\hat X_1\hat Y_2$, $F=-\hat X_2\hat Y_1$, and $H = \hat X_1\hat Y_1-\hat X_2\hat Y_2$, and write
\be
\hat \C[\CM_H] \simeq U(\mathfrak{sl}_2)/(C_2=\frac12(t_\C^2-\epsilon^2))\, ,
\ee
as in equation \eqref{EFH-XY}. The four possible Dirichlet boundary conditions $\CD_\varepsilon$ lead to modules of the form
\be
\hat \CD_\varepsilon^{(H)} = \hat \C[\CM_H]/\hat \C[\CM_H] \CI_\varepsilon\, ,
\ee
where the ideals $\CI_\varepsilon$ are generated by
\be \label{DH-qSQED2}
\begin{array}{r@{\quad}l@{\quad\;\;}l}
\hat\CD_{+-}: & E = -\frac14{\wt \xi}(H+t_\C-\epsilon)(H-t_\C-\epsilon)\,, & F = {\wt \xi}^{-1}\,, \\[.1cm]
\hat\CD_{--}: & E = -\frac12{\wt \xi}(H+t_\C-\epsilon) \,, & F = \frac12{\wt \xi}^{-1}(H-t_\C+\epsilon)\,, \\[.1cm]
\hat\CD_{++}: & E = \frac12{\wt \xi}(H-t_\C-\epsilon) \,, & F = -\frac12{\wt \xi}^{-1}(H+t_\C+\epsilon)\,, \\[.1cm]
\hat\CD_{-+}: & E = {\wt \xi}\,, & \hspace{-.97in}F  = -\frac14{\wt \xi}^{-1}(H+t_\C+\epsilon)(H-t_\C+\epsilon)\,,
\end{array}
\ee
with, respectively, ${\wt \xi} = 1/(c_1c_2)$, ${\wt \xi} = c_1/c_2$, ${\wt \xi} = c_2/c_1$, and ${\wt \xi} = c_1c_2$. These modules perfectly match the modules \eqref{NC-qSQED2} produced by the Coulomb branch image of Neumann boundary conditions, up to some relabelings $(\varepsilon_1,\varepsilon_2)\to (\varepsilon_1,-\varepsilon_2)$ and $(E,F,H)\to (F,E,-H)$, and some signs that can be absorbed in the definition of ${\wt \xi}$.

Following Section \ref{sec:DH-cinf}, we can deform these modules to lowest-weight modules with respect to a given real mass parameter $m_\R$, by sending $c_1$ and $c_2$ to zero or infinity. Suppose that $m_\R<0$. Then the relevant limit for \eqref{DH-qSQED2} is $\wt\xi\to0$. This deforms $\hat \CD_{--}$ and $\hat \CD_{++}$ to Verma modules of lowest weights $t_\C-\epsilon$ and $-t_\C-\epsilon$, respectively, and it deforms $\hat \CD_{+-}$ to a direct sum of the two Verma modules. The direct sum becomes a nontrivial extension as in \eqref{proj-2} if $t_\R=k\epsilon$ for integer $k$. The module $\hat \CD_{-+}$ does not have a regular limit as $\wt\xi\to 0$, corresponding to the fact that $\CD_{-+}$ is not $m_\R$-feasible (it breaks supersymmetry).

In the case of $\hat \CD_{++}$ and $\hat \CD_{--}$, the deformation to a Verma module can equivalently be understood by using a hidden, nonabelian $SU(2)$ flavor rotation to send one of the parameters $c_i$ to zero, as in \eqref{SQED-sym}. For example, if we consider $\hat \CD_{++}$ at $c_2=0$ (or $\wt\xi = 0$), we can describe the module directly by using a basis $|n\rangle := (X_2)^n$. (Note that we can no longer use polynomials in $X_1$ when $c_2=0$.) Following the same procedure as in \eqref{mesons-mu}, we find that
\be \begin{array}{ll}
 E|n\rangle &= 
 -\frac{1}{c_1}n\epsilon(t_\C+n\epsilon)|n-1\rangle\,, \\[.1cm]
 F|n\rangle &= c_1|n+1\rangle\,, \\[.1cm]
 H|n\rangle &=(-t_\C+(2n+1)\epsilon)|n\rangle\,.
\end{array}
\ee
This is the Verma module of lowest weight $-t_\C+\epsilon$.

\subsubsection{SQCD}

Here we follow the notation of the SQCD examples from Section \ref{sec:N}; we consider SQCD with gauge group
 $G=U(K)$ and hypermultiplets $(X^i,Y_i)_{i=1}^N$ in $N$ copies of the fundamental $(X^i)$ and anti-fundamental $(Y_i)$ representations. We will assume $N \geq K$. The set of interesting Dirichlet boundary conditions could be larger in SQCD than it is in SQED, as we can pick our Lagrangian splitting $L$ to break the boundary symmetry $U(K)_\pd$. We will discuss first an example where this does not happen and then assess the possibility to use a $U(K)_\pd$-breaking splitting. 

The first case, which we can denote simply $\CD_{++...+}$, uses the trivial splitting $L$: all scalar fields $Y_i$ in the anti-fundamental representation 
are set to generic constant values at the boundary,
\be Y_i^a\big|_\pd = c_i^a\,. \ee
The $R$-symmetry preserved at the boundary $U(1)_V'$ differs from $U(1)_V$ by a diagonal boundary gauge transformation.
In terms of the meson fields $M_i^j :=  Y_i \cdot X^j = \sum_a Y_i^aX_a^j$ that parameterize the Higgs branch, the boundary condition takes the form
\begin{equation}
M_i^j\big|_\pd = c_i \cdot X^j\big|_\pd\,. \label{DH-SQCD-MX}
\end{equation}
It is clear from this that the boundary operators $X_j$ are fully determined by the bulk mesons $M_i^j$. In fact, they are determined by the set of mesons $M_{(K)}^j$ with $i\leq K$. Indeed, if we denote as $c_{(K)}$ the leading $K\times K$ submatrix of the $K\times N$ matrix $c$ (assumed to be invertible), then we have $X^j = (c_{(K)})^{-1} M_{(K)}^j$. Substituting back into \eqref{DH-SQCD-MX}, we find that the boundary condition requires gauge-invariant operators to obey
\begin{equation}
M_i^j = c_i \cdot (c_{(K)})^{-1} M_{(K)}^j\,, \label{DH-SQCD-MM}
\end{equation}
which simply says that all the mesons can be written in terms of $M_{(K)}^j$.
We should supplement this with the complex moment-map constraint $\sum_j X_a^j c_j^b +t_\C\delta_a^b=0$, which is equivalent to $ \sum_j M_{(K)}^j c^j + c_{(K)} t_\C =0$. This allows us to eliminate the entire $K \times K$ submatrix of mesons with $i\leq K$ and $j\leq K$ in favor of the remaining $K \times (N-K)$ elements of $M^{(K)}_j$. Thus \eqref{DH-SQCD-MM} defines a half-dimensional (in fact, Lagrangian) ideal in the chiral ring $\C[\CM_H]$.

The quantum version of the story proceeds in a similar manner. The module $\hat \CD_{++...+}^{(H)}$ can be given a concrete description in terms of polynomials in the 
$X_j$ with $j>K$. All polynomials can be reconstructed by acting on $1 := |$ with the same $K \times (N-K)$ generators $\hat M^{(K)}_j$ we 
used in the classical analysis. Thus we can recast the module in terms of the ideal of operators which annihilate $|$. Since
\begin{equation}
\hat M_i^j \big| = \;\,:\!\!Y_i \cdot X^j \!\!:\big|= c_i \cdot X^j \big| + \frac{\epsilon}{2} \delta_i^j \big|
\end{equation}
the quantum ideal takes the same linear form as \eqref{DH-SQCD-MM} up to the replacement $M^i_j \to (\hat M^i_j -  \frac{\epsilon}{2} \delta^i_j)|$.

Notice that although we did our analysis with generic $c$, it is always possible to do a global unitary transformation to rotate $c$ 
to $(1_{K \times K}, 0,\cdots,0)$. This shows that the boundary condition preserves an $SU(N_f-K)$ subgroup of the $SU(N_f)$ global symmetry.
The module is reorganized in a similar manner as we saw for the SQED example.

The simplest possibility for a boundary condition breaking $U(K)_\pd$ is to preserves a $U(1)_V'$ that differs from $U(1)_V$ by a diagonal $U(K)_\pd$ generator $(1,\cdots, 1, -1, \cdots,-1)$
with $n$ ``$-1$'' entries. This leads to a Lagrangian splitting that sets to generic constants the first $K-n$ gauge entries $Y^{(K-n)}$ of $Y$ and the last $n$ gauge entries $X_{(n)}$ of $X$. 
In terms of the meson fields, the boundary condition takes the form
\begin{equation}
M_i^j := Y_i \cdot X^j = c^{(K-n)}_i \cdot X_{(K-n)}^j + Y^{(n)}_i \cdot c_{(n)}^j \, .
\end{equation}
These constraints alone do not fully fix the 2d degrees of freedom, though: they are unaffected by coordinated shifts of $X_{(K-n)}^j$ by $A_{(K-n)}^{(n)} \cdot c_{(n)}^j$ and 
of $Y^{(n)}_i$ by $- c^{(K-n)}_i A_{(K-n)}^{(n)}$. The complex moment map constraints
\be
\begin{array}{l@{\qquad\quad}l}
\sum_j X_{(K-n)}^j c^{(K-n)}_j + t_\C = 0 &
\sum_j X_{(K-n)}^j Y^{(n)}_j= 0 \\[.2cm]
\sum_j c_{(n)}^j c^{(K-n)}_j = 0 &
\sum_j c_{(n)}^j Y^{(n)}_j + t_\C = 0
\end{array}
\ee
are also invariant under that shift. Thus, this boundary condition has massless two-dimensional degrees of freedom, and does not belong to the simple class of Dirichlet boundary conditions that we have been studying so far. It is tricky to find choices of $L$ that break $U(K)_\pd$ and do not suffer from this problem. 

\subsection{Coulomb-branch image}
\label{sec:DC}

In order to describe the Coulomb-branch image of a Dirichlet boundary condition, we recall the ``integrable system'' fibration from Section \ref{sec:intsys}
\be \CM_C \to \mathfrak t_\C/W \simeq \C^{\text{rank}(G)}\,, \label{C-fibration} \ee
where the base is parameterized by the expectation values of gauge-invariant polynomials in $\varphi$. The generic fiber is $(\C^*)^{\text{rank}(G)}$, parameterized by abelian monopole operators (or, classically, by $\sigma$ and the dual photons). Dirichlet boundary conditions fix $\varphi\big|_\pd = \varphi_0$ at the boundary, while (classically) leaving $\sigma$ and the dual photons unconstrained; thus the Coulomb-branch image of a Dirichlet boundary condition is supported on a single fiber of \eqref{C-fibration}.

For $c=0$, the fiber may be generic. However, as we turn on vacuum expectation values for some $Y_{L,i}$, the complex fields $\varphi$ and the complex masses $m_\C$ are restricted 
in such a way that the effective complex masses $M_{L,i}$ of the corresponding $Y_{L,i}$ vanish. In the extreme case that $c$ breaks both the boundary gauge symmetry $G_\pd$ and flavor symmetry $G_H$, we must have $m_\C=0$ and $\varphi = 0$, so the Dirichlet boundary condition is supported on the ``most singular fiber'' of the fibration \eqref{C-fibration}, lying above the origin. Singularities of this fiber can be (partially) resolved by turning on real masses $m_\R$, and it is useful to introduce them in the following.

Determining where in the fiber above $\varphi_0$ the support of a Dirichlet boundary condition lies can be tricky. The classical effect of turning on $c$ can be analyzed by looking at the 2d $\CN=(2,2)$ BPS equations for $X_L$, $Y_L$, and $\sigma$. We remind the reader that these equations read 
\be
\CD_1X_L = \CD_1Y_L = 0 \qquad D_1\sigma + g^2\mu_\R = 0 \qquad \mu_\C=0
\ee
as in Appendix \ref{app:2d}. (We have set the complex FI parameter to zero, $t_\C=0$, so that the full Coulomb branch is available in the bulk.) Working in axial gauge $A_1=0$, the equations for $X_L$ and $Y_L$ become
\be \label{MRXY} (\pd_1 + M_\R^L)X_L = 0\,,\qquad (\pd_1 - M_\R^L)Y_L = 0\,, \ee 
\be M_\R^L := \sigma T_L+ m_\R T_L^{(F)}\,, \notag  \ee
where $M_\R^L = M_\R^L(\sigma,m_\R)$ is the effective real mass matrix for the $X_L$ and $-M_\R^L$ the real mass matrix for the $Y_L$. This Hermitian matrix depends on the generator of gauge transformations $\sigma T_L$ in the representation appropriate for the $X_L$, and on the corresponding generator $m_\R T_L^{(F)}$ for $G_H$ flavor symmetry transformations. 

We now ask: for which values $\sigma_\infty$ far from the boundary does there exist a solution of the BPS equations compatible with the boundary values $Y_L\big|_\pd = c$\,? It turns out that a necessary condition is that $c$ lies in the non-positive (non-negative) eigenspace of $M_\R^L(\sigma_\infty)$ for a left (right) boundary condition:
\be \label{Mpos}
\CD_{L,c}^{(C)}\,:\quad \varphi\big|_\pd = \varphi_0\quad\text{and}\quad \begin{cases} c\in \text{non-pos espace of } M_\R^L(\sigma_\infty,m_\R)  & \text{left b.c.} \\
c\in \text{non-neg espace of } M_\R^L(\sigma_\infty,m_\R)  & \text{right b.c.} \end{cases}
\ee
For an abelian theory, or more generally if $L$ preserves $G_\pd$, we can diagonalize $\sigma$ and simplify the constraint to the requirement that the effective real mass of every field with non-zero vev $c$ should be non-negative (non-positive). 

To see where this condition comes from, first observe that the $X_L$ fields should (generically) be set to zero to satisfy $\mu_\C = 0$. The $Y_L$ that are nonzero at the boundary will evolve according to \eqref{MRXY}, and will blow up as $x^1\to \pm\infty$ unless $c$ belongs to the non-negative (non-positive) eigenspace of $M^L_\R$. If \eqref{Mpos} is satisfied, then the $Y_L$ can safely decay to zero. If the $Y_L$ decay, then at infinity $\mu_\R \sim |Y_L|^2$ is negligible, and it is consistent to assume that $\sigma\sim\sigma_\infty$ attains a constant value. We have not proven that given \eqref{Mpos} a unique non-singular solution to the BPS equations \emph{does} exist, but it is conceivable that this is the case. (We will see the solution explicitly in examples below.) 

It is important to note that this analysis is only semi-classical. However, corrections to the Coulomb-branch metric only enter the real moment-map equation $\pd_1\sigma+g^2\mu_\R=0$, whose precise form does not matter. All we only need to know that an asymptotically constant $\sigma_\infty$ is consistent as long as $\mu_\R\to 0$ exponentially fast.

Naively, the real inequalities in \eqref{Mpos} suggest that the support of $\CD_{L,\epsilon}^{(C)}$ simply ends at some real-codimension-one walls in the fiber $\varphi = \varphi_0$. However, the values of $\sigma$ for which $M_\R^L$ has a null eigenvalue are precisely the locations at which
some matter fields become massless and quantum corrections shrink some circle in the torus of dual photons.
This effect can allow the brane $\CD_{L,\epsilon}^{(C)}$ to end smoothly. Thus the intuitive picture is that the fiber 
$\varphi=\varphi_0$ consists of several components and that the Coulomb-branch image $\CD^C_{L,c}$ consists of a subset of the components 
that is selected semi-classically by \eqref{Mpos}.

We can also describe the locus $\CD_{L,\epsilon}^{(C)}$ (at least partially) in terms of the chiral ring. Consider first an abelian theory. Setting complex masses $M_{L,i}$ to zero for nonvanishing $Y_{L,i}$ gives relations of the form $v_A v_{-A}=0$ for 
every cocharacter $A$ such that the charge $Q_{A}^i$ is non-zero.
The classical condition \eqref{Mpos} corresponds to a locus where
\be \label{Mposv}
\text{for all $A$ s.t. $Q_{A,L}^i > 0$}\,, \quad \begin{cases} v_{A} = 0 & \text{left b.c.} \\
v_{-A}  = 0 & \text{right b.c.} \end{cases}
\ee

For nonabelian theories we expect a similar constraint to hold at the abelianized level. An abelianized description, however, may be inappropriate 
for the fiber we are interested in -- for example, at $\varphi=0$.
A better general strategy is to identify how the fiber splits whenever we turn on 
a vacuum expectation value for some specific fields, and to identify which component is selected by the boundary condition for these fields. The final answer should be the intersection of the
constraints associated to each field that receives a vev at the boundary. 

\subsubsection{Effect of real masses and real FI}
\label{sec:DC-mt}

Real masses resolve the Coulomb branch, and it was already natural to include them in the preceding analysis. 
For a given choice of real masses, the condition \eqref{Mpos} may or may not admit solutions. 
Correspondingly, the UV boundary condition $\CD_{L,c}$ is either feasible, with an image $\CD_{L,c}^{(C)}$ on the Coulomb branch; or infeasible, in which case it breaks supersymmetry. 
For an abelian theory, one can systematically identify the components of the resolved fiber associated to a given boundary condition. 
We will discuss this in detail in Section \ref{sec:abel}.

The effect of FI parameters may be complicated.  
If we proceed as we did for Neumann boundary conditions, we can work in the sigma model approximation and look at gradient flows on the Coulomb branch. 
For example, for an abelian theory one would look at the intersection between the condition $v_A=0$ for all $A$ such that $Q_{A,L}^i$ is positive
and the condition associated to gradient flows, $v_A=0$ for all $A$ such that $t_\R \cdot A$ is positive. 

As $\CD_{L,c}^{(C)}$ sits in the fiber above the fixed-point locus and the gradient flow happens within that same fiber, the intersection will 
often lead to noncompact moduli spaces of 2d classical vacua and thus possibly to strong dynamics, 
much as we encountered for Higgs-branch images of Neumann boundary conditions.

\subsubsection{Quantum Coulomb-branch image}
\label{sec:qDC}

In Section \ref{sec:NC-q}, the $\Omega$-background and the corresponding quantization of the Coulomb branch chiral ring $\hat\C[\CM_C]$ was defined using the standard bulk R-symmetry $U(1)_V$. In the presence of a Dirichlet boundary condition, we should instead use the preserved $U(1)_V'$ R-symmetry to define the $\Omega$-background. This has the effect of shifting $\hat \varphi$ and $m_\C$ by half-integer multiples of $\epsilon$, corresponding to the amount by which $U(1)_V'$ differs from $U(1)_V$ by gauge and/or flavor symmetry rotations. This phenomenon is mirror to what we saw in Section \ref{sec:NHiggs}: in the presence of a Neumann boundary condition, the axial anomaly causes the effective value of $t_\C$ that appears in $\hat\C[\CM_H]$ to be shifted from zero by half-integer multiples of $\epsilon$.

Here, we will continue to use the formulas of Section \ref{sec:NC-q}, but must occasionally account for the presence of extra shifts. For example, when a boundary condition $\CD_{L,c}$ breaks all of the flavor symmetry $G_H$, the complex masses $m_\C$ in chiral-ring expressions should not be set to zero, but rather to appropriate half-integer multiplets of $\epsilon$.

For now, we will describe the module $\CD^{(C)}_{L,c}$ corresponding to $\CD_{L,c}$ by the quantization of the 
classical answer, \ie\ as the quotient of the full quantum algebra by an ideal generated by the 
vector corresponding identity operator $1$ or `$|$' at the boundary.
In order for some of the fields $Y_{L,i}$ to have a nonvanishing vev $c_i$ at the boundary, we saw the corresponding complex masses $M_{L,i} = M_{L,i}(\varphi,m_\C)$ had to vanish. In the presence of the $\Omega$-background, this condition is slightly modified: we expect $M_{L,i} - \frac12\epsilon = 0$, where $\frac12$ is the $U(1)_V$ R-charge of $Y_{L,i}$.
Thus the identity vector on the boundary obeys
\be \big(M_{L,i}(\hat \varphi,m_\C) -\tfrac12\epsilon\big)\big| = 0 \qquad \text{(for all nonvanishing $Y_{L,i}$)}\,. \label{qDC-M}\ee
(Alternatively, we have $M_{L,i}(\hat\varphi',m_\C')\big|=0$, where $\hat\varphi',m_\C'$ have been shifted to account for the redefinition of $U(1)_V'$ as discussed above.)

The conditions \eqref{qDC-M} are the quantum equivalent of restricting to the fiber of $\CM_C$ at $\varphi=\varphi_0$. We must supplement them with additional constraints on the monopole operators that select a particular locus in that fiber, as in \eqref{Mposv}. For abelian theories, this means that
\be  v_{A}\big| = 0\qquad \text{for all $A$ s.t. $Q_{A,L}^i < 0$}\,. \label{qDC-ve} \ee
(The notation makes implicit that we are studying a module coming from a right boundary condition. Otherwise, for a left boundary condition, we would want $\big|v_{-A}=0$.) We expect the module $\hat \CD_{L,c}^{(C)}$ to be generated from the identity vector, subject to the relations \eqref{qDC-M} and \eqref{qDC-ve}.

For a nonabelian theory we can formulate similar definitions. Each scalar field with non-zero vev should lead to a constraint 
encoded into an ideal of the full quantum algebra and the final module should be the quotient by the union of these ideals.

The construction in \eqref{qDC-M}-\eqref{qDC-ve} does not extend in a straightforward way to \emph{twisted} modules, corresponding to the insertion in the theory of a vortex line defect at the fixed axis of the $\Omega$-deformation (\cf\ Section \ref{sec:lineC}). Even when a Dirichlet boundary condition breaks all flavor symmetry $G_H$, the insertion of vortex lines should allow us to (effectively) set the complex masses $m_\C$ to arbitrary integer or half-integer multiples of $\epsilon$,
\be m_\C = k\epsilon\,. \ee
However, with generic values of $k$, it is usually impossible to impose conditions \eqref{qDC-M} on the identity vector simultaneously for all $Y_{L,i}$.%
\footnote{Note that the $M_{L,i}$ all commute with each other, so imposing all the conditions simultaneously is equivalent to asking that all classical $M_{L,i}$ vanish for generic $m_\C$ --- but this is impossible if the boundary condition breaks $G_H$ flavor symmetry.}

In order to resolve this puzzle, we need to look more closely at the physical origin of the boundary twisted chiral ring for Dirichlet boundary conditions. 
The full analysis is somewhat lengthy and we postpone it to Section \ref{sec:xD}. The basic idea, though, is rather intuitive: in the presence of a line 
defect, there is no canonical choice of ``identity operator''. Rather, there is a module generated by boundary monopole operators $|A\rangle$ labelled by a charge $A$,
and distinct generators are annihilated by each of the conditions \eqref{qDC-M}. Each of these generators enters the corresponding twisted version of \eqref{qDC-M}--\eqref{qDC-ve}. We will give a brief illustration of this for SQED.

\subsection{Examples}

\subsubsection{SQED}
\label{sec:DC-SQED}

The Coulomb branch of SQED with $N$ hypermultiplets is a singularity $\CM_C \simeq \C^2/\Z_N$, which is deformed by complex masses and resolved by real masses (see Section \ref{sec:NC-SQED}).
Consider the basic Dirichlet boundary conditions $\CD_\varepsilon$ from \eqref{SQED-De}, which are labelled by a sign vector $\varepsilon\in \{\pm\}^N$. Naively, these boundary conditions break both the boundary gauge symmetry $G_\pd$ and the flavor symmetry $G_H$. Thus, to begin, we turn off all complex masses. The image of the boundary condition $\CD_\varepsilon$ is then supported in the fiber of the Coulomb branch at $\varphi = 0$.

\begin{figure}[htb]
\centering
\includegraphics[width=4.5in]{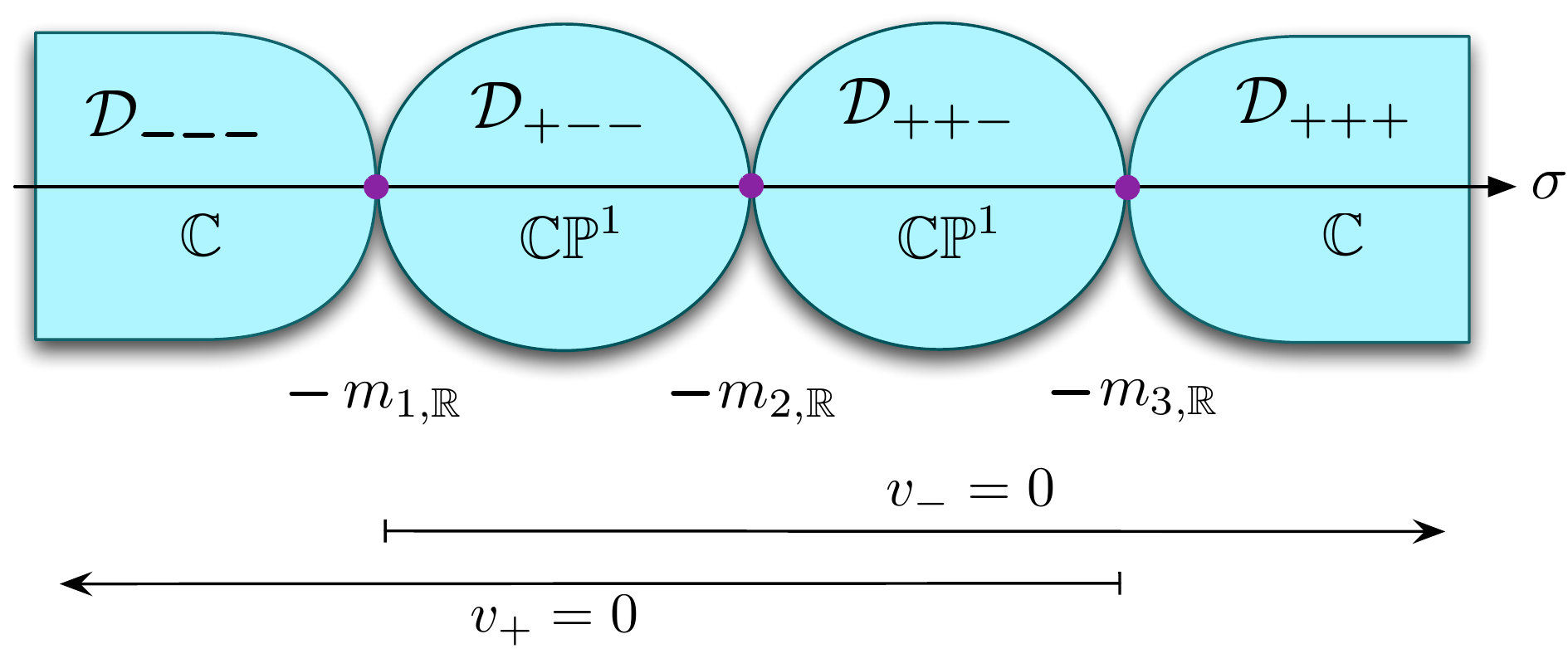}
\caption{The fiber above $\varphi=0$ of the Coulomb branch of SQED with $N=3$ hypers.}
\label{fig:DC}
\end{figure}

The fiber at $\varphi=0$, which we call $\CS_0$, passes through the singularity of $\CM_C$. Upon resolving the singularity with real mass parameters $m_{i,\R}$ (normalized so that $\sum_{i=1}^N m_{i,\R}=0$), the fiber becomes a union of $N-1$ singular divisors $\cp^1$ and two copies of $\C$. Intuitively, the fiber $\CS_0$ is itself a fibration, with the base $\R$ parameterized by $\sigma$ and the fibers $S^1$ parameterized by the dual photon $\gamma$. The dual-photon circle shrinks at the locations where hypermultiplets become massless, \ie\ where $\sigma = - m_{i,\R}$, trapping a string of $\cp^1$'s (Figure \ref{fig:DC}).

Following equation \eqref{Mpos}, the image of the boundary condition $\CD_\varepsilon$ as a right boundary condition must be supported on the locus of $\CS_0$ where
\be (\sigma+m_{i,\R}) > 0 \quad (\varepsilon_i=+)\,,\qquad  \sigma+m_{i,\R} < 0\quad (\varepsilon_i=-)\,. \label{Mpos-SQED} \ee
If the real mass parameters $m_{i,\R}$ are generic, there are exactly $N+1$ sign vectors for which these conditions can be simultaneously satisfied; the others break supersymmetry. For example, if we order $m_{1,\R}>m_{2,\R}>...>m_{N,\R}$, then the feasible sign vectors are of the form $\varepsilon = (++...+--...-)$, with $s$ plus signs followed by $N-s$ minus signs. The resulting image $\CD_\varepsilon^{(C)}$ wraps one of the copies of $\C$ if $s=0$ or $s=N$, and otherwise wraps the $s$-th $\cp^1$.

For non-generic values of the real mass parameters (for example, $m_{i,\R}\equiv 0$), $\cp^1$'s can shrink and the fiber $\CS_0$ becomes singular. In this case, the constraints \eqref{Mpos-SQED} can be satisfied for additional choices of $\varepsilon$ and the resulting boundary conditions $\CD_{\varepsilon}$ will have images supported at singularities.

\begin{figure}[htb]
\centering
\includegraphics[width=4in]{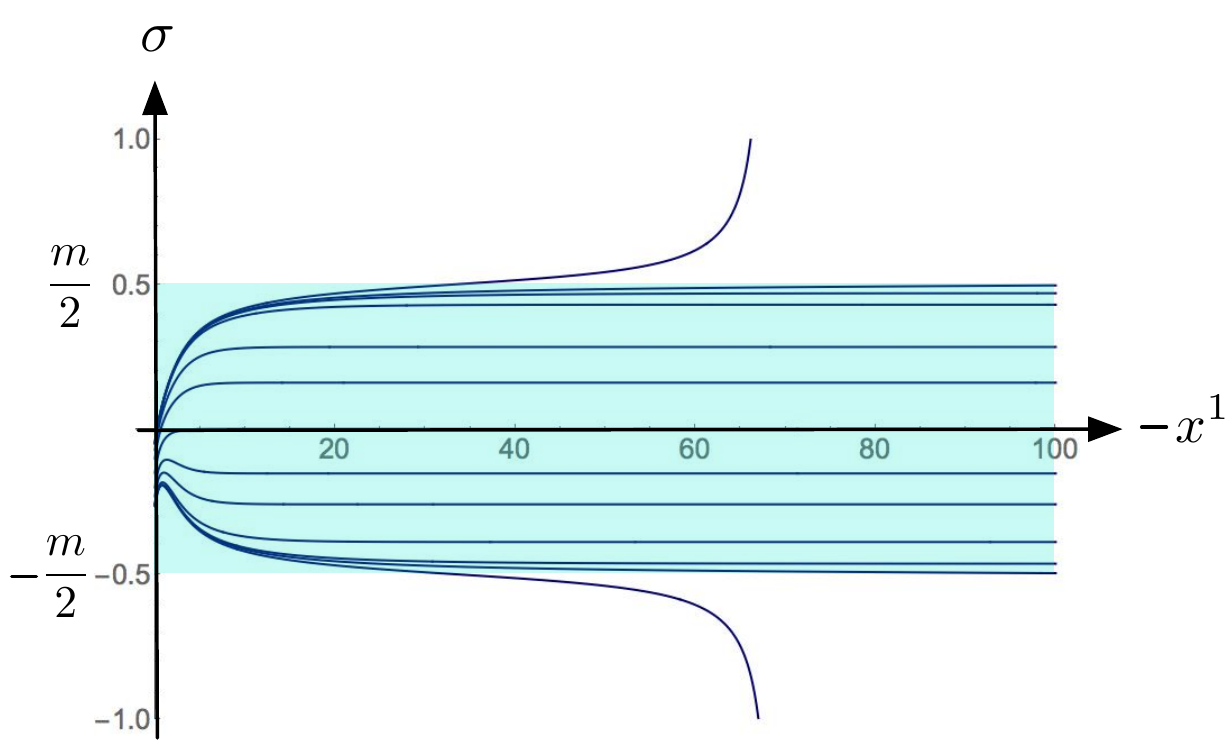}
\caption{Numerical solutions to the BPS equations \eqref{BPS-SQED} with $X_1=Y_2\equiv0$ and $Y_1|_\pd = .5$, $X_2|_\pd = .7$, for $m_\R=1$ and various initial values of $\sigma$. Any asymptotic value $\sigma_\infty$ in the range $\big[-\frac12 m,\frac12 m\big]$ can be attained.}
\label{fig:BPS-SQED}
\end{figure}

In the present example, we can illustrate explicitly that the conditions \eqref{Mpos-SQED} are necessary and sufficient for the existence of a 2d vacuum by solving the 2d $\CN = (2,2)$ BPS equations numerically. As a representative case, let us take $N=2$ and the boundary condition $\CD_{+-}$. We set $(m_{1,\R},m_{2,\R})=(\frac12 m,-\frac12 m)$ with $m>0$. The relevant BPS equations are
\be \label{BPS-SQED}
 \begin{array}{l@{\qquad}l} \pd_1X_1 = -(\sigma+\tfrac m2)X_1 & \pd_1Y_1 = (\sigma+\tfrac m2)Y_1 \\[.1cm]
\pd_1X_2 = -(\sigma-\tfrac m2)X_2 & \pd_1Y_2 = (\sigma-\tfrac m2)Y_2 \end{array}\qquad
g^{-2}\pd_1\sigma = -|\vec X|^2+|\vec Y|^2\,, \ee
together with the boundary conditions
\be
Y_1|_\pd=c_1 \qquad X_2|_\pd=c_2 \qquad \sigma\sim \sigma_{\infty} 
\ee
as $x^1\to-\infty$. It is useful to observe that we can replace $X_i,Y_i$ with their absolute values $|X_i|,|Y_i|$ because the phases of these fields are constant; thus we have a set of five equations in five real variables. If $c_1,c_2\neq 0$, all solutions to these equations blow up at finite $x^1$ unless $X_1$ and $Y_2$ vanish identically. Setting $X_1=Y_2\equiv 0$, we find that there exist regular solutions as long as the initial value $\sigma|_\pd$ is constrained to lie within a small interval close to the origin. (Otherwise the solutions again blow up at finite distance.) The asymptotic value of $\sigma$ lies anywhere in the range $-\frac m2\leq \sigma \leq \frac m2$. At large $x^1$, the solutions have the approximate form 
\be
Y_1\sim e^{(\sigma_\infty+m/2)x^1} \qquad X_2\sim e^{(-\sigma_\infty +m/2)x^1}\, . 
\ee
This example is illustrated in Figure~\ref{fig:BPS-SQED}.

Next, we consider quantization induced by the presence of $\Omega$-background. Recall that with complex masses turned on the Coulomb-branch chiral ring takes the form $v_+v_- = \prod_{i=1}^{N}(\varphi+m_{i,\C})$, and its quantization $\hat\C[\CM_C]$ in $\Omega$-background is given by
\be\textstyle \hat v_+\hat v_- = \prod\limits_{i=1}^N (\hat\varphi+m_{i,\C}-\tfrac12\epsilon)\,,\qquad
\hat v_-\hat v_+ = \prod\limits_{i=1}^N (\hat\varphi+m_{i,\C}+\tfrac12\epsilon)\,,\ee
along with the commutators $[\hat\varphi,\hat v_\pm]=\pm \epsilon \hat v_\pm$. For the $\CD_{\varepsilon}$ boundary condition, we must (naively) set all complex masses to zero, $m_\C=0$, because $G_H$ flavor symmetry is broken. The correction from using $U(1)_V'$ rather than $U(1)_V$ to define the $\Omega$-background modifies this to
\be \label{me-SQED}
 m_{i,\C} = \frac12 \varepsilon_i\, \epsilon\,,
\ee
modulo an overall shift of $\hat \varphi$ that could be used to set $\sum_i m_{i,\C}=0$.

Consider the boundary condition $\CD_{++...+}$. The corrections \eqref{me-SQED} all have the same sign, so they can be absorbed in a shift of $\hat\varphi$, and we simply set $m_{i,\C}\equiv 0$. Following \eqref{qDC-M}--\eqref{qDC-ve}, we find that the identity vector should simply satisfy
\be \big(\hat\varphi-\tfrac12\epsilon\big)\big| = v_-\big| = 0\,,\ee
which  generates a lowest-weight Verma module with states $(v_+)^n\big|$ for $n\geq 0$. In a similar way, $\CD_{--...-}$ leads to a highest-weight Verma module with states $(v_-)^n\big|$.

For any other choice of sign vector containing both $+$'s and $-$'s, it is more convenient not to normalize the complex masses so that $\sum_i m_{i,\C}=0$. Then the prescription \eqref{me-SQED} simply sets 
\be
M_{L,i}(\hat\varphi,m_{i,\C}) - \frac12\epsilon = \varepsilon_i(\hat\varphi+m_{i,\C}) - \frac12\epsilon = \varepsilon_i\hat \varphi
\ee 
for all $i$. Thus we find a module generated by an identity vector that satisfies
\be \hat\varphi\big| = \hat v_+ \big| = \hat v_-\big| = 0\,. \ee
This is a trivial module, which contains only the identity. In summary, we have found
\begin{itemize}
\item $\hat\CD_{++...+}^{(C)}$ and $\hat \CD_{--...-}^{(C)}$ are infinite-dimensional irreducible Verma modules; and
\item all other $\hat \CD_\varepsilon^{(C)}$ are one-dimensional trivial modules.
\end{itemize}

As anticipated, the situation is more complicated if we introduce vortex-line operators in order to set $m_{i,\C} = k_i\epsilon$ with $k_i \in \frac{1}{2}\mathbb{Z}$. So far, the Coulomb-branch images of Dirichlet boundary conditions have resembled the Higgs-branch images of Neumann boundary conditions in all possible ways, and we will argue in Section \ref{sec:abel} that (in the case of abelian theories) these boundary conditions are actually mirror to each other. By analogy with Section \eqref{sec:NH-qSQED}, we therefore expect that at general $k_i$ the Verma modules $\hat\CD_{++...+}^{(C)}$ and $\hat\CD_{--...-}^{(C)}$ remain irreducible Verma modules, while the remaining $\hat \CD_\varepsilon^{(C)}$ (for appropriate $\varepsilon$) become non-trivial irreducible finite-dimensional representations of the algebra $\hat \C[\CM_C]$.
We illustrate how this might come about in the case $N=2$.

For $N=2$ hypermultiplets, let us introduce a vortex-line operator that sets
\be 
(m_{1,\C},m_{2,\C}) = (\frac12 k\epsilon,-\frac12 k\epsilon)
\ee
with $k\geq 1$. The chiral-ring equations are then
\be \hat v_+\hat v_- = (\hat \varphi+\frac{k-1}{2} \epsilon )( \hat \varphi- \frac{k+1}{2}\epsilon )\,,\qquad \hat v_-\hat v_+ = (\hat \varphi+\frac{k+1}{2}\epsilon )(\hat \varphi- \frac{k-1}{2} \epsilon)\,.\ee
Consider the boundary condition $\CD_{+-}$. Following \eqref{qDC-M}, we might be led to consider an identity vector that satisfies separately
\be (\hat\varphi+\frac{k-1}{2}\epsilon )\big| = (\hat\varphi-\frac{k-1}{2}\epsilon) \big| = 0\,, \label{qDC-id} \ee
which is clearly impossible unless $k=1$. 

The best we can do is to consider separate vectors $\big|_\pm$ that are eigenvectors of the operator $\hat\varphi$ with eigenvalues $\pm \frac12(k-1)\epsilon$, both satisfying
\be (\hat\varphi+\frac{k-1}{2}\epsilon )( \hat\varphi-\frac{k-1}{2}\epsilon )\big|_\pm = 0\,. \ee
We cannot impose both constraints $\hat v_+|=\hat v_-|=0$ in \eqref{qDC-ve} on the same vector, but we may require that $\hat v_+\big|_+=0$ and $\hat v_-\big|_-=0$. 

The module generated by $\big|_+$ and $\big|_-$ turns out to be too large: it is the direct sum of completely independent highest-weight and lowest-weight Verma modules. We may reduce it by making an additional identification among vectors in the same eigenspace for $\hat \varphi$. In particular, if we identify
\be \label{SQED-id}
(v_-)^{k-1}\big|_+ \;\sim\; \big|_-\,
\ee
we get precisely the module we are after: the Verma modules truncate to a single $k$-dimensional module, with states $(v_-)^{n}\big|_+$ 
(or equivalently $(v_+)^{k-1-n}\big|_-$) for $0\leq n\leq k-1$. We will argue in Section \ref{sec:xD} that the identification \eqref{SQED-id} is actually prescribed by the physics of boundary monopole operators.

Applying similar reasoning to the other boundary conditions $\CD_\varepsilon$ we find that, for $k\geq 1$:
\begin{itemize}
\item $\hat \CD_{++}$ is an irreducible Verma module of lowest weight (\ie\ eigenvalue of $H=2\hat\varphi$) $k+1$;
\item $\hat \CD_{--}$ is an irreducible Verma module of highest weight $-k-1$;
\item $\hat \CD_{+-}$ is a $k$-dimensional irreducible module;
\item $\hat \CD_{-+}$ does not admit any states.
\end{itemize}
This parallels the classification of $\hat \CN_\varepsilon$ modules in Section \ref{sec:NH-qSQED}.

Finally, we recall that the flavor symmetry $G_H$ is not actually fully broken by Dirichlet boundary conditions for SQED. 
Indeed, as in \eqref{SQED-sym}, we could have rotated the hypermultiplets to set all but one or two of the boundary vevs $c_i$ to zero. This allows some complex masses to be turned on, which do not change the above conclusions about supports of branes or modules, but does make the analysis a bit simpler. We briefly describe this.

Consider (for general $N$) the boundary condition $\CD_{++...+}$. We can rotate the boundary condition to set 
\be
(Y_1,...,Y_N)\big|_\pd = (c,0,...,0) \, ,
\ee
which manifests that there is an unbroken $SU(N-1)$ flavor symmetry. Turning on $N-1$ corresponding complex masses fully eliminates the singularity in the Coulomb branch, and makes the fiber $\CS_0 = \{v_+v_-=0\}$ a union of just two copies of $\C$, parameterized by $v_+$ or $v_-$. Thus, the complex mass deformation has effectively reduced the problem to the case $N=1$. The Coulomb branch image $\CD_{++...+}^{(C)}$ wraps the copy of $\C$ parameterized by $v_+$. Its quantization is the corresponding Verma module. In the limit $m_\C\to 0$, we recover the copy of $\C$ in the more complicated fiber shown in Figure \ref{fig:DC}. Similar arguments apply to $\CD_{--...-}$.

In the case of a sign vector $\varepsilon$ of the form $+++---$, with $s$ plus signs and $N-s$ minus signs, we rotate the boundary condition to the form \eqref{SQED-sym}. Turning on $N-2$ complex masses for the unbroken $S( U(s-1)\times U(N-s-1))$ flavor symmetry deforms the Coulomb branch so that only a $\Z_2$ singularity remains. The fiber $\CS_0$ intersects this $\Z_2$ singularity. Resolving the singularity with a real mass makes the fiber $\CS_0$ a union of two copies of $\C$ and one singular divisor $\cp^1$. Thus, the complex mass deformation has effectively reduced the problem to the case $N=2$. The brane $\CD_{++...+--...-}^{(C)}$ wraps the single $\cp^1$. Its quantization is the expected trivial module.

\subsubsection{SQCD}

Let us assume that $N \geq K$ here. We will consider only Lagrangian splittings that preserve the boundary symmetry $G_\pd$. 
As for SQED, we can use a global symmetry rotation to bring $\varepsilon$ to a $+++---$ form with $s$ plus signs and $N-s$ minus signs.
If $s\geq K$ and $N- s \geq K$ we can reduce $c$ to two $K \times K$ identity matrices 
in the $+$ set of flavors and in the $-$ set of flavors. If not, we will have $s \times s$ and 
$(N - s) \times (N- s)$ identity matrices respectively. 

A non-zero boundary vev $c^i_a$ forces $\varphi_a = m_{i,\C}$ for the corresponding eigenvalue of the complex vectormultiplet scalar. 
As a consequence, the polynomials $Q(z)$ and $P(z)$ have a common factor $z - m_{i,\C}$, 
which must then divide either $U_+(z)$ or $U_-(z)$, depending on the sign of $\varepsilon_i$. 
From the point of view of abelianized variables, this follows simply from the 
observation that $u^{\varepsilon_i}_a$ becomes zero at the boundary. 

If $s\geq K$, $U_-(z)$ ends up being zero, as we are imposing too many constraints on it. 
If $s<K$, we are fixing $s$ roots of $U_-(z)$. Similarly, if $N- s\geq K$, $U_+(z)$ ends up being zero, as we are imposing too many constraints on it. 
If $N-s<K$, we are fixing $N-s$ roots of $U_-(z)$.

In order to understand the Coulomb-branch module, we may start from the quotient by the ideal 
that sets the coefficients of $\hat Q(z)$ to specific values. A full treatment will appear elsewhere.

\section{Exceptional Dirichlet boundary conditions}
\label{sec:xD}

In Section \ref{sec:D}, we considered Dirichlet boundary conditions supplemented by a constraint $Y_L|_\pd = c$ for half of the hypermultiplets in a given Lagrangian splitting. We assumed there that the boundary values $c$ were as generic as possible in order to allow a $U(1)_V'$ R-symmetry to be preserved; typically this meant that both the Higgs-branch flavor symmetry $G_H$ and the boundary global symmetry $G_\pd$ were completely broken. In this section, we consider a second class of ``exceptional'' Dirichlet boundary conditions $\CD_{L,c}$ for which the boundary vevs $c$ still completely break $G_\pd$, but preserve a maximal abelian subgroup 
of the flavor symmetry group $G_H$ (as well as $U(1)_V'$).

These boundary conditions will be compatible with generic complex mass deformations as well as complex FI deformations. The Coulomb-branch flavor symmetry $G_C$ is always preserved by Dirichlet boundary conditions, so complex FI deformations are always possible. A complex mass deformation will have to be accompanied by an appropriate boundary vev $\varphi\big|_\pd  = \varphi_0$ to ensure that the scalar fields $Y_L$ that receive nonzero boundary vevs continue to have zero effective complex mass.

The Higgs and Coulomb-branch images, classical or quantum, of these boundary conditions can be analyzed in the same way as we did for generic 
Dirichlet boundary conditions. The main difference on the Higgs-branch side is that the images will be invariant under the maximal torus of the flavor symmetry group. The main difference on the Coulomb-branch side is the possibility of turning on generic complex mass deformations.

When analyzing quantum Coulomb-branch images in Sections \ref{sec:qDC}--\ref{sec:DC-SQED}, we encountered a puzzle in some examples, regarding the identification of boundary states. It will be important to resolve this puzzle in order to understand exceptional Dirichlet boundary conditions (and their relationship to generic Dirichlet boundary conditions). We do so in Section \ref{sec:bdymon} by more carefully studying the boundary twisted-chiral ring in the presence of a Dirichlet boundary condition, which is generated by boundary monopole operators.

One of our main interests in exceptional Dirichlet boundary conditions is that they provide a candidate for \emph{thimble} boundary conditions. In a theory with isolated massive vacua, a thimble boundary condition $\CB$ mimics a vacuum, in the sense that putting the theory on a half-space $x^1\leq 0$ with $\CB$ at the origin is equivalent (for certain BPS computations) to putting the theory on the whole space $x^1\in \R$ with a fixed vacuum as $x^1\to \infty$. Such boundary conditions for 2d $\CN = (2,2)$ theories appeared in (\eg) \cite{CV-WC, CV-tt*, HIV}; they provide an exceptional collection of generators for the Fukaya-Seidel category \cite{Seidel-Fukaya} of boundary conditions in a massive A-model.
\footnote{
The mathematical notion of a ``thimble'' originated in Picard-Lefschetz theory.} 
In sigma-models to the Higgs and Coulomb branches, the thimble branes are supported precisely on the gradient-flow cycles $\CM_H^>[m_\R^\nu]$, $\CM_C^>[t_\R^\nu]$ that were introduced in Sections \ref{sec:NH-mt}, \ref{sec:NC-mt} to describe the effect of real mass and FI deformations. 

We will argue in Section \ref{sec:thimbles} that thimble boundary conditions can be given a direct definition in the full gauge theory, which is explicitly self-mirror. 
We will also argue that exceptional Dirichlet boundary conditions have precisely the properties expected from such thimble boundary conditions. 
To every vacuum $\nu$ in the presence of real mass and FI deformations $(m_\R,t_\R)$, we associate the data of a particular UV Dirichlet boundary condition
\be (m_\R,t_\R;\nu) \;\to\; (L,c)\,. \label{vacLc} \ee
In particular, the existence of such an association strongly suggests that exceptional Dirichlet boundary conditions are self-mirror: given two mirror gauge theories $\CT,\CT'$, the mirror of $\CD_{L,c}$ associated to a vacuum in $\CT$ should be another exceptional Dirichlet boundary condition $\CD_{L',c'}$ associated to the same vacuum in $\CT'$.

We will find that the quantization of a thimble boundary condition produces some canonical modules. In particular, if the complex parameters $t_\C$ and $m_\C$ that enter the quantization of $\hat\C[\CM_H]$ and $\hat \C[\CM_C]$ (respectively) are generic, the quantization of thimbles produces all possible Verma modules. If the complex parameters are specialized to integer or half-integer values $t_\C\sim k_t\epsilon$ and $m_\C\sim k_m\epsilon$, the description is more subtle. For the Higgs branch, we find that if $k_t$ is chosen proportional to $-t_\R$, with a positive proportionality constant, then right thimble branes lead to Verma modules (also called \emph{standard} modules); while if $k_t$ is proportional to $t_\R$ we get \emph{costandard} modules, which are dual to Vermas/standards. A similar statement holds for the Coulomb branch:
\be \begin{array}{l@{\quad}l@{\quad}l} \text{Higgs branch:} & k_t\sim t_\R \;\Rightarrow\; \text{costandard} & k_t\sim -t_\R \;\Rightarrow \text{standard} \\[.1cm]
\text{Coulomb branch:} & k_m\sim m_\R \;\Rightarrow\; \text{costandard} & k_m\sim -m_\R \;\Rightarrow \text{standard}
\end{array} \ee
We discuss this in greater detail in Section \ref{sec:standards}, with a proof for abelian theories in Section \ref{sec:abel}. In the categories $\CO_H,\,\CO_C$, the standard modules form an exceptional collection with respect to a particular ordering, and the costandard modules form an exceptional collection with respect to the opposite ordering.

\subsection{Boundary monopole operators}
\label{sec:bdymon}

\begin{figure}[htb]
\centering
\includegraphics[width=3.5in]{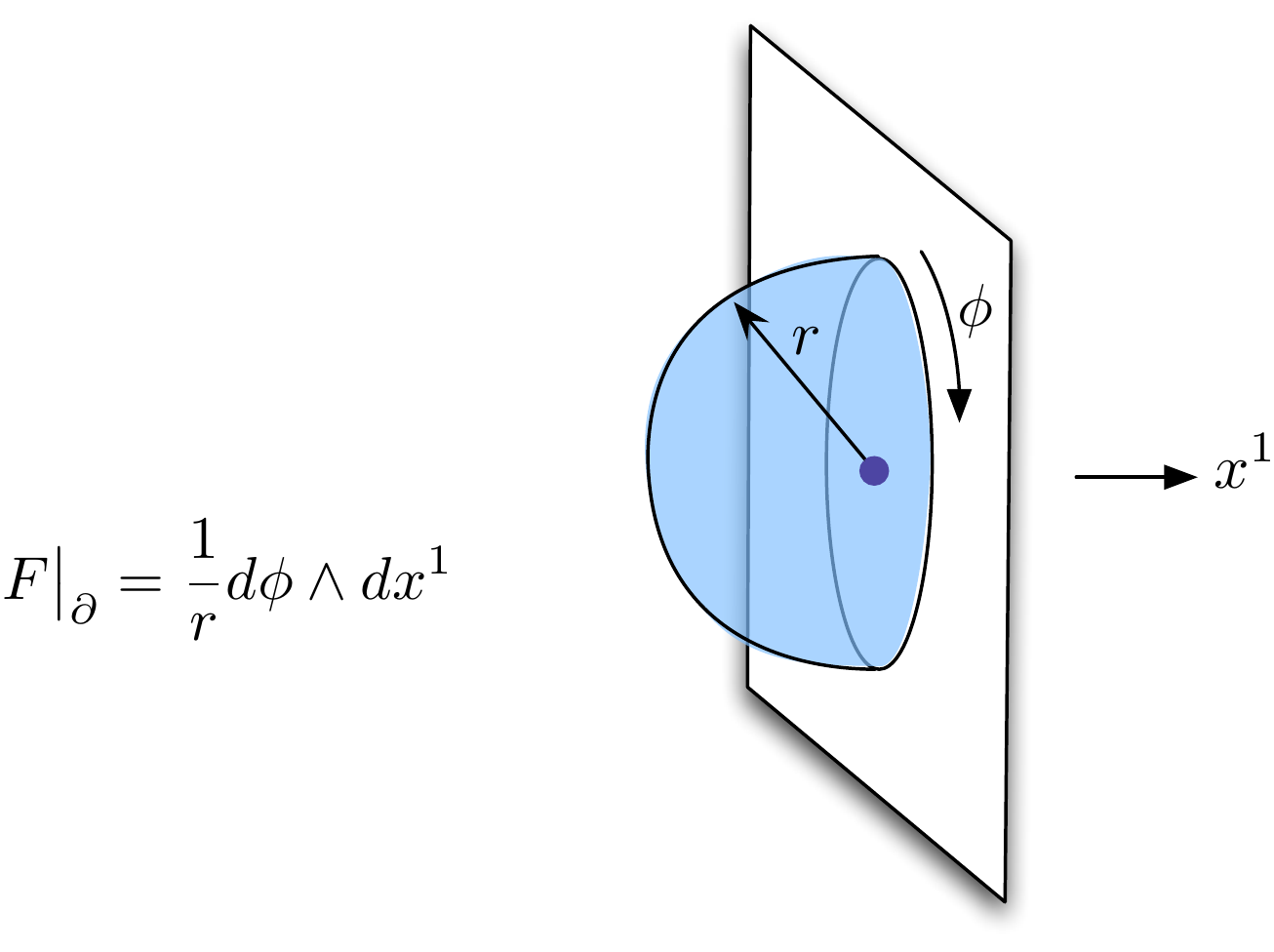}
\vspace{-.2in}
\caption{A boundary monopole-operator configuration}
\label{fig:bdymon}
\end{figure}

Dirichlet boundary conditions are compatible with slicing in half a 
standard BPS monopole singularity, and thus we can simply impose on the gauge field $A$ and scalar $\sigma$ the same singular functional form as for a standard bulk monopole. One can see, for example, that the basic abelian monopole configuration
\be  F \sim *\,d\Big(\frac 1r\Big)\,,\qquad \sigma\sim \frac1r \label{Fmonbdy} \ee
has $F|_\pd \sim r^{-1}d\varphi\wedge dx^1$ (see Figure \ref{fig:bdymon}), which is compatible with the boundary condition  $A_\parallel |_\pd=0$. Alternatively, we can use the semi-classical description $v\sim e^{(\sigma+i\gamma)/g^2}$ for a BPS abelian monopole operator, and recall that Dirichlet boundary conditions for the gauge field imply Neumann for $\sigma+i\gamma$, thus allowing $v$ to exist on the boundary.
In general, a boundary monopole operator has a conserved, quantized flux through the half-sphere that surrounds it, due to the fact that $A_\parallel |_\pd=0$.

Just as BPS monopole operators in the bulk (together with $\varphi$) generate the bulk Coulomb-branch chiral ring $\C[\CM_C]$, the BPS monopole operators bound to a Dirichlet boundary condition generate the boundary twisted-chiral ring.
In an abelian theory, the boundary monopole operators $\v_A$ are labelled by cocharacters $A$, \ie\ subgroups $U(1)_A\subset G$, the same way as bulk monopole operators. The cocharacter specifies how to embed the basic monopole configuration \eqref{Fmonbdy} into $G$. We postulate that if we bring a bulk monopole operator $v_A$ to the boundary we will obtain a boundary monopole $\v_A$, with a relative normalization
\be v_A\big|_\pd \,=\,  \v_A\prod_{i=1}^N (M_{L,i})^{(-Q_{A,L}^i)_+} \label{vw} \ee
for a right boundary condition, while for a left boundary condition we would replace $(-Q_{A,L}^i)_+$ with $(Q_{A,L}^i)_+$. Here $-Q_{A,L}^i$ is the charge of the chiral $Y_{L,i}$ under $U(1)_A$, and $M_{L,i}$ denotes its effective complex mass.
(More accurately, the complex mass is $-M_{L,i}$; we absorb such minus signs in the definition of $\v_A$.)

Note the resemblance of \eqref{vw} to \eqref{NC-abel}, which described the effect of bringing a monopole operator $v_A$ up to a Neumann boundary condition:
\be
 v_A\big|_\pd  = \xi_A\,  \prod_{i=1}^N (M_{L,i})^{(-Q_{A,L}^i)_+}  \tag{\ref{NC-abel}}
\ee
The only essential difference is that $\sigma$ and the dual photon are now dynamical at the boundary, so that $\xi_A \leadsto \v_A$. Indeed, we can justify \eqref{vw} by defining a Dirichlet boundary condition by starting with a Neumann boundary condition and making the complexified 2d FI parameters $t_{2d}$ dynamical --- \ie\ enriching the Neumann boundary condition with an extra set of $(\R\times S^1)$-valued twisted-chiral fields $T$. Following Appendix~\ref{app:2d}, we find that this leads to a twisted superpotential
\be  \wt W_{\rm bdy} = \varphi \,T\,, \ee
which has several effects: it sets $\varphi|_\pd = 0$, imposes Dirichlet boundary conditions on the gauge field (since the boundary field strength is in the boundary $\varphi$ multiplet), and for monopole operators turns \eqref{NC-abel} into \eqref{vw}. Another useful perspective is that 
$T$ is a 2d mirror description of a $(\R\times S^1)$-valued chiral field $\phi$ charged under the 3d gauge symmetry 
which acts as a compensator field, Higgsing the boundary gauge symmetry and converting the Neumann b.c. to a Dirichlet b.c. 

Formula \eqref{vw} also implies that the boundary operator product of the $\v_A$ is simply
\be \v_A \v_B = \v_{A+B}\,, \ee
and that the $U(1)_A$ R-charge of $\v_A$ is $\frac12\sum_i Q_{A,L}^i$. It would be interesting to verify both of these predictions directly. For now, however, we will simply assume \eqref{vw} and use it to find classical and quantum images of Dirichlet boundary conditions. We will see that \eqref{vw} resolves the puzzle with identification of boundary states from Sections \ref{sec:qDC}--\ref{sec:DC-SQED} (see \eqref{SQED-id}), and also produces images of exceptional Dirichlet boundary conditions that are self-mirror.

Upon introducing an $\Omega$-background and reducing the 3d theory to quantum mechanics, each boundary monopole operator $\v_A$ defines a state $|A\rangle$. The state $|0\rangle$ corresponding to the identity operator should be an eigenvector of $\hat\varphi$, with eigenvalue $\varphi_0$. Then $|A\rangle$ must be an eigenvector with eigenvalue $\varphi_0 + A\epsilon$. Generalizing \eqref{vw}, we propose that the action of bulk monopole operators takes the form
\be \hat v_A |B\rangle =  \prod_i [\hat M_{L,i}]^{(-Q_{A,L}^i)_+} |A+B\rangle\,, \label{vw-mod} \ee
where the only $B$ dependence arrises from the choice of eigenvalue for $\hat M_{L,i} = \hat M_{L,i}(\hat\varphi,m_\C)$. This action is consistent with the bulk chiral-ring relations \eqref{Cqring}, the same way as the quantized Neumann boundary conditions \eqref{NC-qabel} from Section \ref{sec:NC-q}.

For nonabelian theories, a full discussion of boundary monopole operators goes beyond the scope of this paper. There are important subtleties to be understood, such as 
how to treat quantum mechanically the continuous choice of possible ways to embed an abelian magnetic charge into the nonabelian gauge field, as the gauge symmetry is broken at the boundary. 

As a working hypothesis in the nonabelian case, we assume that the nonabelian monopole vevs will be expanded in terms of abelianized operators $\v_A$. 
Then we expect to be able to get the image on the boundary of a bulk monopole operator by substituting 
\begin{equation} \label{eq:DirCImage}
v_A =  \v_A\, \frac{\prod_{i=1}^N (M_{L,i})^{(-Q_{A,L}^i)_+}}{\prod_{\text{roots $\alpha$}} (\alpha \cdot \varphi^{\mathrm{ab}})^{(\alpha \cdot A)_+}}
\end{equation}
into the abelianized expression, to be re-grouped somehow into boundary nonabelian operators. 
The quantized version of \eqref{eq:DirCImage} would be
\begin{equation}
\hat v_A |B\rangle =  \frac{\prod_i [\hat M_{L,i}]^{(-Q_{A,L}^i)_+} }{\prod_{\text{roots $\alpha$}} [\alpha \cdot \hat \varphi^{\mathrm{ab}}- \frac{\epsilon}{2}]^{-(\alpha \cdot A)_+}}|A+B\rangle\,.
\end{equation}

\subsection{Coulomb-branch images, revisited}
\label{sec:DC2}

Having described the boundary twisted-chiral ring in the presence of Dirichlet boundary conditions more explicitly, we now reconsider the effect of turning on boundary vevs $c_i$ for some of the $Y_{L,i}$.

Having nonzero $Y_{L,i}$ requires the corresponding complex mass $M_{L,i}$ to vanish. From \eqref{vw} we immediately find that
\be \label{Mposv2}
\begin{cases} v_A\big|_\pd = 0 \qquad \text{$\forall\,A$ s.t. $Q^i_{A,L}>0$} & \text{left b.c.} \\
v_A\big|_\pd = 0 \qquad \text{$\forall\,A$ s.t. $Q^i_{A,L}<0$} & \text{right b.c.}\,.  \end{cases}
\ee
This is the same conclusion we reached from an analysis of $\CN = (2,2)$ BPS equations in \eqref{Mposv}. Conditions \eqref{Mposv2} identify the Coulomb-branch image of a Dirichlet boundary condition as a complex manifold.
We will argue momentarily that turning on $c_i$ actually eliminates the boundary monopole operators $\v_A$ with the wrong sign of $Q_{A,L}^i$, which is a bit stronger than just setting the boundary value of $v_A$ to zero.

In the presence of an $\Omega$-background and (say) a right boundary condition, we expect for each nonzero $c_i$ that the modified complex masses $\hat M_{L,i}(\hat\varphi,m_\C)-\frac12\epsilon$ annihilate the identity state $|0\rangle$. From \eqref{vw-mod}, this implies that all operators $\hat v_A$ with $Q^i_{A,L}<0$ annihilate the identity state as well. Therefore, the states $|A\rangle$ with $Q^i_{A,L}\geq 0$ (including the identity) form a submodule. The quantum Coulomb-branch image of a Dirichlet boundary condition $\CD_{L,c}$ is the intersection of all these submodules,
\be \label{qDC-image}
 \hat\CD_{L,c}^{(C)}\; = \; \text{span}\big\{\,   \text{$|A\rangle$ s.t. $Q^i_{A,L}\geq 0$ for all nonzero $c_i$}\,\big\}\,.\ee
Each $|A\rangle$ is an eigenvector for $\hat\varphi$ with eigenvalue fixed by the conditions $(\hat M_{L,i}-\frac12\epsilon)|0\rangle=0$, and the action of monopole operators on the module is that of \eqref{vw-mod}.
This general result applies to the generic Dirichlet boundary conditions of Section \ref{sec:D} as well as the exceptional ones considered here.

In the case of exceptional Dirichlet boundary conditions for abelian theories, the number of $Y_{L,i}$ that can gain nonzero vevs $c_i$ is exactly the rank $r$ of the gauge group. Of course, this also equals the rank of the cocharacter lattice $A \in \Z^{r}$. The module \eqref{qDC-image} is thus infinite-dimensional, with support on an orthant of the cocharacter lattice. For generic values of $m_\C$ it is an irreducible Verma module, generated from the identity $|0\rangle$ by repeatedly applying ``raising operators'' $v_A$. At special values of $m_\C$ equal to integer or half-integer multiples of $\epsilon$, additional structure can arise.

In the case of generic boundary conditions that break the flavor symmetry $G_H$, all masses $m_\C$ must be set to fixed multiples of $\epsilon$ (as in \eqref{me-SQED}) in order for the constraints $(\hat M_{L,i}-\frac12\epsilon)|0\rangle =0$ to be consistent. The resulting module \eqref{qDC-image} can be finite-dimensional (in fact, trivial), as we already saw in Section \ref{sec:DC-SQED}.

The introduction of vortex-line operators complicates matters, but only slightly. If we force $m^a_\C=k^a\epsilon$ then (in the case of generic Dirichlet b.c.) it may be impossible for all $(\hat M_{L,i}-\frac12\epsilon)$ to annihilate the identity $|0\rangle$. At the same time, in the presence of vortex lines, the ``identity'' is no longer uniquely defined.
 Practically, we proceed by choosing a state $|0\rangle$ that is annihilated by all the $\hat\varphi$. Then if we define integers $\wt k_i$ as the values of complex masses at $\hat\varphi=0$, namely $\hat M_{L,i}(\hat\varphi = 0,m_\C=k\epsilon)-\frac12\epsilon = \wt k_i\epsilon$, we find that $(\hat M_{L,i}-\frac12\epsilon)|A\rangle = 0$ for $A$ such that $Q_{L,A}^i=-\wt k_i$. Moreover, due to \eqref{vw-mod}, such a state $|A\rangle$ is annihilated by $v_B$ with $Q_{L,B}^i<0$; thus the states $|A+B\rangle$ with $Q_{L,B}^i\geq 0$ (or simply $|A\rangle$ with $Q_{L,A}^i\geq -\wt k_i$) form a submodule.
Taking an intersection of submodules as before, we obtain 
 \be \label{qDC-image-k}
 \hat\CD_{L,c}^{(C)}\; = \; \text{span}\big\{\,   \text{$|A\rangle$ s.t. $Q^i_{A,L}\geq -\wt k_i$ for all nonzero $c_i$}\,\big\}\,.\ee
This leads much more directly to the finite-dimensional module discussed around \eqref{SQED-id} in Section \ref{sec:DC-SQED}.

Now, let us come back to the assertion below \eqref{Mposv2} that turning on boundary vevs of the $Y_L$ eliminates some boundary monopole operators. 
Naively, the vev $c$ is part of the F-term data and thus should not affect the twisted F-term data that determines the Coulomb-branch image of a boundary condition,
except for constraining to zero the twisted masses of global symmetries broken by the $c$ vevs. 

This naive intuition is incorrect even in purely two-dimensional systems. Consider for example a 2d free chiral multiplet $\phi$ valued in $S^1 \times \R$. 
By T-duality, it is equivalent to a twisted chiral field $\tilde \phi$ valued in  the dual circle $\tilde S^1 \times \R$. The 2d theory includes both an infinite series of 
chiral operators $\exp (n \phi)$ and of twisted chiral operators $\exp (w \tilde \phi)$, which are represented as twist fields for the original field $\phi$. 
If we add a superpotential $W = e^\phi$, the mirror theory becomes a cigar sigma-model and half of the twisted chiral fields disappear: the operators with negative winding number 
$w$ are singular on the cigar target space \cite{HoriVafa,HoriKapustin}. 

This phenomenon can be explained directly in the $\phi$ theory, without reference to the cigar, by observing that 
a twist field can be BPS only if the theory admits classical BPS solutions of the equations of motion in the neighborhood of the twist field.
The BPS equations for chiral operators take the form 
\be
\partial_{\bar z} \phi = \frac{\partial \ol W}{\partial \ol \phi}
\ee
and do depend on the choice of superpotential. In the neighborhood of a twist field $e^{w\tilde\phi}$ we look for singular solutions with winding number $w$. In the absence of superpotential, $\phi = w \log z$ is a good solution. If we turn on the superpotential, 
for $w \geq 0$ it is still possible to correct the $\phi = w \log z$ solution by subleading terms, but for $w<0$ the superpotential term dominates and we lose the solution. 

In our current setup, the boundary twisted chiral ring consists of boundary monopole operators $\v_A$.
If the vev of some charged chiral $Y_{L}$ is $c$ at the equator of a small hemisphere around the monopole, 
a BPS configuration for the chiral will be divergent or zero at the center of the hemisphere, depending on the sign of the charge $Q_{L,A}$.  If the chiral diverges, the boundary monopole is not actually BPS. 

\subsection{Example: SQED}
\label{sec:xD-SQED}

Consider $G=U(1)$ gauge theory with $N$ hypermultiplets $(X_i,Y_i)$ of charges $(+1,-1)$. The Higgs-branch flavor symmetry is $G_H= U(N)/U(1)$, and we will choose to preserve a maximal torus $\mathbb T_H = \big[U(1)^N\big]/U(1)$ acting diagonally, so that the $i$-th $U(1)$ factor in $\mathbb T_H$ rotates (only) the $i$-th hypermultiplet, with charges $(+1,-1)$.

In total there are $N \times 2^N$ exceptional Dirichlet boundary conditions $\CD_{\varepsilon,j}$, labelled by a Lagrangian splitting (encoded in the sign vector $\varepsilon$ as usual) together with the choice of a single chiral to assign a nonzero boundary vev:
\be \CD_{\varepsilon,j}\,:\qquad \begin{cases} Y_i\big|_\pd = c\,\delta_{ij} & \varepsilon_i = + \\
X_i\big|_\pd = c\,\delta_{ij} & \varepsilon_i = - \end{cases}\,,\qquad \varphi\big|_\pd = -m_\C^j\,.
\ee
All these boundary conditions break the gauge symmetry and preserve a $U(N-1)$ flavor symmetry, which includes the maximal torus $\mathbb T_H$. The effective complex mass of the $i$-th hypermultiplet is $M_i = \varphi +m_{i,\C}$, so in order to turn on $c$ we must have $\varphi|_\pd = -m_{j,\C}$.

As a complex manifold, the Coulomb branch is $v_+v_- = \prod_{i=1}^N (\varphi+m_{i,\C})$. Following \eqref{vw}, \eqref{Mposv2}, the image of the exceptional Dirichlet b.c. (as a right boundary condition) is cut out by
\be  \CD_{\varepsilon,j}^{(C)}\,:\quad v_{-\varepsilon_j} = 0\,,\qquad \varphi = -m_{j,\C}\,.\label{xDC-SQED}\ee 
Thus, for generic values of the complex masses, the image is a copy of $\C$ parametrized by the surviving monopole operator $v_{\varepsilon_j}$. The image only depends on the choice of $j$ and the sign $\varepsilon_j$.

Turning on an $\Omega$-background, we obtain modules from \eqref{vw-mod}, \eqref{qDC-image}. For generic values of the complex masses, the modules are freely generated from the identity vector $|0\rangle$, which satisfies
\be \hat \CD_{\varepsilon,j}^{(C)}\,:\quad \hat v_{-\varepsilon_j} |0\rangle  = 0\,,\qquad \big(\varepsilon_j(\hat\varphi+m_{j,\C})-\tfrac12\epsilon\big)|0\rangle = 0\,. \label{qXDC-SQED} \ee
The states in the module are $\hat v_{\varepsilon_j}^n|0\rangle$ for $n\geq 0$. As the masses are specialized $m_{i,\C}\to k_i\epsilon$, however, the modules acquire more interesting structure that depends on the entire sign vector $\varepsilon$.

We consider the case $N=2$ in greater detail, specializing $(m_{1,\C},m_{2,\C})\to (\frac12 k\epsilon,-\frac12 k \epsilon)$ with $k\in \Z$. Two representative boundary conditions are $\CD_{++,1}$ and $\CD_{+-,1}$ (all the others are related to these by symmetries). In both cases, \eqref{qDC-image} dictates that the corresponding modules have a basis $|A\rangle$ for $A\geq 0$. Also, in both cases $(\hat\varphi+\frac12k\epsilon-\frac12\epsilon)|0\rangle=0$, so $\hat\varphi|0\rangle=\frac12(1-k)\epsilon|0\rangle$ and in general $\hat\varphi|A\rangle = (A+\frac12(1-k))\epsilon|A\rangle$. However, from \eqref{vw-mod}, the module actions are
\be\begin{array}{l@{\quad}l@{\qquad}l}
\hat\CD_{++,1}^{(C)}\,: &   \hat v_+|A\rangle = |A+1\rangle & \hat v_-|A\rangle = A(A-k)\epsilon^2|A-1\rangle \\[.2cm]
\hat\CD_{+-,1}^{(C)}\,: &   \hat v_+|A\rangle = (A-k-1)\epsilon|A+1\rangle & \hat v_-|A\rangle = A\epsilon|A-1\rangle\,.
\end{array}
\ee
Thus, for $\hat\CD_{++,1}^{(C)}$, $\hat v_-$ kills not only the identity $|0\rangle$ but the state $|k\rangle$; this means that if $k\geq 1$ the module $\hat\CD_{++,1}^{(C)}$ has an infinite-dimensional submodule with basis $|A\rangle$ for $A\geq k$. In contrast, for $\hat\CD_{+-,1}^{(C)}$, $\hat v_-$ kills only the identity but $\hat v_+$ kills the state $|k-1\rangle$; this means that if $k\geq 1$ there is a finite-dimensional submodule with basis $|A\rangle$ for $0\leq A< k$.

\begin{figure}[htb]
\centering
\hspace{-.25in}
\includegraphics[width=6.5in]{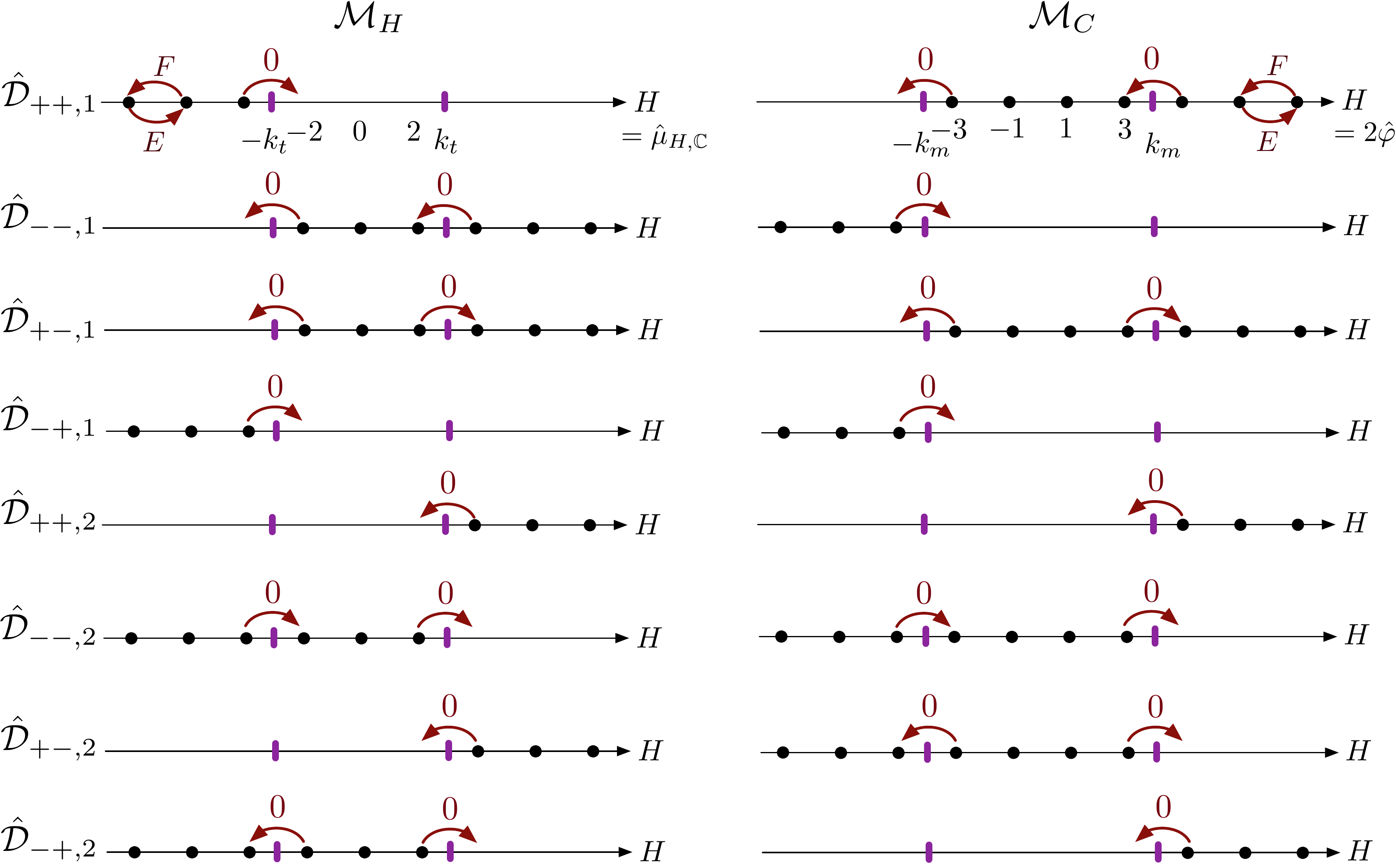}
\caption{Modules for $\hat \C[\CM_H]$ and $\hat \C[\CM_C]$ corresponding to various exceptional Dirichlet boundary conditions in SQED with $N=2$ hypermultiplets. To see the structure of standard (\ie\ Verma) and costandard modules, we set $t_\C = k_t\epsilon$ and $(m_{1,\C},m_{2,\C})=\frac12\epsilon(k_m,-k_m)$ for positive integers $k_t,k_m\geq 1$. For each module, we depict the occupied weight spaces of the Cartan generator $H$ (equal to $\hat X_1\hat Y_1-\hat X_2\hat Y_2$ in $\hat \C[\CM_H]$ and $\hat\varphi$ in $\hat \C[\CM_C]$). The $\hat \C[\CM_H]$ modules with $k_t<0$ correspond to the same pictures modulo the substitution $\hat\CD_{\varepsilon_1,\varepsilon_2;1}\leftrightarrow \hat\CD_{-\varepsilon_2,-\varepsilon_1;2}$, while the $\hat \C[\CM_C]$ modules with $k_m<0$ are described by replacing $\hat\CD_{\varepsilon_1,\varepsilon_2;1}\leftrightarrow \hat\CD_{-\varepsilon_1,-\varepsilon_2,2}$.
}
\label{fig:xD-modules}
\end{figure}

If we identify the quantized chiral ring $\hat \C[\CM_C]$ for $N=2$ with the enveloping algebra $U(\mathfrak sl_2)$ at fixed Casimir $C_2 = \frac12(k^2-1)\epsilon^2$ as in \eqref{C-SQED-Cas}, we may identify $\hat\CD_{++,1}^{(C)}$ as a lowest-weight Verma module (reducible if $k\geq 1$); whereas $\hat\CD_{+-,1}^{(C)}$ is a so-called \emph{costandard} module that coincides with a Verma module only if $k\leq 0$. In a similar way, all the $\CD_{++,j}$ or $\CD_{--,j}$ boundary conditions produce reducible or irreducible Verma modules, while the $\CD_{+-,j}$ and $\CD_{-+,j}$ produce costandard modules. The various possibilities are summarized in Figure \ref{fig:xD-modules}.

Now consider the Higgs branch. The analysis of exceptional Dirichlet boundary conditions is essentially identical to that of generic boundary conditions in Section \ref{sec:DH}.
The chiral ring $\C[\CM_H]$ is generated by the mesons $X_iY_{i'}$, subject to $\sum_i X_i Y_i +t_\C=0$, and we must determine how a given boundary condition fixes the vevs of these operators. For $\CD_{\varepsilon,j}$, we find $X_iY_i=0$ unless $i=j$, so the complex moment-map constraint fixes either $(X_j,Y_j)=(c,-t_\C/c)$ or $(-t_\C/c,c)$, depending on the sign of $\varepsilon_{j}$. The remaining operators $X_iY_{i'}$ vanish if either $i\neq j$ and $\varepsilon_i=-$, or if $i'\neq j$ and $\varepsilon_{i'}=+$. The nonvanishing operators can all be expressed as products of the $N-1$ combinations
\be X_iY_j \quad (\varepsilon_i=+)\qquad\text{or}\qquad X_jY_i \quad(\varepsilon_i=-)\,,\qquad i\neq j\,. \label{XijSQED} \ee
Thus, for generic $t_\C$ the image $\CD_{\varepsilon,j}^{(H)}$ is a copy of $\C^{N-1}$ parameterized by the operators in \eqref{XijSQED}. For example, with $\varepsilon=(++...+)$ and $j=1$ we have $(X_1,Y_1)=(-t_\C/c,c)$ and all other mesons vanishing except the $X_iY_1=X_ic$ for $i>1$.
Note that the Higgs-branch image $\CD_{\varepsilon,j}^{(H)}$ depends on the choice of $j$ and the entire sign vector $\varepsilon$ \emph{except} the component $\varepsilon_j$ (which determined the Coulomb-branch image!).

For the special case $N=2$, we define as usual $E=X_1Y_2,\, F=Y_1X_2,\,H=X_1Y_1-X_2Y_2$ subject to the complex moment-map constraint $\mu_\C = X_1Y_1+X_2Y_2=-t_\C$, whence  $EF = -\frac14(H^2-t_\C^2)$. Then our representative boundary conditions $\CD_{++,1}$ and $\CD_{+-,1}$ have images
\be \label{xDH-SQED}
\begin{array}{l@{\quad}l@{\qquad\Rightarrow\qquad}l}
\CD_{++,1}^{(H)}\,: & Y_1\big|_\pd = c\,,\quad Y_2\big|_\pd = 0 & E=0\,,\quad H = -t_\C\,,\\[.1cm]
\CD_{+-,1}^{(H)}\,: & Y_1\big|_\pd = c\,,\quad X_2\big|_\pd = 0 & F=0\,,\quad H = t_\C\,,
\end{array}\ee
mirroring the Coulomb-branch images \eqref{xDC-SQED}.

Turning on the $\wt\Omega$-background and continuing to work with $N=2$, we identify the quantum chiral ring $\hat\C[\CM_H]$ as a quotient of the algebra generated by the mesons $E=\hat X_1\hat Y_2$, $F=\hat Y_1\hat X_2$, $H=\hat X_1\hat Y_1-\hat X_2\hat Y_2$ by the moment-map constraint $\hat \mu_\C:=\hat X_1\hat Y_1+\hat X_2\hat Y_2+\epsilon = -t_\C$. Recall that the result is the enveloping algebra $U(\mathfrak{sl}_2)$ at fixed Casimir $\frac12(t_\C^2-\epsilon^2)$. In particular, the ring relations are
\be EF = -\tfrac14(H+t_\C-\epsilon)(H-t_\C-\epsilon)\,,\qquad FE=-\tfrac14(H+t_\C+\epsilon)(H-t_C-\epsilon)\,.\ee

To find the module $\hat \CD_{++,1}^{(H)}$, we start with a module for the Heisenberg algebra with basis $X_1^{n_1}X_2^{n_2}\big|$, on which the $Y$'s act as $\hat Y_1=\pd_1+c$ and $\hat Y_2=\pd_2$. Then we quotient by all polynomials of the form $(\hat \mu_\C+t_\C)p(X_1,X_2)$, which allows us to write any vector uniquely as a polynomial in $X_2$ alone and use a basis $|n\rangle:=X_2^n\big|$. For example, in the quotient we have
$(\hat \mu_\C+t_\C)X_1^{n_1}X_2^{n_2}\big| = \big(cX_1+(n_1+n_2+1)\epsilon+t_\C\big)X_1^{n_1}X_2^{n_2}\big| =0$, so
\be X_1^{n_1}X_2^{n_2}\big| = \frac{((n_2+n_1)\epsilon+t_\C)((n_2+n_1-1)\epsilon+t_\C)\cdots (n_2\epsilon+t_\C)}{(-c)^{n_1}}|n_2\rangle
\,.\ee
Acting on the basis $|n\rangle$, we find
\begin{subequations} \label{qxDH-SQED}
\be E|n\rangle = -\tfrac{1}{c}n\epsilon(n\epsilon+t_\C)|n-1\rangle\,,\quad F|n\rangle = c|n+1\rangle\,,\quad
H|n\rangle = -((2n+1)\epsilon+t_\C)|n\rangle\,.  \ee
Therefore, $\hat \CD_{++,1}^{(H)}$ is a highest-weight Verma module, freely generated from the identity $|0\rangle$ (which obeys $E|0\rangle=0$) by acting with $F^n$. If we specialize $t_\C = k\epsilon$ with $k\leq -1$ the Verma module becomes reducible, since the states $|n\rangle$ with $n\geq -k$ form a Verma sub-module.

Repeating the same analysis for the boundary condition $\CD_{+-,1}$, we start with a basis $X_1^{n_1}Y_2^{n_2}\big|$ and quotient by vectors of the form $(\hat \mu_\C+t_\C)X_1^{n_1}Y_2^{n_2}\big| = \big(cX_1+(n_1-n_2)\epsilon+t_\C\big)X_1^{n_1}Y_2^{n_2}\big|$ to obtain a module $\hat \CD_{+-,1}^{(H)}$ with basis $|n\rangle := Y_2^n\big|$. The action is
\be E|n\rangle = \tfrac1c((n+1)\epsilon-t_\C)|n+1\rangle\,,\quad F|n\rangle = -cn\epsilon|n-1\rangle\,,\quad H|n\rangle = (2n\epsilon-t_\C)|n\rangle\,.\ee
\end{subequations}
For generic $t_\C$, this is now an irreducible lowest-weight Verma module, freely generated from the identity (which obeys $F|0\rangle = 0$) by acting with $E^n$. However, if we specialize $t_\C=k\epsilon$ with $k\geq 0$, we obtain a costandard module that contains a finite-dimensional submodule with basis $|n\rangle$ for $0\leq n < k$, just like the module $\hat \CD_{+-}^{(C)}$ on the Coulomb branch.

Altogether, the exceptional boundary conditions $\CD_{++,j}$ and $\CD_{--,j}$ lead to Verma modules on the Higgs branch, while $\CD_{+-,j}$ and $\CD_{-+,j}$ lead to Verma modules that for special values of $t_\C$ may become costandard, with finite-dimensional submodules. We summarize the different possibilities in Figure \ref{fig:xD-modules}. 

\subsection{Exceptional Dirichlet b.c. and thimbles}
\label{sec:thimbles}

One important reason to consider exceptional Dirichlet boundary conditions is that a special class of them flow to ``thimble'' boundary conditions on the Higgs and Coulomb branches. A thimble boundary condition is labelled by a vacuum $\nu$. For certain BPS computations involving BPS objects placed at $x^1<0$,
a thimble boundary condition at $x^1=0$ is equivalent to the bulk theory on the whole half line $x^1>0$, with the corresponding choice of 
vacuum at $x^1 \to \infty$.

In order to study thimble boundary conditions, we turn on real mass and FI deformations $m_\R,t_\R$. We assume for the moment that in the presence of generic deformations the theory has isolated massive vacua ${\nu}$.
Recall from Sections \ref{sec:NH-mt} and \ref{sec:NC-mt} that the 2d $\CN=(2,2)$ BPS equations reduce to (inverse) gradient-flow equations in sigma models, with respect to a real potential
\be h_m = m_\R\cdot \mu_{H,\R}\quad \text{(Higgs branch)}\,,\qquad h_t \approx t_\R\cdot \sigma_{\rm ab} \quad\text{(Coulomb branch)}\,.\ee
These potentials are the real moment maps for the infinitesimal $U(1)_m$ or $U(1)_t$ symmetries associated with a mass or FI deformation.
Thus thimble boundary conditions should be supported on the gradient-flow cycles
\be  \CM_H^<[m_\R^\nu]\,,\; \CM_C^<[t_\R^\nu] \quad \text{(left b.c.)}\,,\qquad
 \CM_H^>[m_\R^\nu]\,,\; \CM_C^>[t_\R^\nu] \quad \text{(right b.c.)}\,. \label{M<>thimble} \ee

In order to identify exceptional Dirichlet boundary conditions with similar properties, we should look for choices of $L$ and $c$ such that the Higgs-branch and Coulomb-branch images mimic \eqref{M<>thimble}. For the Higgs branch, recall that (say) $\CM_H^>[m_\R^\nu]$ is defined by first lifting the flavor symmetry $U(1)_m\subset G_H$ to an (infinitesimal) $U(1)_{m,\nu} \subset G\times G_H$ with generator $m_\R^\nu$, such that the matter fields that get a vev in the vacuum $\nu$ are invariant under $U(1)_{m,\nu}$. (In other words, we find the gauge transformation that compensates for $U(1)_m$ in order to keep the vacuum invariant.) Then, as in Section \ref{sec:NH-mt}, $\CM_H^>[m_\R^\nu]$ is the image on the Higgs branch of the locus where all chirals $X^-_{m^\nu_\R}$ and $Y^-_{m^\nu_\R}$ of negative charge under $U(1)_{m,\nu}$ vanish. We expect to include these fields in the set $Y_L$ that is set to zero by an exceptional Dirichlet boundary condition.

For each remaining hypermultiplet $(X^0_{m^\nu_\R},Y^0_{m^\nu_\R})$ that is neutral under $U(1)_{m,\nu}$, either $X^0_{m^\nu_\R}$ or $Y^0_{m^\nu_\R}$ gets a vev in the vacuum $\nu$. The choice is determined by the signs of the FI parameters $t_\R$, via the real moment-map constraints. We include the neutral chirals that acquire vevs in the set $Y_L$, and set them equal to nonzero constants $c$ at the boundary. For example, we can choose the $c$'s so that
\be \mu_\R(X_L=0,Y_L=c) +t_\R = 0\,. \label{muc} \ee
Thus, given $m_\R,t_\R$ and a vacuum $\nu$, we define a Dirichlet boundary condition $\CD_{L,c}$ with
\be \label{nuLc}
Y_L = \begin{cases} \text{chirals of negative $U(1)_{m,\nu}$ charge} & \text{(set to zero at the boundary)} \\
\text{neutral chirals that get a vev in $\nu$} & \text{(set to $c$ at the boundary)\,.}
\end{cases}
\ee
In addition, the scalar $\varphi\big|_\pd = \varphi_0$ is set to its value in the vacuum $\nu$.

If the gauge group $G$ is abelian, then this is an exceptional Dirichlet boundary condition: the nonzero vevs of $Y_L$ break $G$ (and hence $G_\pd$) completely, while preserving a maximal torus of $G_H$. For non-abelian theories, however, the prescription must be slightly modified. The reason is that a Dirichlet boundary condition only fixes the scalar fields at some distance from the boundary up to \emph{complexified} $G_\C$ gauge transformations, and while \eqref{nuLc} breaks $G$ completely it may preserve some unipotent subgroup $P\subset G_\C$. 
This leads to additional noncompact 2d degrees of freedom on the boundary.
To eliminate these degrees of freedom, we modify \eqref{nuLc} by 1) additionally setting to zero at the boundary all chirals of \emph{positive} $U(1)_{m,\nu}$ charge that are in the $P$-orbit of the vacuum $\nu$, and dually 2) relaxing the boundary condition on (\ie\ not fixing) the chirals of negative $U(1)_{m,\nu}$ that are canonically conjugate to those in (1). We will see an example of this in Section \ref{sec:xD-SQCD}.

Geometrically, the modification has the following description. If a Dirichlet boundary condition restricts chiral fields $(X,Y)$ to lie on some Lagrangian $\CB\subset T^*\C^N$ at the boundary, then it will restrict bulk Higgs-branch vacua to lie on the image of $\CB$ under a complex symplectic quotient
\be \CM_H \simeq T^*\C^N/\!/G_\C\,,\qquad \CB^{(H)}_{\rm bulk} \simeq [\CB\cap (\mu_\C+t_\C=0)]/G_\C\,.\ee
Recall, however, that the full space of vacua of the bulk-boundary system is actually $\CB^{(H)}\simeq \CB\cap (\mu_\C+t_\C=0)$; thus any nontrivial orbits of $G_\C$ on $\CB\cap (\mu_\C+t_\C=0)$ show up as additional 2d degrees of freedom fibered over $\CB^{(H)}_{\rm bulk}$. For example, if $\CB$ preserves a unipotent $P\subset G_\C$, then a $P$-worth of 2d degrees of freedom will sit above every bulk vacuum. The above modification amounts to first replacing $\CB$ by a complex submanifold $\CB'\simeq \CB/P$ that is transverse to all $P$-orbits, and then using the subgroup $P^T$ conjugate to $P$ in $G_\C$ to smear $\CB'$ into a new Lagrangian $\CB'' \simeq P^T\cdot \CB'$. Then the bulk images $\CB^{(H)}_{\rm bulk} = {\CB''}^{(H)}_{\rm bulk}$ coincide, but now boundary degrees of freedom are eliminated.

The Coulomb-branch image of \eqref{nuLc} also has a good chance to match the thimble $\CM_C^{>}[t_\R^\nu]$. Recall that $\CM_C^{>}[t_\R^\nu]$ is characterized in abelian theories as the locus where all monopole operators with negative charge under $U(1)_t$ vanish, \ie\ $v_A=0$ for all $A$ such that $t_\R\cdot A < 0$.
Suppose, therefore, that $t_\R\cdot A < 0$. The real moment-map constraint for the subgroup $U(1)_A\subset G$ takes the form $\sum_i\big(Q_{A,L}^i |X_{L,i}|^2 - Q_{A,L}^i |Y_{L,i}|^2\big)+ t_\R\cdot A = 0$, and restricting this to $(X_L,Y_L)=(0,c)$ we find
\be \textstyle \sum_i Q_{A,L}^i\cdot |c_i|^2 =  t_\R\cdot A\,. \ee
If $t_\R\cdot A < 0$, then $Q_{A,L}^i<0$ for some $i$ with nonvanishing $c_i$. It follows from \eqref{Mposv2} that on the Coulomb-branch image of $\CD_{L,c}$ (for a right boundary condition) we will indeed have $v_A = 0$.

There is an alternative, more physical route to constructing exceptional Dirichlet boundary conditions associated to thimbles.
We may attempt to define a thimble-like boundary condition by varying real masses and FI parameters as functions of the spatial coordinate $x^1$.%
\footnote{
Such configurations are analogous to ``Janus'' configurations of 4d Yang-Mills theory, \cf\ \cite{Bak-Janus, Clark-Janus, GW-Janus, CCV, DGV-hybrid}, or their 2d and 3d cousins as in (\eg) \cite{GW-surface, GGP-walls}.}
Arbitrary variations will preserve 2d $(2,2)$ supersymmetry. We start from a configuration where $m_\R,t_\R$ are close to zero for negative $x^1$ but go to large constant values for positive $x^1$. 
For positive $x^1$, the hypermultiplet scalars will sit close to their vacuum values,
while the gauge group will be Higgsed.
It is thus natural to replace the 
region of positive $x^1$ with a boundary condition that sets the scalar fields to their vacuum values and 
breaks the gauge symmetry at the boundary. The condition that scalar fields should not blow up 
at large positive $x^1$ also forces us to set to zero the appropriate charged scalars, up to complexified gauge transformations.
The result is precisely the same exceptional Dirichlet b.c. we just defined in \eqref{nuLc}.
Strictly speaking, if there are multiple vacua, the construction here produces a direct sum of exceptional Dirichlet boundary conditions, one for each vacuum. We can combine the construction with a projection to a single vacuum in order to recover a single boundary condition.\label{mtJanus}

This definition of boundary conditions in terms of varying parameters
is invariant under mirror symmetry. 
This strongly suggests that exceptional Dirichlet boundary conditions are mirror to other exceptional Dirichlet boundary conditions, as we already saw in the example of Section \ref{sec:xD-SQED}.

Finally, we note that the notion of a thimble can be generalized to situations where vacua are not isolated or massive, but rather correspond to a collection of low-energy sub-theories. In this case, there is a notion of a thimble interface between the full theory and any one of the sub-theories. We would expect that the thimble interface is realized in the UV by an exceptional Dirichlet interface, which lets part of the matter and gauge fields propagate across the interface.

\subsubsection{Thimbles for SQED}

In SQED with $N$ hypermultiplets, at generic values of $(m_\R,t_\R)$ there are $N$ massive vacua $\nu_j$: in each vacuum exactly one of the hypermultiplets $(X_j,Y_j)$ gets a vev. The thimbles $\CM_C^>[t_\R^{\nu_j}]$ depend on the sign of the FI parameter, while the thimbles $\CM_H^>[m_\R^{\nu_j}]$ depend on the charges of the $N-1$ hypermultiplets $(X_i,Y_i)_{i\neq j}$ under $m_\R^{\nu_j}$. Altogether, there are $N\times 2\times 2^{N-1}$ choices that determine a pair of thimbles on the Higgs and Coulomb branches, which via \eqref{nuLc} can be matched with the $N\times 2^N$ exceptional Dirichlet boundary conditions discussed in Section \ref{sec:xD-SQED}.

In the case of $N=2$ hypermultiplets, we expect from \eqref{nuLc} that the right thimbles correspond to the exceptional Dirichlet boundary conditions
\be \label{SQED-thimbles}
\begin{array}{l@{\quad}l@{\,:\;\quad}l} m_\R > 0 & t_\R > 0  &  \CD_{+-,1}, \CD_{++,2} \\
m_\R > 0 & t_\R < 0  &  \CD_{--,1},   \CD_{+-,2} \\
m_\R < 0 & t_\R > 0  &  \CD_{++,1},   \CD_{-+,2} \\
m_\R < 0 & t_\R < 0  &  \CD_{-+,1},   \CD_{--,2}\,.
\end{array}
\ee
On the other hand, left thimbles correspond to the exceptional Dirichlet boundary conditions  
\be \label{SQED-lthimbles}
\begin{array}{l@{\quad}l@{\,:\;\quad}l} m_\R > 0 & t_\R > 0  &  \CD_{++,1},   \CD_{-+,2} \\
m_\R > 0 & t_\R < 0  &  \CD_{-+,1},   \CD_{--,2} \\
m_\R < 0 & t_\R > 0  &  \CD_{+-,1},   \CD_{++,2} \\
m_\R < 0 & t_\R < 0  &  \CD_{--,1},   \CD_{+-,2}\,.
\end{array}
\ee

\subsection{Thimbles and (co)standard modules}
\label{sec:standards}

In the presence of $\Omega$ or $\wt\Omega$ backgrounds, we find that exceptional Dirichlet boundary conditions that correspond to thimbles generically define Verma modules for $\hat \C[\CM_H]$ and $\hat \C[\CM_C]$. Specifically, right boundary conditions produce lowest-weight Verma modules, and left boundary conditions produce highest-weight Verma modules. If the parameters $t_\C$ and $m_\C$ that enter the quantization of  $\hat \C[\CM_H]$ and $\hat \C[\CM_C]$ (respectively) are specialized to integral or half-integral values $t_\C = k_t\epsilon$, $m_\C=k_m\epsilon$, the situation is more subtle. The modules in this case are not always Verma modules, and their behavior depends critically on the values of $k_t,k_m$.

In order to characterize the situation, we introduce a few mathematical notions. For concreteness, we'll work on the Higgs branch.
Recall that a choice of $m_\R \in \mathfrak t_H$ splits the algebra $\hat \C[\CM_H]=\hat \C[\CM_H]_>\oplus \C[\CM_H]_0\oplus \C[\CM_H]_<$ into operators of positive, zero, and negative charge under the corresponding (infinitesimal) flavor symmetry $U(1)_m\subset G_H$, as in \eqref{eq:mdecomp}. A lowest-weight Verma module $V_{\nu}$, also called a \emph{standard} module in the context of Category $\CO$, is freely generated from a single vacuum vector $e_{\nu}$ that is an eigenvector for $\C[\CM_H]_0$ and is annihilated by all of $\hat \C[\CM_H]_<$. (Different vacua are distinguished by their eigenvalues for $\hat \C[\CM_H]_0$.)
Dually, a \emph{costandard} module $\Lambda_{\nu}$ is freely co-generated from $e_\nu$, meaning that every $e_\nu$ is the only state in $\Lambda_{\nu}$ annihilated by all of $\hat \C[\CM_H]_<$, and that $e_{\nu}$ can be reached from every other state by repeatedly applying $\hat \C[\CM_H]_<$ operators. (The formal definition of standard and costandard modules appears in Section \ref{sec:cast}.)

As vector spaces, standard and costandard modules are completely isomorphic. More so, as weight modules, they have the same weight spaces, with the same multiplicities. The difference is that in a standard module the $\hat \C[\CM_H]_<$ operators may occasionally act as zero on states (``null vectors'') other than $e_\nu$, while in a costandard module the $\hat \C[\CM_H]_>$ operators may occasionally act as zero. The modules may be related to one another by combining linear duality $\Lambda_\nu \simeq V_\nu^*$ with an involution of $\hat\C[\CM_H]$ that swaps $\C[\CM_H]_>$ and $\C[\CM_H]_<$ and reverses the sign of $\epsilon$. (See Section \ref{sec:star} for details.)

For example, consider the Higgs-branch modules $\hat \CD_{--,1}^{(H)}$ and $\hat\CD_{+-,1}^{(H)}$ for SQED with $N=2$ hypers in Figure \ref{fig:xD-modules}. These are thimble boundary conditions for the first vacuum corresponding to $m_\R>0,\,t_\R<0$ and $m_\R>0,\,t_\R>0$, respectively, as in \eqref{SQED-thimbles}. The module $\hat \CD_{--,1}^{(H)}$ is standard while $\hat\CD_{+-,1}^{(H)}$ is costandard. In this case, noting that $k_t>0$ in Figure \ref{fig:xD-modules}, we may observe that the costandard module arises when $k_t$ is aligned with $t_\R$, while the standard module arises when $k_t$ is anti-aligned with $t_\R$.%
This behavior turns out to be quite general.

For abelian theories, we will prove in Section \ref{sec:abel-Hmod} that right thimble boundary conditions (depending on $m_\R,t_\R$ and a choice of vacuum) always quantize to costandard modules on the Higgs branch when $k_t\sim t_\R$ and to standard modules when $k_t\sim -t_\R$. Similarly, on the Coulomb branch, thimble boundary conditions become costandard (standard) modules when $k_m \sim m_\R$ ($k_m \sim -m_\R)$. For left thimble b.c., the role of standard and costandard modules is swapped.
We expect to find similar behavior in massive non-abelian theories, though a systematic treatment remains to be performed.%
\footnote{The precise definition of ``aligned'' parameters $k_t\sim \pm t_\R$ will be given in Section \ref{sec:corresp}.}

The physical significance of aligning $k_m,k_t$ with $m_\R,t_\R$ is not obvious from the point of view of exceptional Dirichlet boundary conditions. It becomes clearer when considering pure Neumann or generic Dirichlet boundary conditions. Namely, it is only for $k_t\sim t_\R$ that the same Neumann b.c. have Higgs-branch images that preserve SUSY with and without a twisted $\wt\Omega$-background;  and only for $k_m\sim m_\R$ that the same generic Dirichlet b.c. have Coulomb-branch images that preserve SUSY with and without an $\Omega$-background. We come back to this in Section \ref{sec:corresp}.

If the moduli space of $m_\R, t_\R$ parameters has dimension greater than one, there may be many possible values of $k_m,k_t$ that are neither aligned nor anti-aligned with $m_\R,t_\R$. This leads to modules whose weight spaces coincide with those of $V_\nu$ and $\Lambda_\nu$, but which are only partially standard and partially costandard.

\subsection{Example: SQCD}
\label{sec:xD-SQCD}

Consider SQCD with gauge group $G=U(K)$ and $N\geq K$ hypermultiplets $(X,Y)$ transforming in the fundamental representation of $G$ and the anti-fundamental representation of $G_H=U(N)/U(1)$. The simplest example of an exceptional Dirichlet boundary condition is defined by choosing $Y_L = Y$ and setting $c$ equal to the identity matrix in the first $K$ flavors,
\be \label{xD-SQCD}
Y^T\big|_\pd =
\raisebox{.3cm}{$
\begin{array}{l} \qquad\;\; K \;\;\qquad N-K \\
\left(\begin{array}{cccc@{\;}|@{\;}ccc}
c & 0 &\cdots & 0 & 0 & \cdots & 0 \\
0 & c & & 0 & 0 &\cdots & 0 \\
 &&\ddots & & 0 & \cdots & 0 \\
0 & 0 & & c & 0 & \cdots & 0\end{array}\right) K
\end{array}$}
\ee
This boundary condition fully Higgses the gauge group and preserves a $[U(K)\times U(N-K)]/U(1)$ global symmetry (including the maximal torus of $G_H$). The effective real and complex masses of $Y^a_i$ are $-(\sigma_a-m_{i,\R})$ and $-(\varphi_a-m_{i,\C})$ (where $\sigma$ and $\varphi$ have been diagonalized). In order for the complex masses of the nonzero $Y$ to vanish, we need $\varphi_a\big|_\pd = m_{a,\C}$.

Although it is not obvious, this is a good candidate for a thimble boundary condition. To see it, suppose that the real FI parameter $t_\R$ is positive and that the real masses decrease $m_{1,\R}> m_{2,\R} >\ldots > m_{N,\R}$, and let us try to find a boundary condition whose image is $\CM_H^>[m_\R^\nu]$.
The ``lift'' $m_\R^\nu$ is obtained by requiring effective real masses of nonzero $Y$'s to vanish, \ie\ setting $\sigma_a = m_{a,\R}$. Then the fields $(X^i_a,Y^a_i)$ have effective real mass $(m_{a,\R}-m_{i,\R}, -m_{a,\R}+m_{i,\R})$, respectively. Naively, \eqref{nuLc} tells us to set to zero the fields of negative mass, meaning $X_a^i\big|_\pd = 0$ if $a>i$ and $Y^a_i\big|_\pd = 0$ if $a<i$. Moreover, given the sign of the FI, we set $Y^a_a\big|_\pd=c$. For example, if $K=3$ and $N=5$,
\be \label{wrong}
X\big|_\pd = \left(\begin{array}{ccc@{\;}|@{\;}cc}
 *  & * & * & * & * \\
 0 & * & * & * & * \\
 0 & 0 & * & * & * \end{array}\right)\,,\qquad
Y^T\big|_\pd = \left(\begin{array}{ccc@{\;}|@{\;}cc}
 c & 0 & 0 & 0 & 0 \\
 * & c & 0 & 0 & 0 \\
 * & * & c & 0 & 0 \end{array}\right)\,.
\ee

However, as discussed below \eqref{nuLc}, this boundary condition needs to be modified a little. Notice that \eqref{wrong} is preserved by unipotent complexified gauge transformations of the form
\be g= \begin{pmatrix} 1 & * & * \\ 0 & 1 & * \\ 0 & 0 & 1 \end{pmatrix}\,, \ee
acting as $X\to gX$ and $Y \to Yg^{-1}$. Thus, \eqref{wrong} leads to extra massless 2d degrees of freedom on the boundary, and has a redundancy that we need to remove. We can fully break the complexified gauge symmetry without changing the bulk Higgs-branch image precisely by modifying \eqref{wrong} to the form \eqref{xD-SQCD}.

We see that \eqref{xD-SQCD} together with the complex moment-map constraint $XY+t_\C=0$ forces us to set $X^{(K)}= -(t_\C/c) 1\!\!1_{K\times K}$, where $X^{(K)}$ is the leading $K\times K$ block of the matrix $X$. In the matrix of mesons $M_i^j = (YX)_i{}^j$, the blocks $M^{(K)}_{(N-K)}$ and $M^{(N-K)}_{(N-K)}$ are set to zero, while $M^{(K)}_{(K)} = - t_\C$ and the only nontrivial block $M_{(K)}^{(N-K)} = Y_{(K)} X^{(N-K)}$ can directly be identified with the scalars $X^{(N-K)}$. We obtain a Higgs-branch image isomorphic to a holomorphic Lagrangian $\C^{K(N-K)}$.

The quantum module consists of polynomials in the $X^{(N-K)}$, which we can simply denote as a $K \times (N-K)$ ``$x$'' variables. 
The module action takes the form
\begin{align}
\hat X^{(K)} p(x) &= - \left(t_\C + \tfrac12 N\epsilon\right)1\!\!1_{K \times K}  - \epsilon x \cdot \partial_x p(x)\,, \cr
\hat Y^{(K)} p(x) &= 1\!\!1_{K \times K} \,p(x)\,, \cr
\hat X^{(N-K)} p(x) &= x p(x)\,, \cr
\hat Y^{(N-K)} p(x) &= \epsilon \partial_x p(x)\,.
\end{align}
Then the action of the mesons operators can be computed by first putting them in the schematic order $XY$ 
and then acting on $p(x)$ as described above. In particular, the identity vector $| = 1$ is annihilated by all the meson involving $\hat Y_{(N-K)}$ but also by the whole traceless part of the mesons built from 
$\hat X^{(K)}$ and $\hat Y_{(K)}$. 

The Coulomb-branch image should consist of the locus $U^-(z)=0$. Quantum mechanically we may define a module generated by an identity vector $|$ with the same property. 
For this particular case, in analogy to the $\CD_{+..+,1}$ boundary condition for SQED, we expect a standard Verma module built from such a vector. 

In an abelianized setup, we expect the abelianized module to consist of vectors $|n_1, \cdots, n_K \rangle$ with all $n_a \geq 0$.
Each $\hat u^+_a$ generator should simply raise $n_a$, up to a prefactor 
\begin{equation}
\frac{1}{\prod_{b \neq a}(\hat \varphi_a - \hat \varphi_b)} = \frac{1}{\prod_{b \neq a}(m_a - m_b+(n_a- n_b +1)\epsilon)}\,;
\end{equation}
while the $\hat u^+_a$ generator lowers $n_a$ with a prefactor 
\begin{equation}
\frac{\prod_i (\hat \varphi_a - m_i + \frac{\epsilon}{2})}{\prod_{b \neq a}(\hat \varphi_a - \hat \varphi_b)} = \frac{\prod_i (m_a -m_i+ n_a \epsilon )}{\prod_{b \neq a}(m_a - m_b+(n_a- n_b -1)\epsilon)}\,.
\end{equation}

The action of operators such as the coefficients of $\hat U^+(z)$ on the identity vector $|0,\cdots,0\rangle$ produces complicated expressions with 
coefficients that are rational in $\hat \varphi_a$. It should be possible to given these an interpretation in terms of boundary monopole operators. 
Classically, boundary monopole operators have continuous moduli, corresponding to the embedding of the Dirac singularity into 
the nonabelian gauge fields at the boundary. The vectors $|n_1, \cdots, n_K \rangle$ may correspond to $U(K)$-equivariant fixed points in these moduli spaces.

If we look at thimbles for more general real masses, we may encounter much more complicated examples, where $Y_L=c$ 
will include the whole $Y_{(K)}=c\,1\!\!1_{K\times K}$, as before, but will set to zero the first $s_i$ $Y_{(N-K)}$ and the last $K-s_i$ $X^{(N-j)}$ fields independently for each flavor. 
In analogy with SQED, we expect the Coulomb-branch modules to be equivalent to the standard Verma modules 
away from special values of the masses, but not at the special values where extra complex masses go to zero. 
In the abelianized setting, the $\hat u^+_a$ generators should raise the $n_a$, up to a prefactor 
\begin{equation}
\frac{\prod_i (\hat \varphi_a - m_i - \frac{\epsilon}{2})^{(-\varepsilon_{i,a})_+}}{\prod_{b \neq a}(\hat \varphi_a - \hat \varphi_b)}\,, 
\end{equation}
while the $\hat u^-_a$ generators will include a prefactor 
\begin{equation}
\frac{\prod_i (\hat \varphi_a - m_i + \frac{\epsilon}{2})^{(\varepsilon_{i,a})_+}}{\prod_{b \neq a}(\hat \varphi_a - \hat \varphi_b)}\,.
\end{equation}

\section{Enriched boundary conditions}
\label{sec:en}

It is possible to enrich both Neumann and Dirichlet boundary conditions by adding extra boundary degrees of freedom. We want to preserve 2d $\CN=(2,2)$ supersymmetry, so these boundary degrees of freedom should appear in $\CN = (2,2)$ multiplets. We also generally want to preserve both $U(1)_V$ and $U(1)_A$ R-symmetries. An example of a 2d theory that accomplishes this is a Calabi-Yau sigma model or GLSM with homogeneous superpotential.

The boundary degrees of freedom may be further coupled to bulk hypermultiplets (resp., vectormultiplets) by a boundary superpotential (twisted superpotential and gauging). Such boundary couplings appeared in~\cite{KRS} in the context of 3d $\CN=4$ sigma-models, where they were called ``curvings.'' We will discuss their effect on boundary conditions for gauge theories in Sections \ref{sec:bdyW}--\ref{sec:W-modules}, and then proceed in Section \ref{sec:NC-justified} to use them in order to justify formula \eqref{NC-abel} for the Coulomb-branch image of Neumann boundary conditions.
In Section \ref{sec:Toda}, we study a more interesting application of enriched Neumann boundary conditions in nonabelian gauge theories that is related to the Toda integrable system and the mathematical work of \cite{Teleman-MS}.

\subsection{Effect of boundary superpotentials}
\label{sec:bdyW}

Let us focus first on hypermultiplets. Starting with a Lagrangian splitting and a boundary condition of the form
\be Y_L\big|_\pd = 0\,\qquad \pd_1X_L\big|_\pd = 0\,, \label{YL0}\ee
we can introduce boundary chiral multiplet(s) $\phi$ and superpotential $W_{\rm bdy}(X_L|_\pd,\phi)$. We recall from Appendix \ref{app:hyper} that when writing the bulk 3d theory in 2d $\CN=(2,2)$ language, there is always a \emph{bulk} superpotential of the form $W_{\rm bulk} = \int dx^1 X_L\pd_1 Y_L$. In the presence of a boundary superpotential, the $\CN=(2,2)$ F-terms receive a delta-function contribution that vanishes if the boundary condition \eqref{YL0} is deformed to
\be Y_L\big|_\pd = \begin{cases} {\pd W_{\rm bdy}}/{\pd X_L\big|_\pd} & \text{right b.c.} \\[.1cm]
 -{\pd W_{\rm bdy}}/{\pd X_L\big|_\pd}& \text{left b.c.} \end{cases}\,,\qquad  \frac{\pd W_{\rm bdy}}{\pd \phi} = 0\,.\ee
The boundary condition for $X_L$ is also deformed, in a manner compatible with $\pd W_{\rm bdy}/\pd \phi=0$ and the fact that the F-term for $Y_L$ is $\pd_1\ol{X}_L$.

It is easy to see that a boundary superpotential can be used to deform the initial hypermultiplet boundary condition $Y_L|_\pd = 0$ to
\be (X_L,Y_L)\big|_\pd \subset L_W\,,\qquad (\pd_1X_L,\pd_1Y_L) \subset N^*L_W \label{LW} \ee
for an arbitrary holomorphic Lagrangian $L_W\subset \R^4$. To achieve \eqref{LW}, we simply choose $W_{\rm bdy}(X_L|_\pd,\phi)$ so that after imposing $\pd W_{\rm bdy}/\pd \phi=0$, the function $W_{\rm bdy}$ is a generating function for the Lagrangian $L_W$, \ie\ $L_W$ is the graph of $Y_L = \pd W_{\rm bdy}/\pd X_L$.
After integrating out $\phi$, the generating function may be multivalued.

There are several simple examples of boundary superpotentials that have (in different guises) already shown up in this paper:
\begin{itemize}
\item In  \eqref{WXYPhi}, in the context of Neumann boundary conditions, we used a boundary superpotential $W_{\rm bdy} = X_L|_\pd \phi$ to ``flip'' the $Y_L|_\pd = 0$ boundary condition to a $X_L|_\pd = 0$ boundary condition. Notice that the equation $\pd W_{\rm bdy}/\pd\phi=0$ sets $X_L|_\pd = 0$, while imposing $Y_L|_\pd = \pd W_{\rm bdy}/\pd X_L|_\pd = \phi$ allows $Y_L$ to fluctuate at the boundary. (Such flips work just as well in the presence of Dirichlet boundary conditions on the gauge fields.)

\item A generic Dirichlet boundary condition that sets $\varphi\big|_\pd = \varphi_0$ and $Y_L\big|=c$ can be engineered by starting with a ``pure'' Dirichlet boundary condition $\varphi\big|_\pd = \varphi_0,Y_L\big|=0$ (that preserves both $G_\pd$ and $G_H$ in the language of Section \ref{sec:D}), and adding a linear boundary superpotential
\be W_{\rm bdy} = c\cdot X_L\,.\ee
This generically breaks $G_\pd\times G_H$ symmetry, and deforms $Y_L\big|_\pd=0$ to $Y_L\big|_\pd=c$.

\item An exceptional Dirichlet boundary condition that splits the $Y_L$ into two sets $Y_{L,c}$ and $Y_{L,0}$, with $(Y_{L,c},Y_{L,0})\big|_\pd = (c\neq 0, 0)$ can be obtained from a generic Dirichlet boundary condition with $(Y_{L,c},X_{L,0})\big| = (c,c')$ simply by promoting the $c'$ fields to dynamical boundary chirals $\phi'$. Then the boundary superpotential terms $c'Y_{L,0}\to \phi' Y_{L,0}$ have the effect of flipping the $X_{L,0}$ b.c. to $Y_{L,0}\big|_\pd = 0$ as desired.

\item A more interesting example involves a free chiral $(X,Y)$ with boundary condition $Y|_\pd = 0$ deformed by
\be W_{\rm bdy}(X|_\pd,\phi) = X|_\pd e^{-\phi} + \wt m\phi\,, \label{mirrorD} \ee
where $\phi$ is a chiral multiplet valued in $\R\times S^1$.
We find $Y|_\pd = \pd W/\pd X|_\pd = e^{-\phi}$ and $\pd W/\pd \phi = - X|_\pd e^{-\phi} + \wt m = 0$, whence
\be (XY)\big|_\pd = \wt m\,.\ee
This free-hypermultiplet theory is mirror to a $U(1)$ gauge theory with a hypermultiplet, with the operator $XY$ mapping to the vectormultiplet scalar $\varphi$ in the gauge theory. The deformed boundary condition \eqref{mirrorD} turns out to be mirror a Dirichlet b.c. in the gauge theory that sets $\varphi|_\pd = \wt m$. (The parameter $\wt m$ is a mass parameter in the gauge theory.) We revisit this example in Section~\ref{sec:mirrorwall}.

\end{itemize}
In the presence of Neumann b.c. for the gauge fields, the boundary superpotential must preserve the bulk gauge symmetry. (In the presence of a Dirichlet b.c., this is of course not necessary.) 

The introduction of boundary twisted-chiral multiplets and boundary twisted superpotentials has a similar effect on the bulk vectormultiplet fields. It is simplest to analyze this first in a pure abelian gauge theory. As shown in Appendix \ref{app:2d}, the complex vectormultiplet scalar $\varphi$ is part of a (2,2) twisted-chiral multiplet that includes the 2d gauge field strength. Similarly, the fields $\sigma+iA_1$ are part of (2,2) chiral multiplet that can be T-dualized to a twisted chiral with scalar component $\sigma+i\gamma$. When writing the bulk theory in (2,2) language, there is a bulk twisted superpotential 
\be \wt W_{\rm bulk} \sim \int dx^1\, \varphi\,\pd_1(\sigma+i\gamma)\,, \ee
very much analogous to $W_{\rm bulk}\sim\int dx^1\, X\pd_1 Y$. If we start with a Neumann boundary condition
\be (\sigma+i\gamma)\big|_\pd  = \pm t_{2d}\,,\qquad  \pd_1\varphi\big|_\pd = 0\,, \ee
and deform it with boundary twisted chirals $\eta$ and a twisted superpotential $\wt W_{\rm bdy}(\varphi,\eta)$, then we find
\be (\sigma+i\gamma)\big|_\pd = \begin{cases} t_{2d}-{\pd \wt W_{\rm bdy}}/{\pd \varphi\big|_\pd} & \text{right b.c.} \\[.1cm]
 -t_{2d}+{\pd \wt W_{\rm bdy}}/{\pd \varphi\big|_\pd}& \text{left b.c.} \end{cases}\,,\qquad  \frac{\pd \wt W_{\rm bdy}}{\pd \eta} = 0\,.\ee
 
Some simple examples of twisted superpotential deformations should already be familiar:
\begin{itemize}
\item The 2d FI term itself can be thought of as arising from a twisted boundary superpotential, of the standard 2d form $\wt W_{\rm bdy} = -t_{2d}\varphi$. This deforms $(\sigma+i\gamma)\big|_\pd=0$ to $(\sigma+i\gamma)\big|_\pd=\pm t_{2d}$ as above.

\item To change a Neumann boundary condition to a Dirichlet boundary condition in an abelian theory, the 2d FI term should be promoted to a dynamical field $\R\times S^1$-valued field $t_{2d}\to \eta$. Then $\wt W_{\rm bdy} = \eta\varphi$ imposes $\varphi|_\pd = 0$ and $(\sigma+i\gamma)|_\pd = \eta$, or $v_\pm|_\pd  \sim e^{\pm \eta}$. The fields $e^{\pm \eta}$ are the boundary monopole operators discussed in Section \ref{sec:bdymon}. 

\end{itemize}

\subsection{Deformed modules}
\label{sec:W-modules}

In the presence of a twisted (say) $\wt\Omega$-deformation, boundary conditions produce modules for the quantized Higgs-branch chiral ring $\hat \C[\CM_H]$. To understand the effect of a boundary superpotential on these modules, we start by looking at a purely 2d Landau-Ginzburg model in the  $\wt\Omega$-background.

In a 2d LG model, the $\wt \Omega$-deformation gives a partition function of the form
\begin{equation}
\tilde Z_\gamma = \int_\gamma e^{\tfrac1\epsilon W} \Omega
\end{equation}
where $\Omega$ is the holomorphic top form on the target space and $\gamma$ is a middle-dimensional Lagrangian manifold that encodes the boundary conditions at infinity. 
Expectation values of chiral operators are computed as
\begin{equation}
\langle \hat O(\phi) \rangle = \int_\gamma e^{\frac{W}{\epsilon}} O(\phi)\, \Omega
\end{equation}
In particular, the notion of trivial chiral operator is deformed: rather than setting to zero multiples of $\partial W(\phi)$,
one has to throw away operators for which the right hand side is a total derivative. (The expectation values of such operators vanish.)
For example, if the target space is simply $\C$, then polynomials of the form
\begin{equation}
\partial_{\phi} W(\phi) P(\phi) + \epsilon \partial_{\phi} P(\phi) 
\end{equation}
are set to zero. The space of chiral operators (no longer a ring) is $\C[\phi]/\text{im}(\epsilon\pd_\phi+\pd_\phi W).$

We can define our Higgs-branch module in the same fashion. We will do it explicitly for a theory of free hypermultiplets.
For general gauge theories, one simply needs to project this onto a gauge-invariant subspace (for Neumann b.c.) or to impose complex moment-map constraints (for Dirichlet b.c.).
Suppose we start with a module that consists of polynomials in $X_L$, with the usual action
\be \hat X_L \cdot P(X_L) = X_LP(X_L)\,,\qquad \hat Y_L \cdot P(X_L) = \epsilon \pd_{X_L}P(X_L)\,.\ee
Adding a boundary superpotential $W_{\rm bdy}(X_L,\phi)$ has three effects:
\begin{enumerate}
\item The space of boundary chiral operators is initially enlarged to polynomials $P(X_L,\varphi)$\,;
\item The action of the bulk algebra is conjugated by $\exp(W_{\rm bdy}/\epsilon)$, so that $\hat X_L$ and $\hat Y_L$ act on $P(X_L,\varphi)$ as
\be \hat X_L = X_L\cdot\,,\qquad \hat Y_L = \epsilon\pd_{X_L}  + \pd_{X_L}W_{\rm bdy}\cdot\,; \ee
\item Boundary operators of the form $(\epsilon \pd_\phi + \pd_\phi W)P$ are set to zero. (Such operators generate a submodule, which we must quotient by. Explicitly, we can start with a subspace $N$ transverse to the polynomials divisible by $\pd_\phi W_{\rm bdy}$, and work recursively to bring the image of $\hat Y_{L,i}$ and $\hat X_{L,i}$ back to this subspace, much as in Section \ref{sec:qDH}.)
\end{enumerate}

A simple example of a deformed module already appeared in Section \ref{sec:qDH}. Recall that a generic Dirichlet boundary condition with $Y_L|_\pd = c$ can be constructed by starting with $Y_L|_\pd = 0$ and introducing a boundary superpotential $W_{\rm bdy}= c\cdot X_L$. The resulting module is built out of polynomials in $X_L$, with a conjugated action
\be \hat X_L = X_L\cdot\,,\qquad \hat Y_L = \epsilon\pd_{X_L} + c\cdot\,. \ee

Another simple example is the flip of (say) a boundary condition $Y|_\pd = 0$, implemented by the superpotential $W_{\rm bdy} = X\phi$. The deformed module consists of polynomials $P(X,\phi)$, modulo polynomials of the form $(\epsilon\pd_\phi+X)P(X,\phi)$. We can thus choose the transverse subspace `$N$' to be generated by polynomials $P(\phi)$. The deformed module action is
\be \hat X\, P(\phi) = X P(\phi) \simeq -\epsilon\pd_\phi P(\phi)\,,\qquad \hat Y\,P(\phi) = (\epsilon\pd_X+\phi)P(\phi) = \phi P(\phi)\,.\ee
Thus differentiation and multiplication are reversed in the action of  $(\hat X,\hat Y)$, as we would expect from the flip.

This discussion applies equally well to the ordinary $\Omega$-background and modules for the Coulomb-branch algebra $\hat\C[\CM_C]$. Such modules are deformed exactly the same way by boundary twisted superpotentials.

\subsection{Application 1: Coulomb-branch image of Neumann}
\label{sec:NC-justified}

We can use boundary superpotentials to finally motivate our prescription for the Coulomb-branch image of a Neumann boundary condition \eqref{NC-abel}. Recall that in Section \ref{sec:NC} we postulated that a Neumann b.c. for vectormultiplets supplemented by $Y_L\big|=0$ for hypermultiplets leads to
\begin{equation} \tag{\ref{NC-abel}}
\CN_L^{(C)}\,:\qquad \begin{cases} \ds v_A  = \;  \xi_{-A} \hspace{-.3cm}\prod_{\text{$i$ s.t. $Q_{A,L}^i>0$}} \hspace{-.3cm} M_{L,i}^{|Q_{A,L}^i|} =  \xi_{-A} \prod_{i=1}^N M_{L,i}^{(Q_{A,L}^i)_+}  & \text{left b.c.} \\
\ds v_A  = \;  \xi_{A} \hspace{-.3cm}\prod_{\text{$i$ s.t. $Q_{A,L}^i<0$}} \hspace{-.3cm} M_{L,i}^{|Q_{A,L}^i|}  =  \xi_{A} \prod_{i=1}^N M_{L,i}^{(-Q_{A,L}^i)_+} & \text{right b.c.}
\end{cases}
\end{equation}
We checked there that this image preserves the correct R-symmetries. This formula played a fundamental role later, in Section \ref{sec:DC2}, where we used it to derive a relation between bulk and boundary monopole operators in the presence of a Dirichlet boundary condition.

The idea is simple: \ref{NC-abel} is essentially the only choice which is both compatible with the symmetries of the system and covariant under ``flips'' of the hypermultiplet boundary conditions. Remember that we can flip an $Y=0$ b.c for an hypermultiplet to a $X=0$ b.c. by adding a 2d chiral multiplet which acts as a Lagrange multiplier, with linear superpotential coupling to $X$. That chiral field must have the same gauge charge 
as $Y$ and at a general point in the Coulomb branch will be massive. Integrating the 2d chiral field away, we get a boundary twisted superpotential which shuffles the 
factors on the right hand side of \ref{NC-abel} exactly as expected from the change in $L$. 

We can see this process in detail in an abelian $G=U(1)$ gauge theory with a single hypermultiplet $(X,Y)$ of gauge charge $(Q,-Q)$, with $Q\in \Z$. Suppose that a Neumann boundary condition with $Y|_\pd=0$ sets
\be \hspace{-.3in} Y\big|_\pd=0\,:\qquad v_+\big|_\pd = O_+\,,\qquad v_-\big|_\pd = O_-\,,  \label{YOO} \ee
where $O_\pm$ are some boundary twisted-chiral operators (possible constant). Moreover, if we require that the bulk chiral-ring relation  $v_+v_-=\pm M_X^{|Q|}$ is obeyed, with $M_X = Q\varphi$ being the effective complex mass of $X$, then $O_+O_- = \pm M_X^{|Q|}$. (The sign in the bulk chiral-ring relation is slightly ambiguous, and can be absorbed in the definition of (say) $v_-$.)

We can flip this boundary condition to one with $X|_\pd = 0$ by introducing a boundary chiral $\phi$ and using the usual boundary superpotential $W_{\rm bdy} = X|_\pd \phi$. The chiral $\phi$ must have gauge charge $-Q$ in order for $W_{\rm bdy}$ to preserve gauge symmetry at the boundary. The presence of $\phi$, moreover, induces a 1-loop correction to the boundary \emph{twisted} superpotential \cite{Witten-phases, HIV}
\be \wt W_{\rm bdy} = -t_{2d}\varphi \;\to\; -t_{2d}\varphi + M_\phi (\log M_\phi -1)\,,\ee
where $M_\phi = M_Y = -Q\varphi$ is the effective complex mass of $\phi$. This effective twisted superpotential deforms \eqref{YOO} to
\be 
\hspace{-.3in} X\big|_\pd=0\,:\qquad v_+\big|_\pd = O_+e^{\pd \wt W_{\rm bdy}/\pd \varphi} = O_+(M_Y)^{Q}\,,\qquad v_-\big|_\pd = O_-(M_Y)^{-Q}\,.  
\label{XOO}
\ee

We should require that the RHS of the $v_\pm|_\pd$ boundary conditions in both \eqref{YOO} and \eqref{XOO} are well-defined boundary operators (so no negative powers of $\varphi$ appear). Then it follows that if $Q>0$ we must have $O_+ =\xi$ and $O_- = \xi^{-1}(M_Y)^{|Q|}$ for some constant $\xi$; otherwise, if $Q<0$ we must have $O_+ =\xi (M_Y)^{|Q|}$ and $O_- = \xi^{-1}$. Therefore,
\be 
Y\big|_\pd = 0\,:\quad v_\pm\big|_\pd = \xi^{\pm 1} (M_Y)^{(\mp Q)_+}\,,\qquad
X\big|_\pd = 0\,:\quad v_\pm\big|_\pd = \xi^{\pm 1} (M_Y)^{(\pm Q)_+}\,. 
\label{XYOO} 
\ee
It is natural to identify the constant $\xi$ with the 2d FI parameter,
\be \xi = \exp(t_{2d})\,.\ee
We may write \eqref{XYOO} even more succinctly if we denote as $(X_L,Y_L) = (X,Y)$ or $(Y,-X)$ a Lagrangian splitting of the hypermultiplet, such that $X_L$ has gauge charge $Q_{X,L}$ and effective complex mass $M_{X,L}= -M_{Y,L}$. Then the image of the boundary condition with $Y_L|_\pd=0$ is
\be Y_L\big|_\pd=0\,:\qquad v_A\big|_\pd = (\xi)^A (M_{X,L})^{(-A\cdot Q_{X,L})_+}= \pm (\xi)^A (M_{Y,L})^{(-A\cdot Q_{X,L})_+}\qquad (A\in\Z) \,. \label{XLOO} \ee
(Up to a possible sign that can be absorbed in the monopole operators, it does not matter whether $M_{X,L}$ or $M_{Y,L}$ is used.)

Formula \eqref{XLOO} is a special case of \eqref{NC-abel} (as a right boundary condition) for $G=U(1)$ and a single hypermultiplet. The same argument, though, can be used to derive \eqref{NC-abel} for a general abelian theory with any number of hypermultiplets. For a left boundary condition, the roles of $X_L$ and $Y_L$ fields are simply reversed. (The corrections to $v_\pm\big|_\pd$ induced by a flip as in \eqref{XOO} come with opposite signs.)
For nonabelian gauge theories, we combine the abelianization map with the formula \eqref{NC-abel} to obtain \eqref{NC-nonabel}.

\subsection{Application 2: `Toda boundary condition' for SQCD}
\label{sec:Toda}

\begin{figure}[htp]
\centering
\includegraphics[width=7cm]{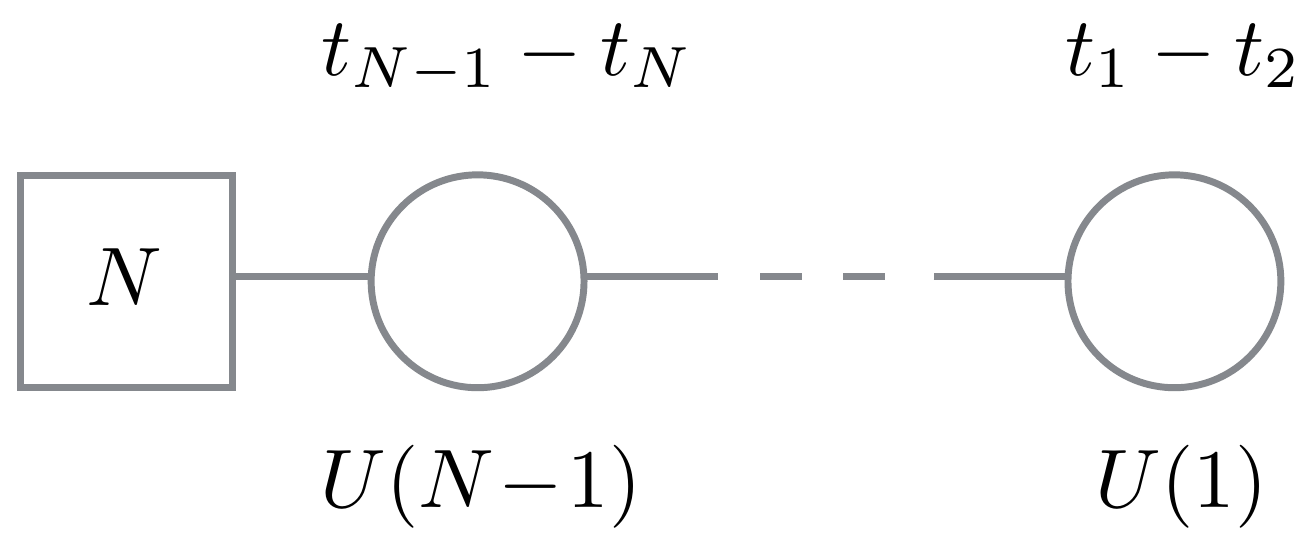}
\caption{The 2d $\CN=(2,2)$ quiver whose Higgs branch is the complete flag variety $U(N) / U(1)^N$. FI parameters (modulo conventional shifts by $i\pi$) are shown above each gauge group.}
\label{fig:2dquiver}
\end{figure}

Nonabelian gauge theories coupled to boundary degrees of freedom can display extremely rich structure. Here we consider one particular example, related to recent work of Teleman \cite{Teleman-MS}: we deform a Neumann boundary condition for pure $U(N)$ super-Yang-Mills by coupling to a 2d $\CN=(2,2)$ triangular quiver gauge theory, as shown in Figure~\ref{fig:2dquiver}. The 2d quiver describes a GLSM whose Higgs branch is the complete flag variety $U(N)/U(1)^N$. We will compute the Coulomb-branch image of this boundary condition.

We denote the 2d complex vectormultiplet scalar at the $j$-th node by $\varphi^{(j)}$ (the bottom component of a twisted chiral multiplet $\Sigma^{(j)}$). To simplify some expressions, we have the convention that $\varphi^{(N)} = \varphi|$ is the boundary value of the 3d complex scalar. We also introduce 2d FI parameters $t_j - t_{j+1} + i\pi$ at the $j$-th two-dimensional node, together with a boundary FI parameter $t_N+i\pi(N-1)$ for the $U(N)$ gauge group. At generic points, the theory is massive with effective twisted superpotential
\begin{multline}
\widetilde\CW = \sum_{j=1}^{N-1} \left[ \sum_{a \neq b}^j \ell(\varphi^{(j)}_a - \varphi^{(j)}_{b}) + \sum_{a=1}^{j-1} \sum_{a'=1}^j \ell(\varphi^{(j-1)}_a - \varphi^{(j)}_{a'}) + (t_j-t_{j+1}+i\pi)  \sum_{a=1}^j  \varphi^{(j)}_a \right] \\ 
+ (t_N+i\pi(N+1)) \sum_{a=1}^N \varphi_a\, .
\end{multline}
where $\ell(s) = s(\log s-1)$ is the one-loop contribution from a massive 2d chiral multiplet with twisted mass $s$.

Let us first concentrate on the 2d quiver gauge theory in isolation. The supersymmetric massive vacua of the 2d quiver are given by
\be
\exp\left( \partial\widetilde\CW / \partial\varphi^{(j)}_a \right) = 1  \qquad j =1,\ldots,N-1 \,,
\ee
which are equivalent to the polynomial equations
\be
Q_{j+1}(z) - e^{t_{j+1}-t_j} Q_{j-1}(z) = Q_j (z) (z - p_{j+1})\,,
\label{QQ}
\ee
where 
\be
Q_j(z) := \prod_{a=1}^j(z-\varphi^{(j)}_a)
\ee
and 
\be
p_j := \frac{\partial \widetilde\CW}{\partial t_j} = \sum_{a=1}^j \varphi^{(j)}_a - \sum_{a'=1}^{j-1} \varphi^{(j-1)}_{a'}\, .
\ee
are the `momenta' conjugate to the 2d FI parameters. We use the convention that $Q_N(z) = Q(z)$ and $Q_0(z)=1$ when appropriate to write the equations uniformly. The twisted chiral ring of the 2d quiver is generated by the coefficients of the polynomials $Q_j(z)$, \ie\ gauge invariant combinations of the complex scalars, subject to the relations~\eqref{QQ}. This is the equivariant quantum cohomology of the complete flag variety~\cite{Kim:aa,Astashkevich:1995aa,Givental:1995aa}. 

We now consider a deformation of Neumann boundary conditions for pure $U(N)$ SYM by adding the above 2d GLSM and using the boundary vectormultiplets to gauge the $U(N)$ symmetry at the final node. The Coulomb-branch image is determined by the 2d twisted chiral ring equations~\eqref{QQ} together with the boundary condition for the monopole operators. Recalling our convention that $u^+_a = v_a^+$ and $u^-_a = (-1)^Nv^-_a$, we find
\be
U^-(\varphi_a) = -e^{t_N} Q_{N-1}(\varphi_a) \,,
\ee
which implies the polynomial equation
\be
U^-(z) = -e^{-t_N} Q_{N-1}(z) \, .
\label{eq:todabc}
\ee

We now want to solve systematically for the scattering data $S(z)$ in terms of the boundary FI parameters $t_j$ and the conjugate momenta $p_j$ on the support of a supersymmetric massive vacuum of the 2d quiver. To do this, we first define polynomials $U_j^-(z)$ for all two-dimensional nodes $j=1,\ldots,N-1$ by the equations 
\be
U_j^-(z) := - e^{-t_j} Q_{j-1}(z) \, .
\label{rec1}
\ee
mirroring equation~\eqref{eq:todabc}. Subsitituting this definition into the twisted chiral ring relations~\eqref{QQ} we find
\be
Q_j(z)  = (z-p_j)Q_{j-1}(z) + e^{t_j} U^-_{j-1}(z) \, .
\label{rec2}
\ee
Equations~\eqref{rec1} and~\eqref{rec2} determine a set of recursion relations that can be solved to find the boundary values of the polynomial $Q(z)$ and $U^-(z)$ in terms of 2d FI parameters $t_j$ and their momenta $p_j$. 

The pair $Q_j(z)$, $U_j^-(z)$ are coprime and can be uniquely completed to a $2 \times 2$ matrix of polynomials $S_j(z)$ with unit determinant by defining polynomials $U^+_j(z)$ and $\tilde Q_j(z)$ by the equations
\be
Q_j(z) \tilde Q_j(z) - U^+_j(z)U^-_j(z) = 1\, .
\label{det}
\ee
Extending the recursion relations~\eqref{rec1} and~\eqref{rec2}, it is straightforward to show that the scattering matrices obey 
\be
S_j(z)
= 
L_j(z) S_{j-1}(z) \, ,
\ee
where
\be
L_j(z) = \begin{pmatrix}
 z-p_j & \, e^{t_j}\\
-  e^{-t_j} & 0
\end{pmatrix} 
\ee
is the 1-particle scattering matrix of the Toda integrable system, with $t_j$ playing the role of the position of the particle and $p_j$ its momenta. It is also the scattering data for one $PSU(2)$ monopole with position $p_j$ and phase $e^{t_j}$. 

The solution of the recursion relation is
\be
S(z) = L_N(z) \cdots L_1(z) 
\ee
which is the Lax matrix of the $N$-body open Toda system, or equivalently, the scattering data for $N$ well-separated $PSU(2)$ monopoles.

Thus our boundary condition encodes a parameterization of the Coulomb branch in terms of a natural collection of Darboux coordinates $(p_j, t_j)$.
Although we cast this result in the language of boundary conditions, it is straightforward to reformulate it and extend it in the language of interfaces between pure 3d $\CN=4$ gauge theories 
with different ranks. We leave the exercise to an enthusiastic reader. 

\section{Abelian theories and mirror symmetry}
\label{sec:abel}

In this section, we take a closer look at half-BPS boundary conditions in abelian theories.
The Higgs and Coulomb branches of abelian theories are hypertoric varieties, whose geometry and quantization have been studied at length in the mathematics literature, \cf\ \cite{Bielawski-hypt, BielawskiDancer, Proudfoot-survey, BLPW-hyp}.
The geometry of hypertoric varieties can be understood using so-called hyperplane arrangements, which play a  role analogous to convex polytopes in toric geometry. We introduce hyperplane arrangements for the Higgs and Coulomb branches from a physical perspective in Sections \ref{sec:hypH} and \ref{sec:hypC}, and show that they provide a systematic, geometric description of chiral rings and the IR images of Neumann and Dirichlet boundary conditions, both classical and quantum.

We have hinted previously that Neumann and generic Dirichlet boundary conditions should be 3d mirrors of each other, while exceptional Dirichlet boundary conditions are self-mirror.
In the case of abelian theories, mirror symmetry is a systematic, combinatorial operation \cite{IS, dBHOY, KS-mirror} that corresponds to Gale duality of Higgs and Coulomb-branch hyperplane arrangements \cite{BLPW-gale, BLPW-hyp}.
We will use hyperplane arrangements to prove that the expected pairs of boundary conditions are in fact mirror to each other, in that their infrared images and quantizations are identical. In Section \ref{sec:mirrorwall}, we will go a step further, defining a ``mirror symmetry interface'' in abelian theories that implements the action of mirror symmetry not just on boundary conditions but on BPS operators of all types.

Throughout this section, we will consider theories with gauge group $G= U(1)^r$ and $N$ hypermultiplets $(X_i,Y_i)$. We make a few simplifying assumptions: 1) that no nontrivial subgroup of $G$ acts trivially on the hypermultiplets (hence $N\geq r$); and 2) that after a generic mass and FI deformation the theory has isolated vacua. Note that (2) is equivalent to saying that a generic (infinitesimal) subgroup $U(1)_m\times U(1)_t$ of the $G_H\times G_C$ flavor symmetry has isolated fixed points on the Higgs and Coulomb branches. Also, (1) implies that the flavor symmetry $G_H$ acting on hypermultiplets has rank $r':=N-r$. 
Since we are only focusing on universal aspects of abelian theories, we will assume that $G_H\simeq U(1)^{r'}$ and $G_C\simeq U(1)^r$ are both abelian (if these groups happen to have a nonabelian enhancement, we will just work with their maximal tori).
We denote the matrices of abelian gauge and flavor charges as $Q = (Q_a{}^i)_{1\leq a\leq r}^{1\leq i \leq N}$ and $q = (q_\alpha{}^i)_{1\leq \alpha\leq r'}^{1\leq i\leq N}$, respectively.

\subsection{Higgs branch}
\label{sec:hypH}

It is convenient to introduce the notation
\be
z_i = X_i Y_i\,, \qquad Z_i = |X_i|^2 - |Y_i|^2 \qquad i = 1,\ldots,N 
\label{eq:zZdef}
\ee
so that the moment-map constraints for the gauge symmetry ($F$ and $D$ terms) are
\be
Q \cdot z + t_\C = 0\,, \qquad  Q \cdot Z + t_\R = 0 \,, \label{Gmom}
\ee
with $t_\C \in \mathfrak g_\C \simeq \C^r$ and $t_\R \in \mathfrak g_\R \simeq \R^r$.
Similarly, the complex and real moment maps for the flavor symmetry become
\be
\mu_{H,\C} = q \cdot z \;\in \C^{r'}\,, \qquad \mu_{H,\R} = q \cdot Z\;\in \R^{r'}\,. \label{GHmom}
\ee

As a simple running example throughout this section, we consider $G=U(1)$ gauge theory with three hypermultiplets of charge $+1$. We focus on a maximal torus $U(1) \times U(1)$ of the full $U(3)/U(1)$ flavor symmetry, such that the gauge and flavor charges matrices are
\be \label{QqSQED}
Q = (1,1,1) \qquad
q = 
\begin{pmatrix}
1 & 0 & 0 \\
0 & 1 & 0
\end{pmatrix} \, .
\ee
The Higgs branch is found by imposing the gauge moment-map constraints
\be
\sum_{i=1}^3 z_i + t_\C = 0 \qquad
\sum_{i=1}^3 Z_i + t_\R = 0
\ee
and dividing by the $U(1)$ gauge symmetry.
For $t_\C=0$, this gives $\CM_H = T^*\cp^2$ with K\"ahler parameter $t_\R$ for the base; for nonzero $t_\C$, we find the usual affine deformation of $T^*\cp^2$. Since this theory is a quiver, the Higgs branch has a nice description as a resolution and/or deformation of the closure of the minimal nilpotent orbit in $\mathfrak{sl}_3$. Nevertheless, in this section we want to understand it in the language of hypertoric geometry, which may also be applied to abelian gauge theories that are not quivers.

\subsubsection{Hyperplane arrangements}

The starting point is to exhibit the Higgs branch as a fibration
\be
\CM_H \longrightarrow \mathbb{R}^{3r'}
\ee
with typical fiber $(S^1)^{r'}$. The base is parametrized by the real and complex moment maps \eqref{GHmom} for the $U(1)^{r'}$ flavor symmetry. This symmetry acts by rotating the $(S^1)^{r'}$ fibers. A particular fiber degenerates on each of the $N$ codimension-three hyperplanes $\CH_i:=\{X_i = Y_i = 0\}$ where one of the hypermultiplets vanishes. 

In our example, the base $\mathbb{R}^6$ is parametrized by the real and complex moment maps for the $U(1)^2$ symmetry, namely $z_1,z_2,Z_1,Z_2$. The fibers are parametrized by, say,
\be
\vartheta_1 = \mathrm{arg}(X_1) - \mathrm{arg}(X_3)\,, \qquad 
\vartheta_2 = \mathrm{arg}(X_2) - \mathrm{arg}(X_3)\,,
\ee
and degenerate  along the three hyperplanes in the base of the fibration
\be
\begin{array}{rll}
\CH_1 & : \quad z_1 = 0 \quad Z_1 = 0  & \quad \text{$\vartheta_1$ degenerates} \\[.1cm]
\CH_2 & : \quad z_2 = 0 \quad Z_2 = 0 & \quad \text{$\vartheta_2$ degenerates} \\[.1cm]
\CH_3 & : \quad z_1+ z_2 = - t_\C  \quad \quad Z_1+ Z_2 = - t_\R & \quad \text{$\vartheta_1-\vartheta_2$ degenerates\,.}
\end{array}
\ee

\begin{figure}[htb]
\centering
\includegraphics[width=6in]{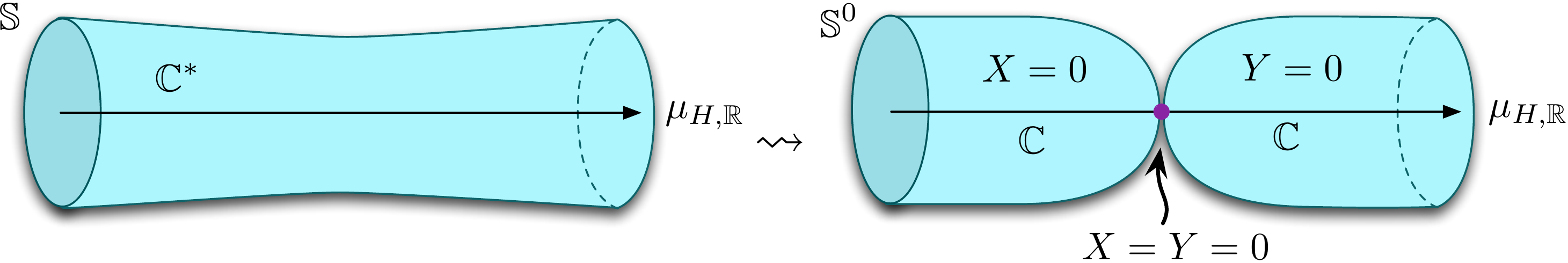}
\caption{The theory of a free hypermultiplet ($G=\oslash,\,N=1$) provides a local model for the behavior of Higgs-branch slices. A generic slice $XY=a$ (left) and the special slice $\S_0=\{XY=0\}$ (right) are shown.
In both cases, the slice is a fibration over $\R$, parameterized by $\mu_{H,\R} = Z = |X|^2-|Y|^2$; the fiber is parameterized by $\frac12(\arg X-\arg Y)$.}
\label{fig:freeslice}
\end{figure}

We next consider holomorphic Lagrangian slices $\S \subset \CM_H$ defined by fixing the complex moment maps for the $U(1)^{r'}$ flavor symmetry. They are fibrations
\be
\S \longrightarrow \mathbb{R}^{r'}
\ee
with the base parametrized by the real moment maps. If a hyperplane $\CH_i$ intersects such a slice, the projection of the intersection to the base $\mathbb{R}^{r'}$ has real codimension one.
In a generic slice, the intersections $\S \cap \CH_i$ are all empty, and the slice has the topology of a cylinder; 
as a complex manifold $\S \simeq (\C^*)^{r'}$. However, whenever there is an intersection $\S \cap \CH_i$, one factor of $\mathbb{C}^*$ degenerates into two cigars $\C \cup \C$ whose tips coincide with the intersection point: see Figure \ref{fig:freeslice}.

We are interested in special slices that intersect multiple hyperplanes. Generically, it is possible to intersect at most $r'$ hyperplanes. We choose a subset $S \subset \{ 1,\ldots,N \}$ of size $r$ such that the corresponding $r\times r$ submatrix $Q^{(S)}$ of the charge matrix $Q$ is nondegenerate. Then there exists a unique slice, denoted $\S_S$, that intersects all of the hyperplanes $\CH_i$ with $i \notin S$. It has the following properties:
\begin{itemize}
\item The common intersection of $\S_S$ and all the hyperplanes $\CH_i$ ($i\notin S$) is a single point $\nu_S\in \CM_H$, which is a vacuum in the presence of generic mass parameters.
\item The hyperplanes cut the slice $\S_S$ into $2^{r'}$ toric varieties.
\item If the submatrix $Q^{(S)}$ of $Q$ is unimodular, then $\nu_S$ is a massive vacuum, the Higgs branch is smooth in a neighborhood of $\nu_S$, and each of the $2^{r'}$ toric varieties is a copy of $\C^{r'}$. Otherwise, there is an orbifold singularity at $\nu_S$.\label{page:unimod}
\end{itemize}
The base of $\S_S$ is cut into $2^{r'}$ orthants by the hyperplanes $\CH_i$ ($i \notin S$). On the base, the two sides of any hyperplane $\CH_i$ are distinguished by either $X_i$ or $Y_i$ getting a vev; we call these the `$+$' and `$-$' sides, respectively. We can then label each orthant (or the toric variety sitting above it) by a sign vector $\varepsilon\in \{\pm\}^{r'}$, such that
\be V_{S,\varepsilon}\,:\quad\text{orthant in $\S_S$ on the $\varepsilon_i$ side of $\CH_i$ for all $i\notin S$}\,. \label{orthantV} \ee
We will often to complete $\varepsilon$ to a full sign vector $\varepsilon=(\varepsilon_1,...,\varepsilon_N)\in \{\pm\}^N$, with the understanding that $V_{S,\varepsilon}$ only depends on $\varepsilon_i$ for $i\notin S$.

\begin{figure}[htb]
\centering
\includegraphics[width=2.5in]{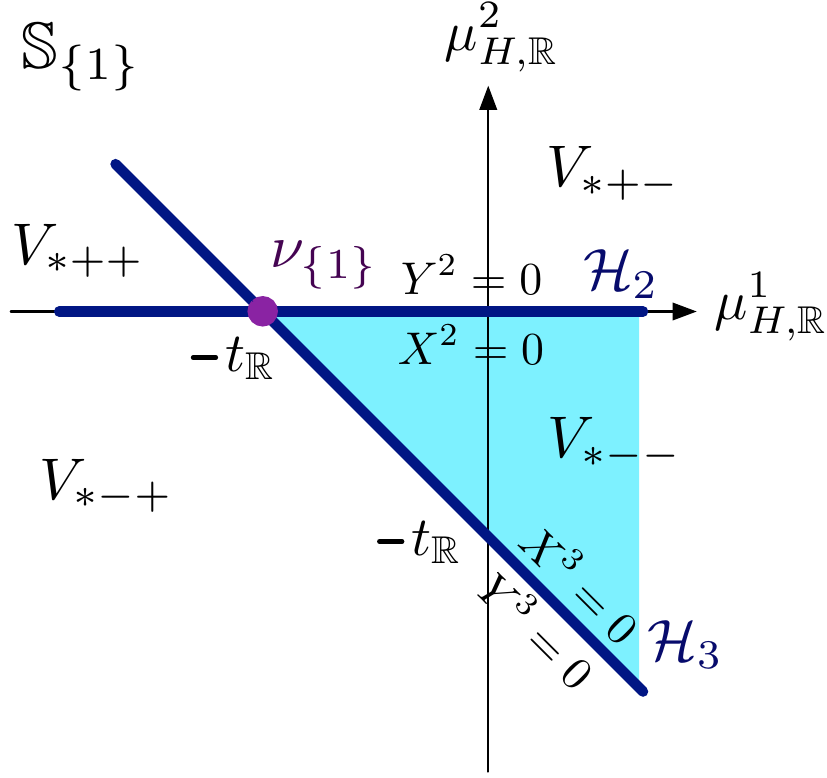}
\caption{The slice $\S_{\{1\}}$ considered in the main text, defined by $z_2=z_3=0$, for real FI parameter $t_\R> 0$, and generic nonzero complex FI. The dark blue lines are the intersections of this slice with the hyperplanes $\CH_2$ and $\CH_3$. The light blue shaded region supports the IR image of the exceptional Dirichlet boundary conditions $\CD_{\pm--,\{1\}}$ (Section \ref{sec:abel-HxD}).}
\label{fig:cp2S1}
\end{figure}

Let us consider the slice $\S_{\{1\}}$ in our running example. This slice must intersect the hyperplanes $\CH_2$ and $\CH_3$, which forces the complex moment maps to equal $z_1 = - t_\C$ and $z_2 = 0$. The base of the slice is $\R^2$, parameterized by the real moment maps $\mu_{H,\R}^1=Z_1$ and $\mu_{H,\R}^2=Z_2$; the hyperplanes $\CH_2$ and $\CH_3$ intersect along $Z_2 = 0$ and $Z_1 +Z_2 = - t_\R$. The intersection of these lines at $Z_1 = -t_\R, Z_2=0$ becomes one of the three massive supersymmetric vacua when masses are turned on. The slice $\S_{\{1\}}$ is cut into four quadrants distinguished by different combinations of $X_2,Y_2.X_2,Y_3$ vanishing.  This is illustrated in Figure~\ref{fig:cp2S1}. Similar comments apply to the slices $\S_{\{2\}}$ and $\S_{\{3\}}$.

So far we have assumed generic complex FI parameters. For special values of the complex FI parameters, more than $r'$ of the hyperplanes $\CH_i$ may intersect a given slice. The extreme case when all complex FI parameters vanish is particularly interesting: there is a canonical slice $\S_0$ defined by $z_i=0$ for all $i=1,\ldots,N$. The canonical slice has the following properties:
\begin{itemize}
\item $\S_0$ intersects all of the hyperplanes $\CH_i$.
\item  The hyperplanes $\CH_i$ cut the base of $\S_0$ into at most $2^N$ convex polytopes.
\item  $\S_0$ itself is cut into toric varieties fibered over the corresponding convex polytopes. The real FI parameters determine the K\"ahler parameters of these toric varieties. 
\end{itemize}
Note that if all real FI parameters vanish, the Higgs branch becomes a singular cone and the hyperplanes all pass through the origin of the canonical slice.

\begin{figure}[htb]
\centering
\includegraphics[width=3in]{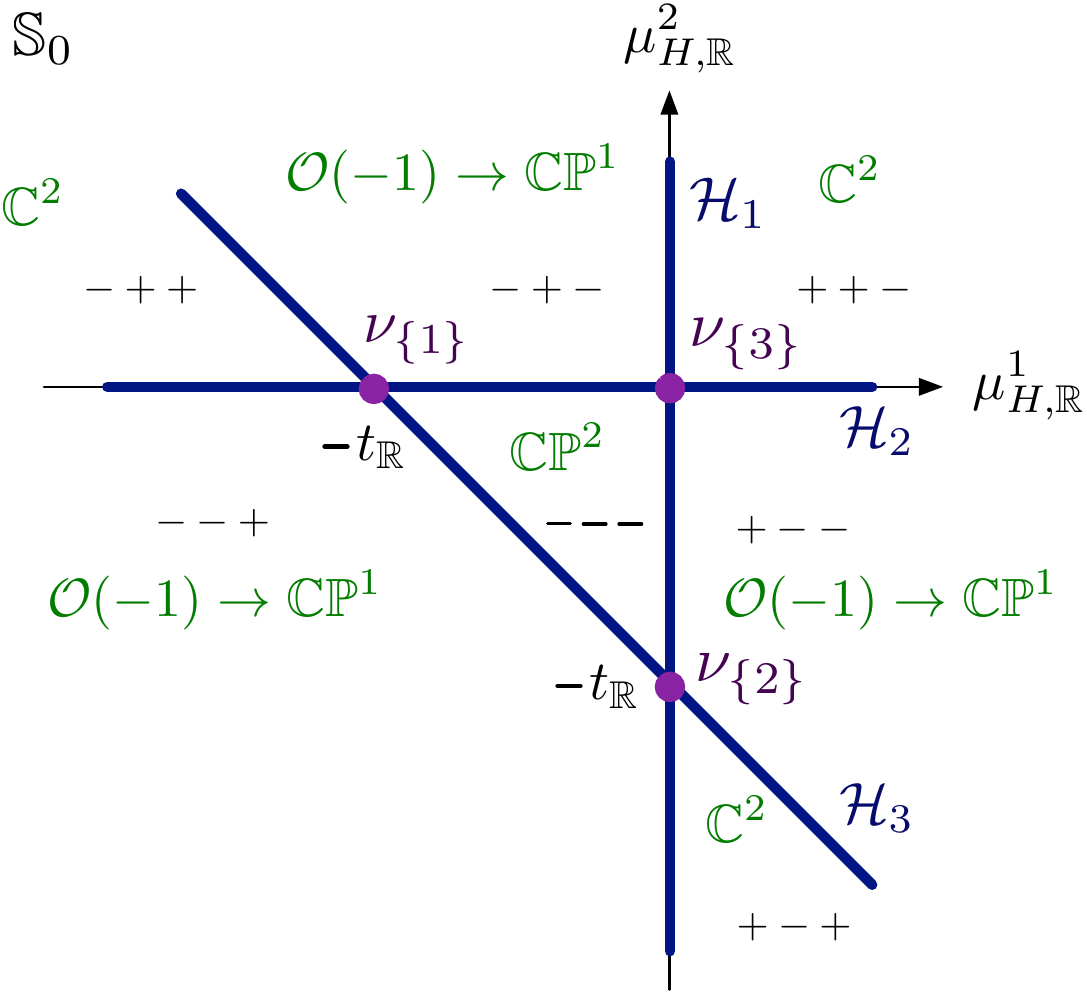}
\caption{The canonical slice $\S_0$ of $T^*\mathbb{CP}^2$ at vanishing complex FI $t_\C=0$, with $t_\R > 0$. As before, dark blue lines show intersection with the hyperplanes $\CH_i$.}
\label{fig:cp2S0}
\end{figure}

Let us illustrate the canonical slice in our example. With $t_\R \neq 0$, the Higgs branch is $T^*\mathbb{CP}^2$. The canonical slice has seven components: the compact $\mathbb{CP}^2$, the conormal bundles to the three projective coordinate hyperplanes $\mathbb{CP}_{\{i\}}^1 \subset \mathbb{CP}^2$, and the conormal bundles to three points $\nu_{\{i\}}$, the intersections of coordinate hyperplanes. Each $\nu_{\{i\}}$ is a vacuum, in which the hypermultiplet $(X_i,Y_i)$ gets a vev. This canonical slice is depicted in Figure~\ref{fig:cp2S0}.

Recall that the two sides of the hyperplane $\CH_i$ can be labelled `$+$' and `$-$' depending on whether $X_i$ or $Y_i$ (respectively) gets a vev. Therefore, each chamber $\Delta_\varepsilon$ in the canonical slice is uniquely labelled by a sign vector $\varepsilon \in \{\pm\}^N$,
\be
\Delta_\varepsilon \;= \;\text{chamber on $\varepsilon_i$ side of each $\CH_i$}\;=\;
\begin{cases}
\; |X_i| \geq 0 \quad Y_i = 0 & \mathrm{if} \quad \varepsilon_i = + \\
\; |Y_i| \geq 0 \quad X_i = 0 & \mathrm{if} \quad  \varepsilon_i = - \\
\end{cases}\, .
\label{eq:chamberdef}
\ee
However, depending on the sign of the real FI parameters, not all of the $2^N$ possible sign vectors correspond to a chamber in the canonical slice. A sign vector that does correspond to a chamber in the canonical slice is called `feasible'. In our example, with $t_\R>0$, the toric varieties associated to the chambers are (see Figure~\ref{fig:cp2S0})
\begin{itemize}
\item $ \Delta_{---} $ : compact base $ \mathbb{CP}^2$.
\item $ \Delta_{+--} $ : conormal bundle to the coordinate hyperplane $\mathbb{CP}_{\{1\}}^1 = \{ Y_1 = 0 \}$.
\item $ \Delta_{++-} $ : conormal bundle to the point $\nu_{\{3\}} = \{ Y_1 = Y_2 = 0\}$.
\item $\Delta_{+++}$ : not feasible.
\end{itemize}
together with obvious permutations. If we had chosen $t_\R<0$ instead, related to the $t_\R>0$ geometry by a hyperk\"ahler flop, we would interchange $+ \leftrightarrow -$ in the above statements.

The orthants $V_{S,\varepsilon}$ still make sense on the canonical slice, but they decompose into a union of chambers. Namely, $V_{S,\varepsilon}$ is a union of all the feasible chambers $\Delta_{\varepsilon'}$ such that $\varepsilon_i=\varepsilon_i'$ for $i\notin S$. Correspondingly, the simple $\C^{r'}$ hypertoric varieties that would be supported on an orthant in $\S_S$ are cut into a union of more interesting ones.

\subsubsection{Chiral Ring}
\label{sec:abel-Hring}

Any gauge-invariant chiral operator is a sum of gauge-invariant monomials in the fields $X_i,Y_i$. Gauge-invariant monomials come in two types.
First, there are the operators $z_i$ defined in~\eqref{eq:zZdef}, which obey the complex moment map equations, $Q \cdot z = - t_\C$. They are neutral under the flavor symmetry $U(1)^{r'}$, and generate a subring $\C[\CM_H]_0 \subset \C[\CM_H]$. (It coincides with the subring $\C[\CM_H]_0$ in \eqref{eq:mdecomp}, given a generic real mass deformation.)

The remaining monomials are charged under the flavor symmetry. To describe them, we introduce another charge matrix $\widetilde Q$ of dimension $r'\times N$ so that 
\be
0\longrightarrow \Z^{r'} \overset{\wt Q^T}{\longrightarrow} \Z^N \overset{Q}{\longrightarrow} \Z^r \longrightarrow 0
\label{eq:exact}
\ee
is an exact sequence of lattices. It will turn out that $\widetilde{Q}$ is the charge matrix of the mirror theory. Having fixed gauge and flavor matrices $Q,q$, a canonical way to choose $\wt Q$ is to set $\left(\begin{smallmatrix} * \\ \wt Q \end{smallmatrix}\right) = \left(\begin{smallmatrix} Q \\ q \end{smallmatrix}\right)^{-1,T}$.
Then
for every element $A \in \mathbb{Z}^{r'}$ of the flavor charge lattice,
\be  \label{wA}
w^A := \prod_{i=1}^N \begin{cases} \, X_i^{\, | \, \wt Q_A^i |} & \wt Q_A^i > 0 \\
 \, Y_i^{\, | \, \wt Q_A^i |} & \wt Q_A^i  < 0 \end{cases}\,,
\ee
with $\wt Q_A := \wt Q^T\cdot A \; \in \mathbb{Z}^N$, is a gauge-invariant monomial. These obey the ring relations
\begin{align} \label{CH-abel}
\hspace{.5in} w^A w^B &= \; w^{A+B}\!\!\!\prod_{\text{$i$ s.t. $\wt Q_A^i \wt Q_B^i  < 0$}} z_i^{\text{min}(\, | \, \wt Q_A^i |, | \, \wt Q_B^i |\,)}  \\ &=\;  w^{A+B}\prod_{1\leq i \leq N}  z_i^{(\wt Q_A^i)_+ + (\wt Q_B^i)_+ - (\wt Q_A^i + \wt Q_B^i)_+} \qquad \text{(equivalently)}\,. \notag
\label{eq:abel-rel-H} 
\end{align}
The $w^A$ and $z_i$ together generate the chiral ring $\C[\CM_H]$.

\begin{figure}[htb]
\centering
\includegraphics[width=3.0in]{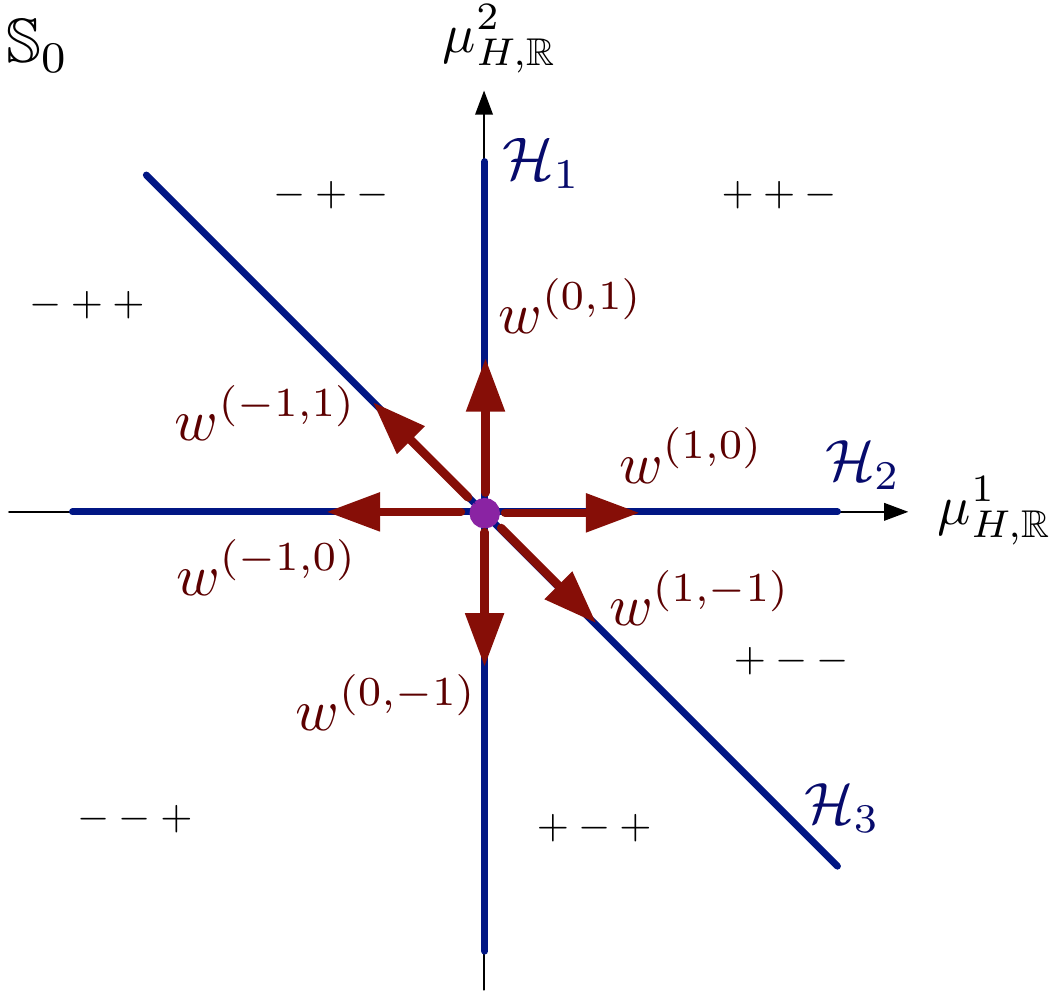}
\caption{The operators $w^A$ that provide linear functions along edges of cones in the canonical slice $\S_0$ (for vanishing real FI parameter).}
\label{fig:wA}
\end{figure}

We can interpret the above generators and relations in terms of the geometry of the canonical slice $\S_0$, in the limit that all FI parameters are set to zero. Recall that all hyperplanes $\CH_j$ then pass through the origin and cut the base $\R^{r'}$ into a union of polyhedral cones. We may identify the base as $\R^{r'}\simeq \R\otimes \Z^{r'}$, so that each charged operator $w^A$ is associated with a ray $\rho(A)$ in the base, in the direction of its flavor charge. Along the ray $\rho(A)$, the function $|w^A|$ increases monotonically from zero. We illustrate this for our example in Figure~\ref{fig:wA}.

Now consider the chiral-ring relations~\eqref{CH-abel}\,: geometrically, they say that the product $w^Aw^B$ is equal to $w^{A+B}$ up to a correction factor for each hyperplane $\CH_j$ that lies between the rays $\rho(A)$ and $\rho(B)$. In particular, if $\rho(A)$ and $\rho(B)$ are contained in a single cone, then there are no corrections. At the opposite extreme, if $B=-A$, then every single hyperplane is crossed and there is a correction factor for every hypermultiplet.

This observation can be used to construct a finite set of generators for the chiral ring: one simply takes the $z^i$ (or the flavor moment maps) together with a finite set of operators $\{w^A\}_{A\in \CA}$ such that the $A\in \CA$ generate the integral lattice inside each of the cones in the canonical slice. (For a proof, see \cite{Hilburn}.) In our example, we take
\be \wt Q = \begin{pmatrix} 1 & 0 & -1 \\ 0 & 1 & -1 \end{pmatrix}\,. \label{SQED-tQ} \ee
Then the finite set of generators is given by $z_1, z_2$ together with
\be
\begin{array}{rl@{\qquad}rl@{\qquad}rl}
 w^{(1,0)} & = X_1Y_3\,,  & w^{(0,1)}  &=X_2Y_3\,, & w^{(1,-1)}&=X_1Y_2, \\[.1cm]
 w^{(-1,0)} & =Y_1X_3\,, & w^{(0,-1)} &=Y_2X_3\,, & w^{(-1,1)}&=Y_1X_2\,.
 \end{array}
\ee
which are illustrated in Figure~\ref{fig:wA}.

\subsubsection{Quantum chiral ring}
\label{sec:abel-qHring}

In the presence of a twisted $\wt\Omega$ background, the Higgs-branch chiral ring is quantized. We review the structure of the quantization, in parallel with the above discussion.

The quantum algebra $\hat \C[\CM_H]$ is obtained by starting with an $N$-dimensional Heisenberg algebra generated by $\hat X_i,\hat Y_i$ with $[\hat Y_i,\hat X_j]=\epsilon\,\delta_{ij}$, then restricting to gauge-invariant operators, and imposing complex moment-map constraints. The gauge-invariant part of the Heisenberg algebra is generated by the normal-ordered operators
\be \hat z_i =\, :\! \hat X_i\hat Y_i\!: \,= \hat X_i\hat Y_i + \tfrac\epsilon2 = \hat Y_i\hat X_i-\tfrac\epsilon2\,, \label{zNO} \ee
which are neutral under the flavor symmetry, and by the monomials 
\be  \hat w^A := \prod_{i=1}^N \begin{cases} \, \hat X_i^{\, | \, \wt Q_A^i |} & \wt Q_A^i > 0 \\
 \, \hat Y_i^{\, | \, \wt Q_A^i |} & \wt Q_A^i  < 0 \end{cases}\,,\qquad A\in \Z^{r'}\,,
\ee
which have flavor charge $A\in\Z^{r'}$. The $\hat z_i$ obey
\be Q\cdot \hat z + t_\C = 0\,. \label{qCH-abel} \ee
and generate a maximal commutative subalgebra $\hat \C[\CM_H]_0\subset \hat \C[\CM_H]$. A concrete basis for $\hat \C[\CM_H]_0$ is given by the flavor moment maps
\be \hat \mu_{H,\C}^\alpha := (q\cdot \hat z)^\alpha\qquad (\alpha=1,...,r')\,. \ee

Taking commutators with flavor moment maps measures the flavor charges of the remaining elements in $\hat C[\CM_H]$,
\begin{subequations} \label{Hqring}
\be [\hat \mu_{H,\C}^\alpha,\hat w^A] = \epsilon\,A^\alpha\,\hat w^A\,. \label{Hqring-a}\ee
There are also additional algebra relations that quantize \eqref{CH-abel},
\begin{equation} \label{Hqring-b}
\hat w^A \hat w^B =   \hspace{-.3cm}  \prod_{\substack{\text{$i$ s.t. $|\wt Q_A^i|\leq |\wt Q_B^i|$,}\\[.05cm] \wt Q_A^i \wt Q_B^i<0}} \hspace{-.3cm}  [\hat z_i]^{-\wt Q_A^i} \;
\hat w^{A+B}
\hspace{-.3cm}  \prod_{\substack{\text{$i$ s.t. $|\wt Q_A^i|> |\wt Q_B^i|$,}\\[.05cm] \wt Q_A^i \wt Q_B^i<0}}  \hspace{-.3cm}  [\hat z_i]^{\wt Q_B^i}
\end{equation}
\end{subequations}
with the usual quantum products
\be [a]^b := \begin{cases} \prod_{i=1}^{b} (a+(i-\frac12)\epsilon) & b>0 \\
 \prod_{i=1}^{|b|} (a-(i-\frac12)\epsilon) & b <0 \\
 1 & b=0\,. \end{cases}   \label{q-exp-abel}
\ee

Altogether, algebra $\hat\C[\CM_H]$ is generated by $\hat z_i,\hat w^A$ subject to \eqref{qCH-abel} and \eqref{Hqring}. A finite set of generators can be obtained exactly as in the classical case: among the infinitely many charged operators, one takes some $\{\hat w^A\}_{A\in \CA}$ such that $A\in \CA$ generate the integral lattice inside each of the cones in the canonical slice $\S_0$ of the hyperplane arrangement.

\subsubsection{Quantum hyperplane arrangements and weight modules}
\label{sec:abel-Hqhyp}

Many UV boundary conditions produce weight modules for the algebras $\hat \C[\CM_H]$ and $\hat\C[\CM_C]$, at least after taking some limits such as $t_{2d}\to \infty$ or $c\to\infty$ from Sections \ref{sec:NC-tinf}, \ref{sec:DH-cinf}. By a weight module for $\hat \C[\CM_H]$, we mean a module $M$ that decomposes $M = \oplus_\lambda M_\lambda$ into finite-dimensional generalized eigenspaces $M_\lambda$ for $\hat \C[\CM_H]_0$, which should be thought of as the Cartan subalgebra of $\hat \C[\CM_H]$.

In the presence of a real mass $m_\R$, we expect that a weight module coming from a right (say) boundary condition is $m_\R$-feasible (preserves supersymmetry) if all the operators $\hat w^A \in \hat\C[\CM_H]_<$ of negative charge (\ie\ $m_\R\cdot A<0$) act locally nilpotently. Specifically, this means that for any vector $v\in M_\lambda$ in a fixed weight space and any negatively charged $\hat w^A$, $(\hat w^A)^nv=0$ for sufficiently large~$n$. We call such a module \emph{lowest-weight} with respect to $m_\R$.

We can give a geometric description of weight modules by introducing \emph{quantum} hyperplane arrangements, following \cite{BLPW-hyp}. In general, the quantum hyperplane arrangement for the Higgs branch is a particular system of lattices embedded in $\C^{r'}$, with the coordinates on $\C^{r'}$ corresponding to eigenvalues of the \emph{complex} flavor moment maps $\hat \mu_{H,\C}$, acting on weight spaces of a putative representation.
The $N$ hyperplanes $\CH_i$ have images in $\C^{r'}$: they are defined to lie along loci where $\hat z_i=0$.
For every maximal intersection $\nu_S \in \C^{r'}$ of the $\CH_i$ with $i\notin S$ (labelled by a subset $S$ of size $r$, just as on page \pageref{page:unimod}), we define an integral lattice
\be \Gamma_S = \nu_S + (\Z+\tfrac12)^{r'} \subset \C^{r'}\,,\ee
such that at each lattice point of $\Gamma_S$ the $\hat z_i$ with $i\notin S$ have half-integer eigenvalues. The significance of $\Gamma_S$ is that any Verma module with a lowest-weight vector corresponding to the vacuum $\nu_S$ must have weight spaces in this lattice. Each $\Gamma_S$ should be considered a quantization of the special slice $\S_S$ of the classical Higgs branch.

Now, recall that modules for $\hat \C[\CM_H]$ are most interesting%
\footnote{Meaning there exist nontrivial maps and extensions among them.} %
when the complex FI parameters are specialized to integral or half-integral values $t_\C=k_t\epsilon$. The specialization is a quantum analogue of setting $t_\C=0$ in the absence of $\Omega$-background; for Neumann boundary conditions it is obligatory. 
The integrality condition $t_\C=k_t\epsilon$ is equivalent to requiring that the lattices $\Gamma_S$ for various $S$ all \emph{coincide}. In this case, the quantum hyperplane arrangement may be restricted to a single lattice $\Gamma\simeq \Z^{r'} \subset \R^{r'}$, identified with the weight lattice of the flavor group. The ambient space $\R^{r'}$, whose coordinates are collections of real eigenvalues for $\hat \mu_{H,\C}$, may be identified with the canonical slice $\S_0$. Each hyperplane $\CH_i \subset \R^{r'}$ lies exactly half-way between lattice points of $\Gamma$, and the relative position of different hyperplanes is fixed by the parameters $k_t$. Thus, the quantum hyperplane arrangement simply becomes a discretized version of the canonical slice $\S_0$ of the Higgs branch, with
\be \begin{array}{ccc}
 \text{real moment maps $\mu_{H,\R}$} & \leadsto & \text{eigenvalues of complex moment maps $\hat \mu_{H,\C}$} \\[.1cm]
 \text{real FI (resolution) params $t_\R$} & \;\leadsto\; & \text{quantized complex FI (quantization) param's $k_t$}
\end{array} \ee

\begin{figure}[htb]
\centering
\includegraphics[width=4.5in]{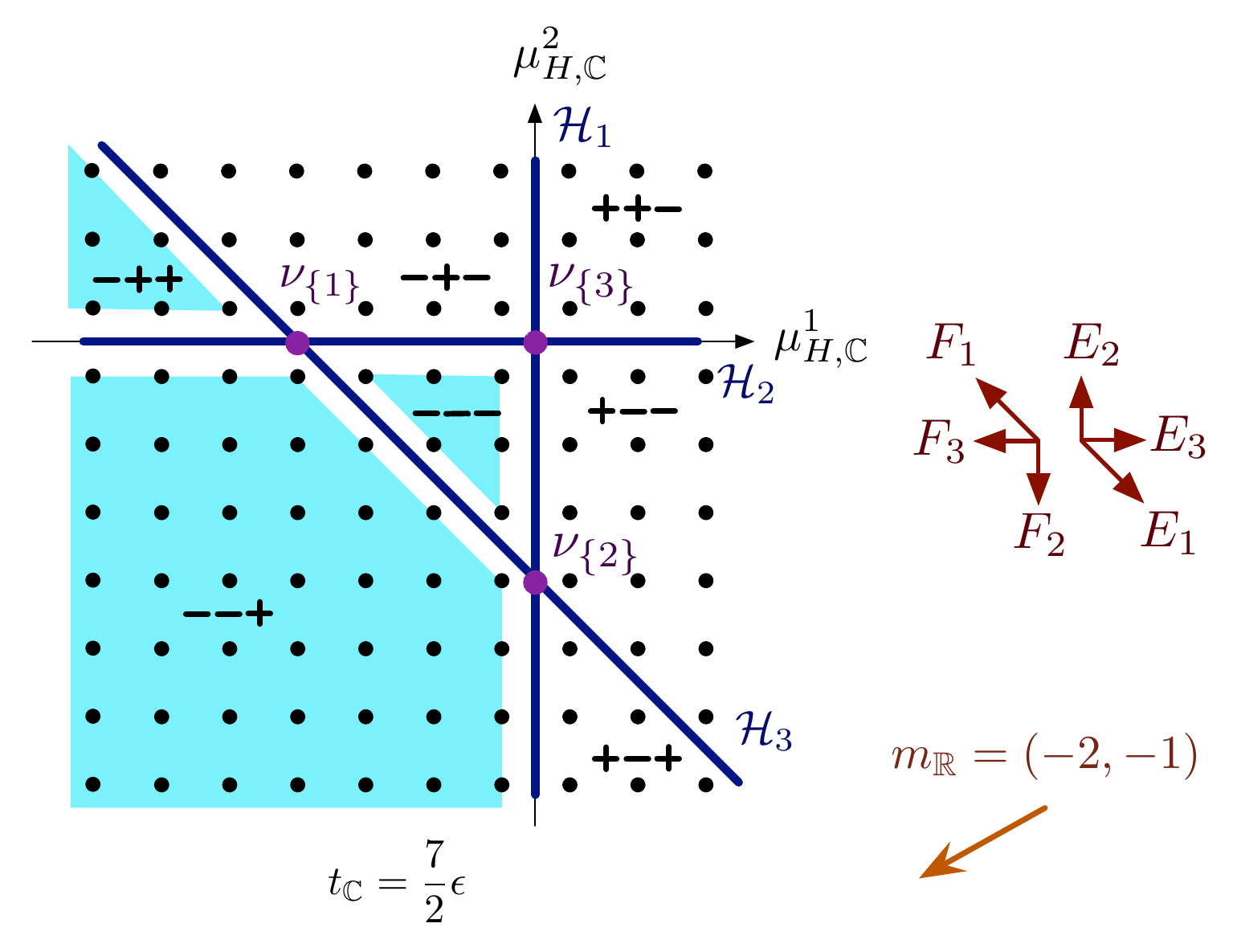}
\caption{The quantum hyperplane arrangement for SQED with three hypers, with quantized FI parameter $t_\C = \frac72\epsilon$. The charged operators $\hat w^A$ (equal to $E_i$ or $F_i$) map one weight space to another, along the same directions that appeared classically in Figure \ref{fig:wA}.}
\label{fig:cp2q}
\end{figure}

For our running example of SQED with three hypermultiplets, $\hat\C[\CM_H]$ may be identified with a quotient of the universal enveloping algebra of $\mathfrak{sl}_3$, by setting the Chevalley-Serre generators to be (say)
\be \label{EFH3}
\begin{array}{c@{\quad}c@{\quad}c} E_1 = \hat X_1\hat Y_2 = \hat w^{(1,-1)}\,, &  F_1 = \hat X_2\hat Y_1= \hat w^{(-1,1)}\,,& H_1 = \hat X_1\hat Y_1-\hat X_2\hat Y_2  = \hat z_1-\hat z_2 \\[.1cm]
 E_2 = \hat X_2\hat Y_3=\hat w^{(0,1)}\,, & F_2 = \hat X_3\hat Y_2 = \hat w^{(0,-1)} \,,& H_2 = \hat X_2\hat Y_2 -\hat X_3\hat Y_3 = \hat z_2-\hat z_3\,. \end{array}
\ee 
We can also introduce $E_3=\frac1\epsilon[E_1,E_2]=\hat X_1\hat Y_3=\hat w^{(1,0)}$ and $F_3 = \frac1\epsilon[F_2,F_1]=\hat w^{(-1,0)}$.
Recall that the flavor moment maps are $\hat\mu_{H,\C}^1=\hat z_1$, $\hat \mu_{H,\C}^2=\hat z_2$, and the gauge constraint is $\hat z_1+\hat z_2+\hat z_3+t_\C=0$. Specializing $t_\C \in (\Z+\frac12)\epsilon$, the quantum hyperplane arrangement takes the form shown in Figure \ref{fig:cp2q}: it looks like the weight lattice of $\mathfrak{sl}_3$.

In this quotient of the enveloping algebra $U(\mathfrak{sl}_3)$, the Casimir operators are fixed. A short calculation shows that they are both fixed in terms of $t_\C$:
\be C_2 = \tfrac23 (t_\C)^2 - \tfrac32\epsilon^2\,,\qquad C_3 = C_2\big(\tfrac13 t_\C + \tfrac32\epsilon\big)\,. \label{Cas-sl3} \ee
These are the values that the Casimirs would take in the $n$-th symmetric power of the antifundamental representation if $t_\C = (n+\frac32)\epsilon$ and the $n$-th symmetric power of the fundamental if $t_\C = -(n+\frac32)\epsilon$.

The charged operators $\hat w^A \in \hat\C[\CM_H]$ (labelled by weights $A$ of the flavor group $G_H$) act on a weight module by take one weight space to another. Thus, having identified the lattice(s) $\Gamma_S$ with the weight lattice of $G_H$, we see that a lattice point $p$ is mapped by $\hat w^A$ to another lattice point with coordinates $p+A$. This is the quantum analogue of the linear functions $w^A$ pointing along rays in Figure \ref{fig:wA}.
Moreover, due to the ring relation $\hat w^A\hat w^{-A} = \prod_i [\hat z_i]^{-\wt Q_A^i}$, the operators $\hat w^A$ can act as zero if and only if they cross one of the hyperplanes in the quantum arrangement. Therefore, any weight module must be supported on (\ie\ have nontrivial weight spaces inside) some union of complete chambers of the arrangement.

For example, at quantized $t_\C=k_t\epsilon$, the irreducible weight modules of $\hat \C[\CM_H]$ are precisely supported on chambers $\Delta_\varepsilon$ of the quantum arrangement. The chambers are labelled by a sign vector $\varepsilon$, exactly the same way as on the canonical slice $\CS_0$. We denote by $\hat \Delta_\varepsilon$ the irreducible module supported on $\Delta_\varepsilon$.

Similarly, for each orthant $V_{S,\varepsilon}$ of the quantum arrangement as in \eqref{orthantV}, there is a Verma module $\hat V_{S,\varepsilon}$. It is freely generated from an identity state $|0\rangle$ that satisfies
\be \label{modV-def}
 \begin{array}{ll} \hat z_i|0\rangle = \tfrac12\varepsilon_i\epsilon|0\rangle \qquad  & \text{for all $i\notin S$} \\[.1cm]
\hat w^A|0\rangle = 0 & \text{for all $A$ pointing out of $V_{S,\varepsilon}$}\,. \end{array}
\ee
The state $|0\rangle$ lies in the weight space closest to the origin of the orthant, and may be identified with the classical vacuum $\nu_S$ at the origin itself.
The module is reducible if and only if additional hyperplanes $\CH_i$ ($i\in S$) intersect $V_{S,\varepsilon}$.

In the presence of a real mass $m_\R$, it is useful to introduce a linear function
\be  \hat h_m := m_\R\cdot \hat \mu_{H,\C} \label{def-hathm} \ee
on the quantum hyperplane arrangement,
which simply measures the charge of each weight space.
It is analogous to the classical Morse function $h_m=m_\R \cdot \mu_{H,\R}$ on the Higgs branch.
The lowest-weight (highest-weight) modules with respect to $m_\R$ are precisely those supported on chambers such that $\hat h_m$ is bounded below (above). For example, in Figure \ref{fig:cp2q} the lowest-weight modules for $m_\R=(-2,-1)$ must be supported on some union of the three shaded chambers.

\subsection{Higgs branes and modules}
\label{sec:abel-Hmod}

We now use the formalism of hyperplane arrangements to systematically describe the IR images of various boundary conditions.

\subsubsection{Neumann boundary conditions}
\label{sec:abel-HN}

A basic Neumann boundary condition (Section \ref{sec:defN}) requires a Lagrangian splitting $L$ of the hypermultiplets. For abelian theories, the splitting can be encoded in a choice of sign vector $\varepsilon = (\varepsilon_1,\ldots,\varepsilon_N) \in \{\pm\}^N$, so that the boundary condition $\CN_\varepsilon$ sets
\be 
\CN_\varepsilon\,:\quad \text{Neumann b.c. for gauge multiplets and} \quad \begin{cases} Y_j | = 0 & \mbox{if} \quad \varepsilon_j = + \\  X_j |= 0& \mbox{if} \quad \varepsilon_j = - \end{cases}\,. 
\ee
Since complex FI parameters necessarily vanish for Neumann boundary conditions, the only interesting slice of the Higgs branch is the canonical slice $\S_0$. Then the analysis of Section \ref{sec:NHiggs} shows that the Higgs-branch image of $\CN_\varepsilon$ is precisely the toric component of $\S_0$ with base polytope $\Delta_\varepsilon$,
\be \CN_\varepsilon \quad\leadsto\quad \CN_\varepsilon^{(H)} = \text{Toric}(\, \Delta_\varepsilon \, )\,.\label{NeL}\ee
The boundary condition breaks supersymmetry in the IR unless the chamber $\Delta_\varepsilon$ is feasible, for a given choice of real FI parameters.

Thus, in our example of SQED with three hypermultiplets, seven of the eight possible boundary conditions have images on the feasible chambers in Figure~\ref{fig:cp2S0} (for, say, $t_\R > 0$); and the eighth breaks supersymmetry.

Turning on real masses $m_\R \in \mathfrak t_H \simeq \R^{r'}$ introduces a potential on the Higgs branch given by (Section \ref{sec:NH-mt})
\be h_m = m_\R \cdot \mu_{H,\R}\,. \label{hm-abel} \ee
This is the real moment map for a particular (infinitesimal) $U(1)_m$ subgroup of $G_H$.
On the base of any slice $\mathbb S$, in terms of the coordinates $\mu_{H,\R}^\alpha$, $h_m$ is just a linear function; and $m_\R$ itself can be interpreted as a direction (the gradient of $h_m$) in the slice. For generic $m_\R$, the critical points of $h_m$ coincide with the massive vacua of the theory, which lie at maximal intersections of hyperplanes.

The gradient-flow cycles on the Higgs branch $\CM_H^\lessgtr[m_\R]$ that were first described in Section \ref{sec:NH-mt} are precisely the toric components of $\S_0$ on which $h_m$ is \emph{bounded},
\be \begin{array}{l} \CM_H^<[m_\R]\;:\qquad \text{union of $\Delta_\varepsilon$'s s.t. $h_m\big|_{\Delta_\varepsilon} < \infty$} \\
\CM_H^>[m_\R]\;:\qquad \text{union of $\Delta_\varepsilon$'s s.t. $h_m\big|_{\Delta_\varepsilon} > -\infty$}
\end{array} \ee
From the analysis of Section \ref{sec:NH-mt}, we expect that a right (left) boundary condition $\CN_\varepsilon$ preserves supersymmetry if the intersection of its image with $\CM_H^<[m_\R]$ ($\CM_H^>[m_\R]$) is compact. We called the corresponding boundary conditions $m_\R$-feasible.
Notice that when images of boundary conditions are restricted to the slices $\S$, having a compact intersection with $\CM_H^<[m_\R]$ ($\CM_H^>[m_\R]$) is itself equivalent to being supported on chambers that are bounded from below (above).

\begin{figure}[htb]
\centering
\includegraphics[width=3.7in]{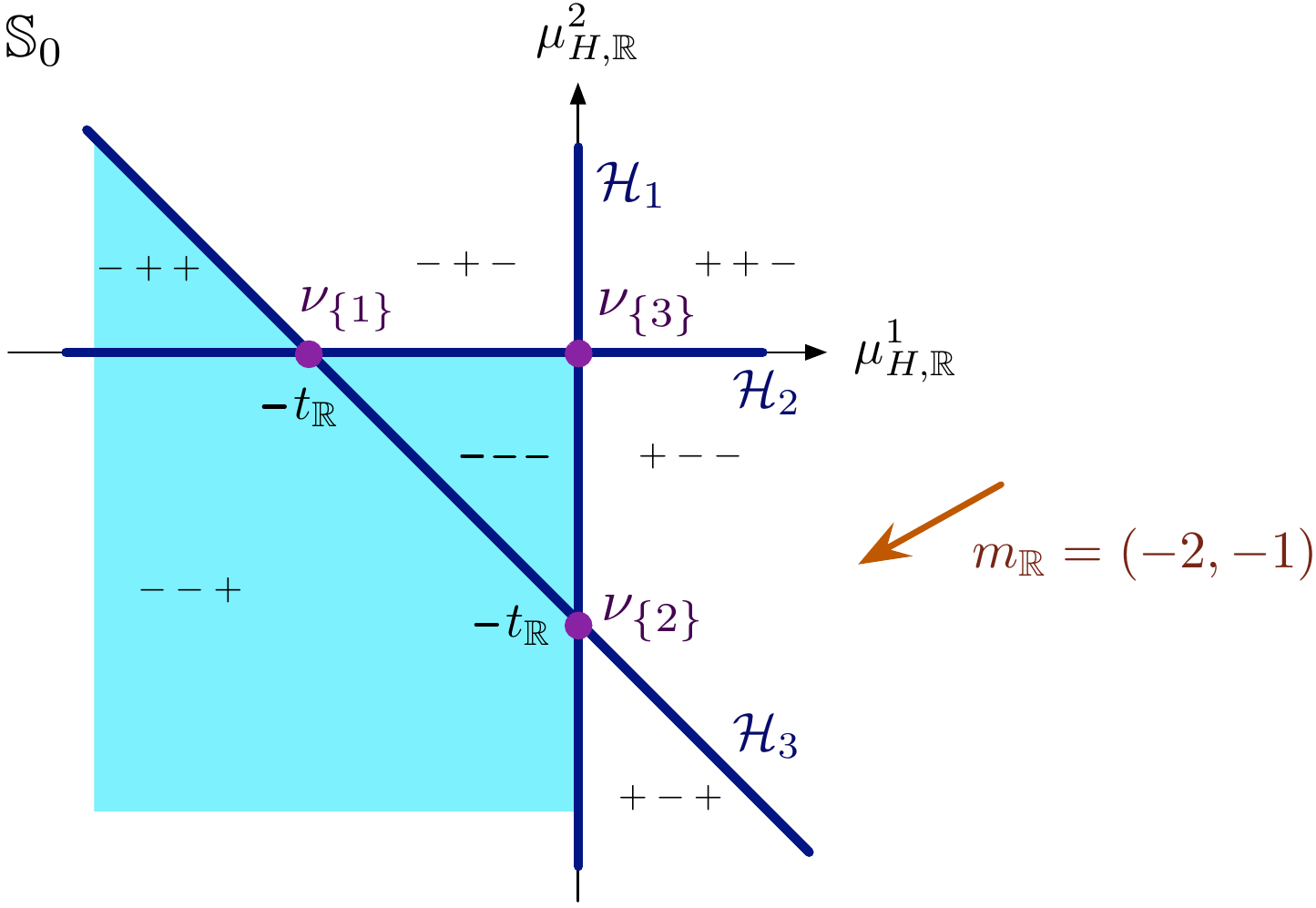}
\caption{Chambers bounded from below in the slice $\S_0$ of the Higgs branch for SQED with $N=3$, supporting supersymmetric right boundary conditions in the presence of~$m_\R$.}
\label{fig:cp2bdd}
\end{figure}

In our example, if we choose $t_\R>0$ and $m_\R = (-2,-1)$ so that $h_m = -(2Z_1+Z_2)$, we find exactly three chambers that are both bounded and feasible (Figure \ref{fig:cp2bdd}). They support the IR images of the boundary conditions $\CN_{---}$, $\CN_{--+}$, and $\CN_{-++}$. These are the conormal bundles to Schubert cells in $T^*\cp^2$ with respect to a specific choice of flag.

After turning on the $\wt \Omega$ background, Neumann boundary conditions define modules for the quantized operator algebra $\hat \C[\CM_H]$. The quantization depends on $t_\C=k_t\epsilon$.
Specifically, a Neumann boundary condition $\CN_\varepsilon$ for an abelian theory produces an irreducible module $\hat \CN_\varepsilon^{(H)}$ for the algebra $\hat\C[\CM_H]$, whose weight spaces have multiplicity one and are in 1-1 correspondence with the internal lattice points of the chamber $\Delta_\varepsilon$ in the quantum arrangement. Each state in the module represents a particular chiral operator on the boundary. If the chamber $\Delta_\varepsilon$ is not $k_t$-feasible in the quantum arrangement, then the boundary condition breaks supersymmetry, in the sense that there exist no chiral operators with appropriate gauge charges (depending on $k_t$) that survive at the boundary.

Consider again our SQED example with (say) $k_t = \frac72$. We find
\begin{itemize}
\item $\hat\CN_{---}^{(H)}\simeq \hat\Delta_{---}$: the finite-dimensional $\bar{\mb 6}$ of $\mathfrak{sl}_3$, generated from a lowest-weight vector by $F_2$ and $F_3$;
\item $\hat\CN_{--+}^{(H)}\simeq \hat \Delta_{--+}$: an infinite-dimensional irreducible representation that is a quotient of two Verma modules, generated from a lowest-weight vector by $F_1$ and $F_2$;
\item $\hat\CN_{+-+}^{(H)}\simeq \hat\Delta_{+-+}$: an irreducible Verma module, freely generated from a lowest-weight vector by $F_1$ and $F_3$;
\end{itemize}

\begin{figure}[htb]
\centering
\includegraphics[width=1.7in]{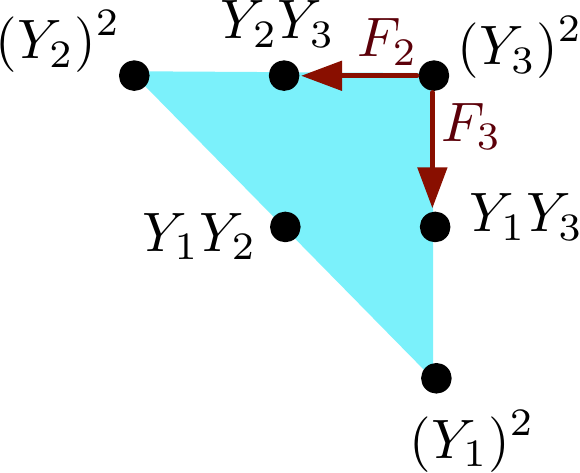}
\caption{Weight spaces for $\hat\CN_{---}^{(H)}$\,.}
\label{fig:N---}
\end{figure}

\noindent together with four other infinite-dimensional irreducible modules of a similar form.
Note that in all these cases setting $k_t = \frac72$ requires the introduction of a Wilson loop in addition to the usual R-symmetry redefinition to avoid the axial anomaly. For example, for $\hat \CN_{---}^{(H)}$, the R-symmetry redefinition alone would set $t_\C = \frac32\epsilon$, and an additional Wilson line of charge $2$ is required to achieve $t_\C = \frac72\epsilon$. The weight spaces of the module $\hat\CN_{---}^{(H)}$ each correspond to a boundary chiral operator formed from the $Y_i$ and with total gauge charge $2$, which can exist at the end of the Wilson line; the operators are shown in Figure \ref{fig:N---}. For general $t_\C = (n+\frac32)\epsilon$, the weight spaces of $\hat\CN_{---}^{(H)}$ correspond to $Y_1^{n_1}Y_2^{n_2}Y_3^{n_3}$, with $\sum_i n_i = n$.

The three modules $\hat\Delta_{---}$, $\hat\Delta_{--+}$, $\hat\Delta_{+-+}$ above are all lowest-weight with respect to $m_\R=(-2,-1)$. Namely, if we decompose the bulk algebra according to $m_\R$-charge
\be \begin{array}{ll} \hat \C[\CM_H] &= \hat \C[\CM_H]_< \oplus \hat \C[\CM_H]_0\oplus \hat \C[\CM_H]_> \\[.1cm]
 & =\langle E_1,E_2,E_3\rangle \oplus \langle H_1,H_2\rangle \oplus \langle F_1,F_2,F_3\rangle\,, \end{array} \ee
the operators $E_1,E_2,E_3$ of negative charge all act nilpotently. We see from the quantum arrangement that the function $\hat h_m = m_R\cdot \hat \mu_{H,\C}$ as in \eqref{def-hathm} is bounded from below on the support of these modules. This is the quantum analogue of $m_\R$-feasibility.

\subsubsection{Exceptional Dirichlet boundary conditions}
\label{sec:abel-HxD}

Exceptional Dirichlet boundary conditions in abelian theories are labelled by a sign vector $\varepsilon$ and a subset $S\subset \{1,...,N\}$ of size $r$ such that the corresponding $r\times r$ submatrix $Q^{(S)}$ of gauge charges is nondegenerate. The boundary condition sets
\be \label{abel-xD}
 \CD_{\varepsilon,S} \,:\quad \begin{cases} Y_i\big| = c_i   & \varepsilon_i = + \\
 X_i\big| = c_i & \varepsilon_i = -\end{cases}\quad(i\in S)\,,\qquad  \begin{cases} Y_i\big| = 0   & \varepsilon_i = + \\
 X_i\big| = 0 & \varepsilon_i = -\end{cases}\quad(i\notin S)\,,
\ee
with nonzero $c_i$, together with the usual $\varphi\big|_\pd=\varphi_0$ in order allow the hypers with $i\in S$ to get vevs. This fully breaks the gauge symmetry and preserves a $G_H$ flavor symmetry at the boundary.

The classical Higgs-branch image of this boundary condition is easy to describe, at least at generic values of complex FI parameters $t_\C$. The image is confined to the slice $\S_S$ that intersects the $r'$ hyperplanes $\CH_i$ with $i\notin S$. Recall that the hyperplanes cut the base of the slice into orthants $V_{S,\varepsilon}$ \eqref{orthantV}, and cut the slice itself into $2^{r'}$ copies of $\C^{r'}$. The image of $\CD_{\varepsilon,S}$ is simply the copy of $\C^{r'}$ fibered over the orthant $V_{S,\varepsilon}$,
\be \CD_{\varepsilon,S}\quad\leadsto\quad \CD_{\varepsilon,S}^{(H)} \;=\;\text{toric}(V_{S,\varepsilon}). 
\ee
The image depends only on the signs $\varepsilon_i$ for $i\notin S$.

In our running example of SQED, the images of $\CD_{+--,\{1\}}$ and $\CD_{---,\{1\}}$ coincide, and are shown in Figure \ref{fig:cp2S1}.

We similarly expect that the module $\hat \CD_{\varepsilon,S}$ is a Verma module $\hat V_{S,\varepsilon}$. At least, this should be the result at generic $t_\C$. At quantized values of $t_\C$, extra structure may appear, which depends on the signs $\varepsilon_i$ with $i\in S$ (\ie\ on which chirals are given boundary vevs). To clarify the situation, we take a moment to study the boundary chiral ring and its quantization in the presence of an exceptional Dirichlet boundary condition. In the process, we identity the mirrors of the boundary monopole operators from Section~\ref{sec:bdymon}.

With a Dirichlet boundary condition $\CD_{\varepsilon,S}$, the chiral operators that can fluctuate on the boundary are $X_i$ for $\varepsilon_i=+$ and $Y_i$ for $\varepsilon_i=-$. Let us introduce the notation
\be (X_{\varepsilon,i},\, Y_{\varepsilon,i}) := \begin{cases} (X_i,\,Y_i) & \varepsilon_i = + \\ (Y_i,-X_i) & \varepsilon_i = - \end{cases}\,,\ee
so that the fluctuating fields are $X_{\varepsilon,i}$. (In the previous Section \ref{sec:N}, \ref{sec:D}, \ref{sec:xD}, we would have called these $X_{L,i}$.) The $X_{\varepsilon,i}$ are not all independent, due to the complex moment-map constraints
\be (\mu_{a,\C} +t_{a,\C})\big|_\pd = \sum_{i\in S} Q_a{}^i c_i X_{\varepsilon,i}\big|_\pd + t_{a,\C} = 0\,, \ee
or schematically $(Q^{(S)})\cdot (cX_\varepsilon) +t_\C = 0$. Since $Q^{(S)}$ is nondegenerate, all the $X_{\varepsilon,i}$ with $i\in S$ are fixed in terms of the $t_\C$. Thus the boundary chiral ring is generated by the $X_{\varepsilon,i}$ with $i\notin S$. These operators parameterize the image $\CD_{\varepsilon,S}^{(H)}\simeq \C^{r'}$ described above.

If we further assume that $Q^{(S)}$ is unimodular (so the Higgs branch is smooth around the vacuum $\nu_S$, as on page \pageref{page:unimod}), then we may equivalently take as generators for the boundary chiral ring the operators
\be \label{abel-defw}
\mb w^A := \prod_{i\in S}c_i^{-\varepsilon_i\wt Q_A^i} \prod_{i\notin S}  (X_{\varepsilon,i})^{\varepsilon_i \wt Q_A^i} \qquad\text{for $A\in \Z^{r'}$ s.t. $\varepsilon_i\wt Q_A^i \geq 0$ $\forall\, i\notin S$}\,.
\ee
Note that these only make sense if $\varepsilon_i\wt Q_A^i \geq 0$ for all $i\notin S$.%
\footnote{To see that all the $X_{\varepsilon,i}$ can indeed be expressed in terms of the $\mb w^A$, we use the fact that $Q^{(S)}$ is unimodular to perform a change of basis on the gauge and flavor charges so that $Q^{(S)}=1\!\!1_{r\times r}$; then in the exact sequence \eqref{eq:exact}, we can choose $\wt Q$ so that its submatrices with $i\in S$ and $i\notin S$ are $\wt Q^{(S)}=0$ and $\wt Q^{(\notin S)}=1\!\!1_{r'\times r'}$. In this case, either
$\mb w^A \sim \delta^{Ai}X_{\varepsilon,i}$ or $\mb w^{-A} \sim \delta^{Ai}X_{\varepsilon,i}$.} %
The boundary OPE is $\mb w^A\mb w^B = \mb w^{A+B}$.

In our running example, we expect that the boundary condition $\CD_{+--,\{1\}}$ has $Y_2$ and $Y_3$ as unconstrained operators generating the boundary chiral ring. The canonical prescription in \eqref{abel-defw} reproduces this result: we have
\be \begin{array}{l@{\qquad}l}\mb w^{(1,0)} = c^{-1}Y_3\,, & \mb w^{(-1,0)} = c Y_3^{-1}\,, \\
\mb w^{(0,1)} = Y_2^{-1}Y_3\,, & \mb w^{(0,-1)} = Y_2 Y_3^{-1}\,, \\
\mb w^{(1,-1)} = c^{-1}Y_2\,, & \mb w^{(-1,1)} = c Y_2^{-1}\,,\end{array}
\ee
and only keep $\mb w^{(1,0)} = c^{-1}Y_3$ and $\mb w^{(1,-1)} = c^{-1}Y_2$, since they are the operators with $\varepsilon_i \wt Q^i\cdot A\geq 0$ for $i=2,3$.
The analysis for $\CD_{---,\{1\}}$ is identical, with $c\to c^{-1}$.

The restriction of bulk chiral operators to the boundary is also easy to calculate. First, we have 
\be z_i\big|_\pd  =  0\qquad (i\notin S)\,, \qquad z_i\big|_\pd = -(Q^{(S)})^{-1}{}_i{}^a t_{a,\C}\qquad (i\in S)\,, \label{abel-zbdy} \ee
where for $i\in S$ the boundary vevs are determined by the moment-map constraint $Q^{(S)}\cdot z\big|_\pd + t_\C = 0$. For operators $w^A$ with flavor charges, we find
\be w^A\big|_\pd \,=\, \prod_i (z_i)^{(-\ol\varepsilon_i \wt Q_A^i)_+} \mb w^A\,, \label{abel-wbdy} \ee
with $(a)_+ = \max(a,0)$ as usual, and a new sign vector $\ol\varepsilon$ defined as
\be \ol\varepsilon_i = \begin{cases} -\varepsilon_i & i\in S \\ \varepsilon_i & i\notin S\,.\end{cases} \label{bare} \ee
Formula \eqref{abel-wbdy} bears a striking resemblance to \eqref{vw}. Indeed, in abelian theories the $\mb w^A$ are the Higgs-branch mirrors of boundary monopole operators. Together, \eqref{abel-zbdy} and \eqref{abel-wbdy} imply that $w^A\big|_\pd = 0$ if $\varepsilon_i \wt Q_A^i < 0$ for any $i\notin S$.

Upon introducing the $\wt\Omega$-background, each nontrivial boundary operator $\mb w^A$ defines a state $|A\rangle$. We obtain a (left) module with basis
\be \hat \CD_{\varepsilon,S} \,:\qquad |A\rangle\qquad\text{for $A\in \Z^{r'}$ s.t. $\varepsilon_i \wt Q_A^i\geq 0$ $\forall\,i\notin S$}\,. \label{abel-qxD} \ee 
In terms of our general analysis of Dirichlet boundary conditions on the Higgs branch from Section \ref{sec:qDH}, we can construct this module by starting with a module for the Heisenberg algebra with basis $p(X_{\varepsilon,i})\big|$, then quotienting by all states of the form $(\hat\mu_C + t_\C)p(X_{\varepsilon,i})\big|$. The states $|A\rangle = \mb w^A\big| = \prod_{i\in S}c_i^{-\varepsilon_i\wt Q_A^i} \prod_{i\notin S}  (X_{\varepsilon,i})^{\varepsilon_i \wt Q_A^i}\Big|$ are transverse to the orbits of the equivalence relation, and constitute a basis for the quotient module.

It is a straightforward combinatorial exercise to work out the action of the bulk algebra on $|A\rangle$. For uncharged operators we find
\be  \hat z_i|0\rangle = \tfrac12\varepsilon_i\, \epsilon |0\rangle\,,\qquad  \hat z_i |A\rangle = \big(\tfrac12\varepsilon_i + \wt Q_A^i\big)\,\epsilon |A\rangle\qquad (i\notin S) \label{abel-qxD-z} \ee
whereas for $i\in S$ the eigenvalue of $\hat z_i$ acting on $|A\rangle$ is fixed by the moment-map relations $(\hat \mu_{a,\C}+t_{a,\C})|A\rangle = (\sum_{i} Q_a{}^i \hat z_i +t_\C)|A\rangle = 0$. The charged operators then act as
\begin{align} \label{abel-qxD-w}
 \hat w^A|B\rangle  &= \prod_i (\ol \varepsilon_i)^{(\wt Q_A^i)_+} [\ol \varepsilon_i \hat z_i]^{(-\ol\varepsilon_i \wt Q_A^i)_+}|A+B\rangle  \\ &= \prod_{\text{$i$ s.t. $\ol\varepsilon_i\wt Q_A^i <0$}} [\hat z_i]^{-\wt Q_A^i}|A+B\rangle\,.\notag \end{align}
Again, this mirrors the Coulomb-branch relation \eqref{vw-mod}.

\begin{figure}[htb]
\centering
\includegraphics[width=6in]{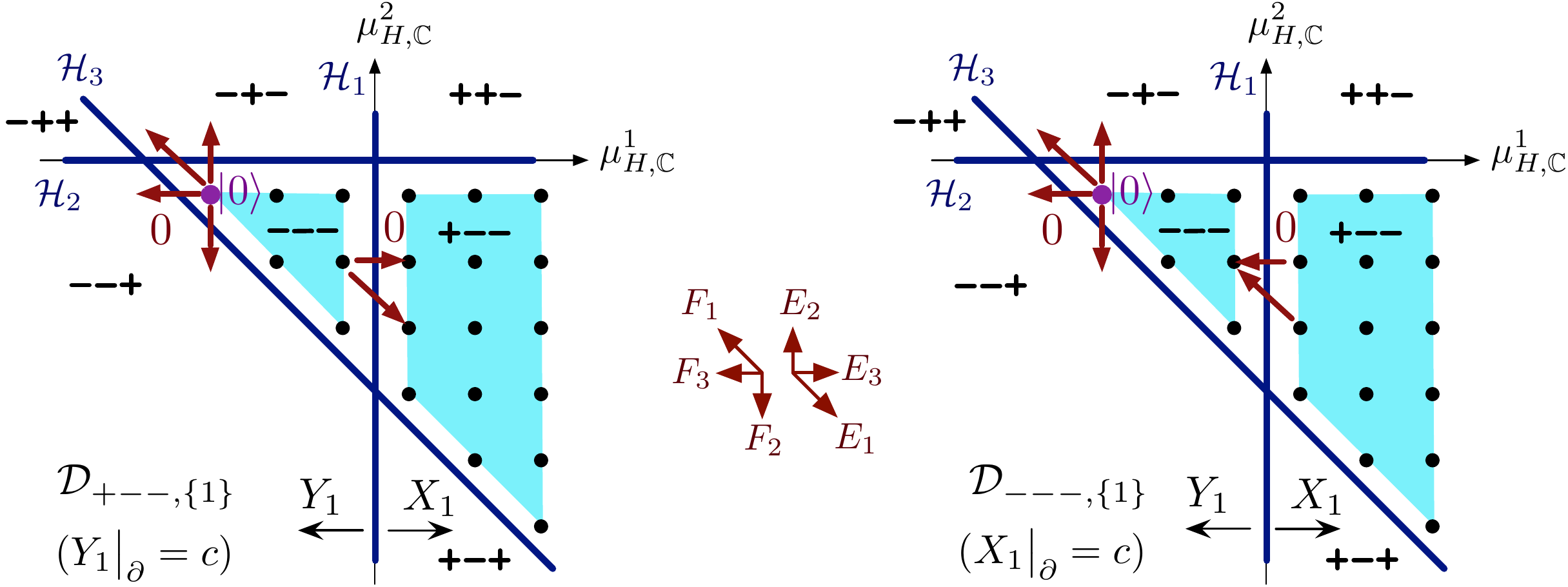}
\caption{The Higgs-branch modules defined by exceptional Dirichlet boundary conditions $\hat \CD_{\pm--,\{1\}}$ in SQED. For generic $t_\C$ they are both irreducible Verma modules with weight spaces supported in the shaded orthant. When $t_\C=k_t\epsilon$ with $k_t\geq\frac32 $ ($t_\C = \frac72\epsilon$ is shown) the modules decompose as an extension of smaller irreducible modules, supported on individual chambers inside the orthant.}
\label{fig:cp2-Verma}
\end{figure}

The action \eqref{abel-qxD-w} implies that all operators $\hat w^A$ with $\varepsilon_i \wt Q_A^i<0$ for some $i\notin S$ annihilate the identity $|0\rangle$. These are simply the operators for which $A$ points out of the orthant $V_{S,\varepsilon}$. Moreover, as long as FI parameters $t_\C$ are generic, the action of $\hat w^A$ for $A$ pointing into the orthant is never zero. Thus, comparing \eqref{abel-qxD-z} to \eqref{modV-def}, we find that the exceptional Dirichlet boundary condition precisely produces the Verma module $\hat V_{S,\varepsilon}$,
\be \hat \CD_{\varepsilon,S}^{(H)}\;\simeq \; \hat V_{S,\varepsilon}\,. \ee
The states $|B\rangle$ fill out the orthant $V_{S,\varepsilon}$ in the quantum hyperplane arrangement, and the module is irreducible. We depict the modules $\hat \CD_{\pm--,\{1\}}$ for SQED in Figure \ref{fig:cp2-Verma}: the operators $E_2,F_1,F_2,F_3$ all annihilate the identity $|0\rangle$, and the modules are freely generated from the identity by $E_1,E_3$.

If the FI parameters are fixed to quantized values, the module $\hat \CD_{\varepsilon,S}^{(H)}$ may no longer be irreducible, and it may not be a Verma module. In terms of the quantum hyperplane arrangement, we deduce from \eqref{abel-qxD-w} that an operator $\hat w^A$ acts as zero when crossing any hyperplane $\CH_i$ if $\ol\varepsilon_i\wt Q_A^i< 0$. This means
\begin{itemize}
\item $\hat w^A$ is zero if it moves us out of the orthant $V_{S,\varepsilon}$ (as before); and
\item $\hat w^A$ is also zero if it crosses $\CH_i$ ($i\in S$) toward the $-\ol\varepsilon_i=\varepsilon_i$ side of the hyperplane, \ie\ the side where $X_{\varepsilon,i}$ could (classically) get a vev.
\end{itemize}
From the second property, we see that $\hat \CD_{\varepsilon,S}^{(H)}$ is reducible if and only if additional hyperplanes $\CH_i$ ($i\in S$)  intersect the orthant $V_{S,\varepsilon}$.

In our example, we consider again the modules $\hat \CD_{\pm--,\{1\}}$ in Figure \ref{fig:cp2-Verma} and set $t_\C = k_t\epsilon$, $k_t\in \Z+\frac12$. Both modules are supported on the same orthant of the quantum arrangement, which is intersected by the hyperplane $\CH_1$ so long as $k_t \geq  \frac32$. (If $k_t\leq \frac12$, then $\CH_1$ does not intersect the interior of the orthant, and $\hat \CD_{\pm--,\{1\}}$ are automatically irreducible Verma modules.)
In $\hat \CD_{---,\{1\}}$, the operators $F_1$ and $F_3$ act as zero when crossing $\CH_1$, because they move toward the $\varepsilon_1=-$ side. Thus $\hat \CD_{---,\{1\}}$ is a reducible Verma module, freely generated from the identity by $E_1$ and $E_3$, which has an irreducible submodule  $\hat\Delta_{+--}$ and a finite-dimensional irreducible quotient $\hat \CD_{---,\{1\}}/\hat\Delta_{+--} \simeq \hat\Delta_{---}$. 
In contrast, in $\hat \CD_{+--,\{1\}}$ we find that $E_1,E_3$ are zero when crossing $\CH_1$, while $F_1,F_3$ are nonzero. Thus $\hat \CD_{+--,\{1\}}$ has $\hat\Delta_{---}$ as an irreducible submodule and $\hat\Delta_{+--}$ as an irreducible quotient. It is a costandard module rather than a Verma module.

In general, the module $\hat \CD_{\varepsilon,S}$ at quantized values of $t_\C$ will be a successive extension of the irreducible modules supported on the chambers inside the orthant $V_{S,\varepsilon}$. In other words, there is a filtration by submodules
\be \hat \CD_{\varepsilon,S}^{(H)} = M_n\supset M_{n-1}\supset\ldots\supset M_1\supset M_0 = \oslash\,.\label{xD-filter} \ee
such that each quotient $M_a/M_{a-1}$ is irreducible, supported on one of the chambers. The order in which the chambers appear depends on the $k_t$ (they determine how hyperplanes $\CH_i$ $i\in S$ intersect the orthant) and on the signs $\varepsilon_i$ for $i\in S$. 

To emphasize the individual modules that are successively extended to build $\hat \CD_{\varepsilon,S}^{(H)}$ in the filtration \eqref{xD-filter} we will write 
\be \hat \CD_{\varepsilon,S}^{(H)} = \big[\, M_n/M_{n-1}\,\big|\, M_{n-1}/M_{n-2}\,\big|...\big|\,M_2/M_1\,\big|\, M_1\,\big]. \label{xD-comp} \ee
When the subquotients $M_a/M_{a-1}$ are irreducible, as they are in this case, this filtration is called a \emph{composition series} of $\hat \CD_{\varepsilon,S}^{(H)}$ and the modules $M_a/M_{a-1}$ are called the \emph{composition factors} of $\hat \CD_{\varepsilon,S}^{(H)}$.

\subsubsection*{Thimbles, standards, and costandards}

Sometimes $\hat \CD_{\varepsilon,S}$ is a Verma (\emph{a.k.a.} standard) module even at quantized $t_\C$, meaning that it is freely generated from $|0\rangle$. We claimed in Section \ref{sec:standards} that this would be the case whenever we associated an exceptional Dirichlet boundary condition to a vacuum $\nu$ and a choice of $(m_\R,t_\R)$ using \eqref{nuLc}, and the quantized value of $t_\C$ was aligned with $-t_\R$. We can now prove this claim for abelian theories.

We first express \eqref{nuLc} geometrically, producing an assignment
\be (\nu;m_\R,t_\R) \qquad\leadsto\qquad (\varepsilon,S)\,. \label{nueS} \ee
We first set $t_\C=0$ and consider the classical canonical slice $\S_0$, in which $t_\R$ determines the relative positions of hyperplanes and $m_\R$ determines a direction. The vacuum $\nu=\nu_S$ lies at an intersection of $N-r$ hyperplanes, which define the subset $S$ (namely, $i\notin S$ iff $\CH_i$ intersects $\nu$).
Remember that every hyperplane $\CH_i$ has positive and negative sides, on which $X_i$ and $Y_i$ (respectively) can get vevs. The sign vector $\varepsilon$ is fixed by requiring that
\begin{itemize}
\item the vector $m_\R$ (or rather the potential $h_m$) is bounded from below on the orthant $V_{S,\varepsilon}$, which fixes $\varepsilon_i$ for $i\notin S$\,;
\item for $i\in S$, $\nu$ lies on the $\ol \varepsilon_i=-\varepsilon_i$ side of $\CH_i$ (this depends on $t_\R$).
\end{itemize}

Now consider the module $\hat \CD_{\varepsilon,S}$, with $\varepsilon,S$ associated to $(\nu;m_\R,t_\R)$ in this way. We choose quantized values of the complex FI parameters $t_\C = k_t\epsilon$. In the corresponding quantum hyperplane arrangement, the identity state $|0\rangle$ is a lattice point adjacent to the vacuum $\nu_S$. We argued above that the operators $\hat w^A$ that cross hyperplanes $\CH_i$ ($i\in S$) toward their $\varepsilon_i$ sides act as zero. Thus, in order for $\hat \CD_{\varepsilon,S}$ to remain a Verma module, the $\varepsilon_i$ sides of the these hyperplanes must all point toward the identity $|0\rangle$. This is true precisely if
\be k_t \sim -t_\R\,. \ee

Note that while $m_\R,t_\R$ are continuous, the assignment $(\nu;m_\R,t_\R)\leadsto (\varepsilon,S)$ only depends on $m_\R,t_\R$ in a piecewise constant manner. Thus the spaces of mass and FI parameters are divided into chambers on which the assignment is constant. What we mean by $k_t \sim -t_\R$ is that $k_t$ is in the same chamber as $-t_\R$.%
\footnote{A precise discussion of chambers and alignment of parameters appears in Sections \ref{sec:shuffle} and \ref{sec:corresp}.}

In the opposite regime $k_t\sim +t_\R$, then the $\varepsilon_i$ sides of the hyperplanes $\CH_i$ ($i\in S$) all point away from the identity $|0\rangle$. In this case, $\hat \CD_{\varepsilon,S}$ is not a Verma module but a costandard module, as defined in Section \ref{sec:standards}: it is freely co-generated by $|0\rangle$.

\subsubsection{Generic Dirichlet boundary conditions}
\label{sec:abel-HD}

The data of a generic Dirichlet boundary condition is simply encoded in a sign vector $\varepsilon\in \{\pm\}^N$. We set
\be \CD_{\varepsilon}:\quad \varphi\big|_\pd = 0\,,\qquad \begin{cases} X_i\big|_\pd = c_i & \varepsilon_i = - \\
Y_i\big|_\pd = c_i & \varepsilon_i = +\,,\end{cases} \label{Dec}
\ee
with all $c_i$ nonzero. An easy calculation shows that, at the boundary, the generators of the chiral ring satisfy
\be \hspace{0in} \CD_\varepsilon^{(H)}\,:\qquad w^A\big|_\pd = \tilde \xi^A  \prod_{1\leq i\leq N} z_i^{(\varepsilon_i\wt Q_A^i)_+}\Big|_\pd \qquad (\forall\;A\in \Z^{r'})\,, \label{Dec-H} \ee
where as usual $(x)_+ = \max(x,0)$ and we have introduced
\be \wt\xi^\alpha := \prod_{1\leq i\leq N} c_i^{-\varepsilon_i \wt Q_\alpha^i}\qquad (1\leq \alpha\leq r')\,,
\ee
so that $\wt\xi^A = \prod_{\alpha} (\wt\xi^\alpha)^{A^\alpha} =  \prod_{i} c_i^{-\varepsilon_i\wt Q_A^i}$.
The $\wt\xi^\alpha$ are independent gauge-invariant monomials in the $c_i^{\pm 1}$ that were introduced heuristically in Section \ref{sec:DH}. They are the only combinations of the $c_i$ that can appear in chiral-ring equations. As long as $t_\C$ is generic, \eqref{Dec} defines the image $\CD_{\varepsilon}^{(H)}$ of the Dirichlet boundary condition on the Higgs branch.

In an $\wt \Omega$-background, the boundary condition \eqref{Dec} assures that the identity state `$|$' obeys
\begin{align} \label{Dec-Hq} 
\hat w^A\big| \;&=\; \tilde \xi^A  \prod_{\text{$i$ s.t. $\varepsilon_i\wt Q_A^i> 0$}} [\hat z_i]^{-\wt Q_A^i}\Big| \\
 &= \;\wt \xi^A \prod_{1\leq i\leq N} (\varepsilon_i)^{(-\wt Q_A^i)_+}  [\epsilon_i \hat z_i]^{-(\varepsilon_i\wt Q_A^i)_+}\big| \qquad \text{(equivalently)}\,, \hspace{-.5in} \notag
\end{align}
with the usual convention for $[z]^b$ \eqref{q-exp-abel}. The relations \eqref{Dec-Hq} define a left ideal $\CI_\varepsilon$ in the algebra $\hat \C[\CM_H]$. As described at the end of Section \ref{sec:qDH}, the module $\hat \CD_{\varepsilon,c}^{(H)}$ has the abstract form
\be \hat \CD_{\varepsilon}^{(H)} \simeq \hat \C[\CM_H]/\CI_\varepsilon\,.\ee

For example, for SQED with three hypermultiplets, the algebra $\hat C[\CM_H]$ is generated by the operators $E_i,F_i,H_i$ described in \eqref{EFH3}, or equivalently $E_i,F_i$ and the $\hat z_i$ subject to $\hat z_1+\hat z_2+\hat z_3+t_\C=0$.
The eight basic Dirichlet boundary conditions produce ideals
\be \label{DforSQED}
\hspace{-.5in}
\begin{array}{l@{\quad}l}
\hat \CD_{+++}^{(H)}:\; \begin{array}{l@{\quad}l}
 E_1 = \frac{\tilde\xi_1}{\tilde \xi_2}  (z_1-\frac\epsilon2)  &   F_1 = \frac{\tilde\xi_2}{\tilde \xi_1} (z_2-\frac\epsilon2) \\[.1cm]
 E_2 = \tilde \xi_2 (z_2-\frac\epsilon2)  &   F_2 =  \tilde\xi_2^{-1} (z_3-\frac\epsilon2)  \\[.1cm]
 E_3 = \tilde \xi_1 (z_1-\frac\epsilon2) &   F_3 =  \tilde\xi_1^{-1} (z_3-\frac\epsilon2)  \end{array}
& \hat \CD_{---}^{(H)}:\; \begin{array}{l@{\quad}l}
 E_1 = \frac{\tilde\xi_1}{\tilde \xi_2}   (z_2+\frac\epsilon2)  &   F_1 = \frac{\tilde\xi_2}{\tilde \xi_1}  (z_1+\frac\epsilon2) \\[.1cm]
 E_2 = \tilde \xi_2 (z_3+\frac\epsilon2)  &   F_2 =  \tilde\xi_2^{-1} (z_2+\frac\epsilon2)  \\[.1cm]
 E_3 = \tilde \xi_1 (z_3+\frac\epsilon2) &   F_3 =  \tilde\xi_1^{-1} (z_1+\frac\epsilon2)  \end{array} \\[1cm]
\hat \CD_{++-}^{(H)}:\; \begin{array}{l@{\quad}l}
 E_1 = \frac{\tilde\xi_1}{\tilde \xi_2}  (z_1-\frac\epsilon2)  &   F_1 = \frac{\tilde\xi_2}{\tilde \xi_1} (z_2-\frac\epsilon2) \\
 E_2 = \tilde \xi_2 (z_2-\frac\epsilon2)(z_3+\frac\epsilon2)  &   F_2 =  \tilde\xi_2^{-1} \\
 E_3 = \tilde \xi_1 (z_1-\frac\epsilon2)(z_3+\frac\epsilon2) &   F_3 =  \tilde\xi_1^{-1}  \end{array}
& \hat \CD_{--+}^{(H)}:\; \begin{array}{l@{\quad}l}
 E_1 = \frac{\tilde\xi_1}{\tilde \xi_2}  (z_2+\frac\epsilon2)  &   F_1 = \frac{\tilde\xi_2}{\tilde \xi_1} (z_1+\frac\epsilon2) \\
 E_2 = \tilde \xi_2  &   F_2 =  \tilde\xi_2^{-1} (z_2+\frac\epsilon2)(z_3-\frac\epsilon2)   \\
 E_3 = \tilde \xi_1  &   F_3 =  \tilde\xi_1^{-1} (z_1+\frac\epsilon2)(z_3-\frac\epsilon2)  \end{array} \\[.9cm]
\hat \CD_{+-+}^{(H)}:\; \begin{array}{l@{\quad}l}
 E_1 = \frac{\tilde\xi_1}{\tilde \xi_2}  (z_1-\frac\epsilon2)(z_2+\frac\epsilon2)  &   F_1 = \frac{\tilde\xi_2}{\tilde \xi_1}  \\
 E_2 = \tilde \xi_2  &   \hspace{-1.5cm} F_2 =  \tilde\xi_2^{-1} (z_2+\frac\epsilon2) (z_3-\frac\epsilon2)  \\
 E_3 = \tilde \xi_1 (z_1-\frac\epsilon2) & \hspace{-1.5cm}   F_3 =  \tilde\xi_1^{-1} (z_3-\frac\epsilon2)  \end{array}
& \hat \CD_{-+-}^{(H)}:\; \begin{array}{l@{\quad}l}
 E_1 = \frac{\tilde\xi_1}{\tilde \xi_2}    &   \hspace{-1.4cm}F_1 =  \frac{\tilde\xi_2}{\tilde \xi_1} (z_1+\frac\epsilon2)(z_2-\frac\epsilon2) \\
 E_2 = \tilde \xi_2 (z_2-\frac\epsilon2)(z_3+\frac\epsilon2)  &   F_2 =  \tilde\xi_2^{-1}   \\
 E_3 = \tilde \xi_1 (z_3+\frac\epsilon2) &   F_3 =  \tilde\xi_1^{-1} (z_1+\frac\epsilon2)  \end{array} \\[.9cm]
\hat \CD_{-++}^{(H)}:\; \begin{array}{l@{\quad}l}
 E_1 = \frac{\tilde\xi_1}{\tilde \xi_2}  &   F_1 = \frac{\tilde\xi_2}{\tilde \xi_1} (z_1+\frac\epsilon2)  (z_2-\frac\epsilon2) \\
 E_2 = \tilde \xi_2 (z_2-\frac\epsilon2)  &   F_2 =  \tilde\xi_2^{-1} (z_3-\frac\epsilon2)  \\
 E_3 = \tilde \xi_1 &   F_3 =  \tilde\xi_1^{-1} (z_1+\frac\epsilon2) (z_3-\frac\epsilon2)  \end{array}
& \hat \CD_{+--}^{(H)}:\; \begin{array}{l@{\quad}l}
 E_1 = \frac{\tilde\xi_1}{\tilde \xi_2}  (z_1-\frac\epsilon2)(z_2+\frac\epsilon2)  &   F_1 = \frac{\tilde\xi_2}{\tilde \xi_1}  \\
 E_2 = \tilde \xi_2 (z_3+\frac\epsilon2)  &   F_2 =  \tilde\xi_2^{-1} (z_2+\frac\epsilon2)  \\
 E_3 = \tilde \xi_1 (z_1-\frac\epsilon2) (z_3+\frac\epsilon2) &   F_3 =  \tilde\xi_1^{-1}  \end{array} \\[.9cm]
\end{array}
\ee

The classical images $\CD_\varepsilon^{(H)}$ are not contained in any slice $\S$ of the Higgs branch, and the modules $\hat \CD_\varepsilon^{(H)}$ are not obviously weight modules. Thus, naively, it does not seem that hyperplane arrangements are relevant here. However, if we take particular limits that send to $\wt \xi^\alpha$ to zero or infinity as in Section \ref{sec:DH-cinf}, the classical images \emph{do} become supported on slices. Correspondingly, if we allow infinite changes of basis involving particular Laurent series in $\wt\xi^\alpha$ or $(\wt\xi^\alpha)^{-1}$, the modules become isomorphic to weight modules. This phenomenon was first explored in the Coulomb-branch examples of Section \ref{sec:NC-qSQED}, and then more briefly for the Higgs branch in \ref{sec:DH-SQED}. We proceed to explain how the limits should work in general abelian theories, deferring some details to \cite{Hilburn}.

The limits we are interested in for right boundary conditions correspond to introducing parameters $m_\R$ and applying an infinite gradient flow with respect to $h_m=m_\R\cdot \mu_{H,\R}$. These limits have the same effect as setting
\be \wt\xi^\alpha = e^{\lambda\, m_\R^\alpha}\, \wt\xi^\alpha_0\qquad\text{or}\qquad \wt\xi^A = e^{\lambda(m_\R\cdot A)}\wt\xi_0^A \ee
for a real scaling factor $\lambda$, and sending $\lambda\to \infty$. (For left boundary conditions, one should send $\lambda\to 0$ instead.)

In the limit $\lambda\to\infty$, the classical relations \eqref{Dec-H} split into two sets, depending on the sign of $m_\R\cdot A$\,:
\be w^A\big|_\pd = 0 \qquad (m_\R\cdot A<0)\;;\qquad \prod_i z_i^{(\varepsilon_i\wt Q_A^i)_+}\big|_\pd = 0 \qquad (m_\R\cdot A > 0)\,.\ee
We assume that $m_\R$ is generic, so that as $A$ ranges over any finite set of generators for the chiral ring from Section \ref{sec:abel-Hring}, either $m_\R\cdot A<0$ or $m_\R\cdot A>0$. (Geometrically, this means that if we think of $m_\R$ as a direction in a hyperplane arrangement, it is not parallel to any hyperplane.) 

The equations for the $z_i$ have a finite number of solutions. Correspondingly, the support of $\CD_\varepsilon^{(H)}$ becomes restricted to a finite number of slices $\S$ in the Higgs branch, as we wanted. Each solution is characterized by the vanishing of $r'$ of the $z_i$'s, and hence can be labelled by a subset $S$ of size $r$.  A bit of further analysis shows that the solutions are in 1-1 correspondence with subsets $S$ with the
special property that the potential $h_m=m_\R\cdot \mu_{H,\R}$ is bounded from below on the orthant $V_{S,\varepsilon}\subset \S_S$. Therefore, $\CD_\varepsilon^{(H)}$ is supported on a union of the slices $\S_S$ for such $S$. The equations for the $w^A$ simply say that $w^A$ vanishes if $m_\R$ decreases in the direction $A$. Therefore, the image $\CD_\varepsilon^{(H)}$ is supported precisely on the orthants $V_{S,\varepsilon}$,
\be \CD_\varepsilon^{(H)}\;\overset{\lambda\to\infty}{\longrightarrow}\; \bigcup_S \;\text{toric}(V_{S,\varepsilon})\qquad \text{s.t. $h_m$ bounded below on $V_{S,\varepsilon}$}\,.\hspace{-.5in} \label{Dlimit} \ee
This image is manifestly $m_\R$-feasible, or empty.

\begin{figure}[htb]
\hspace{-.25in}
\includegraphics[width=6.5in]{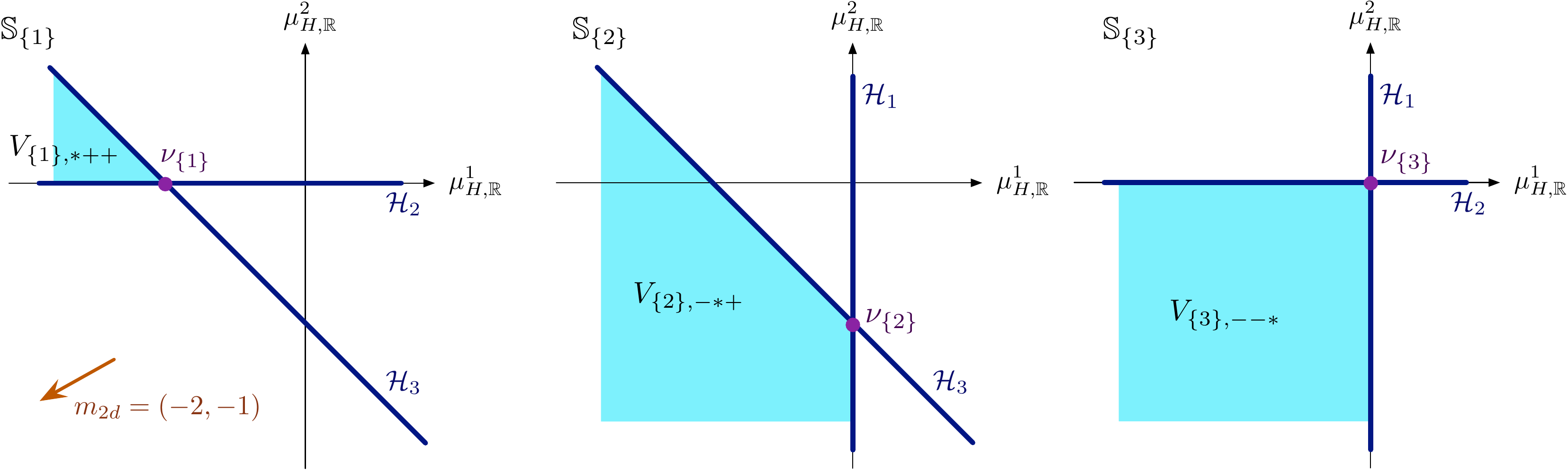}
\caption{The three orthants that contribute to the support of $\CD_\varepsilon^{(H)}$ when $\mtwod=(-2,-1)$.}
\label{fig:cp2D}
\end{figure}

In our SQED example, suppose we choose $m_\R = (-2,-1)$. This means that as $\lambda\to\infty$ we send $\wt\xi_1\to 0$, $\wt\xi_2\to 0$, and $\wt \xi_1/\wt\xi_2\to 0$.
There are three orthants that can potentially contribute to the support of Dirichlet boundary condition in the limit $\lambda\to \infty$, namely $V_{\{1\},*++}$, $V_{\{2\},-*+}$, and $V_{\{3\},--*}$. As shown in Figure \ref{fig:cp2D}, these are the orthants on which $m_\R$ (or more accurately the linear function $h_{2d}=m_\R\cdot \mu_{H,\R}$) is bounded from below. We very quickly deduce that
\be \label{DlimitSQED}
\begin{array}{l@{\qquad}l} \CD_{+++}^{(H)} = V_{\{1\},*++} & \CD_{---}^{(H)} = V_{\{3\},--*} \\[.1cm]
 \CD_{++-}^{(H)} = \oslash & \CD_{--+}^{(H)} = V_{\{2\},-*+}\,\cup\, V_{\{3\},--*} \\[.1cm]
 \CD_{+-+}^{(H)} = \oslash & \CD_{-+-}^{(H)} = \oslash \\[.1cm]
 \CD_{-++}^{(H)} = V_{\{1\},*++}\,\cup\,V_{\{2\},-*+} & \CD_{+--}^{(H)} = \oslash\,,
 \end{array}
\ee
simply by matching the potential orthants with the sign vector of the Dirichlet boundary condition. To verify that the process makes sense, consider (say) $\CD_{-++}$: by consulting \eqref{DforSQED} we see that sending $\lambda\to \infty$ forces $E_1=E_2=E_3=0$ and $z_1z_2=z_3=z_1z_3=0$. The latter has two solutions $z_1=z_3=0$ (so $S=\{2\}$) and $z_2=z_3=0$ (so $S=\{1\}$). For each of these solutions we choose an orthant where $E_1=E_2=E_3=0$, giving $V_{\{1\},*++}$ and $V_{\{2\},-*+}$, respectively. Thus $\CD_{-++}^{(H)} = V_{\{1\},*++}\,\cup\,V_{\{2\},-*+}$. Notice that taking $\lambda\to \infty$ pushes the images of some boundary conditions (such as $\CD_{++-}^{(H)}$) to infinity on the Higgs branch, giving empty support in \eqref{DlimitSQED}.

For modules, one sensible way to take the ``limit'' $\lambda\to \infty$ is to find an isomorphism between $\hat\CD_\varepsilon^{(H)}$ and a weight module, in such a way that the factors $\wt \xi^A$ appearing in the isomorphism have $\mtwod\cdot A$ bounded from above. 
This generalizes the notion of working over formal Laurent series from Section \ref{sec:NC-qSQED}. 
Equivalently, we may ask for an isomorphism between $\hat \CD_{\varepsilon}^{(H)}$ and a \emph{lowest}-weight module with respect to $m_\R$. 

If the complex FI parameters $t_\C$ are generic, then the result of this isomorphism can be achieved more directly by just sending $\lambda\to\infty$ exactly as in the classical case. We find that $\hat \CD_\varepsilon^{(H)}$ becomes a direct sum of irreducible Verma modules
\be \hat\CD_\varepsilon^{(H)}\;\overset{\lambda\to\infty}{\longrightarrow}\; \bigoplus_S \;\hat V_{S,\varepsilon}\qquad \text{s.t. $h_m$ bounded below on $V_{S,\varepsilon}$}\,,\hspace{-.5in} \label{qDlimit} \ee
which is the naive quantization of \eqref{Dlimit}.

If $t_\C=k_t\epsilon$ is quantized, the naive limit \eqref{qDlimit} is no longer correct: the different Verma modules in the direct sum begin interacting with one another. We defer the full explanation of this phenomenon to \cite{Hilburn}, simply postulating the result here. We expect that rather than being a direct sum, $\hat \CD_\varepsilon^{(H)}$ is an iterated extension of Verma modules; in other words there is a filtration by submodules
\be \hat \CD_\varepsilon^{(H)}  \overset{\lambda\to\infty}{\longrightarrow} M_n \supset \ldots M_1\supset M_0 = \oslash\label{qDextension} \ee
such that each successive quotient $M_a/M_{a-1}$ is isomorphic to one of the $\hat V_{S,\varepsilon}$ in \eqref{qDlimit}. In particular, the operators $\hat z_i$ can no longer be diagonalized, but rather acquire generalized weight spaces in which they act with nontrivial Jordan blocks.

The order in which the $\hat V_{S,\varepsilon}$ appear as quotients in \eqref{qDextension} is dictated by $m_\R$ and by~$k_t$. To each $\hat V_{S,\varepsilon}$ we can associate a vacuum $\nu_S$, the origin of the orthant $V_{S,\varepsilon}$ in the quantum hyperplane arrangement. Also recall that $m_\R$ defines a function $\hat h_m=m_\R\cdot \hat \mu_{H,\C}$ on the quantum arrangement. Then $\hat V_{S,\varepsilon}$ appears after $\hat V_{S',\varepsilon}$ if $\hat h_m(\nu_S) < \hat h_m(\nu_{S'})$. Using the notation of \eqref{xD-comp}, we may write
\be \hat\CD_\varepsilon^{(H)} \overset{\lambda\to\infty}{\longrightarrow} \big[\, \hat V_{S_n,\varepsilon}\,\big|... \big| \hat V_{S_2,\varepsilon}\,\big|\, \hat V_{S_1,\varepsilon}\,\big] \ee
where $\hat V_{S_i,\varepsilon}$ are the modules in \eqref{qDlimit}, in decreasing order of $\hat h_m(\nu_{S_i})$.

In our SQED example, if we choose $t_\C = k_t \epsilon$ with $k_t$ a half-integer and $m_\R==(-2,-1)$ as before, then the order of subsets (or equivalently, of vacua) is
\be \{3\}<\{2\}<\{1\}\quad (k_t> \tfrac12)\,,\qquad  \{1\}<\{2\}<\{3\}\quad (k_t< -\tfrac12)\,. \ee
Let us take $k_t>\frac12$ as usual. Then the modules defined by Dirichlet boundary conditions are isomorphic to lowest-weight modules (with respect to $m_\R$) of the form
\be \label{qDlimitSQED}
\begin{array}{c@{\qquad}c}
 \hat \CD_{+++}^{(H)} \simeq \hat V_{\{1\},*++} & \hat \CD_{---}^{(H)}\simeq \hat V_{\{3\},--*} \\[.1cm]
 0\to \hat V_{\{2\},-*+} \to \hat \CD_{-++}^{(H)} \to \hat V_{\{1\},*++}\to 0 &
  0\to \hat V_{\{3\},--*} \to \hat \CD_{--+}^{(H)} \to \hat V_{\{2\},-*+}\to 0\,. \end{array}
\ee
The remaining four boundary conditions produce modules that are not isomorphic to any lowest-weight modules with respect to $m_\R=(-2,-1)$.

\begin{figure}[htb]
\centering
\includegraphics[width=4in]{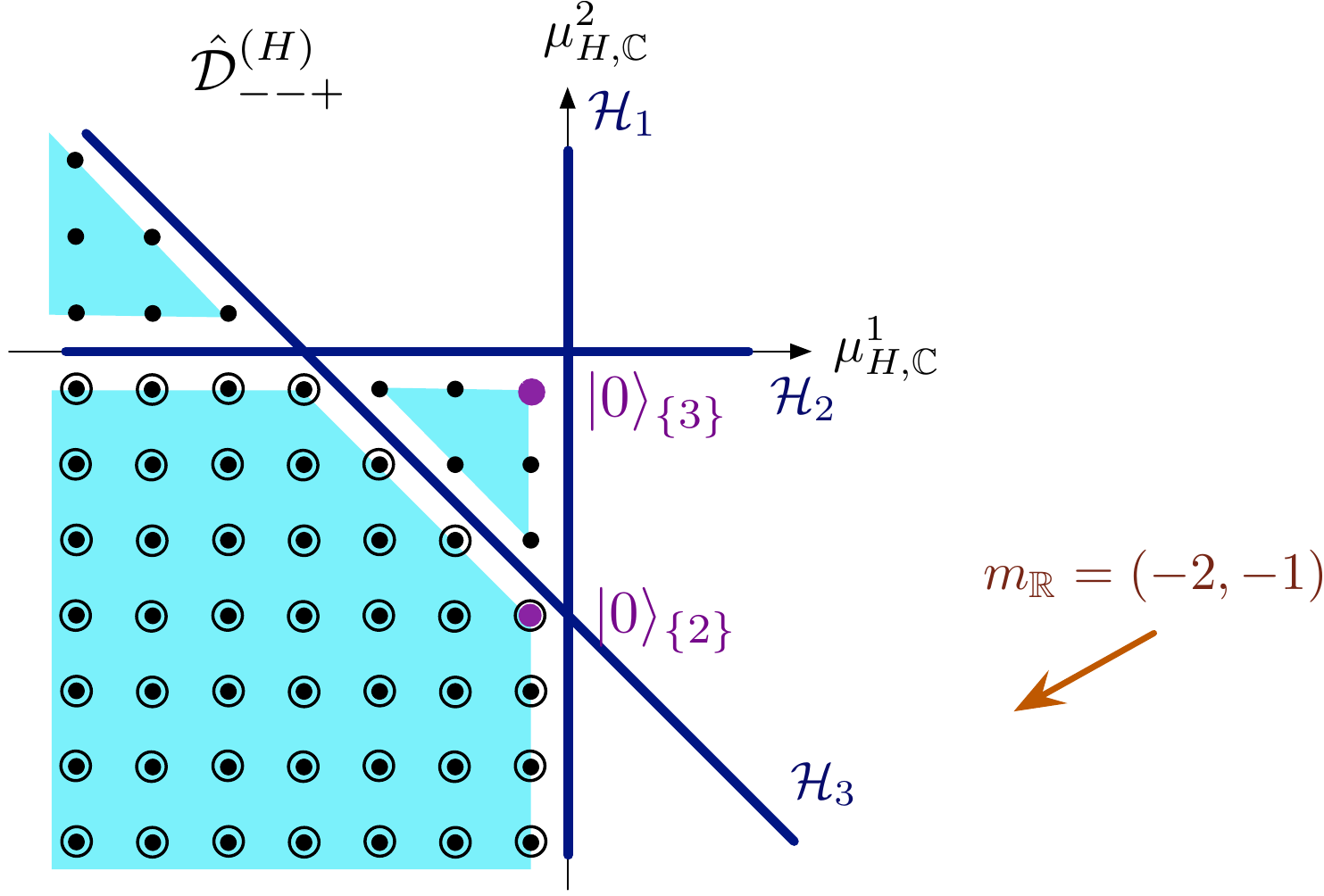}
\caption{The (generalized) weight spaces in the lowest-weight module isomorphic to $\hat \CD_{-++}^{(H)}$. Weight spaces of dimension two are depicted as circled lattice points.}
\label{fig:cp2qD}
\end{figure}

Let us look at $\hat \CD_{--+}^{(H)}$ in slightly more detail. The two Verma modules $\hat V_{\{2\},-*+}$ and $\hat V_{\{3\},--*}$ are freely generated from vectors $|0\rangle_{\{2\}}$ and $|0\rangle_{\{3\}}$ (respectively), which should obey
\be (\hat z_1+\tfrac\epsilon2)|0\rangle_{\{2\}}=(\hat z_3-\frac12)|0\rangle_{\{2\}}=0\,,\qquad
(\hat z_1+\tfrac\epsilon2)|0\rangle_{\{3\}}=(\hat z_2+\frac12)|0\rangle_{\{3\}}=0\,. \label{D-++V} \ee
In the naive $\lambda\to \infty$ of $\hat \CD_{--+}^{(H)}$, the identity state `$|$' satisfies $(\hat z_1+\tfrac\epsilon2)\big| = (\hat z_2+\tfrac\epsilon2)(\hat z_3-\tfrac\epsilon2)\big| = 0$ (reading off from \eqref{DforSQED}), so if we set
\be |0\rangle_{\{2\}} = (\hat z_3-\tfrac\epsilon2)\big|\,,\qquad |0\rangle_{\{3\}} = (\hat z_2+\tfrac\epsilon2)\big|\,. \label{D-++V12} \ee
the expected relations \eqref{D-++V} will be satisfied. The naive $\lambda\to\infty$ limit simply produces a direct sum $\hat V_{\{2\},-*+}\oplus \hat V_{\{3\},--*}$, whose weight spaces are depicted in Figure \ref{fig:cp2qD}. The more careful procedure of establishing an isomorphism between $\hat \CD_{--+}^{(H)}$ and a lowest-weight module leads leads to a module with the same weight spaces but with a modified action of bulk operators. In particular, acting on two-dimensional weight spaces, the $\hat z_i$ are no longer diagonal but have nontrivial Jordan blocks.

\subsection{Coulomb branch}
\label{sec:hypC}

The Coulomb branch of an abelian gauge theory can also be described using hyperplane arrangements, in a manner analogous to the preceding Higgs-branch discussions. In the infrared (\ie\, at infinite gauge coupling) the Coulomb branch is a hypertoric variety, equivalent to the Higgs branch of a mirror abelian theory.
Turning on a finite gauge coupling smoothly deforms the metric of the Coulomb branch to a generalized Taub-NUT metric, while preserving the topology and complex structure. Thus even at finite coupling many features of the Coulomb branch are encapsulated in hyperplane arrangements.

As a running example in this section and the next, we will consider $G=U(1)^2$ gauge theory with $N=3$ hypermultiplets of gauge and flavor charges
\be Q = \begin{pmatrix} 1 & 0 & -1 \\ 0 & 1 & -1 \end{pmatrix}\,,\qquad q = \begin{pmatrix} 0 & 0 & 1 \end{pmatrix}\,.\ee
This turns out to be the mirror of the SQED with $N=3$. The Higgs-branch flavor symmetry is $G_H=U(1)$, while the topological Coulomb-branch symmetry is $G_C = U(1)^2$. Thus there is a single set of mass parameters $(m_\R,m_\C)$ and there are two sets of FI parameters $(t_{a,\R},t_{a,\C})_{a=1,2}$.
The effective masses of the three hypermultiplets (which play the same role as $Z_i,z_i$ did for the Higgs branch) are
\be \begin{array}{l@{\qquad}l}
 M_\R^1 = \sigma^1\,,  &  M_\C^1 = \varphi^1\,, \\[.1cm]
 M_\R^2 = \sigma^2\,,  &  M_\C^2 = \varphi^2\,, \\[.1cm]
 M_\R^3 = m_\R-\sigma^1-\sigma^2\,,  &  M_\C^3 = m_\C- \varphi^1 -\varphi^2\,.
\end{array} \ee

\subsubsection{Hyperplane arrangements}
\label{sec:abel-Hhyp}

Our starting point is a description of the Coulomb branch as a fibration
\be
\CM_C \longrightarrow \mathbb{R}^{3r}
\ee
with typical fiber $(S^1)^r$. The base of the fibration is parametrized by the real and complex vectormultiplet scalars, which are the moment maps for the $G_C\simeq U(1)^r$ topological symmetry,
\be
\mu^a_{C,\R} = \sigma^a \qquad \mu^a_{C,\C}=\varphi^a \, .
\ee
The fibers are parametrized by the dual photons $\gamma^a$, which are rotated by $G_C$. Due to a standard 1-loop correction (\cf\ \cite{SW-3d, IS}), one fiber degenerates on each of the $N$ hyperplanes where the effective real and complex masses of each hypermultiplet vanish, $\CH^i = \{ M_\R^i = M_\C^i = 0\}$. We recall that
\be
M^i_\R = \sigma\cdot Q^i + m_\R\cdot q^i\,,\qquad M^i_\C = \varphi\cdot Q^i+m_\C \cdot q^i\,. \label{Mabel}
\ee

Consider our running example with $G=U(1)^2$. The Coulomb branch is an $(S^1)^2$ fibration over $\R^6$, with the base parameterized by $(\sigma^a,\varphi^a)_{a=1,2}$. The dual-photon circles degenerate along the three hyperplanes
\be
\begin{array}{rll}
\CH^1 & : \quad \sigma^1 = 0 \quad & \varphi^1 = 0   \\[.1cm]
\CH^2 & : \quad \sigma^2= 0 \quad & \varphi^2 = 0   \\[.1cm]
\CH^3 & : \quad \sigma^1+\sigma^2 = m_\R \qquad & \varphi^1+\varphi^2 = m_\C \, .
\end{array}
\ee
When $m_\C=0$, we recognize this as a topological description of $T^*\cp^2$.

As before, we are interested in slices $\mathbb S$ defined by fixed values of the complex moment maps $\varphi_a$. Such slices are fibrations 
\be
\mathbb S \longrightarrow \R^r
\ee
with fiber $(S^1)^r$ and base parameterized by the real moment maps $\sigma_a$. A generic slice does not intersect any of the hyperplanes and has the topology of $(\C^*)^r$. However, given a subset $S\in \{1,...,N\}$ of size $r$ such that the corresponding submatrix $Q^{(S)}$ is nondegenerate, we can choose $\varphi_a$ such that $M_\C^i=0$ for all $i\in S$. The corresponding slice $\S^S$ intersects the hyperplanes $\CH^i$ with $i\in S$ in real codimension one. The common intersection of all these hyperplanes on $\S^S$ is a single point $\nu_S$ --- it is the same vacuum that we described previously on the Higgs branch, which becomes massive if $Q^{(S)}$ is unimodular and generic FI parameters are turned on. The slice $\mathbb S^S$ is a union of $2^r$ toric varieties; if $Q^{(S)}$ is unimodular, they all have topology $\C^r$.

On the base of a slice $\CS_S$, each hyperplane $\CH^i$ ($i\in S$) has positive and negative sides, distinguished by $M_\R^i>0$ and $M_\R^i<0$, respectively. Thus the $2^r$ orthants of $\CS_S$ are each labelled by a sign vector,
\be V^{S,\varepsilon}\;:\quad \begin{array}{l}\text{orthant in $\S^S$ on the $\varepsilon_i$ side of $\CH_i$ for all $i\in S$\,,} \\ \text{\ie\ $\varepsilon_iM_\R^i>0$ for all $i\in S$}\,. \end{array} \ee

\begin{figure}[htb]
\centering
\includegraphics[width=5.8in]{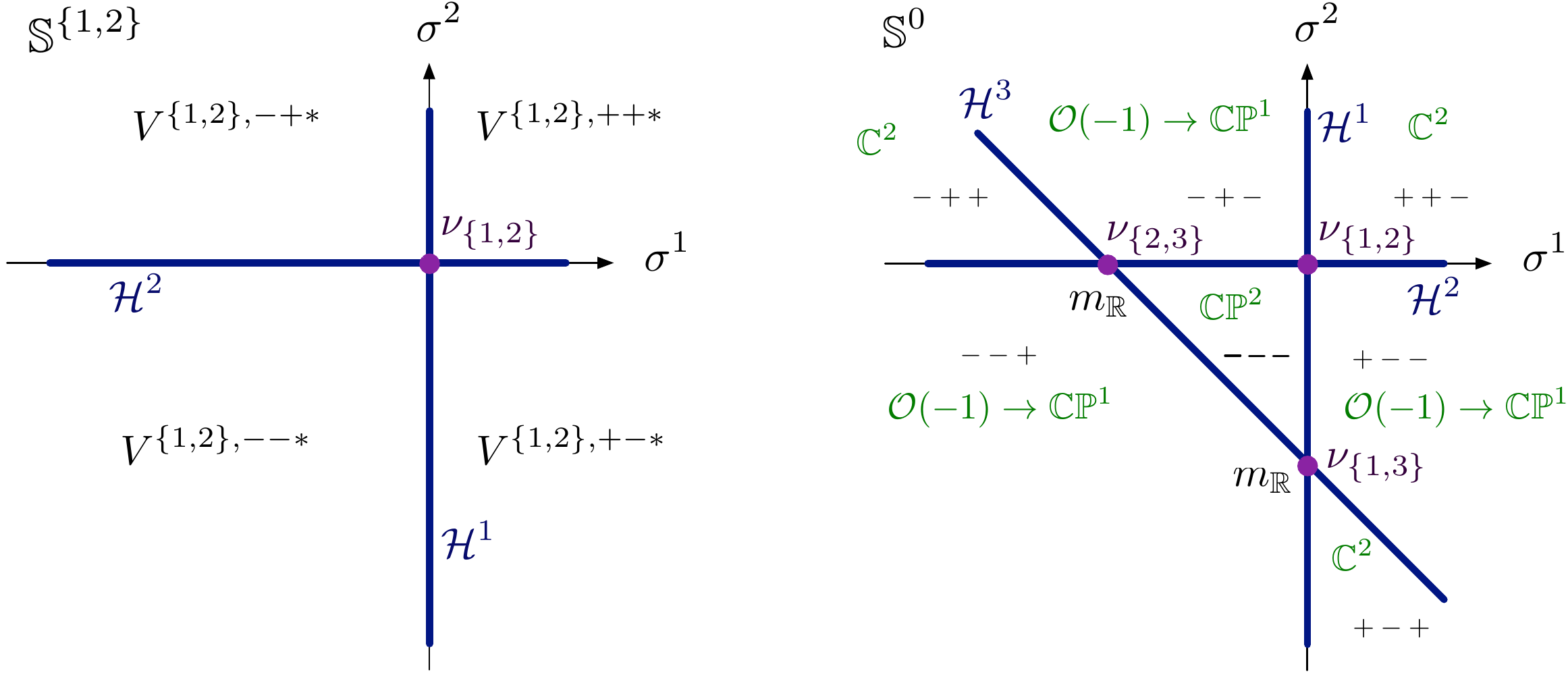}
\caption{Slices in the Coulomb branch for our $G=U(1)^2$ example: on the left the slice $\varphi^1=\varphi^2=0$ at generic complex mass $m_\C$; on the right, the canonical slice $\varphi^1=\varphi^2=0$  at $m_\C=0$. The real mass is negative, $m_\R<0$.}
\label{fig:cp2-C}
\end{figure}

Let us illustrate this in our example for the slice $\mathbb S^{\{1,2\}}$ defined by $\varphi^1=\varphi^2=0$. This is an $(S^1)^2$ fibration over $\R^2$, with the base parameterized by $(\sigma^1,\sigma^2)$. The two hyperplanes $\CH^1=\{\sigma^1=0\}$ and $\CH^2=\{\sigma^2=0\}$ intersect the slice. One factor $S^1\subset (S^1)^2$ degenerates along $\CH^1$ and the other along $\CH^2$, turning the slice into a union of four copies of $\C^2$, fibered over the four octants in the base. The hyperplane arrangement on the base is shown in Figure \ref{fig:cp2-C}.

When all complex masses are zero, there is a canonical slice $\mathbb S^0$ of the Coulomb branch defined by $\varphi_a=0$ for all $a=1,\ldots,r$. The canonical slice intersects all $N$ hyperplanes $\CH^i$ and is a union of toric varieties.  The hyperplanes cut the base into chambers $\Delta^\varepsilon$, which are labelled by sign vectors $\varepsilon\in\{\pm\}^N$ such that
\be \Delta^\varepsilon\,:\quad  \begin{array}{l}\text{chamber in $\S^0$ on the $\varepsilon_i$ side of $\CH_i$\,,} \\ \text{\ie\ $\varepsilon_iM_\R^i>0$ for all $i$}\,. \end{array}   \label{DeC} \ee
Again, the chambers where this condition has a nonempty solution are called feasible, or more precisely $m_\R$-feasible; feasibility depends on the choice of real masses. 

We illustrate the canonical slice for our example on the right of Figure \ref{fig:cp2-C}. The slice contains a union of a compact $\cp^2$ (fibered over $\Delta^{---}$), three copies of $\CO(-1)\to \cp^1$, and three copies of $\C^2$. For $m_\R<0$ ($m_\R>0$), the only infeasible chamber is $\Delta^{+++}$ ($\Delta^{---}$). This arrangement looks identical to the Higgs-branch hyperplane arrangement for SQED with $N=3$ hypermultiplets in Figure \ref{fig:cp2S0}, at positive FI parameter.

Turning on real FI parameters generates a real (super)potential on the Coulomb branch, of the form
\be h_t = t_\R\cdot \mu_{C,\R} \approx t_\R \cdot \sigma\,. \label{ht-abel} \ee
This is the real moment map for an infinitesimal subgroup $U(1)_t\subset G_C$ of the Coulomb-branch flavor symmetry, specified by $t_\R$.
On the base of the fibration $\CM_C\to \R^{3r}$ (and the base of any slice), $h_t$ is clearly a linear function. Its gradient defines a direction on each slice, which we simply refer to as $t_\R$.
The critical points of $h_t$, which are fixed points of $U(1)_t$, are the supersymmetric vacua $\nu_S$ of the theory. As discussed above, they lie at maximal intersections of $r$ hyperplanes $\CH^i$, $i\in S$.

\subsubsection{Mirror map}
\label{sec:abel-Cmirror}

The Coulomb-branch hyperplane arrangement for our $U(1)^2$ theory above looks identical to the Higgs-branch hyperplane arrangement for SQED with $N=3$ hypermultiplets. This is not a coincidence.

Suppose that we are studying the Coulomb branch of an abelian theory $T$, and want to exhibit it as the Higgs branch of another abelian theory $\wt T$. By counting dimensions of the moduli spaces, we see that if $T$ has gauge group $G=U(1)^r$ and $N$ hypermultiplets then $\wt T$ should have gauge group $\wt G = U(1)^{r'}=U(1)^{N-r}$ and $N$ hypermultiplets. By matching the structure of the Coulomb-branch hyperplane arrangement in $T$ with the Higgs-branch arrangement in $\wt T$, we find that gauge and flavor charge matrices must be related as
\begin{subequations}\label{mirrormap}
\be \begin{pmatrix} \wt q \\ \wt Q \end{pmatrix} = \begin{pmatrix} Q \\ q\end{pmatrix}^{-1,T}\,, \ee
along with $(\wt t_\R,\wt t_\C) = (-m_\R,-m_\C)$. This assures that if the moment maps for flavor symmetries are identified as $(\wt \mu_{H,\R},\wt \mu_{H,\C}) = (\sigma,\varphi)$, then the effective masses $(M_\R^i,M_\C^i)$ in theory $T$ map to the combinations $(\wt Z_i,\wt z_i)$ as in \eqref{eq:zZdef} in theory $\wt T$. Since we have identified the flavor symmetry $\wt G_H$ with $G_C$, we also have $(\wt m_\R,\wt m_\C) = (t_\R,t_\C)$. Thus, altogether
\be (\wt t,\wt m) = (-m,t)\,.\ee
\end{subequations}

In our example, we found that the Coulomb branch of a $G=U(1)^2$ theory with ${\footnotesize \begin{pmatrix}Q\\q\end{pmatrix} = \begin{pmatrix} 1 & 0 & -1 \\ 0 & 1 & -1 \\ 0 & 0 & 1 \end{pmatrix}}$ is equivalent to the Higgs branch of SQED, which has ${\footnotesize \begin{pmatrix}\wt q\\\wt Q\end{pmatrix} = \begin{pmatrix} 1 & 0 & 0 \\ 0 & 1 & 0 \\ 1 & 1 & 1 \end{pmatrix}}$ as in \eqref{QqSQED}.  These matrices obey (\ref{mirrormap}a). We also saw that in order to match resolution parameters we had to set $m_\R = -\wt t_\R$.

If we were not keeping track of resolutions and symmetries, we could translate (\ref{mirrormap}a) into a statement about gauge charges alone. The relation simply says that $\wt Q Q^T = Q\wt Q^T=0$, and more precisely that these two matrices fit into an exact sequence \eqref{eq:exact}. This relation among gauge charges was first derived in \cite{dBHOY}.

The particular form of the mirror map above is adapted to make the Coulomb branch of $T$ (including its resolutions and symmetries) resemble the Higgs branch of $\wt T$. Of course, the Higgs branch of $T$ also resembles the Coulomb branch of $\wt T$. However, since $(Z_i,z_i) = (-\wt M_\R^i,-\wt M_\C^i)$, the hyperplane arrangements corresponding to the Higgs branch of $T$ also resembles the Coulomb branch of $\wt T$ are not quite identical; rather, they are related by a reflection through the origin.

\subsubsection{Chiral ring}
\label{sec:abel-Cring}

The mirror map \eqref{mirrormap} was used in \cite{BDG-Coulomb} to derive the Coulomb-branch chiral ring in an abelian theory. The map of chiral operators is
\be v_A = \wt w^A\,,\qquad \varphi = \mu_{H,\C}\,,\ee
leading to the usual chiral-ring relations $v_Av_B = v_{A+B}\prod_i (M_\C^i)^{(Q_A^i)_++(Q_B^i)_+-(Q_{A+B}^i)_+}$ and their quantization \eqref{Cqring}.
Here $A\in \Z^r$ is identified (equivalently) as either a weight of the flavor group $G_C$ or a cocharacter of the gauge group $G$. Recall that $Q_A^i = \sum_a A^a Q_a{}^i$ is the charge of the $i$-th hypermultiplet under a subgroup $U(1)_A\subset G$ specified by the cocharacter $A$.

\begin{figure}[htb]
\centering
\includegraphics[width=3in]{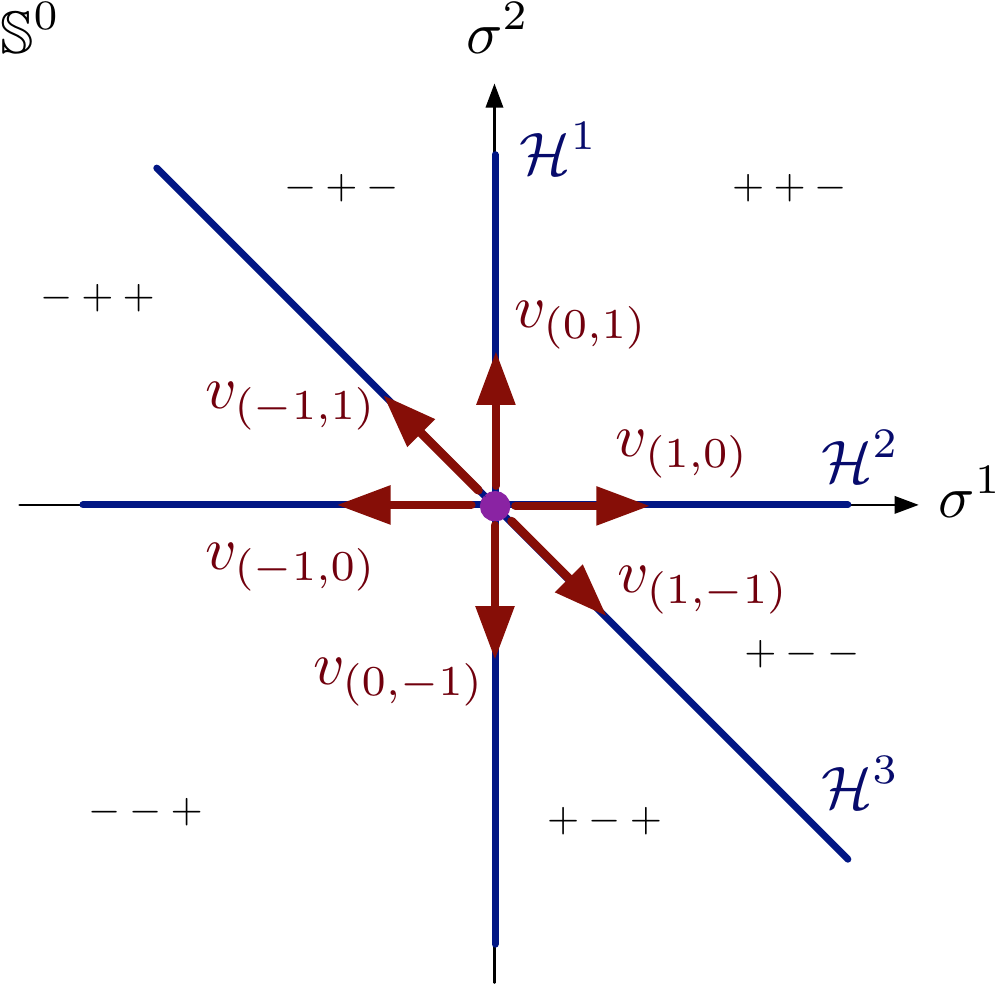}
\caption{Monopole operators parametrizing rays on the canonical slice.}
\label{fig:vA}
\end{figure}

The mirror map of chiral operators together with the Higgs-branch discussion of Section \ref{sec:abel-Hring} imply that
\begin{itemize}

\item On a special slice $\mathbb S^S$, we have $v_Av_{-A}=0$ for all monopole operators such that $Q_A^i\neq 0$ for some $i\in S$. Specifically, if $Q_A^i> 0$ then $v_A$ (resp. $v_{-A}$) vanishes on the negative (positive) side of the hyperplane $\CH^i$, $i\in S$.

\item On the canonical slice $\mathbb S^0$, we have $v_Av_{-A}=0$ for all $A\neq 0$. If $Q_A^i>0$ for any $i$ then $v_A$ (resp. $v_{-A}$) vanishes on the negative (positive) side of the hyperplane~$\CH^i$.

\item If we turn off both real and complex masses, then the canonical slice $\mathbb S^0$ is a cone. Rays $\rho(A) = \R_{\geq 0}\cdot A$ in the base of $\S^0$ are parameterized by monopole operators $v_A$ (Figure \ref{fig:vA}).

\item At vanishing real and complex mass, we can embed the lattice $\Z^r$ in the base of the canonical slice $\mathbb S^0$, identifying the base as $\Z^r\otimes \R$. The hyperplanes $\CH^i$ cut $\Z^r$ into a union of positive sublattices. A finite set of generators for the Coulomb-branch chiral ring $\C[\CM_C]$ is given by the $\varphi^a$ together with monopole operators $\{v_A\}_{A\in \CA}$ such that the set $\CA$ is a union of positive bases for the sublattices of $\Z^r$ cut out by hyperplanes. The ring relations $v_Av_B = v_{A+B}\prod_i(M_\C^i)^{(...)}$ contain a factor $M_\C^i$ for every hyperplane that lies between $A,B\in \Z^r$.

\item The finite set of generators for $\C[\CM_C]$ lifts to a set of generators for the quantum algebra $\hat\C[\CM_C]$.

\end{itemize}

In our $G=U(1)^2$ example, the chiral ring is generated by $\varphi^1,\varphi^2$ and the six monopole operators shown in Figure \ref{fig:vA}. They satisfy relations such as $v_{(0,1)}v_{(1,0)}=v_{(1,1)}$ (no hyperplanes in between), $v_{(1,0)}v_{(-1,1)} = M_\C^1\,v_{(0,1)}=\varphi^1\,v_{(0,1)}$ (hyperplane $\CH^1$ in between), and $v_{(1,0)}v_{(-1,0)} = M_\C^1M_\C^3 = \varphi^1(m_\C-\varphi^1-\varphi^2)$ (hyperplanes $\CH^1,\CH^3$ in between).

\subsubsection{Quantum hyperplane arrangements}
\label{sec:abel-Cqhyp}

The quantized chiral ring $\hat \C[\CM_C]$ of the Coulomb branch was described in \eqref{Cqring}, and is simply the mirror of the Higgs-branch ring from Section \ref{sec:abel-qHring}. We repeat the definition here for convenience: the generators are the complex scalars $\{\hat \varphi^a\}_{a=1}^r$ and monopole operators $\{\hat v_A\}_{A\in \Z^r}$; and the relations are
\begin{subequations}\label{abel-Cqring}
\be [\hat\varphi^a,\hat v_A] = \epsilon\,A^a\,\hat v^A\,, \ee
\be
 \hat v_A \hat v_B = \hspace{-.3cm}  \prod_{\substack{\text{$i$ s.t. $| Q_A^i|\leq | Q_B^i|$,}\\[.05cm] \wt Q_A^i  Q_B^i<0}} \hspace{-.3cm}  [\hat M_\C^i]^{- Q_A^i}\;
  \hat v_{A+B} 
   \prod_{\substack{\text{$i$ s.t. $| Q_A^i|> | Q_B^i|$,}\\[.05cm]  Q_A^i  Q_B^i<0}}  \hspace{-.3cm}  [\hat M_\C^i]^{ Q_B^i}\,.
\ee
\end{subequations}

We can visualize weight modules for $\hat \C[\CM_C]$ by using quantum hyperplane arrangements, essentially the same way as for the Higgs branch (Section \ref{sec:abel-Hqhyp}). The quantum hyperplane arrangements are constructed on the weight lattice $\Z^r$ of the topological symmetry group $G_C$, embedded into $\R^r$. The coordinates on $\R^r$ are weights of the $\hat \varphi^a$.
Since the $\hat \varphi^a$ are the moment maps for $G_C$, each lattice point can be identified with a weight space for the action of the commutative (Cartan) subalgebra $\hat \C[\CM_C]_0$ generated by the $\hat \varphi^a$. The monopole operators $\hat v_A$ map one weight space to another in the direction $A$.

The hyperplanes in the quantum arrangement are defined by $\CH^i = \{\hat M_\C^i=0\} = \{Q^i\cdot \hat\varphi + q^\alpha\cdot m_\C=0\}$. Their relative positions are determined by the complex masses~$m_\C$.

Just as in the Higgs-branch setup, one generally encounters \emph{multiple} quantum arrangements $\Gamma^S$, one for each classical vacuum $\nu_S$, labelled by a maximal intersection of $r$ hyperplanes. In the special case that the complex masses $m_\C = k_m\varepsilon$ are appropriately quantized, all the lattices $\Gamma^S$ coincide with each other and we can speak about a single, canonical quantum hyperplane arrangement.

\begin{figure}[htb]
\centering
\includegraphics[width=4in]{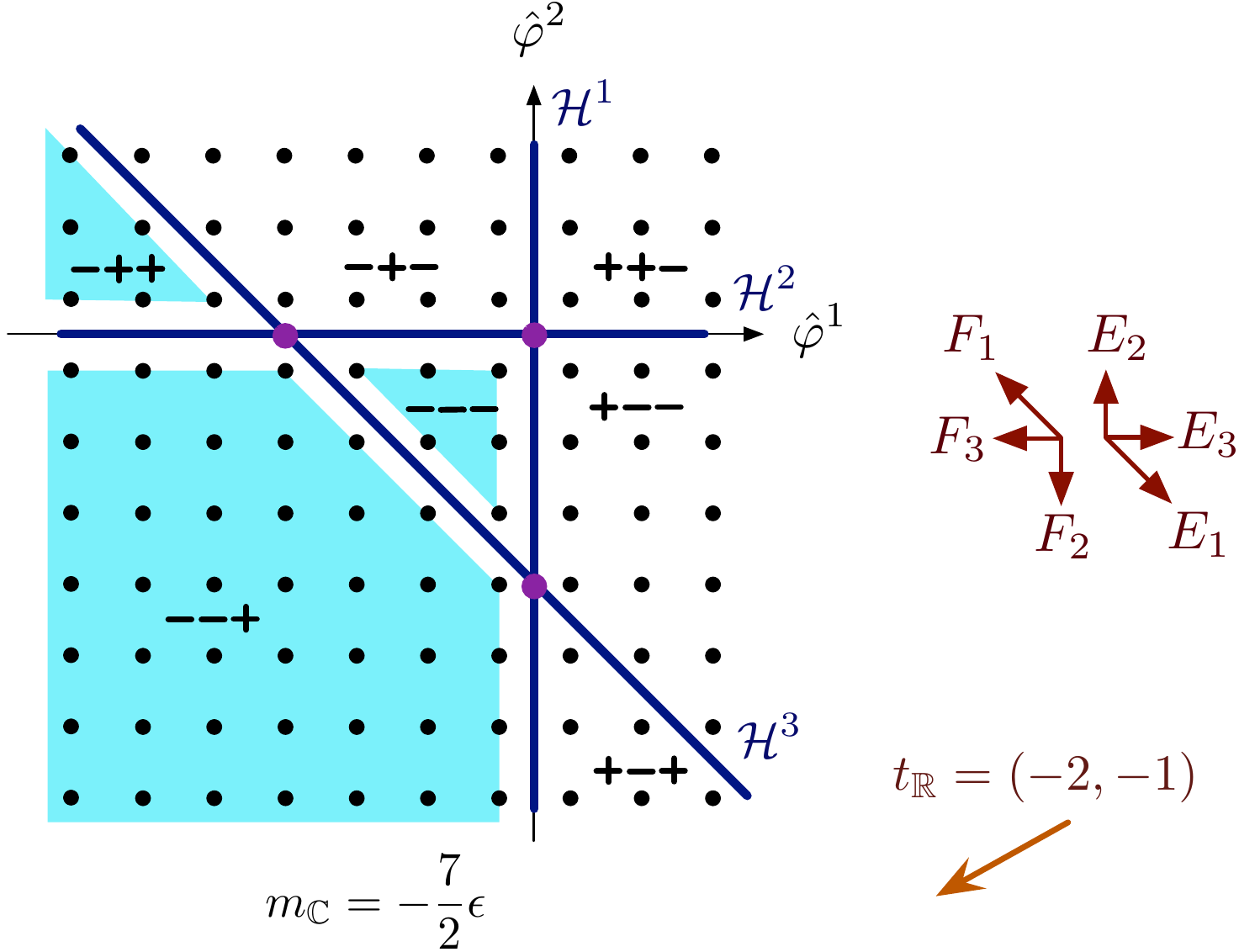}
\caption{The quantum hyperplane arrangement for the $G=U(1)^2$ theory, with quantized mass parameter $m_\C = -\frac72\epsilon$.}
\label{fig:cp2q-C}
\end{figure}

In our $G=U(1)^2$ example, the quantum algebra may be identified as a central quotient of $U(\mathfrak{sl}_3)$, with generators (say)
\be \begin{array}{l@{\quad}l@{\quad}l@{\qquad}l}
 E_1 = \hat v_{(1,-1)}\,, & E_2 = \hat v_{(0,1)}\,, & E_3 = \hat v_{(1,0)}\,, & H_1 =
 \hat M_\C^1-\hat M_\C^2=\hat \varphi^1-\hat\varphi^2\,, \\[.2cm]
  F_1 = \hat v_{(-1,1)}\,, & F_2 = \hat v_{(0,-1)}\,, & F_3 = \hat v_{(-1,0)}\,, & H_2 = 
  \hat M_\C^2-\hat M_\C^3=\hat\varphi^1+2\hat\varphi^2-m_\C\,.
\end{array}
\ee
The Casimirs $C_2,C_3$ are fixed as in \eqref{Cas-sl3}, subject to the replacement $t_\C\to-m_\C$. The quantum hyperplane arrangement at $m_\C = -\frac72\epsilon$ is shown in Figure \ref{fig:cp2q-C} (compare Figure~\ref{fig:cp2q}).

We expect that the Coulomb-branch images of right (resp., left) boundary conditions break supersymmetry unless the moment map $h_t$ in \eqref{ht-abel} is bounded from below (resp. above) (Section \ref{sec:NC-mt}). We called the boundary conditions with bounded $h_t$ $t_\R$-feasible. Similarly, we expect $t_\R$-feasible boundary conditions to define lowest-weight left-modules (resp. highest-weight right modules) for the quantized ring $\hat \C[\CM_C]$. In the case of right b.c., this means that all monopole operators $\hat v_A\in \hat\C[\CM_C]$ with negative charge $t_\R\cdot A<0$ act nilpotently on any fixed weight space in the module. In terms of the quantum hyperplane arrangement, lowest-weight modules are supported on chambers where the ``quantum'' function
\be \hat h_t = t_\R\cdot \hat \mu_{C,\C} = t_\R\cdot \hat \varphi \ee
is bounded from below. These chambers are shaded in Figure \ref{fig:cp2q-C}.

\subsection{Coulomb branes and modules}
\label{sec:abel-Cmod}

The Coulomb-branch images of boundary conditions in abelian theories were all analyzed in Sections \ref{sec:N}-\ref{sec:xD}. We can identify them fairly quickly with various chambers in hyperplane arrangements.

\subsubsection{Dirichlet boundary conditions}
\label{sec:abel-CD}

The generic Dirichlet boundary condition $\CD_\varepsilon$ in an abelian theory was described in \eqref{Dec}. In addition to $\varphi\big|_\pd=0$, all complex masses must vanish $m_\C=0$ in order for all the hypermultiplets to get vevs. Thus the Coulomb-branch image of a Dirichlet boundary condition is restricted to the canonical slice $\S^0$. From the semi-classical analysis of BPS equations in \eqref{Mpos}, we find that the image of a right (left) boundary condition is supported on the part of the slice with $\varepsilon_i M_\R^i\geq 0$ ($\varepsilon_i M_\R^i\leq 0$). Thus
\be \CD_\varepsilon \quad\leadsto\quad \CD_\varepsilon^{(C)} = \begin{cases} \text{toric}(\Delta^\varepsilon) & \text{right b.c.} \\
\text{toric}(\Delta^{-\varepsilon}) & \text{left b.c.} \end{cases} \ee

The quantization of a Dirichlet boundary condition was described most precisely in Section \ref{sec:DC2}, using boundary monopole operators. In the presence of a Dirichlet boundary condition and an $\Omega$-background, the complex masses must be quantized $m_\C^\alpha = k^\alpha \epsilon$. Following Section \ref{sec:DC2}, 
we find that $\CD_\varepsilon$ as a right boundary condition gives rise to a module $\hat\CD_\varepsilon^{(C)}$ with states $|B\rangle$ that satisfy
\be \hat\varphi^a|B\rangle  = B^a|B\rangle\,,\qquad \hat v_A|B\rangle = \prod_{\varepsilon_iQ_A^i<0}[\hat M_\C^i]^{-Q_A^i}|A+B\rangle =  \prod_{i} [\varepsilon_iM_\C^i]^{(-\varepsilon_iQ_A^i)_+}|A+B\rangle
\label{abel-CDaction} \ee
(The two expressions for $\hat v_A|B\rangle$ are equivalent up to a sign that can be absorbed in the definition of $|B\rangle$) The boundary states $|B\rangle$ are labelled either by points in the lattice $\Z^{r'}$ or a half-integer shift of this lattice, \ie\ a torsor. As in \eqref{qDC-image-k}, these boundary states are constrained so that the eigenvalue of $(\varepsilon_i\hat M_\C^i-\frac12\epsilon)$ on $|B\rangle$, namely $(\varepsilon_i Q_B^i+\varepsilon_iq^i\cdot k-\frac12)\epsilon$, is positive for all $i$. This identifies $\hat\CD_\varepsilon^{(C)}$ as the irreducible module whose nonzero weight spaces are the lattice points inside the chamber $\Delta^\varepsilon$ in the quantum hyperplane arrangement:
\be   \hat\CD_\varepsilon^{(C)} \;\simeq\;  \hat \Delta^\varepsilon\,. \label{abel-CDmod} \ee
It follows from \eqref{abel-CDaction} that any monopole operators that would take a state outside this chamber act as zero.

\subsubsection{Exceptional Dirichlet boundary conditions}
\label{sec:abel-CxD}

An exceptional Dirichlet boundary condition $\CD_{\varepsilon,S}$ is labelled by a sign vector and a subset $S$ of size $r$ \eqref{abel-xD}. It gives boundary vevs only to hypermultiplets with $i\in S$, and correspondingly must set
\be M_\C^i\big|_\pd = 0\quad(i\in S)\qquad\Rightarrow\qquad \varphi\big|_\pd = -(Q^{(S)})^{-1}\cdot q \cdot m_\C\,,\ee
where $Q^{(S)}$ is the $r\times r$ submatrix of $Q$ with columns $i\in S$. Thus the Coulomb-branch image of $\CD_{\varepsilon,S}$ is supported on the special slice $\S^S$. Following the analysis of BPS equations in Section \ref{sec:DC} (or from Section \ref{sec:DC2}) we find that
\be \CD_{\varepsilon,S} \quad\leadsto\quad \CD_{\varepsilon,S}^{(C)} = \begin{cases} \text{toric}(V^{S,\varepsilon}) & \text{right b.c.} \\
\text{toric}(V^{S,-\varepsilon}) & \text{left b.c.} \end{cases} \ee

The relation between bulk and boundary monopole operators is
\be v_A\big|_\pd = \prod_{\text{$i$ s.t. $\varepsilon_iQ_A^i<0$}} (M_\C^i)^{|Q_A^i|}\,\mb v_A\,,\qquad {}_\pd\big|v_A = \mb v_A\prod_{\text{$i$ s.t. $\varepsilon_iQ_A^i>0$}} (M_\C^i)^{|Q_A^i|}\,, \ee
on the right and left sides.
Thus, for a right (left) boundary condition, $v_A|_\pd = 0$ (${}_\pd|v_{-A} = 0$) if $\varepsilon_iQ_A^i<0$ for any $i\in S$.

Quantization produces a right module $\hat \CD_{\epsilon,S}^{(C)}$ with states $|B\rangle$ corresponding to lattice points in the interior of the orthant $V^{S,\varepsilon}$ of the quantum hyperplane arrangement. Concretely, there is a single state $|B\rangle$ for each $B\in \Z^r$ such that
\be \varepsilon_i Q_B^i = \varepsilon_i B\cdot Q^i \geq 0 \qquad \forall\; i\in S\,. \ee
In particular, there is an identity state $|0\rangle$ that satisfies $(\hat M_\C^i-\frac12\varepsilon_i\,\epsilon)|0\rangle= 0$ for all $i\in S$, which fixes the eigenvalues of $\hat\varphi^a$. The identity corresponds to the lattice point in $V^{S,\varepsilon}$ closest to the intersection of hyperplanes $\CH^i$ ($i\in S$).
The remaining states satisfy $\hat \varphi^a|B\rangle = (\hat \varphi^a+B^a)|0\rangle$. The monopole operators act as
\be\hspace{-.5in}  \hat \CD_{\epsilon,S}^{(C)}\,:\qquad \hat v_A|B\rangle = \prod_{\text{$i$ s.t. $\varepsilon_iQ_A^i<0$}} [\hat M_\C^i]^{-Q_A^i}|A+B\rangle\,, \label{abel-qCxD} \ee
and in particular act as zero when moving out of the orthant $V^{S,\varepsilon}$.

For generic $m_\C$, $\hat \CD_{\epsilon,S}^{(C)}$ is the irreducible Verma module $\hat V^{S,\varepsilon}$ generated by acting freely on the identity state $|0\rangle$ with monopole operators. For quantized $m_\C = k_m \epsilon$, the module will become reducible if the orthant $V^{S,\varepsilon}$ is intersected by additional hyperplanes $\CH^i$ with $i\notin S$. It follows from \eqref{abel-qCxD} that
\begin{itemize}
\item The monopole operators $\hat v_A$ act as zero when they cross any hyperplane $\CH^i$ from the $\varepsilon_i$ side toward the $-\varepsilon_i$ side. In particular, they act as zero when moving out of the orthant $V^{S,\varepsilon}$.
\end{itemize}
Therefore, much as in the case of exceptional Dirichlet b.c. on the Higgs branch \eqref{xD-filter}, the module $\hat \CD_{\epsilon,S}^{(C)}$ has a filtration
\be \begin{array}{ll} \hat \CD_{\epsilon,S}^{(C)} &= M_n\supset M_{n-1}\supset\ldots\supset M_1\supset (M_0 = \oslash) \\[.1cm]
 &= \big[\, M_n/M_{n-1}\,\big|\, M_{n-1}/M_{n-2}\,\big|...\big|\,M_2/M_1\,\big|\, M_1\,\big]\,,
 \end{array} \label{xD-filter2} \ee
such that each quotient $M_a/M_{a-1}$ is an irreducible module $\hat \Delta^{\varepsilon'}$ supported on one of the chambers in the orthant $V^{S,\varepsilon}$.

\subsubsection*{Thimbles, standards, and costandards}

We expect certain exceptional Dirichlet boundary conditions $\CD_{\varepsilon,S}$ to have images that are thimbles on the Higgs and Coulomb branches, and whose quantizations are standard (Verma) or costandard modules.
In terms of Coulomb-branch data, the association between a massive vacuum $\nu$ and parameters $m_\R,t_\R$ (which label a thimble) and the UV boundary condition
\be (\nu;m_\R,t_\R)\qquad \leadsto\qquad (\varepsilon,S) \label{nueS-C} \ee
is implemented by
\begin{itemize}
\item Choosing $S$ so that the vacuum $\nu_S$ lies at the intersection of hyperplanes $\CH^i$ ($i\in S$) on the special slice $\S^S$ of the Coulomb branch;
\item Choosing $\varepsilon_i$ ($i\in S$) so that the potential $h_t^C=t_\R\cdot \sigma$ is bounded from below on the orthant $V^{S,\varepsilon}$;
\item Choosing $\varepsilon_i$ ($i\notin S$) so that the vacuum $\nu_S$ lies on the $\varepsilon_i$ side of $\CH^i$ in in the slice $\S^S$\,.
\end{itemize}

These criteria are equivalent to the geometric Higgs-branch criteria given below \eqref{nueS}. Indeed, $h_m$ is bounded from below on the orthant $V_{S,\varepsilon}$ if and only if $\nu_S$ lies on the $\varepsilon_i$ side of $\CH^i$ in $\S^S$ (for $i\notin S$); and $h_t^C$ is bounded from below on $V^{S,\varepsilon}$ if and only if $\nu_S$ lies on the $-\varepsilon_i$ side of $\CH_i$ in $\S_S$ (for $i\in S$). The easiest way to see these equivalences is to order the hypermultiplets so that $S=\{1,...,r\}$ and its complement is $\ol S=\{r+1,...,N\}$, and to reparameterize the gauge and flavor group so that
\be \begin{array}{cc} & S\quad\;\; \ol S \\[-.1cm]
\begin{pmatrix} Q\\ q\end{pmatrix} = &\begin{pmatrix} I_{r\times r} & * \\ 0 & I_{r'\times r'} \end{pmatrix} \end{array}\qquad
\begin{array}{cc} & S\quad\;\; \ol S \\[-.1cm]
\begin{pmatrix} \wt q\\ \wt Q\end{pmatrix} = &\begin{pmatrix} I_{r\times r} & 0 \\ * & I_{r'\times r'} \end{pmatrix} \end{array} \ee
The (right) boundary condition $\CD_{\varepsilon,S}$ whose Higgs and Coulomb-branch images are both thimbles for given $m_\R,t_\R$ has $\varepsilon_i=\delta_i^a\text{sign}(t_{a,\R})$ ($i\in S$) and $\varepsilon_i=\delta^{i-r}_{\alpha} \text{sign}(m_\R^\alpha)$ ($i\notin S$)\,. On the other hand, at the vacuum $\nu_S$ we have $Z_i=0$ ($i\notin S$) and $M_\R^i=0$ ($i\in S$), which implies $\mu_{H,\R},\sigma\equiv 0$, and in turn $Z_i=- \delta_i^a t_{a,\R}$ ($i\in S$) and $M_\C^i = \delta^{i-r}_{\alpha} m_\R^\alpha$ ($i\notin S$)\,. Thus the vacuum lies on the $-\varepsilon_i$ side of hyperplanes $\CH_i$ ($i\in S$) on the Higgs-branch slice, and on the $\varepsilon_i$ side of $\CH^i$ ($i\notin S$) on the Coulomb-branch slice.

As for modules, we follow the same reasoning as in Section \ref{sec:abel-HxD} to conclude that if we introduce an $\Omega$-background with quantized mass parameters $m_\C = k_m\epsilon$, the module $\hat \CD_{\varepsilon,S}^{(H)}$ (with $\varepsilon,S$ determined by \eqref{nueS-C}) is 
\be \begin{array}{ll}\text{standard/Verma} & \text{if $k_m \sim -m_\R$\,,} \\[.1cm] \text{costandard} & \text{if $k_m\sim m_\R$\,.} \end{array}\ee

\subsubsection{Neumann boundary conditions}
\label{sec:abel-CN}

Finally, we come to Neumann boundary condition $\CN_\varepsilon$. Following Section \ref{sec:NC}, we find that the classical images of left and right boundary conditions are cut out by the holomorphic equations
\be \CN_\varepsilon^{(C)}\,:\quad v_A\big|_\pd = \xi_A \prod_{\text{$i$ s.t. $\varepsilon_i Q_A^i<0$}} (M_\C^i)^{|Q_A^i|}\big|_\pd\,,\qquad
{}_\pd\big| v_A = \xi_A^{-1} \prod_{\text{$i$ s.t. $\varepsilon_i Q_A^i>0$}} {}_\pd\big|\,(M_\C^i)^{|Q_A^i|}\,.
\label{abel-CN} \ee
Upon turning on the $\Omega$-background, we find a right module (say) for $\hat\C[\CM_C]$, generated from an identity state `$|$' that satisfies relations
\be \label{abel-qCN}
\hat \CN_\varepsilon^{(C)}\,:\quad
\hat v_A\big| = \xi_A \prod_{\text{$i$ s.t. $\varepsilon_i Q_A^i<0$}} [\hat M_\C^i]^{-Q_A^i}\big|\,.
\ee
This should be compared to the Higgs-branch image of a Dirichlet boundary condition \eqref{Dec-Hq} from Section \ref{sec:abel-HD}; the two formulas are identical after applying the mirror map $\hat w^A,\hat z_i,\wt \xi^A\to \hat v_A,\hat M_\C^i,\xi_A$ and sending $\varepsilon\to -\varepsilon$.

In the presence of nonzero $t_\R$, it is natural to deform the boundary conditions by an infinite gradient flow with respect to $h_t$, as first discussed in Section \ref{sec:NC-tinf} (and in parallel to the abelian Higgs-branch discussion of Section \ref{sec:abel-HD}). For right boundary conditions, the deformation is achieved by rescaling
\be \xi_A = e^{\lambda(t_\R\cdot A)} \xi_{A,0}\,, \ee
and sending $\lambda\to\infty$. (For left b.c., one should send $\lambda\to 0$ instead.)

In this limit, the Coulomb-branch image of a right boundary condition satisfies
\be v_A\big|_\pd =0 \quad (t_\R\cdot A <0)\,;\qquad \prod_i (M_\C^i)^{(-\varepsilon_i Q_A^i)_+}\big|_\pd = 0\quad (t_\R\cdot A > 0)\,.\ee
More precisely (by reasoning similar to Section \ref{sec:abel-HD}) the support of $\CN_\varepsilon^{(C)}$ becomes a union of toric varieties on which the potential $h_t=t_\R\cdot\sigma$ is bounded from below:
\be \CN_\varepsilon^{(C)} \;\overset{\lambda\to\infty}{\longrightarrow}\; \bigcup_{S} \text{toric}(V^{S,-\varepsilon})\qquad \text{s.t. $h_t$ bounded below on $V^{S,-\varepsilon}$.} \label{abel-CN-limit} \ee
Similarly, when complex mass parameters $m_\C$ are generic, the module \eqref{abel-qCN} splits into a direct sum of irreducible lowest-weight Verma modules
\be \hat \CN_\varepsilon^{(C)}  \;\overset{\lambda\to\infty}{\longrightarrow}\; \bigoplus_{S} \hat V^{S,-\varepsilon}\qquad \text{s.t. $\hat h_t$ bounded below on $V^{S,-\varepsilon}$.} \label{abel-qCN-limit} \ee

For quantized values of complex masses $m_\C=k_m\epsilon$, the limit $\lambda\to\infty$ must be taken carefully, as explained in Section \ref{sec:NC-tinf} (also Section \ref{sec:NC-SQED}). We expect that the module $\hat \CN_\varepsilon^{(C)}$ becomes a successive extension of Verma modules,
\be \hat\CN_\varepsilon^{(C)} \overset{\lambda\to\infty}{\longrightarrow} \big[\, \hat V_{S_n,-\varepsilon}\,\big|... \big| \hat V_{S_2,-\varepsilon}\,\big|\, \hat V_{S_1,-\varepsilon}\,\big]\,. \ee
where $\{S_i\}$ are the subsets appearing in \eqref{abel-qCN-limit}. The ordering is such that $S_i$ occurs before $S_j$ if $\hat h_t(\nu_{S_i})>\hat h_t(\nu_{S_j})$.

\subsection{Mirror symmetry}
\label{sec:abel-MS}

The explicit description of the Higgs and Coulomb-branch images of UV boundary conditions in abelian theories allows us to propose an explicit mirror map of boundary conditions. Let us take two theories $T,\wt T$ with charge matrices and parameters related as in \eqref{mirrormap}, namely
\be  \begin{pmatrix} \wt q \\ \wt Q \end{pmatrix} = \begin{pmatrix} Q \\ q\end{pmatrix}^{-1,T}\,,\qquad  (m,t) = (-\wt t,\wt m)\,. \ee
Recall that this makes the Coulomb-branch hyperplane arrangement of $T$ identical to the Higgs-branch arrangement of $\wt T$, but relates the Higgs-branch arrangement of $T$ to the inverse (reflection through the origin) of the Coulomb-branch arrangement of $\wt T$. Then the mirror map of boundary conditions is 
\be \label{bcmap}
 \begin{array}{l@{\qquad}l}
(\CN_\varepsilon,\;\CD_\varepsilon,\;\CD_{\varepsilon,S}) \simeq (\wt\CD_{-\varepsilon},\;\wt \CN_{\varepsilon},\;\wt \CD_{-\ol \varepsilon,\ol S}) & \text{right b.c.} \\[.2cm]
(\CN_\varepsilon,\;\CD_\varepsilon,\;\CD_{\varepsilon,S}) \simeq (\wt\CD_{\varepsilon},\;\wt \CN_{-\varepsilon},\;\wt \CD_{\ol \varepsilon,\ol S}) & \text{left b.c.}\,,
\end{array} \ee
in the sense that these lead to identical IR images and modules on both Higgs and Coulomb branches. Here we use the notation $\ol S$ for the complement of $S$, and
\be \ol\varepsilon_i = \begin{cases} -\varepsilon_i & i\in S \\ \varepsilon_i & i\notin S \end{cases}\,.\ee
as in \eqref{bare}.

To illustrate the equivalence, consider the classical images of right boundary conditions. For theory $T$ we have
\be
\begin{array}{c@{\;}|@{\;}c@{\quad}c}
& \CM_H & \CM_C  \\\hline
\CN_\varepsilon & \Delta_\varepsilon & v_A\big| = \xi_A\prod_i (M_\C^i)^{(-\varepsilon_iQ_A^i)_+}  \\[.2cm]
\CD_\varepsilon & w^A = \wt\xi^A \prod_i (z_i)^{(\varepsilon_i\wt Q_A^i)_+} & \Delta^\varepsilon  \\[.2cm]
\CD_{\varepsilon,S} & V_{S,\varepsilon} & V^{S,\varepsilon}\,,
\end{array}
\ee
whereas for theory $\wt T$,
\be
\begin{array}{c@{\;}|@{\;}c@{\quad}c}
& \wt\CM_C & \wt \CM_H \\\hline
\wt \CD_{-\varepsilon} & \Delta^{-\varepsilon} & \wt w^A\big| = \wt \xi^A \prod_i (\wt z_i)^{(-\varepsilon_i Q_A^i)_+} \\[.2cm]
\wt\CN_\varepsilon & \wt v_A = \xi_A \prod_i (\wt M_\C^i)^{(-\varepsilon_i\wt Q_A^i)_+} & \Delta_\varepsilon \\[.2cm]
\wt \CD_{-\ol\varepsilon,\ol S} & V^{\ol S,-\ol\varepsilon} = V^{\ol S,-\varepsilon} & V_{\ol S,-\ol\varepsilon}=V_{\ol S,\varepsilon}\,.
\end{array}
\ee
The $\CM_C$ and $\wt \CM_H$ images match exactly, while the $\CM_H$ and $\wt \CM_C$ images match with an expected inversion $\varepsilon \to -\varepsilon$.

\subsection{The mirror symmetry interface}
\label{sec:mirrorwall}

Suppose that we are given two 3d $\CN=4$ gauge theories $T$, $\wt T$ that are mirror to each other. It is not obvious \emph{a priori} that every UV boundary condition in $T$ should admit a mirror UV boundary condition in $\wt T$, such that the IR images of the boundary condition and its mirror are identical. One way to ensure the existence of mirror boundary conditions is to produce a mirror-symmetry interface, namely a BPS interface between mirror gauge theories that will flow to the almost-trivial interface in the IR, which simply exchanges Higgs and Coulomb data of the IR SCFT's. Then one may formally construct mirrors of boundary conditions by colliding them with the 
mirror symmetry interface, assuming the different RG flows involved in the process commute. 

\begin{figure}[htb]
\centering
\includegraphics[width=2.8in]{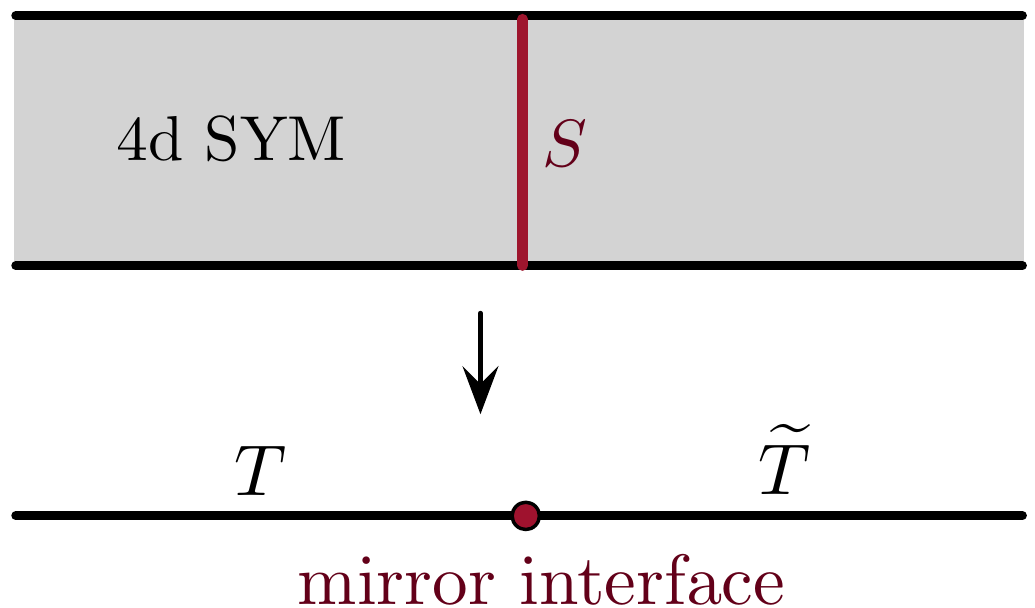}
\caption{Reducing an S-duality interface in 4d $\CN=4$ SYM to a 3d mirror-symmetry interface.}
\label{fig:Sdual}
\end{figure}

In the case of 3d gauge theories that arise from segment compactifications of 4d $\CN=4$ SYM, the existence of a mirror-symmetry interface can be proven by acting with S-duality \cite{GW-Sduality}. In the 4d UV description, the desired mirror-symmetry interface arises from an S-duality wall stretched along the segment (Figure \ref{fig:Sdual}). Such an interface can be engineered (somewhat non-constructively) by representing the S-duality wall in the far UV as a smooth Janus configuration for the 4d gauge coupling. This construction would be explicit if one could find the precise description of the intersections between the S-duality wall and the endpoints of the segment.

In the case of abelian theories $T,\wt T$, we can follow a different approach. We already know the explicit mirror map of chiral and twisted-chiral operators (Sections \ref{sec:abel-Cmirror}, \ref{sec:abel-Cring}), and can simply try to write down a 2d (2,2) interface theory with appropriate couplings to $T$ and $\wt T$ in order to reproduce this mirror map. We do so below in two steps, starting with a simple example. Our construction is closely related to two-dimensional Hori-Vafa mirror symmetry~\cite{HoriVafa}.
We can then verify that the putative interface theory also reproduces the explicit mirrors of boundary conditions from Section \ref{sec:abel-MS}.

\subsubsection{Example: U(1) + 1 hyper}

Consider a gauge theory $T$ with $G=U(1)$ and one hypermultiplet $(X,Y)$ of charge $Q=1$. 
The Higgs branch of this theory is trivial: imposing the moment map constraints $XY + t_\C=0,\,|X|^2-|Y|^2+t_\R=0$ and dividing by the $U(1)$ gauge symmetry leaves a point. The Coulomb branch, with chiral-ring relations $v_+v_- = \varphi$, is a copy of $\mathbb{C}^2$. There is a topological symmetry $G_C=U(1)_t$ rotating the monopole operators $v_\pm$ with charge $\pm1$.

The mirror theory $\wt T$ is simply a free twisted hypermultiplet $(\wt X , \wt Y) = (v^+,v^-)$. In the conventions of Section \ref{sec:abel-Cmirror}, the flavor symmetry $\wt G_H=U(1)$ should be identified with $G_C$, so that $\wt m_\C = t_\C$. The mirror map also sets $(M_\C,z)=(\wt z,-\wt M_\C)$, in other words 
$(\varphi,XY)=(\wt X\wt Y,-\wt m_\C)$.

We want to construct a mirror-symmetry interface that implements this identification. We will build the interface as a deformation of the right Neumann boundary condition $\CN_+$ (\ie\ $Y|_\pd=0$) for the $U(1)$ gauge theory and the left boundary condition $\wt \CN_{-}$ (\ie\ ${}_\pd|\wt X=0$) for the twisted hypermultiplet.
On the interface itself we introduce a 2d chiral multiplet $\phi$ valued in $\C/2\pi i\Z \simeq \R\times S^1$, and its T-dual, a twisted-chiral multiplet $\wt \phi$, also valued in $\C/2\pi i\Z$. We would like to identify
\begin{itemize}
\item $G=U(1)$ as the translation symmetry of $\phi$, or the winding symmetry of $\wt \phi$\,;
\item $G_C = \wt G_H =U(1)_t$ as the translation symmetry of $\wt \phi$, or the winding symmetry of~$\phi$\,. 
\end{itemize}
To this end, we introduce superpotential and twisted-superpotential couplings at the interface,
\be W_{\rm int} = X|_\pd\, e^{-\phi} - \wt m_\C\, \phi\,,\qquad \wt W_{\rm int} = {}_\pd|\wt Y\,e^{\wt \phi}-\varphi|_\pd\, \wt \phi\,. \label{WWU1} \ee
The first (exponential) terms in the superpotentials require $e^{\phi},e^{\wt \phi}$ to have the desired charges under $G$ and $G_C=\wt G_H$. The second (bilinear) terms break these symmetries explicitly whenever $\wt m_\C=t_\C$ or $\varphi|_\pd$ are nonzero. (The same breaking occurs dynamically on the moduli space of the 3d theories $T,\wt T$.)

We determine the effect of superpotentials \eqref{WWU1} on the boundary conditions by using the methods of Section \ref{sec:en}. First, the F-terms for $\phi,\wt\phi$ imply that at the interface
\be X|_\pd\,e^{-\phi} = -\wt m_\C\,,\qquad {}_\pd|\wt Y\,e^{\wt \phi} = \varphi|_\pd\,. \label{U1XY1}\ee
In addition, boundary (twisted) F-terms for $X$ $(\wt Y)$ impose
\be  Y|_\pd = \frac{\pd W_{\rm int}}{\pd X|_\pd} = e^{-\phi}\,,\qquad  {}_\pd|\wt X = \frac{\pd W_{\rm int}}{\pd ({}_\pd|\wt Y)} = e^{\wt \phi}\,, \label{U1XY2} \ee
so that altogether
\be XY|_\pd = -\wt m_\C = t_\C\,,\qquad \varphi|_\pd = {}_\pd|\wt X\wt Y\,,\ee
which is the first part of the mirror map. Similarly, following Sections \ref{sec:NC}, \ref{sec:NC-justified}, we find that the deformed Neumann boundary condition in theory $T$ implies that the monopole operators satisfy
\be v_+|_\pd = e^{-\pd \wt W_{\rm int}/\pd\varphi|_\pd} = e^{\wt \phi} = {}_\pd|\wt X\,,\qquad v_-|_\pd = M_\C\, e^{+\pd \wt W_{\rm int}/\pd\varphi|_\pd}  = \varphi|_\pd e^{-\wt \phi} = {}_\pd|\wt Y\,.\ee
Thus the interface implements the full mirror-symmetry transformation.

Note that, unfortunately, this description of the interface is intrinsically non-Lagrangian: both the 2d chiral $\phi$ and its T-dual $\wt \phi$ are involved in the couplings \eqref{WWU1}. 

As a simple check, let us reproduce some of the mirror pairs of boundary conditions from Section \ref{sec:abel-MS}.
Consider a right b.c. $\wt \CD_+$ (generic Dirichlet) for the twisted hypermultiplet theory $\wt T$, which sets $\wt Y|_\pd =c \neq 0$. After colliding (from the left) with the mirror symmetry interface, we arrive at the $U(1)$ gauge theory $T$ on a half-space, coupled to a 2d theory with superpotentials
\be W_{\rm int} = X|_\pd \, e^{-\phi}\,,\qquad \wt W_{\rm int} = c\,e^{\wt \phi} - \varphi|_\pd\,\wt \phi\,.\ee
The superpotential $\wt W_{\rm int}$ is precisely of the form encountered in Hori-Vafa mirror symmetry \cite{HoriVafa}. The exponential term $c\,e^{\wt \phi}$ with a constant, nonvanishing coefficient has the effect of removing the $e^\phi$ operator and promoting $\eta = e^{-\phi}$ to a $\C$-valued chiral field. We can simply integrate out this field from $W_{\rm int}$, finding that its F-term imposes $X|_\pd=0$. We can also integrate out $\wt \phi$ to find an effective twisted superpotential $\wt W_{\rm int} = -\varphi|_\pd(\log(c/\varphi|_\pd)-1)$, which has the effect of setting $v_+|_\pd = c\varphi|_\pd$ and $v_-|_\pd = c^{-1}$.
Altogether, we find that the gauge theory $T$ effectively has a right Neumann b.c. $\CN_-$, with effective 2d FI parameter $t_{2d} = \log c$. Similar manipulations show that colliding the interface with a right boundary condition $\wt \CD_-$ for $\wt T$ produces an effective boundary condition $\wt \CN_+$ for $T$ (with $t_{2d}=\log c$); thus $\CN_\varepsilon\simeq \wt \CD_{-\varepsilon}$ as expected.

Conversely, suppose we have an exceptional Dirichlet b.c. $\CD_{-,\{1\}}$ for $T$ on the left, which breaks $U(1)$ gauge symmetry, setting ${}_\pd|X=c$ and ${}_\pd|\varphi=0$. Now $\wt m_\C = t_\C$ may be generic. Colliding with the mirror-symmetry interface (from the right) produces a Neumann b.c. for $\wt T$ coupled to 2d fields $\phi,\wt\phi$ with
\be W_{\rm int} = c\,e^{-\phi}-\wt m_\C\,\phi\,,\qquad \wt W_{\rm int} = {}_\pd|\wt Y\, e^{\wt \phi}\,.\ee
By the same argument as above, $\wt\eta=e^{\wt \phi}$ becomes a $\C$-valued field, and its (twisted) F-term sets ${}_\pd|\wt Y=0$. The F-term for $\phi$ also fixes $e^{-\phi}=-\wt m_\C/c$. Thus, we effectively find a left b.c. for $\wt T$ that simply sets ${}_\pd|\wt Y=0$; this can be identified as exceptional Dirichlet $\wt D_{+,\oslash}$, in agreement with the general formula $\CD_{\varepsilon,S}\simeq \wt \CD_{\ol\varepsilon,\ol S}$ (left b.c.).

\subsubsection{General abelian theory}

The basic example above indicates how to proceed for a general abelian gauge theory. Suppose that $T$ and $\wt T$ are mirror theories as in Section \ref{sec:abel-Cmirror}, with
\be \begin{array}{c|@{\;}c@{\quad}c}
 & T & \wt T \\\hline
 \text{hypermultiplets} & (X_i,Y_i)_{i=1}^N & (\wt X_i,\wt Y_i)_{i=1}^N \\[.1cm]
 \text{gauge group} & G=U(1)^r & \wt G = U(1)^{N-r} \\[.1cm]
 \text{flavor symmetry} & \multicolumn{2}{c}{ G_H \simeq \wt G_C^* =  U(1)^{r'} } \\
  & \multicolumn{2}{c}{ G_C \simeq \wt G_H = U(1)^r }
\end{array}
\ee
with $r'=N-r$ as usual. The gauge and flavor charges of the respective hypermultiplets are related as in \eqref{mirrormap}, ${Q\choose q}^{-1,T}={\wt q\choose \wt Q}$, and masses and FI parameters satisfy $(m,t)=(-\wt t,\wt m)$. The mirror map for Higgs and Coulomb-branch chiral operators identifies
\be (M_\C^i,z_i) = (\wt z_i,-\wt M_\C^i)\,,\qquad v_A = \wt w^A\,,\qquad w^A = \wt v_{-A}\,, \label{chiralmap} \ee
where as usual $M_\C = Q\cdot \varphi+q\cdot m_\C$ and $\wt M_\C = \wt Q\cdot \wt \varphi+\wt q\cdot \wt m_\C$ are effective complex masses; $z_i=X_iY_i$ and $\wt z_i=\wt X_i\wt Y_i$;  $v_A,\wt v_A$ are the usual monopole operators in the two theories; and $w^A,\wt w^A$ are defined as in \eqref{wA}, namely
\be \label{wAwA}
w^A = \prod_{i=1}^N \begin{cases} \, X_i^{\, | \, \wt Q_A^i |} & \wt Q_A^i > 0 \\
 \, Y_i^{\, | \, \wt Q_A^i |} & \wt Q_A^i  < 0 \end{cases}\,,\qquad
 \wt w^A = \prod_{i=1}^N \begin{cases} \, \wt X_i^{\, | \,  Q_A^i |} & \wt Q_A^i > 0 \\
 \, \wt Y_i^{\, | \,  Q_A^i |} & \wt Q_A^i  < 0 \end{cases}\,.
\ee
The relations \eqref{chiralmap} comprise a full set of generators for the Higgs and Coulomb-branch chiral rings --- thus an interface that implements these relations will necessarily implement the correct mirror map for all chiral operators.

To construct the mirror-symmetry interface, we first choose any Lagrangian splitting for the hypermultiplets in $T$ and the \emph{same} splitting for the hypermultiplets in $\wt T$ (with respect to the relation \eqref{mirrormap} between gauge charges). By default, we will take the splittings $(X_i,Y_i)$ and $(\wt X_i,\wt Y_i)$. Then we place $T$ (resp., $\wt T$) on the half-line $x^1\leq 0$ ($x^1\geq 0$), with boundary conditions $\CN_{++...+}$ ($\wt \CN_{--...-}$) at $x^1=0$. A priori, the theories on the two half-lines do not interact with each other. We then deform these two Neumann b.c. by coupling to $N$ 2d chiral fields $\phi_i$ and their T-duals $\wt \phi_i$, both valued (with appropriate normalization) in $\C/2\pi i\Z$. The couplings are encoded in a superpotential and a twisted superpotential at the interface:
\be W_{\rm int} = \sum_{i=1}^N\big( X_i|_\pd \, e^{-\phi_i}-{}_\pd|\wt M_\C^i \,\phi_i\big)\,,\qquad
\wt W_{\rm int} = \sum_{i=1}^N\big( {}_\pd|\wt Y_i \,e^{\wt\phi_i} - M_\C^i|_\pd\,\wt\phi_i\big)\,.\ee

In order to check the relations \eqref{chiralmap}, we again use the results of Section \ref{sec:en}. The F-terms for $\phi,\wt\phi$ and the boundary F-terms for the hypermultiplets imply that at the interface (we drop the $|_\pd$ and ${}_\pd|$ to simplify notation):
\be Y_i = e^{-\phi_i}\,, \qquad X_i\, e^{-\phi_i}=-\wt M_\C^i\,;\qquad
 \wt X_i=e^{\wt \phi_i}\,,\qquad \wt Y_i\,e^{\wt \phi_i} = M_\C^i\,,\ee
which immediately gives the first part of the mirror map $(M_\C^i,z_i)=(\wt z_i,-\wt M_\C^i)$.

The part of the mirror map involving flavor-charged operators is slightly trickier. We recall that the pure Neumann b.c. $\CN_{++...+}$ sets $v_A|_\pd = \prod_i (M_\C^i)^{(-A\cdot Q^i)_+}$ (up to an overall sign). Following Section \ref{sec:en}, we find that the 2d superpotentials deform this to
\begin{align} v_A &= \prod_i (M_\C^i)^{(-A\cdot Q^i)_+}  \exp\Big( -\sum_aA_a \frac{\pd \wt W_{\rm int}}{\pd \varphi^a}\Big)  \notag \\
 &= \prod_i (M_\C^i)^{(-A\cdot Q^i)_+} \prod_i e^{A\cdot Q^i \wt \phi_i} \\
 &= \wt w^A \hspace{1.5in}\text{(at the interface).} \notag
\end{align}
Similarly, the pure Neumann b.c. $\wt \CN_{--...-}$ sets ${}_\pd|\wt v_A = \prod_i (\wt M_\C^i)^{(-A\cdot \wt Q^i)_+}$, which gets deformed by the superpotential $W_{\rm int}$ to the desired ${}_\pd|\wt v_A = w^{-A}|_\pd$.

All these relations among chiral operators have an immediate extension to quantum algebras, in the presence of an $\Omega$-background or $\wt\Omega$-background. The mirror map of quantized chiral rings is just \eqref{chiralmap} with `hats' on the operators. The prescription of Section \ref{sec:W-modules} shows that the desired relations are indeed implemented by the mirror-symmetry interface.

Finally, one can check that collision with the mirror-symmetry interface produces the mirror map \eqref{bcmap} of boundary conditions (and the respective map of modules). The procedure is a direct extension of our analysis above for $U(1)$ theory with a hyper (relying, in particular, on Hori-Vafa mirror symmetry) so we leave this as an exercise for the reader.

\section{Towards symplectic duality}
\label{sec:SD}

In this final section, we reconnect to some of the mathematical ideas from the introduction. In particular, we attempt to relate the physics of boundary conditions in 3d $\CN=4$ gauge theories to symplectic duality.

Many of the mathematical ingredients of symplectic duality have already appeared in our story. 
As presented in \cite{BPW-I, BLPW-II}, symplectic duality involves two categories $\CO,\CO^!$ associated to a pair of symplectic manifolds $\CM,\CM^!$ with some very special properties that make them ``conical symplectic resolutions.''%
\footnote{Conical symplectic resolutions, their quantization, and the associated categories have been studied in many other works. Relatively recent examples include \cite{McGN-der, McGN-t, BDMN, Losev-quantizations, Losev-CatO, KWWY, Webster}. As mentioned in the introduction, the basic ideas go back to work of Bernstein-Gel'fand-Gel'fand \cite{BGG} and Beilinson-Bernstein \cite{BeilinsonBernstein} on categories of highest-weight modules for simple Lie algebras.} %
Most of the properties required of $\CM$ and $\CM^!$ in the mathematical literature match natural properties of the Higgs and Coulomb branches of a 3d $\CN=4$ gauge theory that a) flows to an $\CN=4$ conformal theory in the infrared, and b) is fully massive in the presence of generic mass and FI deformations. We review these properties in Section \ref{sec:con}. We will then identify $\CM=\CM_H$ and $\CM^!=\CM_C$ as the Higgs and Coulomb branches of a gauge theory, for some fixed choice of complex structures.

From an algebraic perspective, the next step in defining the categories $\CO,\CO^!$ is to construct a deformation quantization of the rings of functions $\C[\CM]$, $\C[\CM^!]$ (that is equivariant with respect to the $\C^*$ action in property \ref{prop-contract} of Section \ref{sec:con}). Mathematically, the quantizations depend on a \emph{period}, which is a class in $H^2(\CM,\C)$. Physically, the most direct way to obtain these quantizations is to turn on $\wt \Omega$ or $\Omega$ backgrounds, as described in Sections \ref{sec:qNHiggs} and \ref{sec:NC-q}. This produces noncommutative operator algebras
\be \hat \C[\CM_H]_{t_\C}\,,\qquad \hat \C[\CM_C]_{m_\C}\,. \label{quant-mt} \ee
These algebras depend on complex FI parameters $t_\C$ and masses $m_\C$, which we identify with the periods.
Recall that the $t_\C\in H^2(\CM_H,\C)$ and  $m_\C\in H^2(\CM_C,\C)$ as desired (\cf\ \eqref{mt-isom} below).

In order for the setup to be compatible with the many kinds of boundary conditions we study in this paper, including those that break flavor symmetries $G_H$ and $G_C$, the parameters  $t_\C=k_t\epsilon$ and $m_\C=k_m\epsilon$ should be quantized in integer or half-integer multiplets of the Omega-deformation parameter $\epsilon$ (\cf\ Sections \ref{sec:lineH}, \ref{sec:lineC}). In this case, we denote the operator algebras as
\be \hat \C[\CM_H]_{k_t}\,,\qquad \hat \C[\CM_C]_{k_m}\,. \label{quant-kk} \ee

One then defines $\CO,\CO^!$ as categories of lowest-weight modules for the quantum algebras \eqref{quant-kk}.%
\footnote{The modules considered in the mathematics literature are usually ``highest-weight'' rather than ``lowest-weight.'' This is purely a matter of convention. With the definitions of weights given in this, it is more natural for us to consider lowest-weight modules.} %
Mathematically, making sense of ``lowest weight'' requires the choice of a $\C^*$ action on the algebras, induced from Hamiltonians $\C^*$ action on $\CM,\CM^!$ with isolated fixed points, as in property \ref{prop-ham} of Section \ref{sec:con}. Physically, we again know what to do. For $\CM_H$ (following Section \ref{sec:NH-mt}), we turn on a real mass $m_\R$ corresponding to a generic infinitesimal subgroup $U(1)_m\subset G_H$ of the flavor group. It grades the operator algebra $\hat \C[\CM_H]_{k_t}$, and we take
\be \CO_H = \CO_H^{k_t,m_\R}\, := \, \{\text{left $\hat \C[\CM_H]_{k_t}$-modules that are $m_\R$-lowest-weight}\}\,, \label{defOH} \ee
meaning that any operators of positive $U(1)_m$ charge act nilpotently. Similarly, for the Coulomb branch we turn on real FI parameters $t_\R$ corresponding to $U(1)_t\subset G_C$, and define
\be \CO_C = \CO_C^{k_m,t_\R}\, := \, \{\text{left $\hat \C[\CM_C]_{k_t}$-modules that are $t_\R$-lowest-weight}\}\,. \label{defOC} \ee

The lowest-weight restriction in  \eqref{defOH}--\eqref{defOC} is natural from the perspective of boundary conditions.
Indeed, in the presence of generic $m_\R$ and an $\wt \Omega$ background with complex FI parameter $t_\C = k_t\epsilon$, we expect the Higgs-branch image of any boundary condition to either a) break supersymmetry; or b) produce a module in $\CO_H^{k_t,m_\R}$. 
(In order to produce lowest-weight rather than Whittaker-like modules, it may be necessary to apply an infinite gradient flow, as in Sections \ref{sec:NC-tinf}, \ref{sec:DH-cinf}.) Similarly, in the presence of generic $t_\R$ and an $\Omega$ background with $m_\C=k_m\epsilon$, the Coulomb-branch image of any boundary condition will either break SUSY or produce a module in $\CO_C^{k_m,t_\R}$.

This immediately begs the question: if we start with a single UV boundary condition $\CB$ and consider its images $\hat\CB_H$, $\hat\CB_C$ on (quantized) Higgs and Coulomb branches, can we get a meaningful correspondence of objects in $\CO_H^{k_t,m_\R}$ and $\CO_C^{k_m,t_\R}$? We propose that the answer is yes, provided that quantization and isometry parameters are aligned:
\be k_t \sim t_\R\,,\qquad k_m\sim m_\R\,. \label{alignmt} \ee
(The precise definition of `$\sim$' appears in Section \ref{sec:order}.)
Heuristically, this alignment of classical and quantum parameters is motivated by asking that the \emph{same} UV boundary conditions preserve SUSY both in the presence and absence of Omega backgrounds. 
We describe the resulting correspondence of modules explicitly in Section \ref{sec:corresp}, in the case of abelian gauge theories. We explain how it agrees with the predictions of symplectic duality.

From the perspective of Omega backgrounds in 3d, it is not at all obvious how to obtain an equivalence of \emph{categories} (in fact, of derived categories) $\CO_H$ and $\CO_C$, rather than a mere correspondence of some objects in them. There exist two fundamental impediments to doing so.

First, in order to make sense of $\CO_H$ and $\CO_C$ as categories (rather than just sets of modules), we need to define morphisms between the objects they contain.  Mathematically, the morphisms are linear maps between modules preserving the action of $\hat\C[\CM_H]$ or $\hat \C[\CM_C]$. Physically, however, there is no way to realize such maps in an Omega background: an Omega background effectively reduces a 3d theory to 1d quantum mechanics, eliminating (naively) the possibility of having maps/transitions between boundary conditions.

The second impediment is that the $\wt \Omega$ and $\Omega$ backgrounds that quantize the Higgs and Coulomb branches are defined using completely different supercharges. Thus, even if categories $\CO_H,\CO_C$ could be made sense of, it is not physically clear why they should be dual to one another.

We propose to overcome both obstacles by using a slightly different realization of categories $\CO_C$ and $\CO_H$, as categories of A-branes in a two-dimensional theory $\CT_{2d}$ obtained by a careful compactification of a 3d theory on a circle (Section \ref{sec:3d2d}).
The theory $\CT_{2d}$ has $\CN=(4,4)$ supersymmetry and admits an entire $\cp^1\times \cp^1$ family of topological twists compatible with our boundary conditions.
The twists at $(0,1)\in \cp^1\times \cp^1$ and at $(1,0)$ effectively lead to massive A-models on the original 3d Higgs and Coulomb branches, respectively. By a result of Nadler and Zaslow \cite{NadlerZaslow} (originating in work of Kapustin and Witten \cite{KapustinWitten}), the categories of branes in these theories are equivalent to the derived module categories $\CO_H$ and $\CO_C$ when $\CM_H$ and $\CM_C$ are cotangent bundles. We expect this equivalence to hold for more general $\CM_H$ and $\CM_C$ as well.

The statement of symplectic duality now translates to the conjecture that we can move \emph{smoothly} within the family of topological twists of $\CT_{2d}$, from the A-model at $(0,1)$ to the A-model at $(1,0)$, without encountering any phase transitions --- in particular, without changing the spaces of morphisms (boundary-changing operators) in the categories of boundary conditions. While this is still a highly nontrivial conjecture, it is now a well-formed physical statement that can be directly tested and stands some chance of being correct.
 It also leads to some interesting predictions.

As we will explain in Section \ref{sec:derO}, the most interesting path between the Higgs- and Coulomb-branch A-models passes through the topological twist $(0,0)\in \cp^1\times \cp^1$. At this point, the topologically twisted theory $\CT_{2d}$ can be viewed as a B-model, in two different ways. 
If the path through the point $(0,0)$ is indeed smooth, then we expect that it should be possible to relate both categories $\CO_H,\CO_C$ involved in symplectic duality to a category of B-branes. Moreover, in contrast to the A-models at $(0,1)$ and $(1,0)$ or to either Omega-background in 3d, the B-model at $(0,0)$ preserves \emph{both} $U(1)_A$ and $U(1)_V$ R-symmetries of our original 3d theory. This suggests the existence of an extra global symmetry, or an extra grading, in the categories $\CO_H,\CO_C$. Such a grading has played an essential role in the mathematical definition of symplectic/Koszul duality, starting from the earliest examples of \cite{BGS}; nevertheless, it has also been notoriously difficult to define. 
It is promising that the extra grading occurs naturally when considering families of 2d topological twists.

Some other advantages of studying the B-type twist of $\CT_{2d}$ were discussed back in Section \ref{sec:structure} of the Introduction. For example, many functors that act on categories $\CO_C$ and $\CO_H$ --- including functors that braid mass and FI parameters, as well as Koszul duality itself --- are uniformly realized as wall-crossing transformations in $\CT_{2d}$. We explain this idea in Section \ref{sec:WC}. In Section \ref{sec:LG}, we briefly describe the two-dimensional mirror of $\CT_{2d}$, which is a Landau-Ginzburg model whose superpotential has appeared in physical constructions of knot homology.

\subsection{Conical symplectic resolutions}
\label{sec:con}

Here we review the properties that are usually required of conical symplectic resolutions $\CM,\CM^!$ in the literature on symplectic duality (in particular \cite{BPW-I, BLPW-II}), and how these properties correspond to physics of 3d $\CN=4$ gauge theories that flow to CFT's and admit fully massive deformations. Each property manifests itself in slightly different ways in the physical and mathematical descriptions. Most strikingly, the natural physical description of moduli spaces involves \emph{hyperk\"ahler} geometry, while the natural mathematical description involves \emph{complex algebraic} geometry. Here the translation between the two pictures is not very difficult, though it will become much more involved once we consider categories.%
\footnote{In the related setting of the geometric Langlands correspondence (and its physical origin), the dictionary between hyperk\"ahler and algebraic geometry is extremely nontrivial \cite{ElliottYoo}.}

\begin{enumerate}
\item $\CM$ and $\CM^!$ must be resolutions of complex symplectic cones $\CM_0,\CM_0^!$.
Correspondingly, in a 3d $\CN=4$ that flows to a CFT, the Higgs and Coulomb branches are hyperk\"ahler cones in the absence of mass and FI deformations.
The conical structure simply reflects scale invariance of the CFT. Thus we are led to identify $\CM,\CM^!$ with $\CM_H,\CM_C$, for some fixed choice of complex structures on the latter. In the fixed complex structures, the Higgs and Coulomb branches become complex symplectic manifolds as desired. Resolution corresponds to turning on real FI's $t_\R$ (for $\CM_H$) and real masses $m_\R$ (for $\CM_C$). 

\item $\CM_0$ and $\CM_0^!$ are usually required to be affine, meaning that they are completely determined by their rings of holomorphic functions --- they are cut out of $\C^d$ (for some $d$) by the polynomial relations in their rings of functions $\C[\CM_0]$, $\C[\CM_0^!]$.
Mathematically, one would express this as $\CM_0= \text{Spec}\,\C[\CM_0]$. 
This translates physically to requiring that the moduli space of the CFT is fully captured by the vevs of chiral operators --- it is \emph{not} clear why this should always be true, but it does hold in all known examples.

\item\label{prop-contract} $\CM,\CM^!$ each admits a $\C^*$ action that coincides with the contracting action on $\CM_0,\CM_0^!$ and acts on the holomorphic symplectic form with weight 2.%
\footnote{More general weights are occasionally studied in the mathematical setup. In 3d $\CN=4$ gauge theories, however, the only possible weight is 2.} %
Physically, we know that the $SU(2)_H\times SU(2)_C$ R-symmetry group acts via metric isometries of the cones $\CM_H^{(0)}$, $\CM_C^{(0)}$, while rotating the $\cp^1$'s of complex structures. A $U(1)_H\times U(1)_C$ subgroup preserves any given choice of complex structures, while rotating the phases of the complex symplectic forms with weight 2, as desired. Upon turning on real FI and mass parameters to resolve the branches, this $U(1)_H\times U(1)_C$ subgroup is preserved. Moreover, any $U(1)$ isometry of a K\"ahler manifold is automatically promoted to a $\C^*$ complex (but not metric) isometry, matching the mathematical description of the symmetry.%
\footnote{Viewing $\CM_H$ (say) in a fixed complex structure as a K\"ahler manifold, the $U(1)$ that preserves the complex structure is Hamiltonian. The $U(1)$ isometry is promoted to a $\C^*$ by using gradient flow with respect to its real moment map $\mu_\R$. Explicitly, letting $\omega=Ig$ denote the K\"ahler form and metric, and letting  $V=\omega^{-1}d\mu_\R$ denote the vector field that generates $U(1)$, the complexification is $V+g^{-1}d\mu_\R = (\omega^{-1}+g^{-1})d\mu_\R$.\label{foot:C*}}

\item The resolutions $\CM,\CM^!$ are (usually) required to be smooth. Correspondingly, in a physical theory that admits enough FI and mass deformations to make it fully massive, the Higgs (Coulomb) branch can always be fully resolved by turning on generic FI (mass) parameters. The basic idea behind this relationship is that any singularities on (say) the Higgs branch should correspond to massless degrees of freedom on the Coulomb branch, and vice versa.

\item\label{prop-ham} Both $\CM$ and $\CM^!$ are (usually) required to admit $\C^*$ actions that \emph{preserve} the complex symplectic forms and have isolated fixed points. Physically, the existence of these actions is tied to the existence of mass and FI parameters that make the theory fully massive. Indeed, a choice of real masses $m_\R$ that makes $\CM_H$ (say) massive is equivalent to a choice of subgroup $U(1)_m\subset G_H$ in the Higgs-branch flavor group that has isolated fixed points (the vacua). Similarly, a choice of $t_\R$ that makes the Coulomb branch massive is the same as a subgroup $U(1)_t\subset G_C$ with isolated fixed points. In complex geometry, these $U(1)$'s are again promoted to $\C^*$'s. Since they are flavor symmetries, they preserve the full hyperk\"ahler structure -- they are tri-Hamiltonian.

\item A pair $\CM,\CM^!$ involved in symplectic duality has
\be \text{dim}\,H^2(\CM,\R) = \text{rank}\,G^!\,,\qquad \text{dim}\,H^2(\CM^!,\R) = \text{rank}\,G\,,\ee
where $G$ and $G^!$ are the groups of (complex) Hamiltonian isometries of $\CM,\CM^!$. Physically, we simply have
\be \label{mt-isom}
 \begin{array}{llll} \mathfrak t_C &=  \{\text{space of FI parameters}\} &= \{\text{space of $\CM_H$ resolutions}\} &= H^2(\CM_H,\R) \\[.1cm]
\mathfrak t_H &=  \{\text{space of mass parameters}\} &= \{\text{space of $\CM_C$ resolutions}\} &= H^2(\CM_C,\R)\,, 
\end{array}
\ee
where $\mathfrak t_C,\mathfrak t_H$ are the real Cartan subalgebras of the flavor groups $G_C,G_H$.

\item Though we will not need it here, the pairs $\CM_0$ and $\CM_0^!$ involved in symplectic duality have also been observed to admit stratifications that are in 1--1 order-reversing bijection. Physically, these stratifications come from mixed branches in the moduli space. Concretely, the Higgs branch may contain conical ``strata'' $[\CM_H^{(0)}]_{G'} \subset \CM_H^{(0)}$ along which a continuous subgroup $G'\subset G$ of the gauge group remains unbroken. Along each such stratum, the fields of a $G'$ vectormultiplet may get expectation values, so a partial Coulomb branch $[\CM_C^{(0)}]_{G'} \subset \CM_C^{(0)}$, with quaternionic dimension equal to $\text{rank}(G')$. Both the Higgs and Coulomb branches can be expressed as disjoint unions of such strata
\be \CM_H^{(0)} = \bigsqcup_{G'\subset G} [\CM_H^{(0)}]_{G'}\,,\qquad \CM_C^{(0)} = \bigsqcup_{G'\subset G} [\CM_C^{(0)}]_{G'}\,.  \ee
Taking closures, we have $[\CM_H^{(0)}]_{G'} \subset \ol{ [\CM_H^{(0)}]_{G''}}$ and $[\CM_C^{(0)}]_{G''} \subset \ol{  [\CM_C^{(0)}]_{G'}}$ if and only if $G''\subset G'$ (this is what is meant by order-reversing bijection). The full moduli space of the 3d $\CN=4$ theory takes the form
\be \CM_{\rm full} = \bigsqcup_{G'\subset G}  [\CM_H^{(0)}]_{G'}\times  [\CM_H^{(0)}]_{G'}\,, \ee
where the closure of the component with $G'={id}$ is the standard Higgs branch, the closure of the component with $G'=G$ is the standard Coulomb branch, and all other components are known as mixed branches.

\end{enumerate}

Notice that the match between the physical and mathematical properties is not perfect, but is very close. In some cases, the physical properties already come with some nontrivial predictions. For example, from the physics of flavor symmetries and associated mass/FI deformations, it follows that the $\CM_H$ can be fully resolved if and only if $\CM_C$ admits a $U(1)$ action with isolated fixed points, and vice versa. Thus, the generalization of symplectic duality to singular $\CM$ must necessarily involve non-isolated fixed loci of the Hamiltonian $\C^*$ action on $\CM^!$.

\subsection{The cast of modules}
\label{sec:cast}

Physically, we use $\Omega,\wt \Omega$ backgrounds to quantize the algebras of local operators $\hat \C[\CM_H]_{k_t}$, $\hat \C[\CM_C]_{k_m}$, and we find that in the presence of generic $m_\R,t_\R$ any right boundary condition that preserves SUSY defines a lowest-weight module in the categories $\CO_H^{k_t,m_\R},\CO_C^{k_m,t_\R}$, as in \eqref{defOH}--\eqref{defOC}.
The fact that $\CM_H,\CM_C$ are conical symplectic resolutions as in Section \ref{sec:con} implies that the categories $\CO_H^{k_t,m_\R},\CO_C^{k_m,t_\R}$ have a great deal of additional structure. In particular, by \cite[Thm 5.12]{BLPW-II}, they are so-called \emph{highest-weight categories} \cite{CPS}.%
\footnote{We use the standard terminology of ``highest-weight'' categories here even though, in our natural conventions, $\CO_H^{k_t,m_\R},\CO_C^{k_m,t_\R}$ would more properly be called ``lowest-weight'' categories. Throughout this section, the various properties of modules induced by an order on the vacua are actually written in our natural lowest-weight conventions.} %
Also, conjecturally, they are \emph{Koszul} categories \cite{BGS}.

In this section, we want to explain a bit of this additional structure, and how it fits in with the physics of boundary conditions. The basic point to make is that $\CO_H^{k_t,m_\R}$ (or $\CO_C^{k_m,t_\R}$) is generated by any one of \emph{six} fundamental, finite collections of modules: simples, standards, costandards, projectives, injectives, and tiltings. The objects in each collection are indexed by vacua $\nu$ of the underlying 3d $\CN=4$ theory --- which we know can be thought of as $m_\R$-fixed points of $\CM_H$ or $t_\R$-fixed points of $\CM_C$. Moreover, the \emph{ordering} of vacua given by the moment map $h_m$ on $\CM_H$ (or $h_t$ on $\CM_C$) leads to certain constraints among the morphisms in each family.

We have already encountered some of these families in the study of boundary conditions. We will now describe each of them more systematically and in the process explain what it means to be a highest-weight category. The Koszul property will be revisited in Section~\ref{sec:tale}.

\subsubsection{Orders, walls, and chambers}
\label{sec:order}

A central notion in the definition of a highest-weight category is a partially ordered set of ``weights'' $\Pi$, whose elements $\nu$ index various special collections of modules. Physically, $\Pi$ is the set of isolated massive vacua in a 3d $\CN=4$ theory with real parameters $m_\R,t_\R$ turned on. We would like to explain why this set is ordered.

Recall that the vacua $\nu$ can be viewed equivalently as either the critical points of a real moment map $h_m = m_\R\cdot \mu_\R^H$ on the Higgs branch or a real moment map $h_t = t_\R\cdot \mu_\R^C$ on the Coulomb branch. As long as $m_\R$ and $t_\R$ are generic, the critical values $h_m(\nu)$ and $h_t(\nu)$ are all distinct, and we can define an order
\begin{subequations} \label{order-vac}
\be \nu < \nu' \qquad \Leftrightarrow \qquad h_m(\nu) < h_m(\nu')\,, 
 \label{order-vacH}
 \ee
or
\be \nu < \nu' \qquad \Leftrightarrow \qquad h_t(\nu) < h_t(\nu')\,.
 \label{order-vacC} \ee
\end{subequations}

The two orders (\ref{order-vac}a-b) necessarily coincide. One way to see this is to observe that the critical values $h_m(\nu)$ and $h_t(\nu)$ both coincide with a single set of real central charges
\be h_\nu(m_\R,t_\R) = h_m(\nu) = h_t(\nu)\,, \ee
which arise as effective background Chern-Simons couplings in a vacuum of the 3d $\CN=4$ theory (Appendix \ref{app:central}). These central charges govern the tension of half-BPS domain walls. For each fixed vacuum $\nu$, they are bilinear in \emph{both} $m_\R$ and $t_\R$.  An explicit formula for $h_\nu(m_\R,t_\R)$ in abelian theories is given in \eqref{abel-central}.

In the mathematics of highest-weight categories, the set $\Pi$ is only partially ordered. To obtain a partial order on the vacua, one says that $\nu<\nu'$ if and only if the RHS of \eqref{order-vac} are satisfied \emph{and} there exists a half-BPS domain wall interpolating between $\nu$ and $\nu'$. Since the BPS equations for 2d $\CN=(2,2)$ supersymmetry descend to gradient flow for $h_m$ on the Higgs branch and $h_t$ on the Coulomb branch (Sections \ref{sec:NH-mt}, \ref{sec:NC-mt}, Appendix \ref{app:hW}), this additional requirement is equivalent to the existence of a gradient flow between vacua $\nu$ and $\nu'$, on either branch.

The full space of mass and FI parameters is cut into chambers by codimension-one walls  $\mathbb W_{\nu,\nu'}$  labelled by pairs of distinct vacua
\be \mathbb W_{\nu,\nu'} := \{(m,t)\in \mathfrak t_H\times \mathfrak t_C\;\,\text{s.t.}\;\, h_\nu(m,t)=h_{\nu'}(m,t)\}\,. \label{walls} \ee
These walls are the loci in parameter space where the tension of a putative half-BPS domain wall goes to zero. Within each chamber, the order of the vacua is constant.
(One could alternatively say that there is a wall $\mathbb W_{\nu,\nu'}$ if and only if there actually exists a half-BPS domain wall between $\nu$ and $\nu'$. Then within each chamber the partial order of the vacua would be constant. We will not use this refined notion of walls here, and we will will generally use orders rather than partial orders.)

We say that a pair $(t,m)$ is \emph{generic} if it lies in the complement of the walls \eqref{walls}, \ie\ if all critical points of $h_m$ and $h_t$ are isolated and all critical values are distinct. We say that generic parameters $(t,m)\sim (t',m')$ are \emph{aligned} if they lie inside the same chamber of parameter space.

At any point in this discussion, we could have replaced the real parameters $m_\R$, $t_\R$ with the quantized parameters $k_m$, $k_t$ that appear in the definitions of quantum algebras and modules. We simply identify the space of quantized parameters with a sublattice in the space of real parameters. It then makes sense to say that $k_t\sim t_\R$ are aligned (at fixed $m_\R$), or that $k_m\sim m_\R$ are aligned (at fixed $t_\R$).

In a 2d compactification of a 3d $\CN=4$ theory, the loci \eqref{walls} describe some of the walls of marginal stability, corresponding to massless 2d solitons that come from compactifying domain walls. We will come back to this later.

\begin{figure}[htb]
\hspace{-.5in}\includegraphics[width=7in]{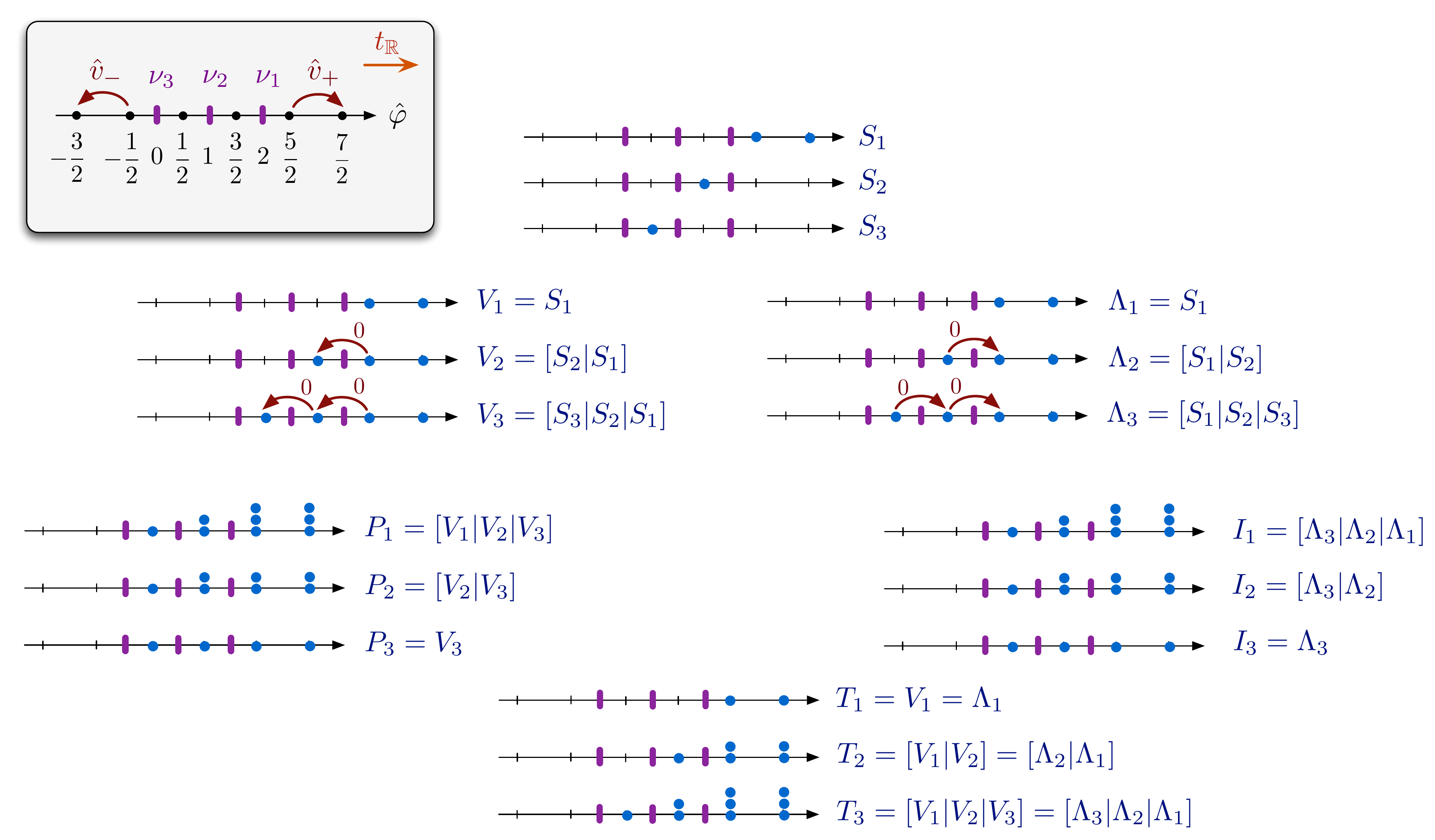}
\caption{The six sets of modules that generate category $\CO_C$ for the Coulomb branch of SQED (\ie\ $U(1)$ gauge theory) with $N_f=3$ hypermultiplets of charge $+1$. We have chosen $t_\R=1$ and $k_m=(-2,-1)$, and conventions for gauge/flavor charges are as in \eqref{QqSQED} on page \pageref{QqSQED}. The Coulomb branch itself is a resolution of the $\C^2/\Z_3$ singularity, and the quantum algebra is $\hat v_\pm\hat v_\mp = (\hat\varphi\mp\tfrac\epsilon2)(\hat\varphi+(k_{m,1}\mp\tfrac12)\epsilon)(\hat\varphi+(k_{m,2}\mp\tfrac12)\epsilon)$,\; $[\hat\varphi,\hat v_\pm]=\pm\epsilon v_\pm$. The real FI parameter corresponds to a potential $\hat h_t = t_\R\hat\varphi$, with respect to which these modules are lowest-weight. Within each generating set, the modules are labelled by the three vacua of the massive theory $(\nu_1,\nu_2,\nu_3)$.
In the figure we use $n$ stacked blue dots to depict a weight space of dimension~$n$.}
\label{fig:cast}
\end{figure}

\subsubsection{Simple modules}
\label{sec:simplemod}

The first property of a highest-weight category is that it is Noetherian and Artinian, which imples that every module has a finite composition series (\cf \, \eqref{xD-comp}). In particular, every module is a finite iterated extension of irreducible modules $S_\nu$, otherwise known as \emph{simple} modules.
Moreover, one requires that there is a partially ordered set $\Pi$ indexing the simple modules, and that for all $\nu,\nu' \in \Pi$
\be
\text{Hom}(S_{\nu}, S_{\nu'}) = \delta_{\nu,\nu'}\C\,.
\ee
The set $\Pi$ is the set of vacua of the theory, ordered (or partially ordered) as explained above.

In an abelian theory, the simples for (say) the Higgs branch are supported on chambers of the quantum hyperplane arrangement that are $k_t$-feasible and on which $h_m$ is bounded from below. We could call these chambers $\Delta_\nu$, labeling them by the vacua $\nu$ lying at the $\hat h_m$-minimal points of the chambers. Equivalently, we may introduce a quantum moment map $\hat h_m$ as in \eqref{def-hathm} and evaluate it on lowest-weight vectors of the modules $S_{\nu}$ to define the ordering. 

In abelian theories, all the simple modules in $\CO_H^{(k_t,m_\R)}$ (resp. $\CO_C^{(k_m,t_\R)}$) are realized as images of Neumann (resp., generic Dirichlet) boundary conditions in the UV. The simple modules on the Coulomb branch of SQED with three hypermultiplets are shown in Figure \ref{fig:cast}.

In nonabelian theories, pure Neumann boundary conditions that preserve the full $G$ gauge symmetry are not enough to produce all the simple modules on (say) the Higgs branches. It appears necessary (and sufficient) to consider a larger family of mixed Neumann-Dirichlet boundary conditions that preserve a maximal torus of $G$. We will investigate this elsewhere.

\subsubsection{Standard modules}
\label{sec:stdmod}

The second property of a highest-weight category is that for each $\nu \in \Pi$ there is a \emph{standard module} $V_{\nu}$ equipped with a surjection
\be V_\nu \to\hspace{-.3cm}\to S_\nu\,, \ee \label{std-map}
such that the composition series for the kernel of \eqref{std-map} contains only $S_{\nu'}$ with $\nu' > \nu$.
More generally, the standard modules have a composition series of the form
\be V_{\nu_i} \,=\, \big[\, S_{\nu_i}\,\big|\, S_{\nu_{j_1}} \,\big|\, ... \,\big|\, S_{\nu_{j_n}}\,\big]\,, \label{VS} \ee
with $S_{\nu}$ appearing before $S_{\nu'}$ if and only if $\nu\leq \nu'$. (Notation is as in \eqref{xD-comp}. A given $S_\nu$ may appear more than once here.)
Thus the relation between standard and simple modules is ``triangular.''

It is easy to see that the standard modules generate any highest-weight category and the properties of projective modules which we will discuss in Section \ref{sec:projmod} imply that the standard modules form an \emph{exceptional collection} with respect to the ordering on $\Pi$. This means there only exist maps and extensions%
\footnote{Recall that for any objects $A$, $B$ in an abelian category (such as a category of modules), $\text{Ext}^0(A,B) = \text{Hom}(A,B)$; and $\text{Ext}^n(A,B)$ is the group of extensions of $A$ by $B$ of length $n$; for example the elements of $\text{Ext}^1(A,B)$ are exact sequences of the form $0\to B\to C\to A \to 0$.} %
among standards in a particular order:
\be \text{Ext}^n(V_{\nu},V_{\mu}) = 0  \quad\text{if $\nu < \mu$}\,;\quad \text{Ext}^n(V_{\nu},V_{\nu}) = \C\,\delta_{n,0}\,. \label{std-order}\ee

It is possible to give uniform construction of standard modules in (say) $\CO_H^{(k_t,m_\R)}$. Let $A = \hat \C[\CM_H]_{k_t}$ denote the quantized algebra of operators on the Higgs branch, and let $A_<, A_0,A_>$ denote the subalgebras of operators with negative, zero, and positive charge (respectively) under the global symmetry $U(1)_m\subset G_H$ generated by $m_\R$. Consider the quotient
\be  B = A_0 / (A_0\cap A_>A_<)\,. \label{AB} \ee
Since $A_0\cap A_>A_<$ is a two-sided ideal, $B$ is again an algebra; in fact, it is just a quantization of the algebra of functions on the $U(1)_m$-fixed locus $\CM_H^0$ of $\CM_H$. This fixed locus is exactly the collection of vacua. In the notation of Section \ref{sec:NH-mt}, we would write $\CM_H^0 = \bigcup_{\nu} \CM_H^0[m_\R^{\nu}] = \bigcup_{\nu} \{\nu\}$. We find that
\be B \simeq \bigoplus_{\nu} \C e_{\nu}\,, \ee
where the generators $e_{\nu}$ of the algebra obey  $e_{\nu}e_{\nu'} = \delta_{\nu, \nu'}e_{\nu}$.
Let $\C e_{\nu}$ denote the 1-dimensional left module for $B$ generated by $e_{\nu}$. It can be upgraded to a left module for $A_{\leq 0} = A_<\oplus A_0$ simply by setting $A_<\cdot e_{\nu}=0$\,.

Then the \emph{standard lowest-weight $A$-module} $V_{\nu}$ is defined as the ``induced'' module
\be V_{\nu} := A \otimes_{A_{\leq 0}} \C e_{\nu}\,. \label{def-Vnu} \ee
Here the tensor product instructs us to take all elements $a\otimes e_{\nu} \in A\otimes \C e_{\nu}$, modulo the relation $(aa')\otimes e_{\nu} = a\otimes (a'e_{\nu})$ for all $a'\in A_{\leq 0}$. The algebra $A$ acts on such elements by multiplication on the left. Intuitively, the module $V_{\nu}$ is freely generated by acting with $A_>$ on a single vector $e_{\nu}$ that is an eigenvector for $A_0$ and is annihilated by all of $A_<$. The construction \eqref{def-Vnu} generalizes the standard definition of Verma modules for Lie algebras.%
\footnote{The process of induction \eqref{def-Vnu} might be given a physical interpretation using the variation of $m_\R,t_\R$ as functions of $x^1$ that was described on page \pageref{mtJanus}.}

We first met standard modules in Section \ref{sec:xD}. We gave in Section \ref{sec:thimbles} a prescription for associating an exceptional Dirichlet boundary condition $\CD_{L,c}$ to any vacuum $\nu$, such that the classical Higgs and Coulomb-branch images of $\CD_{L,c}$ would be thimble branes attached to $\nu$. We conjectured that the quantized images would be standard modules whenever parameters $k_t\sim -t_\R$ (or $k_m\sim -m_\R$) were anti-aligned.
Physically, one would expect that any (IR) boundary condition in a massive theory can be ``built'' by suitably composing thimble branes. This expectation remains to be made precise in three-dimensional theories, but its two-dimensional analogue has been well studied, \cf~\cite{HIV}.  At a rough level, the property \eqref{std-order} of being an exceptional collection  can be understood by considering half-BPS domain walls between vacua.
The space $\text{Ext}^0(V_{\nu},V_{\nu'})$ is generated by half-BPS domain walls on $\R\times \R^2$ that interpolate between $\nu$ at $x^1\to-\infty$ and $\nu'$ at $x^1\to\infty$, which can exist only if $\nu \geq \nu'$.

In abelian theories, the standard modules are easy to describe in terms of quantum hyperplane arrangements. For the Higgs (Coulomb) branch, each standard $V_{\nu}$ is supported on the orthant of the hyperplane arrangement whose origin is the maximal intersection of hyperplanes labeled by $\nu$ on which $\hat h_m$ ($\hat h_t$) is bounded from below. A simple module $S_{\nu'}$ is contained in the composition series for $V_{\nu}$ if and only if the chamber $\Delta_{\nu'}$ is contained in the orthant for $V_{\nu}$. For example, on the Coulomb branch of SQED with three hypermultiplets, the three standard modules are depicted in Figure~\ref{fig:cast}.

\subsubsection{Projective modules}
\label{sec:projmod}

Recall that a module $P$ is projective if and only if it is maximally extended; that is, for any other module $M$,
\be \text{Ext}^n(P,M)=0\,,\qquad n\geq 1\,. \label{proj-exact}\ee
In a highest-weight category every standard module $V_{\nu}$ is required to have an indecomposable projective cover
\be P_{\nu}  \to\hspace{-.3cm}\to V_{\nu} \,. \ee
Moreover, it is required that $P_{\nu}$ admits a \emph{standard filtration} with respect to the reverse ordering of vacua; in other words, each $P_{\nu_i}$ is a successive extension of standard modules
\be P_{\nu_i} \,=\, \big[\, V_{\nu_i}\,\big|\, V_{\nu_{j_1}} \,\big|\, ... \,\big|\, V_{\nu_{j_n}}\,\big]\,, \label{PV} \ee
with $V_\nu$ appearing before $V_{\nu'}$ if and only if $\nu\geq \nu'$. A given standard module may appear more than once.
The quotient $V_{\nu_i}$  is called the head of the filtration.
 (Again, notation is as in \eqref{xD-comp}.)

In an abelian theory, a standard module $V_{\nu'}$ appears in the standard filtration for $P_{\nu}$ if and only if the orthant supporting $V_{\mu'}$ contains $\nu$. We will argue in Section \ref{sec:corresp} that every projective module in category $\CO_{H}^{k_t,m_\R}$ ($\CO_C^{k_m,t_\R}$) of an abelian theory can be obtained as the image of a pure Dirichlet (pure Neumann) boundary condition.

\subsubsection{Costandard modules}
\label{sec:costdmod}

The axioms defining a highest-weight category actually imply the existence of \emph{costandard modules} $\Lambda_{\nu}$ whose behavior is dual to that of $V_{\nu}$. For example, dual to \eqref{std-map} there is an inclusion
\be
S_{\nu} \hookrightarrow \Lambda_\nu\,,
\ee
and more generally
\be \Lambda_{\nu_i} \,=\, \big[\, S_{j_n}\,\big|\, ...\,\big|\, S_{\nu_{j_1}} \,\big|\, S_{\nu_{i}}\,\big]\,, \label{LambdaS} \ee
with the same simples as in \eqref{VS}, but in the opposite order.
The costandard modules form an exceptional collection with respect to the opposite ordering 
\be \text{Ext}^n(\Lambda_{\nu},\Lambda_{\nu'}) = 0  \quad\text{if $\nu >\nu'$}\,;\quad \text{Ext}^n(\Lambda_{\nu},\Lambda_{\nu}) = \C\,\delta_{n,0}\,. \label{costd-order}\ee

The costandard modules in $\CO_H^{(k_t,m_\R)}$ are constructed using the dual of the tensor product \eqref{def-Vnu}. Let $A_<,A_0,A_>$, $B$, and $\C e_{\nu}$ be as in \eqref{AB}. Then one sets 
\be \Lambda_{\nu} \,:=\, \text{Hom}_{A_{\geq 0}}(A,\C e_{\nu})\,. \label{def-costd}\ee
As a vector space, $\Lambda_{\nu}$ is simply the space of all maps $f :A\to\C e_{\nu}$ that commute with the left action of $A_{\geq 0}$. This space has a left action of $A$, given by $a\cdot f(-)=f(- \cdot a)$.

We conjectured in Section \ref{sec:thimbles} (and later proved for abelian theories) that the exceptional Dirichlet b.c. $\CD_{L,c}$ associated to a vacuum $\nu$ produces costandard modules on both Higgs and Coulomb branches so long as the parameters $k_t\sim t_\R$, $k_m\sim m_\R$ are aligned. All costandard modules arise this way. The relative sign in the alignment of parameters for standard and costandard modules accounts for the reversal in \eqref{costd-order}.

\subsubsection{Injective modules}

An injective module $I$ is defined by the property that for any module $M$,
\be  \text{Ext}^n(M,I) = 0\,,\qquad n\geq 1\,. \label{ind-exact}\ee
In a highest-weight category each costandard module has an injective hull 
\be \Lambda_\nu \hookrightarrow
 I_\nu \ee
that behaves dually to $P_{\nu}$. In particular each injective module admits a \emph{costandard filtration}
\be I_{\nu_i} \,=\, \big[\, \Lambda_{j_n}\,\big|\, ...\,\big|\, \Lambda_{\nu_{j_1}} \,\big|\, \Lambda_{\nu_{i}}\,\big]\,, \label{i-filter} \ee
with the same vacua as in \eqref{PV}, but in opposite order. The submodule $\Lambda_{\nu_i}$ is called the \emph{tail} of the filtration.

The injective modules in categories $\CO_{H}^{k_t,m_\R}$ and $\CO_C^{k_m,t_\R}$, are dual to the projectives. For example just as there is a unique indecomposable projective module $P_{\nu}$ sitting in the sequence $P_\nu\to\hspace{-.3cm}\to V_\nu\to\hspace{-.3cm}\to S_\nu$  there is a unique indecomposable injective module $I_\nu$ sitting in the sequence
\be S_\nu \hookrightarrow
\Lambda_\nu \hookrightarrow
I_\nu\,.  \label{IS} \ee 
All indecomposable injectives arise this way, and they generate the category.

None of the UV boundary conditions considered in this paper seem to have images that generically coincide with injective modules.

\subsubsection{Tilting modules}

Finally, the indecomposable tilting modules $T_\nu$ in a highest-weight category are characterized by having \emph{both} a standard filtration with tail $V_{\nu}$ and a costandard filtration with head $\Lambda_{\nu}$. Any module that is both projective and injective is automatically tilting, though the converse is far from true.

While neither $\text{Ext}^{n}(T,M)$ nor $\text{Ext}^{n}(M,T)$ vanish in general when $n\geq 1$ (as they do for projectives and injectives, respectively), tilting modules have the property that
\be \begin{cases} \text{Ext}^{n\geq 1}(T, M) = 0 & \text{if $M$ admits a standard filtration} \\[.1cm]
\text{Ext}^{n\geq 1}(M, T) = 0 & \text{if $M$ admits a costandard filtration} \end{cases}
\label{tilt-exact}
\ee
In particular, for any two tilting modules, $\text{Ext}^n(T_\nu,T_{\nu'})=0$ if $n\geq 1$.

In abelian theories, a standard module $V_{\nu'}$ appears in the standard filtration of $T_{\nu}$ if and only if the orthant $V_{\nu}$ contains $\nu'$. Just like projective modules, we will argue that all tilting modules occur as images of Dirichlet (Neumann) boundary conditions on the Higgs (Coulomb) branches.

Heuristically, the tilting modules in (say) $\CO_{H}^{k_t,m_\R}$ are related to projectives by reversing the sign $k_t\to-k_t$ of the quantization parameter, much the same way that costandards are related to standards. They turn out to play a central role in the physical realization of symplectic duality.

\subsection{A tale of many functors}
\label{sec:tale}
In the previous section, we described six families of modules that each generate the category $\CO_H^{k_t,m_\R}$ (or $\CO_C^{k_m,t_\R}$).
These families of modules come with canonical quotient and inclusion maps, which can be summarized as
\be  \raisebox{-.6in}{$\includegraphics[width=1.8in]{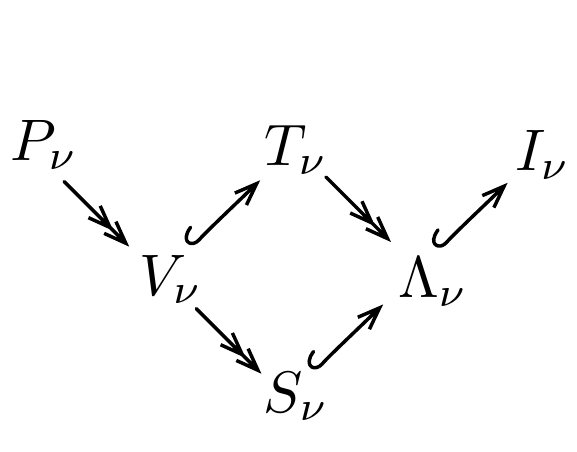}$} \label{rels1} \ee
Moreover, every module in this diagram is related to a collection of modules sitting below it by constructing an iterated extension. For example, $P_\nu$ and $T_\nu$ are both extensions of collections of Vermas that include $V_\nu$. The extensions all occur in a particular order, dictated by the ordering of vacua. Thus, if we view \eqref{rels1} as a graph, every edge in the graph represents a \emph{triangular} relationship of modules.

For many applications, including symplectic duality and (physically) the study of boundary conditions in compactified 2d theories, categories of modules are not quite enough: one must extend $\CO_H^{k_t,m_\R}$ and $\CO_C^{k_m,t_\R}$ to derived categories $D^b\CO_H^{k_t,m_\R}$ and $D^b\CO_C^{k_m,t_\R}$. We briefly recall that the objects of the derived category $D^b\CO_H^{k_t,m_\R}$ (say) are \emph{complexes} of modules of $\CO_H^{k_t,m_\R}$, considered modulo quasi-isomorphism, \ie\ two complexes are deemed isomorphic if there is a map from one to the other that preserves homology.

The derived categories $D^b\CO_H^{k_t,m_\R}$ and $D^b\CO_C^{k_m,t_\R}$ are, understandably, quite complicated. We may, however, summarize quite a few of their properties and auto-equivalences by extending the diagram \eqref{rels1}. The derived category turns out to look like
\be  \raisebox{-1in}{$\includegraphics[width=5.4in]{O-rels-D}$} \label{rels2} \ee
Each dot here represents a family of objects in the derived category labelled by the vacua $\nu$; and each edge represents a triangular relationship between these families. Particularly nice equivalences between categories $\CO$ exchange the various families of modules and hence induce symmetries of the diagram. We proceed to describe a few of them, and in the process justify the diagram itself. We focus on the Higgs branch; the corresponding functors for the Coulomb branch are identical.

\subsubsection{Highest-weight equivalences}
\label{sec:star}

An exact equivalence between highest-weight categories $\CC_1$ and $\CC_2$ is called a  \emph{highest-weight equivalence} if it sends standard modules to standard modules and hence induces an order-preserving bijection between the weights for $\CC_1$ and $\CC_2$. Since all exact equivalences must also preserve simples, projectives, and injectives we see that highest-weight equivalences identify the diagrams \eqref{rels2} for different categories.

One example of a highest-weight equivalence is the functor that takes a module $M$ in $\CO_H^{k_t,m_\R}$ with general $m_\R$-eigenspace decomposition $M = \oplus_\alpha M_{\alpha}$ to its restricted dual $M^{\star} = \oplus_\alpha \text{Hom}_{\C}(M_\alpha, \C)$  \cite[Sec. 4.2]{Losev-quantizations}.
Note that $M^{\star}$ is a right $\hat \C[\CM_H]_{k_t}$-module but using the natural isomorphism $\hat \C[\CM_H]_{-k_t} \cong \hat \C[\CM_H]_{k_t}^{\text{op}}$ we can view $M^{\star}$ as an object of $\CO_H^{-k_t,-m_\R}$. Since $\star$ reverses the order of arrows it is a highest weight equivalence
\[
\star: \CO_H^{k_t,m_\R} \to (\CO_H^{-k_t,-m_\R})^{\text{op}}.
\]
where $(\CO_H^{-k_t,-m_\R})^{\text{op}}$ is the \emph{opposite category} of $\CO_H^{-k_t,-m_\R}$. Thus we can identify the diagrams of $\CO_H^{k_t,m_\R}$ and $(\CO_H^{-k_t,-m_\R})^{\text{op}}$.

Recall that the opposite category $\CC^{op}$ of a category $\CC$ has the same objects as $\CC$ but the morphism spaces are reversed. If $\CC$ is highest weight, then $\CC^{op}$ is highest weight with respect to the opposite order. The standards in $\CC$ become the costandards in $\CC^{op}$. In fact the diagram \eqref{rels1} for $\CC^{op}$ is a vertical reflection of the diagram for $\CC$. For this reason we think it is natural to represent $\star$ as a reflection of diagram \eqref{rels2} about a vertical axis.

\subsubsection{Shuffles, twists, and braiding}
\label{sec:shuffle}

Given any two values $k_t,k_t'$ of the quantization parameter for the Higgs branch with integral difference, there is a covariant functor relating the derived categories
\be  \Phi^{k_t',k_t}\,:\; D^b\,\CO_H^{k_t,m_\R} \to  D^b\,\CO_H^{k_t',m_\R}\,. \label{ptwist} \ee
In the mathematics literature it is sometimes known as a \emph{twisting} functor \cite[Sec. 8.1]{BLPW-II}. Similarly, given any two values $m_\R,m_\R'$ with integral difference there is a \emph{shuffling} functor
\be \Psi^{m_\R',m_\R}\,:\; D^b\,\CO_H^{k_t,m_\R} \to  D^b\,\CO_H^{k_t,m_\R'}\,. \label{pshuffle} \ee
Both of these functors have been proven to be equivalences of derived categories, as long as the parameters $k_t,k_t'$ and $m_\R,m_\R'$ are all generic \cite[Prop. 6.32]{BPW-I} \cite[Thm. 7.3]{Losev-quantizations}.

If $(k_t,m_\R)$ and $(k_t',m_\R')$ belong the same chamber in parameter space, in the sense of Section \ref{sec:order}, the twisting and shuffling actions are fairly trivial. In contrast, the twists and shuffles that cross the walls \eqref{walls} from one chamber to another combine to generate a generalized braid action on the derived $D^b\CO_H$. When we (conjecturally) identify category $D^b\CO_H$ with a category of boundary conditions in a 2d B-model in Section \ref{sec:2d-SD}, we will find that the twisting and shuffling actions correspond to ordinary wall crossing transformations. In the 2d theory, masses, FI parameters, and the central charges $h_\nu(m,t)$ are all complexified. Then the generalized braid action can succinctly be described as an action of the fundamental group of the complexified space
\be \mathfrak t_H^\C\times \mathfrak t_C^\C - \big(\cup_{\nu , \nu'} \mathbb W_{\nu,\nu'}^\C\big) \label{param-braid}\ee
on $D^b\CO_H$, where $\mathfrak t_H^\C$, $\mathfrak t_C^\C$ are the complex Cartans of the flavor symmetry groups $G_H$, $G_C$.

When the flavor groups $G_H$, $G_C$ are non-abelian, the respective Weyl groups $W_H$, $W_C$ also act on $\mathfrak t_H^\C\times \mathfrak t_C^\C$, permuting the walls. One then arrives at a categorical action of the fundamental group of 
\be \big[\mathfrak t_H^\C\times \mathfrak t_C^\C - \big(\cup_{\nu , \nu'} \mathbb W_{\nu,\nu'}^\C\big)\big]/(W_H\times W_C) \ee
on each $D^b\CO_H^{k_t,m_\R}$.

In the mathematics literature, it is well known that twisting and shuffling separately give commuting braid actions. (These are the braid actions that have played a central role in knot homology, as discussed briefly at the end of the Introduction and in Section \ref{sec:LG}.) A new prediction from our physical picture is that both actions are controlled by a \emph{single} set of central charges $h_\nu(m,t)$.

One consequence of this idea is that the the transformations that send $t \mapsto -t$ and $m \mapsto -m$ cross exactly the same walls in parameter space, since the both send $h_\nu(m,t)$ to $-h_\nu(m,t)$. Thus one might guess that the \emph{long twist} $\Phi=\Phi^{-k_t,k_t}$ and the \emph{long shuffle} functor $\Psi = \Psi^{-m_\R,m_\R}$ act the same way on $D^b\CO_H$. 
Mathematically, this doesn't quite make sense because $D^b\CO_H^{k_t,m_\R}$ is mapped to $D^b\CO_H^{-k_t,m_\R}$ and to $D^b\CO_H^{k_t,-m_\R}$ by $\Phi$ and $\Psi$, respectively.  The best that one can hope for is that there is a highest weight equivalence $D^b\,\CO_H^{-k_t,m_\R} \to D^b\,\CO_H^{k_t, -m_\R}$ intertwining the two functors. Indeed, this is almost exactly what happens: Losev has shown that $\Phi$ and $\Psi^{-1}$ are both Ringel dualities and hence are intertwined by a highest weight equivalence \cite{Losev-quantizations}.

\subsubsection{Ringel dualities} \label{sec:ringel}

A Ringel duality $R:\CC_1 \to \CC_2$ is an equivalence of highest-weight categories that restricts to an exact equivalence between the subcategories $\CC_1^{V}$ and $\CC_2^{\Lambda}$ of objects admitting standard and costandard filtrations, respectively. Such a functor reverses the order of weights/vacua $\nu$, and sends the families $(V_{\nu}, P_{\nu}, T_{\nu})$ to $(\Lambda_{\nu}, T_{\nu}, I_{\nu})$. It corresponds to a horizontal \emph{shift} of the diagram \eqref{rels2}:
\be \raisebox{-.5in}{$\includegraphics[width=4in]{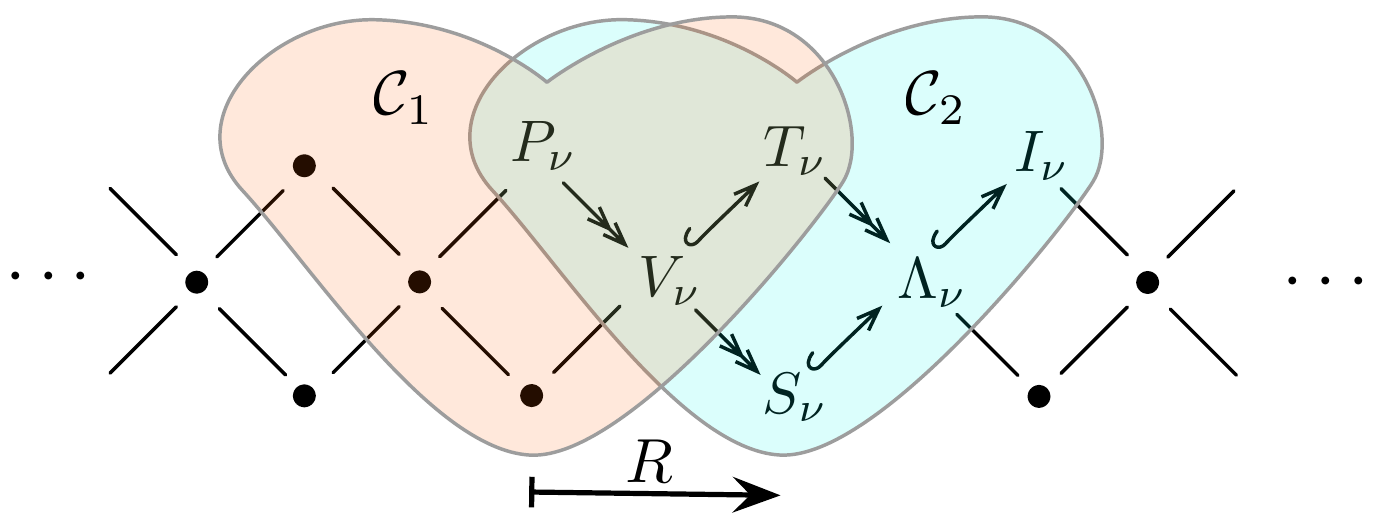}$} \ee

Ringel duality send the remaining families of modules $(S_\nu,\Lambda_\nu,I_\nu)$ to nontrivial complexes in the derived category, denoted by dots $\bullet$ in the diagram.
By starting with the six basic collections of modules in category $\CO_H^{k_t,m_\R}$ and repeatedly applying a Ringel duality, we obtain infinitely many collections of objects in the derived category that all have the same sort of triangular relationships as the original modules.

Notably, the long twist $\Phi$ and inverse long shuffle $\Psi^{-1}$ from the above are both Ringel dualities.
Another sort of Ringel duality $\CD$, corresponding to the composition of a shift $R$ and the restricted dual $\star$, also appears in \cite{Losev-quantizations, BLPW-gale}. For example, \cite[Prop. 7.5]{Losev-quantizations} considers the homological duality
\[\CD = \text{Ext}_{\hat \C[\CM_H]_{k_t}}^{*+\frac{1}{2}\text{dim}_{\C} \, \CM_H}(-,\hat \C[\CM_H]_{k_t}): D^b\CO_H^{k_t,m_\R} \to D^b (\CO_H^{-k_t,m_\R})^{\text{op}}. \]
Just as in the discussion of the restricted dual $\star$ we have used the equivalence between right $\hat \C[\CM_H]_{k_t}$-modules and left $\hat \C[\CM_H]_{-k_t}$-modules with the opposite highest-weight structure. It is natural to think of $\CD$ as a reflection about a shifted vertical axis in \eqref{rels2}.

\subsubsection{Serre functor}

Applying the long-twist or long-shuffle twice acts trivially on the parameters $k_t,m_\R$. However, both of these functors correspond to a non-trivial braiding in the derived category --- a non-trivial monodromy in the parameter space \eqref{param-braid}. Indeed, the results of Losev \cite{Losev-quantizations} mentioned above imply that up to homological shifts
\be \CS \cong \Phi^{k_t,-k_t}\circ\Phi^{-k_t,k_t} \cong (\Phi^{-k_t,k_t}\circ\Phi^{k_t,-k_t})^{-1} \ee
where $\CS$ is the Serre functor for the category $D^b\CO_H^{k_t,m_\R}$. The functor $\CS$ is characterized up to homological shift and isomorphism by the property that
\be \text{Ext}^*(M,N) \simeq \text{Ext}^*(N,\CS(M)) \ee
for any objects $M$ and $N$.

\subsubsection{Koszul duality}
\label{sec:Koszul}

For our purposes, the most interesting functor acting on the derived category $D^b\CO_H^{k_t,m_\R}$ is Koszul duality.
It corresponds to a reflection of \eqref{rels2} about the \emph{horizontal} symmetry axis (denoted $!$); it is a covariant functor that exchanges $S_\nu\leftrightarrow T_\nu$, while preserving both standard $V_\nu$ and costandard $\Lambda_\nu$ modules.

At first glance, a functor with these properties may sound very exotic. Reflecting \eqref{rels2} about a horizontal axis means that the functor must exchange the roles of extensions and quotients in the various triangular relationships among modules. This is actually possible in a derived category, if one is willing to allow the functor to change the category's homological grading. Then, for example, an extension $\alpha\in\text{Ext}^1(M,N)$ between two objects might map to a standard homomorphism $\alpha^!\in\text{Ext}^0(M^!,N^!) = \text{Hom}(M^!,N^!)$ between dual objects, inducting a quotient $N^!/\alpha^!(M^!)$. 

Of course, if homological gradings change, they must do so in a controlled manner. In the standard definition of Koszul duality \cite{BGS} (\cf\ \cite{MOS}), one first introduces an additional ``internal'' grading (\ie\ a non-homological grading) on the categories $\CO_H^{k_t,m_\R}$ and $D^b\CO_H^{k_t,m_\R}$. Let us call this internal grading $\rho$, and the homological grading $\eta$. Then Koszul duality shifts the homological grading by the internal grading, while reversing the sign of the internal grading,
\be \eta^! = \eta + \rho\,,\qquad \rho^! = -\rho\,. \label{gradings}\ee

The internal grading used in defining Koszul duality must satisfy some very special properties, whose role in the physics of boundary conditions has not yet been fully understood. We will not describe them in detail here. One interesting implication of these properties is that the derived endomorphism algebras of simple, tilting, and projective objects in category $\CO_H^{k_t,m_\R}$ are all quadratic algebras --- meaning that they are generated in degree one (with respect to an appropriate grading) and all relations among generators appear in degree two. For example, for tilting objects the endomorphisms are ordinary maps $\alpha\in \text{Hom}(T_\nu,T_{\nu'})$ with $\eta=0$; and one requires that any such map is a composition of elementary maps with $\rho=1$, and that relations among the elementary maps are quadratic. In contrast, for simple objects there are no ordinary maps but rather extensions. One requires that all extensions are generated by elementary extensions $\beta\in \text{Ext}^1(S_\nu,S_{\nu'})$ with $\eta=1$, $\rho=-1$, satisfying quadratic relations. Koszul duality exchanges the quadratic algebras $\text{Hom}^*(\oplus_\nu T_\nu, \oplus_\nu T_\nu)$ and $\text{Ext}^*(\oplus_\nu S_\nu,\oplus_\nu S_\nu)$, subject to the shifts \eqref{gradings}.

If an internal grading with the desired properties exists in category $\CO_H^{k_t,m_\R}$, the category is called Koszul. Establishing the existence of such a grading turns out to be highly non-trivial, both mathematically and physically! Mathematically, existence has been proven only in some special cases, such as parabolic and singular blocks of the BGG category $\CO$ \cite{BGS}, hypertoric varieties \cite{BLPW-gale}, and type A quiver varieties \cite{Webster, RSVV, SVV}.

\begin{figure}[htb]
\centering
\includegraphics[width=4.8in]{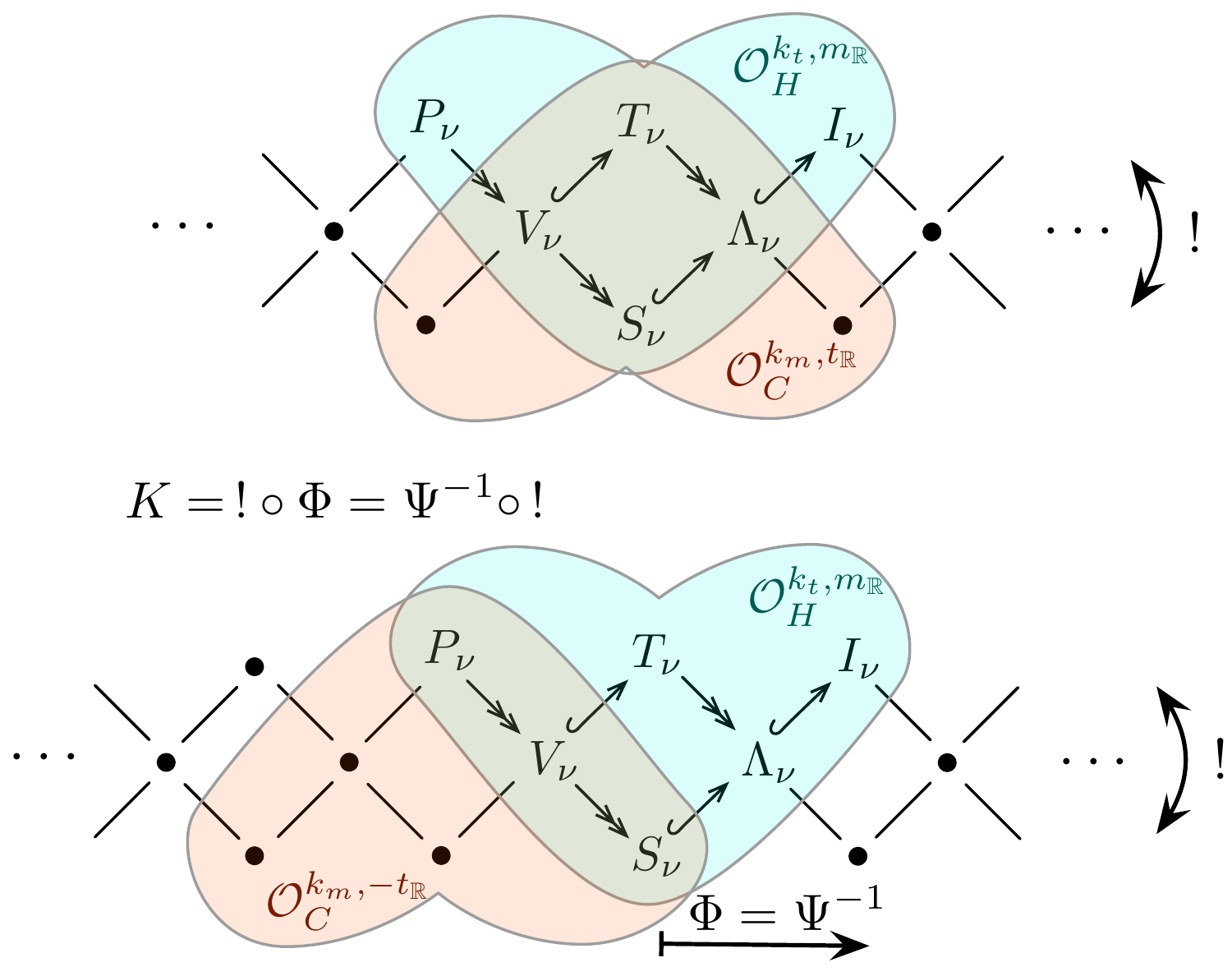}
\caption{Different versions of Koszul duality.}
\label{fig:Koszul}
\end{figure}

When a suitable internal grading exists and Koszul duality can be defined, \cite{BLPW-II} conjecture that the Koszul-dual of the category $D^b\CO_H^{k_t,m_\R}$ (with its shifted gradings) can naturally be identified with category $\CO$ for a symplectic-dual manifold. We of course expect this to be the Coulomb branch. Specifically, in our present conventions, we expect
\be \big(D^b\CO_H^{k_t,m_\R}\big)^! \simeq D^b\CO_C^{k_m,t_\R}\,\qquad  \text{for $(k_t,k_m)\sim (t_\R,m_\R)$},\ee
in such a way that the Koszul-duals of simples in $\CO_H^{k_t,m_\R}$ are identified with tiltings in $\CO_C^{k_m,t_\R}$, and so forth. We depict this relation graphically in Figure \ref{fig:Koszul}.

We will revisit the physical meaning of the gradings $\eta,\rho$ in Section \ref{sec:2d-SD}. After compactifying to two dimensions, we will identify the gradings with charges for the $U(1)_C\times U(1)_H\subset SU(2)_C\times SU(2)_R$ R-symmetries that are unbroken by BPS boundary conditions. From the perspective of a Higgs-branch sigma-model, we will find
\begin{subequations}\label{erHC}
\be \eta = C\,,\qquad \rho = H-C\,;  \ee
whereas from the perspective of a Coulomb-branch sigma-model we will find
\be \eta^! = H\,,\qquad \rho^! = C-H\,. \ee
\end{subequations}
This implies Koszul-duality relation \eqref{gradings}.

For readers that who wish to explore the mathematical literature, we should remark that in Braden-Licata-Proudfoot-Webster \cite[Sec. 10]{BLPW-II} the definition of symplectic duality involves a Koszul duality that reverses the order on vacua and sends $(S_\nu,V_\nu,P_\nu)\mapsto(I_\nu,\Lambda_\nu,S_\nu)$. Such a duality
\be K: D^b\CO_H^{k_t,m_\R} \to D^b\CO_C^{k_m,-t_\R} \ee
is obtained by the formula $K = \, ! \circ \Phi_{-k_t,k_t} \cong (\Psi_{t_\R,-t_\R})^{-1} \circ \, !$. The last isomorphism is meant to be interpreted up to grading shift and is a particular example of the fact that Koszul duality is expected to intertwine twisting and shuffling functors.

In more generality, Mazorchuk-Ovsienko-Stroppel \cite{MOS} have shown that a positively graded category has three different dual categories, each one consisting of linear complexes of either projective, injective, or tilting modules. The three different duality functors are intertwined by Ringel duality just as in the example above.

\subsection{Warmup: a symplectic correspondence}
\label{sec:corresp}

In order to reproduce a small part of the Koszul-duality map between Higgs- and Coulomb-branch categories, we may follow the procedure outlined on page \pageref{alignmt}, and depicted graphically back in Figure \ref{fig:Bflow} of the introduction. Namely, we choose many different UV boundary conditions $\CB$ for a 3d $\CN=4$ gauge theory, and, by turning on $\Omega$ and twisted $\wt\Omega$ backgrounds, use them to define many pairs of modules $(\hat\CB_C,\hat \CB_H)$ for the quantized Coulomb- and Higgs-branch algebras. This leads to a non-categorical ``symplectic correspondence'' between pairs of objects in the module categories $\CO_C$ and~$\CO_H$.

To make the correspondence concrete, we must relate the quantization parameters $k_t,k_m$ for the Higgs- and Coulomb-branch algebras to real parameters $t_\R,m_\R$. To this end, we align
\be k_m\sim m_\R\,,\qquad k_t\sim t_\R \label{alignmt2} \ee
as described in Section \ref{sec:order}. For example, we could fix generic $k_m,k_t$, and simply set $m_\R=k_m$, $t_\R=k_t$. Then we obtain a correspondence between modules $\hat \CB_C\in \CO_C^{k_m,t_\R}$ and $\hat \CB_H\in \CO_H^{k_t,m_\R}$.

The identification \eqref{alignmt2} is motivated by the requirement that boundary conditions preserve supersymmetry in an $\Omega$ (or $\wt\Omega$) background if and only if they preserve supersymmetry in its absence. For example, the Higgs-branch image of a Neumann b.c. $\CN_L$ is supported on a particular submanifold $\CN_L^{(H)}$ of the Higgs branch, the image of the Lagrangian subspace $L$ under the hyperk\"ahler quotient $\CM_H = \C^{2N}\cap(\mu_\R=t_\R,\mu_\C=0)/G$. We called the boundary condition ``$t_\R$-feasible'' if the image $\CN_L^{(H)}$ was non-empty, \ie\ if supersymmetry was preserved. This condition depends on the chamber that $t_\R$ lies in. When $k_t\sim t_\R$, the module $\hat \CN_L^{(H)}$ will be nonempty if and only if $\CN_L^{(H)}$ is feasible. Similarly, it follows from the analysis in Section \ref{sec:DC} and \ref{sec:DC2} that, when $k_m\sim m_\R$,
 the module $\hat\CD_{L,c}^{(C)}$ defined by a generic Dirichlet boundary condition will be nonempty if and only if the Lagrangian $\CD_{L,c}^{(C)}\subset \CM_C$ is nonempty.

Assuming \eqref{alignmt2}, we proceed to describe pairs $(\hat \CB_C,\hat \CB_H)$ of corresponding modules for abelian theories, taking the parent UV boundary condition $\CB$ to be either pure Neumann, generic Dirichlet, or exceptional Dirichlet. These three families of boundary conditions were already analyzed in detail in Section \ref{sec:abel}, so we have mainly to apply our previous results. We find (see below for proofs and examples):
\begin{itemize}
\item Every $m_\R$-lowest-weight simple module $S_\nu$ for the Higgs-branch algebra $\hat \C[\CM_H]_{k_t}$ is the image of a pure Neumann b.c. $\CN_\varepsilon$ (for an appropriate choice of sign vector~$\varepsilon$). The corresponding Coulomb-branch module is of generalized Whittaker type that deforms (under infinite gradient flow) to the $t_\R$-lowest-weight tilting module $T_\nu$.

\item Similarly, every $t_\R$-lowest-weight simple module $S_\nu$ for the Coulomb algebra $\hat \C[\CM_C]_{k_m}$ is the image of a generic Dirichlet b.c., and corresponds on the Higgs branch to the $m_\R$-lowest-weight tilting module $T_\nu$.

\item The exceptional Dirichlet boundary conditions $\CD_{\varepsilon,S}$ that are assigned to vacua as in \eqref{nueS} define lowest-weight costandard modules $\Lambda_\nu$ for both the Coulomb- and Higgs-branch algebras. All costandard modules arise this way. \end{itemize}
We thus find that we can reproduce the part of the Koszul-duality map (Figure \ref{fig:Koszul}) involving simple, costandard, and tilting modules,
\be \raisebox{-.5in}{$\includegraphics[width=1.5in]{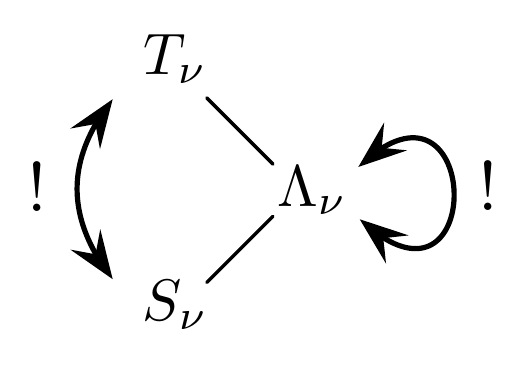}$}
\ee

We expect a similar correspondence to hold for nonabelian theories. However, in nonabelian theories, one must go (slightly) beyond the basic families of boundary conditions studied in this paper to capture all simple and tilting modules. We will explore this elsewhere.

The above claims about modules in abelian theories mostly follow from Section~\ref{sec:abel}. In particular, the statement that exceptional Dirichlet b.c. $\CD_{\varepsilon,S}$ define costandard modules $\Lambda_\nu$ on both Higgs and Coulomb branches already appeared in Sections \ref{sec:abel-HxD} and \ref{sec:abel-CxD}. The claims about tilting modules require an additional argument, as follows.

\subsubsection*{Simples on $\CM_H$ $\leftrightarrow$ tiltings on $\CM_C$}

We consider a $G=U(1)^r$ gauge theory with $N$ hypermultiplets, and use the same notation and formalism as in Section \ref{sec:abel}. We assume that no continuous subgroup of $U(1)^r$ acts trivially (so the quaternionic dimension of the Higgs branch is $N-r$). We also assume (as everywhere in this section) that the Higgs and Coulomb branches can be fully resolved, with a finite number of isolated, massive vacua in the presence of generic real mass and FI parameters. In the Higgs-branch hyperplane arrangement, the massive vacua $\nu_S = \cap_{i\notin S}\CH_i$ are located at the simultaneous intersections of $N-r$ hyperplanes, labelled by a subset $S\in\{1,...,N\}$ of size $r$. On the Coulomb branch, the same vacua are located at the complementary intersections $\nu_S = \cap_{i\in S}\CH^i$.

Since parameters $t_\R\sim k_t$ and $m_\R\sim k_m$ are aligned, the quantum hyperplane arrangements have the same topology as the classical ones. We will not distinguish between the two below, with the understanding that ``a module supported on a chamber $\Delta$'' refers to the quantum arrangement; and ``a vacuum $\nu_S$'' refers both to a maximal intersection of hyperplanes in the classical arrangement and a weight space closest to that intersection in the quantum arrangement.

Given a vacuum $\nu_S$ labelled by a subset $S$ of size $r$, it is useful to define a sign vector $\varepsilon(S;H)$ by
\be \label{def-eH} \begin{cases} \text{$\nu_S$ lies on the $\varepsilon_i(S;H)$ side of $\CH_i$ in the Higgs arrangement for $i\in S$} \\[.1cm]
 \text{$m_\R$ is a positive linear combination of $\varepsilon_i(S;H)\wt Q_i$ for $i\notin S$\,.}
\end{cases}
\ee
The first condition ensures that the chamber $\Delta_{\varepsilon(S;H)}$ in the Higgs arrangement is $t_\R$-feasible (nonempty) and has $\nu_S$ as a vertex, while the second condition ensures that $h_m$ is bounded below on $\Delta_{\varepsilon(S;H)}$, attaining its minimum value at $\nu_S$. (To understand the second condition, note that $(\wt Q_i^\alpha)_{\alpha=1}^{N-r}$ is a vector perpendicular to the hyperplane $\CH_i$ in the Higgs arrangement, pointing toward the positive side of this hyperplane.)

Similarly, we may define a sign vector $\varepsilon(S;C)$ by
\be \label{def-eC} \begin{cases} \text{$\nu_S$ lies on the $\varepsilon_i(S;C)$ side of $\CH^i$ in the Higgs arrangement for $i\notin S$} \\[.1cm]
  \text{$t_\R$ is a positive linear combination of $\varepsilon_i(S;C) Q_i$ for $i\notin S$\,.}
\end{cases}
\ee
The definition ensures that the chamber $\Delta^{\varepsilon(S;C)}$ in the Coulomb-branch arrangement is $m_\R$-feasible (nonempty), and that $h_t$ is bounded from below on the chamber, attaining its minimum at the vertex $\nu_S$.

Now, let us choose any massive vacuum $\nu_S$ and consider the Neumann b.c. $\CN_{\varepsilon(S;H)}$. By construction, its Higgs-branch image is supported on the chamber $\Delta_{\varepsilon(S;H)}$. Upon quantization, it defines the lowest-weight simple module
\be \hat \CN_{\varepsilon(S;H)}^{(H)} = S_{\nu_S}\; \in \CO_H^{k_t,m_\R}\,.\ee
Clearly all simple modules are realized this way.

We would like to show that the quantum Coulomb-branch image is a tilting module
\be \hat \CN_{\varepsilon(S;H)}^{(C)}=T_{\nu_S}\; \in \CO_C^{k_m,t_\R}\,, \label{N-tilting} \ee
labelled by the same vacuum. The result follows from a combination of elementary geometric observations.

We first claim that the sign vectors $\varepsilon(S;H)$ and $\varepsilon(S;C)$ defined above satisfy
\be    \varepsilon(S;C) = \ol{\varepsilon(S;H)}\,,  \label{eCH} \ee
where the `bar' means that the signs for $i\in S$ are negated, as in \eqref{bare}. We may understand this as follows. To determine $\varepsilon_i(S;H)$ for $i\in S$, we first solve the equations $\sum_{i\in S} Q_a{}^iZ_i + t_a = 0$ (for all $a$) to obtain the values of $Z_i$ ($i\in S$) at the vacuum $\nu_S$ (the $Z_{i\notin S}$ are automatically zero there). This fixes $\varepsilon_i(S;H) = \text{sign}(Z_i)$ (for $i\in S$). Then, recalling that the dual charge vector $\wt Q_i$ is the positive normal vector to each $\CH_i$ passing through $\nu_S$, we determine the remaining signs by finding the unique linear combination satisfying $\sum_{i\notin S} \delta^i \wt Q_i = m$, and setting $\varepsilon_i(S;H)=\text{sign}(\delta^i)$ ($i\notin S$). Similarly, on the Coulomb branch we solve $\sum_{i\notin S} \wt Q^\alpha M^i = m^\alpha$ (since $M^{i}=0$ for $i\in S$) to determine the values of $M^i$ at $\nu_S$; and we find a unique linear combination of normal vectors such that $\sum_{i\in S}\beta_i Q^i = t$. Then $\varepsilon_i(S;C)=\text{sign}(M^i)$ (for $i\notin S$) and $\varepsilon_i(S;C)=\text{sign}(\beta_i)$ (for $i\in S$). The pairs of equations we solve for the Higgs and Coulomb branches are identical, subject to the identification $(Z_i,\delta^i)=(-\beta_i,M^i)$. The relation \eqref{eCH} follows.

Next, let us choose a vacuum $\nu_S$ on the Coulomb branch and describe the associated tilting module $T_{\nu_S}$. Let $V^{\nu_S}$ denote the unique orthant of the Coulomb-branch arrangement whose origin lies at $\nu_S$ and on which $h_t$ is bounded below. Let $\hat V^{\nu_S}$ denote the corresponding Verma module. As discussed in Section \ref{sec:cast}, the tilting module $T_{\nu_S}$ is a successive extensions of all Verma modules whose lowest weights are contained in~$\hat V^{\nu_S}$. (The ordering of the extension is uniquely determined by $h_t$.) 
Let $\varepsilon(S;C)$ label the bounded chamber with $h_t$-lowest point $\nu_S$ as above.
In terms of the hyperplane arrangement, a straightforward analysis shows that a Verma module supported on an orthant $V^{S',\varepsilon'}$ appears in the composition series for $T_{\nu_S}$ if and only if 1) $h_t$ is bounded from below on $V^{S',\varepsilon'}$; and 2) $\varepsilon_i'=\tilde\varepsilon_i(S)$ (for $i\in S$, $i\in S'$), while $\varepsilon_i'=-\tilde\varepsilon_i(S)$ (for $i\notin S$, $i\in S'$). In turn this implies that the composition series for $T_{\nu_S}$ contains precisely the Verma modules supported on chambers
\be  V^{S',-\ol{\varepsilon(S;C)}} \qquad \text{for all $S'$ s.t. $h_t$ is bounded below}\,. \ee
Using \eqref{eCH}, we can re-express this as
\be  V^{S',-\varepsilon(S;H)} \qquad \text{for all $S'$ s.t. $h_t$ is bounded below}\,. \label{VinT} \ee

Now we come back to Neumann boundary conditions. For every vacuum $\nu_S$, the Neumann boundary condition $\CN_{\varepsilon(S;H)}$ defines a Whittaker-like module $\hat \CN_{\varepsilon(S;H)}^{(C)}$ for the Coulomb-branch algebra, as in \eqref{abel-qCN}. Following Section \ref{sec:abel-CN}, it can be deformed to an extension of Verma modules $\hat V^{S',-\varepsilon(S)}$ for all $S'$ such that $h_t$ is bounded below on $V^{S',-\varepsilon(S)}$. Since this condition is identical to \eqref{VinT}, we arrive at the desired result~\eqref{N-tilting}.

Finally, we remark that there exists another concise, geometric description of the Verma modules appearing in \eqref{VinT}. Given a vacuum $\nu_S$ that appears as the $h_t$-lowest point of $\Delta_{\varepsilon(S;H)}$ on the Higgs branch, let $\nu'_j$ be the vacua at the vertices of $\Delta_{\varepsilon(S;H)}$. Then the orthants $V^{S',-\varepsilon(S;H)}$ in \eqref{VinT} are precisely the bounded orthants whose origin lies at the vacua $\nu'_j$ on the Coulomb branch. In other words, the vertices of $\Delta_{\varepsilon(S;H)}$ label the Verma modules in $T_{\nu_S}$. The proof follows from elementary arguments similar to those above.

\subsubsection*{Tiltings on $\CM_H$ $\leftrightarrow$ simples on $\CM_C$}

A repetition of the above argument in the case of generic Dirichlet boundary conditions to show that simple modules for the Coulomb-branch algebra correspond to tiltings for the Higgs-branch algebra. In particular, given any vacuum $\nu_S$, it follows from Section \ref{sec:abel-CD} that the Dirichlet boundary condition $\CD_{\varepsilon(S;C)}$ defines the module
\be \hat \CD_{\varepsilon(S;C)}^{(C)} = S_{\nu_S}\;\in \CO_C^{k_m,t_\R}\,.\ee
Its Higgs-branch image, described in Section \ref{sec:abel-HD}, is a successive extension of Verma modules supported on chambers
\be V_{S',\varepsilon(S,C)} = V_{S',\ol{\varepsilon(S,H)}} \qquad \text{for all $S'$ s.t. $h_m$ is bounded below}\,. \ee
From \eqref{eCH}, we have $V_{S',\varepsilon(S,C)} = V_{S',\ol{\varepsilon(S,H)}}$, and we identify $V_{S',\ol{\varepsilon(S,H)}}$ as the Verma modules appearing in the composition series of the tilting module $T_{\nu_S}$. Thus,
\be \hat \CD_{\varepsilon(S;C)}^{(H)} = T_{\nu_S}\;\in \CO_H^{k_t,m_\R}\,.\ee

\subsubsection{Example: SQED}
\label{sec:corresp-SQED}

As an example of the correspondence between Higgs- and Coulomb-branch modules, we consider $G=U(1)$ gauge theory with three hypermultiplets of charge $+1$. The resolved Higgs branch is $T^*\cp^2$ (this was the recurring example in the first half of Section \ref{sec:abel}), and the Coulomb branch resolves the $\C^2/\Z_3$ singularity (\cf\ Section \ref{sec:NC-SQED}). We use the same notation and conventions as in Section \ref{sec:abel}, with gauge and flavor charges \eqref{QqSQED} and dual charges \eqref{SQED-tQ}. We take $t_\R=k_t=7/2$ and $m_\R=k_m=(-2,-1)$.

\begin{figure}[htb]
\centering
\includegraphics[width=6in]{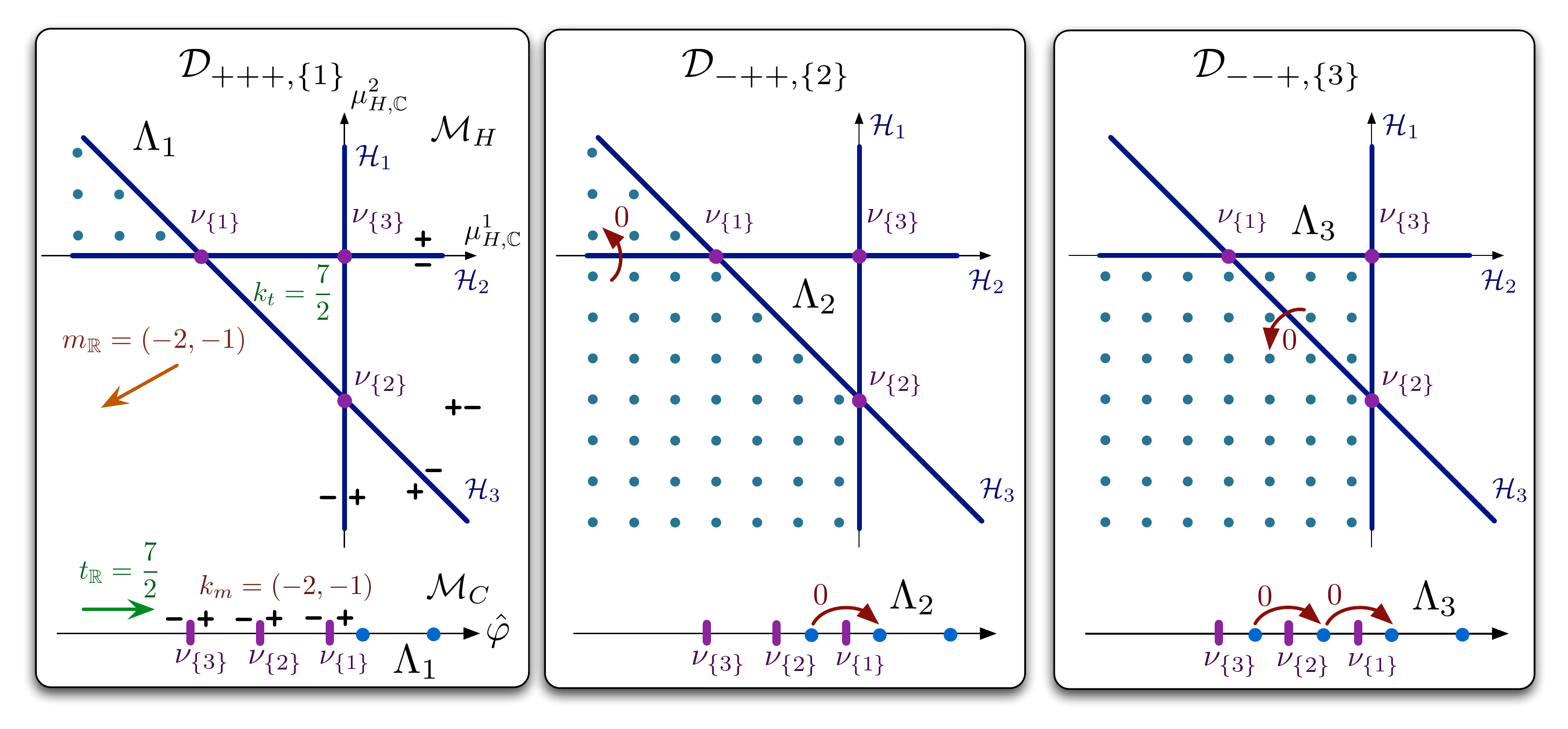}
\caption{Correspondence of lowest-weight Higgs- and Coulomb-branch modules defined by exceptional Dirichlet b.c. for SQED.}
\label{fig:corresp-xD}
\end{figure}

\begin{figure}[htb]
\centering
\includegraphics[width=6in]{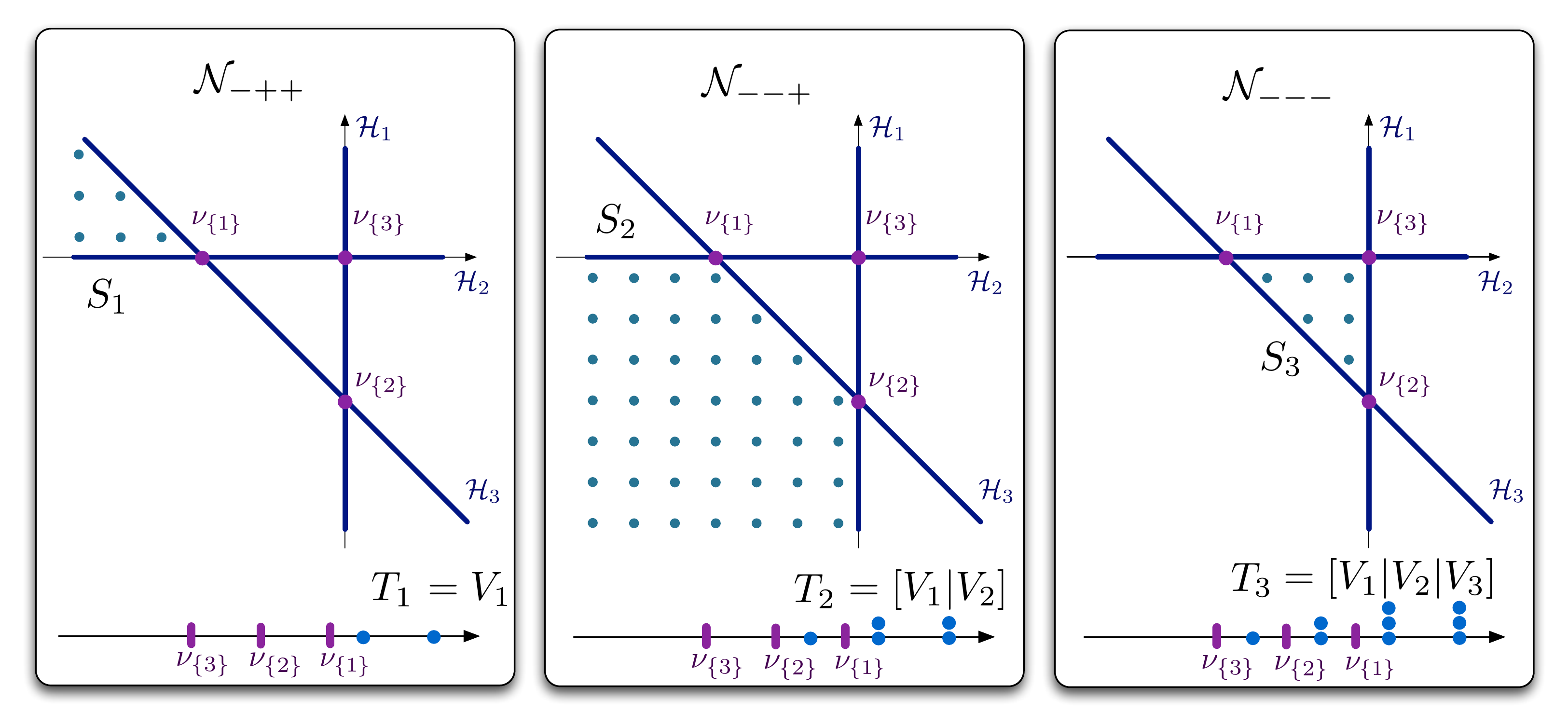}
\caption{Correspondence of lowest-weight modules defined by Neumann b.c. for SQED.}
\label{fig:corresp-N}
\end{figure}

\begin{figure}[htb]
\centering
\includegraphics[width=6in]{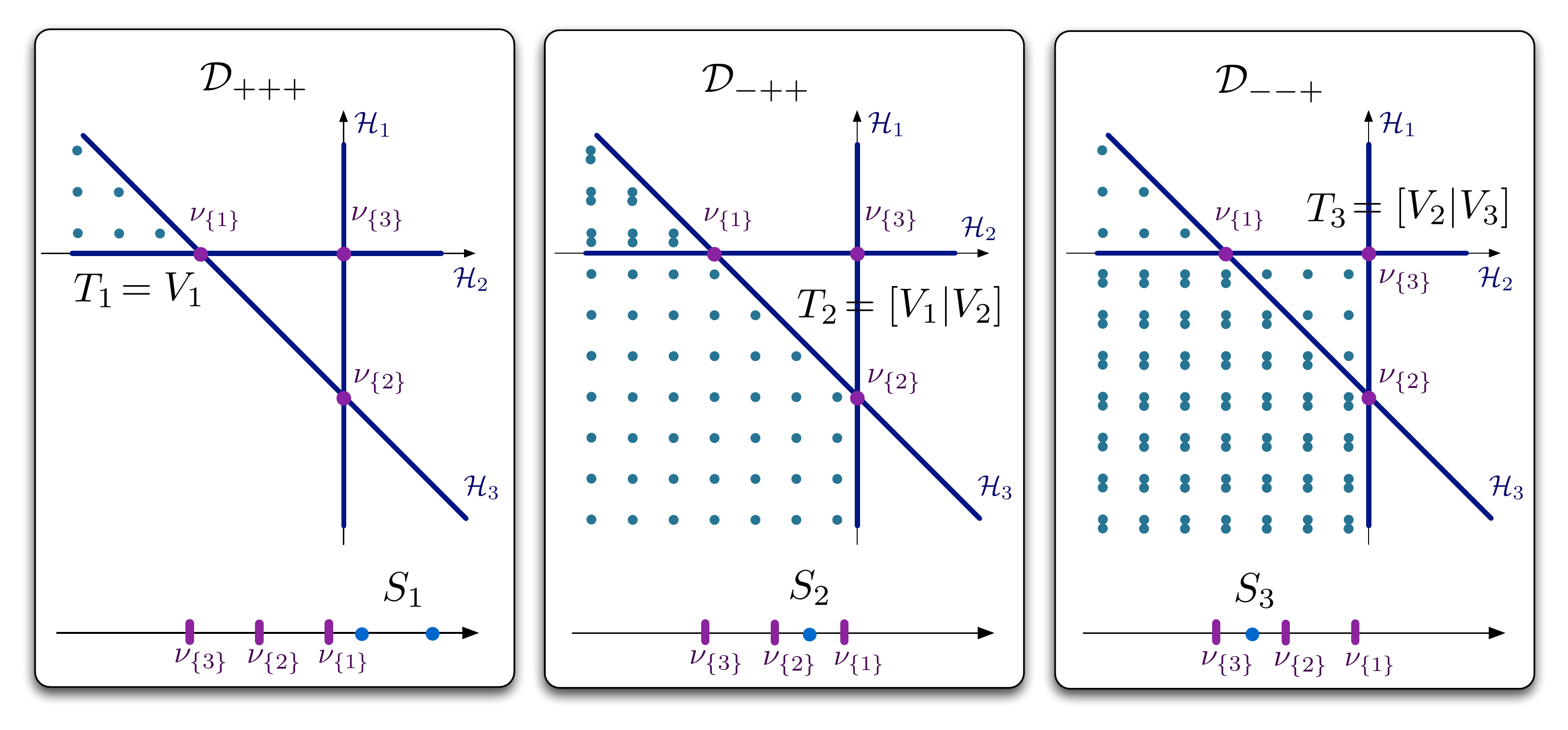}
\caption{Correspondence of lowest-weight modules defined by generic Dirichlet b.c. for SQED.}
\label{fig:corresp-D}
\end{figure}

There are three massive vacua $\nu_{\{1\}}$, $\nu_{\{2\}}$, $\nu_{\{3\}}$, and thus three simple, three tilting, and three costandard modules for both the Higgs- and Coulomb-branch algebras. The UV boundary conditions that realize these various modules are shown in Figures \ref{fig:corresp-xD}--\ref{fig:corresp-D}. For each module, we depict the nontrivial weight spaces by dots, with the number of dots equal to the dimension of the weight space.

\subsubsection{Central charges}
\label{sec:abel-central}

In abelian theories, we can also give an explicit description of the central charges $h_\nu(m_\R,t_\R)$ assigned to vacua, as in Section \ref{sec:order}.

Consider a vacuum $\nu_S$ in an abelian theory, labelled by a subset $S$ of size $r$. From the perspective of the Higgs branch, the central charge in this vacuum is $h_m(\nu_S) = m_\R\cdot\mu_{H,\R}|_{\nu_S} = m_\R \cdot q\cdot Z|_{\nu_S}$. At the vacuum, $Z_i=0$ for $i\notin S$, and the nonvanishing $Z_i$ are determined from the equations $Q\cdot Z + t = 0$. Letting $Q^S = \{Q_a^i\}_{1\leq a\leq r}^{i\in S}$ and $q^S = \{q_\alpha^i\}_{1\leq \alpha\leq N-r}^{i\in S}$ denote the blocks of the gauge and flavor charge matrices corresponding to $i\in S$, we find
\be h_{\nu_S}(m_\R,t_\R) = h_m(\nu_S) = -m_\R \cdot q^S(Q^S)^{-1}\cdot t_\R\,. \label{abel-central}\ee
More explicitly, $h_{\nu_S}(m_\R,t_\R)=-\sum_{a,\alpha,i\in S}m_\R^\alpha q_\alpha{}^i[(Q^S)^{-1}]_i{}^a t_\R^a\,.$ Equivalently, on the Coulomb branch, the central charge to be given by $h_t(\nu_S) = \sigma\cdot t_\R|_{\nu_S}$. At the vacuum we have $M^i=\sigma\cdot Q^i+m_\R\cdot q^i=0$ for $i\in S$, whence $\sigma|_{\nu_S} = -m_\R q^S (Q^S)^{-1}$. Therefore, $h_t(\nu_S)=h_m(\nu_S)$, as expected.

We see from \eqref{abel-central} that every vacuum $\nu_S$ defines an $(N-r)\times r$ matrix $q^S(Q^S)^{-1}$ that allows the mass and FI parameters to be contracted. We expect this to arise as a matrix of effective Chern-Simons couplings of the 3d $\CN=4$ theory in the vacuum $\nu_S$. As discussed briefly in Appendix \ref{app:central}, these determine domain-wall central charges.

In the SQED example of Section \ref{sec:corresp-SQED}, with charge matrices \eqref{QqSQED}, the three vacua lead to matrices
\be q^{\{1\}}(Q^{\{1\}})^{-1} = \begin{pmatrix} 1 \\ 0\end{pmatrix}\,,\qquad  q^{\{2\}}(Q^{\{2\}})^{-1} = \begin{pmatrix} 0 \\ 1\end{pmatrix}\,,\qquad  q^{\{3\}}(Q^{\{3\}})^{-1} = \begin{pmatrix} 0 \\ 0\end{pmatrix}\,,\ee
and central charges
\be h_{\{1\}} = -m_{1,\R}t_\R\,,\qquad h_{\{2\}} = -m_{2,\R}t_\R\,,\qquad h_{\{3\}} = 0\,.\ee
For the values of $m_\R,\,t_\R$ used in Figures \ref{fig:corresp-xD}--\ref{fig:corresp-D}, we thus find $(h_{\{1\}},h_{\{2\}},h_{\{3\}}) = (7,\frac72,0)$. This may readily be checked in either the Coulomb or Higgs arrangements.

\subsection{Compactification to 2d}
\label{sec:3d2d}

The correspondence of modules described above is only a small part of symplectic duality. As we have already emphasized, symplectic duality involves an equivalence of \emph{categories} $\CD^b\CO_H$ and $\CD^b\CO_C$ --- or, more precisely, \emph{graded} versions of these categories as in \eqref{gradings}. The categories $\CD^b\CO_H$ and $\CD^b\CO_C$ strongly resemble categories of boundary conditions in a two-dimensional theory. Therefore, in order to establish a physical basis for symplectic duality, we are led to consider compactifications of a 3d $\CN=4$ theory $\CT_{3d}$ to two dimensions.

We are interested in a setup where we turn on both real masses $m_\R$ and real FI parameters $t_\R$ in $\CT_{3d}$ and compactify on a circle 
of radius $R$. The result is a massive theory with two-dimensional $\CN=(4,4)$ supersymmetry and two unbroken R-symmetries, 
the  $U(1)_H\times U(1)_C\subset SU(2)_H\times SU(2)_C$ that preserve $m_\R$ and $t_\R$.%
\footnote{In the $R\to 0$ limit, the bulk 2d $(4,4)$ theory actually has independent left- and right-moving R-symmetries. However, boundary conditions will only preserve diagonal combinations of the left and right symmetries, which may be identified with the 3d R-symmetry $U(1)_H\times U(1)_C$.} %
We will refer to this theory as the ``unreduced'' theory. 
At sufficiently small energies, this theory behaves as a two-dimensional massive theory, with massive 
particles and solitons that carry a variety of charges: KK momentum, Higgs-branch and Coulomb-branch flavor charges, 
and possibly a topological charge associated to the choice of vacua on the two sides of a soliton. 

In a truly two-dimensional  $(4,4)$ theory one may define various categories of boundary conditions by picking a $(2,2)$ subalgebra commuting with at least one 
unbroken R-charge and applying the standard machinery of topological twists. It is not completely obvious that such a construction would work 
directly on a three-dimensional theory compactified on a circle of finite radius. In principle, it may be possible to give a low energy construction of such categories 
through the web formalism introduced in \cite{GMW}, which constructs categories of branes from a sort of topological low-energy effective Lagrangian for the BPS particles of the theory.     

In any case, the unreduced theory is ``too big'' for our purposes: the categories $\CD^b\CO_H$ and $\CD^b\CO_C$ appear to be associated to true two-dimensional theories, 
non-linear sigma models with targets $\CM_H$ or $\CM_C$. We can get to such theories by a careful $R\to 0$ limit. In order to understand this limit, we need to 
keep track of four important mass scales:
\begin{itemize}
\item the KK scale $R^{-1}$, which controls masses of particles with KK momentum (\ie\ nontrivial Fourier modes around the circle);
\item the scales of real mass and FI deformations $m_\R, t_\R$, which control masses of particles charged under Higgs- and Coulomb-branch isometries, respectively; and
\item the mass scale of topological solitons that come from BPS domain walls wrapping the compactification circle, which is of order $R\, m_\R t_\R$.
\end{itemize}
As reviewed in Appendix \ref{app:central}, all these scales appear in the central charges of the supersymmetry algebra, as twisted masses for the 
KK, flavor and topological charges of the unreduced theory.%
\footnote{The three-dimensional gauge couplings $g_{YM}^2$ also provide a fifth mass scale; however, it does not enter the central charges and we can assume that it is very large in order for 3d mirror symmetry to be valid.}

A naive dimensional reduction $\CT_{3d}$ to a 2d $(4,4)$ gauge theory corresponds to a limit where $R$ is taken to $0$ 
while $t_\R$ is sent to infinity, so that the 2d FI parameters $t^{2d}_\R = R t_\R$ remain finite. The real masses $m_\R$ remain finite and coincide with the real part of twisted masses in the 2d theory. (As usual, the real masses are complexified by the holonomies of a background flavor gauge field on the compactification circle.)
The mass scale of wrapped BPS domain walls also remains finite and is controlled by $t^{2d}_\R m_\R$. 
As the 2d gauge coupling goes to infinity in the limit, we should really think in terms of the mass-deformed 2d sigma model with target $\CM_H[t^{2d}_\R]$, which is a well understood, asymptotically free theory.
We can call this limit ``Higgs-branch reduction''. 

Each boundary condition $\CB$ that we defined in the 3d gauge theory has an image in the 2d sigma model on $\CM_H[t^{2d}_\R]$ --- we expect it to be a brane supported on the holomorphic Lagrangian submanifold $\CB_H$, the Higgs-branch image of $\CB$. Later on we will sharpen this relation, but it is 
well known that such boundary conditions can be associated to D-modules.  

This is of course promising, but we immediately hit a snag: as the real masses and FI parameters play a symmetric role in $\CT_{3d}$, 
it is clear that the ``Higgs-branch reduction'' is not the only limit one may take. If we keep the real FI parameters $t_\R$ fixed and send the real masses to infinity in such a way that $m^{2d}_\R = R m_\R$ is finite, three dimensional 
mirror symmetry indicates that the result will be a mass-deformed 2d sigma model with target  $\CM_C[m^{2d}_\R]$. 
We can call this limit ``Coulomb-branch reduction''.

These two limits are very different from each other and do \emph{not} allow us to predict a full duality 
between the 2d sigma models with target $\CM_H[t^{2d}_\R]$ and $\CM_C[m^{2d}_\R]$. It may be possible, of course, to look for protected quantities 
in the unreduced theory which are independent of $R$ and have a faithful image in both 2d sigma models. This is the simplest way one may imagine 
to give a physical justification for symplectic duality: build some category of boundary conditions in the unreduced theory which would be unaffected by either 
Higgs- or Coulomb-branch reductions and thus would be isomorphic to some (sub)categories of boundary conditions in the two sigma models. 

Before exploring that avenue, it is useful to observe that there is a more general limit one may consider:
we may send both $m_\R$ and $t_\R$ to infinity as $R \to 0$, while keeping $R m_\R t_\R$ finite. 
This ``full reduction'' can be thought of as a combination of the Higgs- and Coulomb-branch reductions. 

For example, we may introduce the Higgs branch and Coulomb branch mass scales 
$\Lambda_H$ and $\Lambda_C$, so that $m_\R \sim \Lambda_H$ and $t_\R \sim \Lambda_C$, 
and scale $R$ as $\mu \Lambda_H^{-1}\Lambda_C^{-1}$ in order to fix the mass scale of wrapped BPS domain walls
to be of order $\mu$. Then the Higgs- and Coulomb-branch reductions correspond to sending either 
$\Lambda_H$ or $\Lambda_C$ to infinity, while the full reduction sends both to infinity. We will call the fully reduced theory $\mathcal T_{2d}$.

\begin{figure}[htb]
\centering
\includegraphics[width=4.5in]{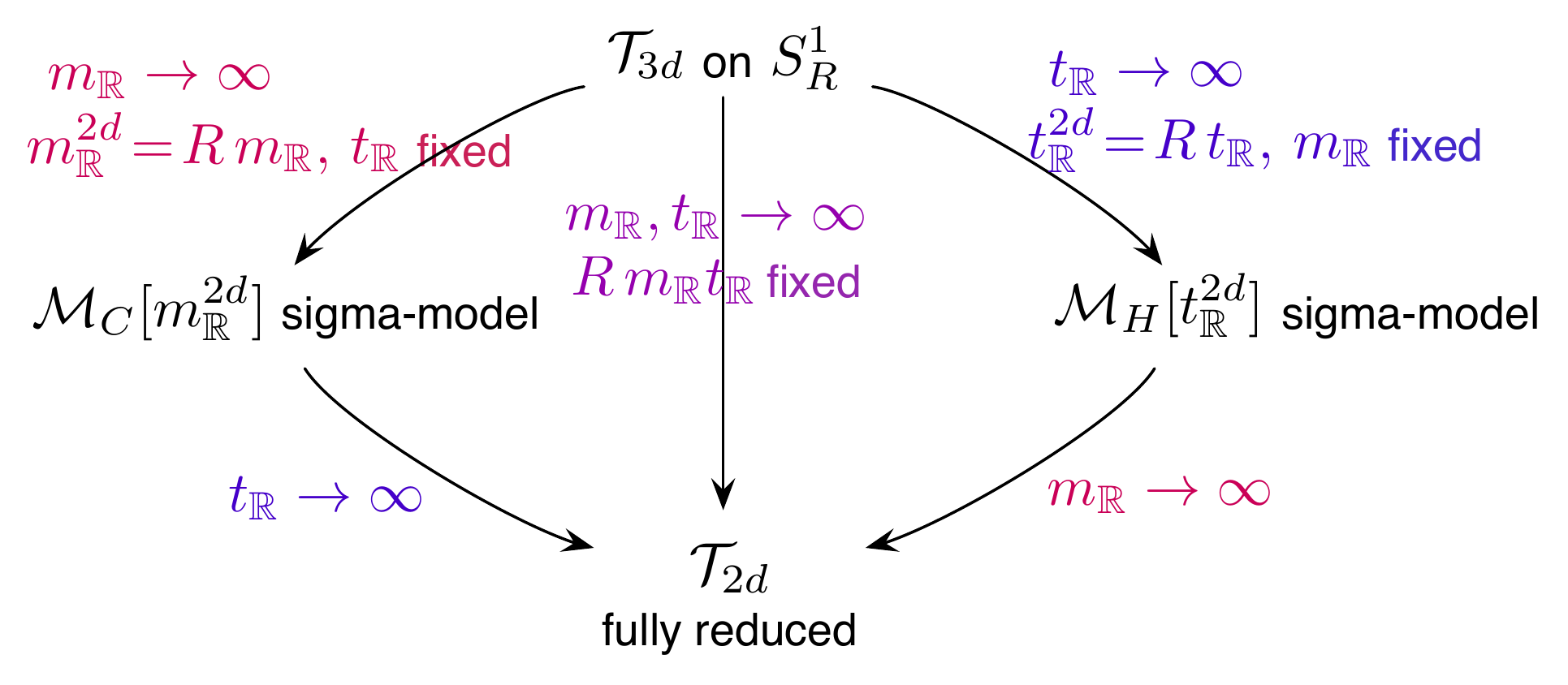}
\caption{Reductions to two dimensions.}
\label{fig:2dlimits}
\end{figure}

We expect these various limits to commute (Figure \ref{fig:2dlimits}). In particular, the fully reduced massive $(4,4)$
theory should admit both a description as the limit of the $\CM_H[t^{2d}_\R]$ 2d sigma model as $m_\R$ is sent to infinity at constant $m_\R t^{2d}_\R$ 
and as a limit of the $\CM_C[m^{2d}_\R]$ 2d sigma model as $t_\R$ is sent to infinity at constant $t_\R m^{2d}_\R$. 
This is a true 2d duality statement: the two mass-deformed sigma models flow to the same 2d theory 
in the limit where the mass deformations are sent to infinity while the resolution parameters are sent to zero in such a way as to keep the topological central charges finite. 

This offers an alternative route to symplectic duality: one may hope to define categories of boundary conditions in the two
massive sigma models that are unaffected by the scaling limit, and end up as the same category of boundary conditions in the fully reduced 
2d theory. This is a weaker requirement than asking for a well-defined category in the unreduced theory invariant under the Higgs- and Coulomb-branch reductions. 
 
It is interesting to track the effect of these reductions on the BPS spectrum of the theory. 
The Higgs-branch reduction removes from the spectrum every particle or soliton which carries KK or Coulomb branch 
flavor charges. Because of the BPS bound, it also removes non-BPS excitations with such charges, of course. 
Similarly, the Coulomb-branch reduction removes from the spectrum every particle or soliton which carries KK or Higgs branch 
flavor charges. The full reduction removes from the spectrum every particle or soliton which carries KK, Coulomb, or Higgs-branch 
flavor charges, leaving only solitons with topological charges. \footnote{These statements have to be understood in the light of wall-crossing. 
As the appropriate central charges increase in magnitude, one may encounter a sequence of walls of marginal stability. As these central charges come to dominate the 
mass of the particles which carry the corresponding quantum numbers, the spectrum splits into ``light'' and ``heavy'' particles and 
the light spectrum stabilizes. At this point the heavy states can be dropped. }

In $(2,2)$ non-linear sigma models, BPS particles and operators with isometry flavor charges only appear 
in B-model calculations and not in A-model calculations. Naively, this suggests that A-model categories may be essentially unaffected by 
the Higgs or Coulomb-branch reductions and thus may be isomorphic in the unreduced theory (if defined), in the 2d sigma models 
and in the fully reduced theory. As categories of D-modules often appear as an economical description of 
A-brane categories, this naive expectation makes the setup very promising. 

Unfortunately, the naive expectation cannot be true. It would lead to a direct relation between the D-module categories on the Higgs and Coulomb branches, 
which is not the correct statement of symplectic duality: the two categories are expected to be isomorphic only after the 
extra hidden grading is restored. In the next section we will find the crucial snag: the A-twists of the two sigma models 
correspond to different topological twists of the unreduced or fully reduced theories.

\subsection{2d twists and symplectic duality}
\label{sec:2d-SD}

In any of the $R\to0$ limits of Section \ref{sec:3d2d}, our 3d $\CN=4$ gauge theory reduces to a 2d $\CN=(4,4)$ theory that admits a large family of topological twists, and corresponding categories of boundary conditions. We would like to relate these categories to those appearing in symplectic duality.

By ``topological twist'' here we simply mean a choice of nilpotent supercharge $Q$ in flat space with the property that all translations are $Q$-exact (\cf\ Appendix \ref{app:twist}).
This is enough to define an associated category,
whose objects are boundary conditions $\CB$ for the theory on $\R\times \R_+$ that preserve $Q$, and whose morphism spaces $\text{Hom}(\CB,\CB')$ are the $Q$-cohomologies of spaces of local operators at a junction of $\CB$ and $\CB'$.

The half-BPS boundary conditions that we have studied throughout the paper preserve four of the eight supercharges of the 2d $\CN=(4,4)$ algebra, which can be combined to form topological supercharges of the form
\be Q_{\zeta,\zeta'} = Q^{++} + \zeta' Q^{+-} + \zeta Q^{-+} + \zeta\zeta' Q^{--}\,,\qquad \zeta,\zeta'\in \C\cup\{\infty\}\,. \label{Qzz-text} \ee
The two indices $\pm,\pm$ indicate charges for the $U(1)_H$ and  $U(1)_C$ R-symmetries, respectively. Thus, a  half-BPS boundary condition in the physical theory will define an object in the category associated to $Q_{\zeta,\zeta'}$ for all $\zeta,\zeta'$.

If a given $Q_{\zeta,\zeta'}$ transforms with nonzero charge under an R-symmetry, then this R-symmetry will provide a ``homological'' or ``fermion number'' grading in the associated category.
This is the situation we are interested in. 
Each morphism space $\text{Hom}(\CB,\CB')$ will split into sectors of fixed R-charge, the $Q_{\zeta,\zeta'}$-cohomology groups. Mathematically, we find what is called a dg (differential graded) category.
It is clear from \eqref{Qzz-text} that $Q_{\zeta,\zeta'}$ transforms under an R-symmetry if and only if at least one of $\zeta,\zeta'$ equals $0$ or $\infty$.

Similarly, if $Q_{\zeta,\zeta'}$ is \emph{invariant} under an R-symmetry, then the corresponding category of branes will have an additional ``internal'' or ``flavor'' grading. In particular, each cohomology group in $\text{Hom}(\CB,\CB')$ gains such a grading. This is only possible if both $\zeta,\zeta'$ equal $0$ or $\infty$, \ie\ if our supercharge is one of  $Q^{++}$, $Q^{+-}$, $Q^{-+}$, and $Q^{--}$. The supercharge $Q^{++}$ may be further distinguished by the property that its cohomology contains local bulk operators that are holomorphic (as opposed to anti-holomorphic) functions on both Higgs and Coulomb branches, in our standard complex structure.

In order to make sense of Koszul duality (Section \ref{sec:Koszul}), both homological and internal gradings must be present. This naturally leads us to consider the topological twist with respect to $Q_{0,0} = Q^{++}$ as a candidate for symplectic duality.
In Appendix \ref{app:twist}, we find that this twist leads to a B-model with respect to both Higgs and Coulomb branches, in our standard complex structure. In particular, if we consider a 2d reduction to a $\CM_H$ sigma-model as on the RHS of Figure \ref{fig:2dlimits}, the $Q^{++}$ twist will be a B-model with homological grading $\eta = C$ (coming from $U(1)_C$, under which the fermions of the sigma-model are charged) and internal grading $\rho=H-C$ (coming from the anti-diagonal of $U(1)_H\times U(1)_C$). Conversely, if we consider a 2d reduction to a $\CM_C$ sigma-model, we get a B-model with homological grading $\eta=H$ and internal grading $\rho = C-H$. This perfectly reproduces the structure in \eqref{gradings}.

In order to match other features of symplectic duality, this picture requires three additional modifications:\label{mod-cat}
\begin{enumerate}
\item To talk about an actual duality, we need to be considering a single 2d theory. As explained in Section \ref{sec:3d2d}, the 2d sigma-models with target $\CM_H$ and target $\CM_C$ are \emph{not} the same theory. However, they can be deformed to a common theory $\CT_{2d}$ by additionally sending $m_\R\to\infty$ or $t_\R\to\infty$, respectively.

\item The category associated to any fixed topological supercharge such as $Q^{++}$ contains many boundary conditions that are quarter-BPS, and do not preserve any other $Q_{\zeta,\zeta'}$. For symplectic duality, we are only interested in boundary conditions that are half-BPS and preserve the entire family of supercharges $Q_{\zeta,\zeta'}$. We should always restrict ourselves to subcategories generated by such boundary conditions.

For example, in a B-model with target $\CM_H$, generic quarter-BPS boundary conditions correspond to holomorphic vector bundles supported on any holomorphic cycles $\CM_H$. (The B-model category is $D^b\text{Coh}(\CM_H)$.) The half-BPS subcategory we are interested in is generated by \emph{flat} vector bundles supported on holomorphic \emph{Lagrangian} cycles. We will always implement such a restriction.

\item In order to find module categories resembling $D^b\CO_H$ (resp. $D^b\CO_C$), we will need to deform the B-model supercharge $Q_{0,0}=Q^{++}$ to $Q_{1,0}$ (resp. $Q_{0,1}$). 
We will discuss this in Section \ref{sec:derO}.
Symplectic duality then rests on the conjecture that the category of half-BPS boundary conditions for the fully reduced theory $\CT_{2d}$ is unchanged under these deformations.

Notice that the twists $Q_{1,0}$ and $Q_{0,1}$ only preserve a single R-symmetry, and thus their categories only have a homological grading. This matches the the state of affairs in the mathematical description of symplectic duality: naively, categories $D^b\CO_H$ and $D^b\CO_C$ only have a homological grading, and one must work hard to find a hidden internal grading as well. For us, the internal grading is manifest in the $Q_{0,0}$ category, and gets transported to the $Q_{1,0}$ and $Q_{0,1}$ categories.

\end{enumerate}

\subsubsection{B-models with twisted masses}
\label{sec:massiveB}

The presence of generic nonzero $t_\R$ and $m_\R$ makes the 2d sigma-models with target $\CM_H$ or $\CM_C$ massive. The categories of branes in these theories, for any twist preserving an R-symmetry, may then be studied using techniques of \cite{GMW}.%
\footnote{Much of \cite{GMW} is presented from the perspective of A-type boundary conditions in a massive Landau-Ginzburg model. However, the formalism is completely general, and applies equally well to a massive 2d (2,2) theory with B-type boundary conditions that preserve a vectorial R-symmetry.} %
In particular, the morphism spaces $\text{Hom}(\CB,\CB')$ can be constructed directly from the spectrum of BPS solitons in the theory.

One expects on general grounds that the category of branes in the B-model with (say) target $\CM_H$ will be graded by Higgs-branch isometries. This can be seen very explicitly from the analysis of \cite{GMW}. Namely, the 2d (4,4) sigma-model has quarter-BPS solitons that descend from 3d particles charged under Higgs-branch isometries. These solitons preserve the B-model supercharge $Q^{++}$ (\cf\ the discussion around \eqref{q-BPS1}), and thus contribute to the morphism spaces in the B-model category.

The real mass $m_\R$ enters the B-model as a twisted mass (Appendix \ref{app:mt}). When $m_\R$ is generic, the solitons charged under any Higgs-branch isometry will have mass of order $m_\R$. As $m_\R\to\infty$, these solitons decouple from the spectrum. Therefore, we can heuristically understand the effect of sending $m_\R\to\infty$ as ``removing'' charged morphisms from the B-model category. 

A more refined analysis of twisted masses along the lines of \cite{GMW} will be presented in Section \ref{sec:WC}. One actually finds that, when $m_\R$ is large, only solitons with non-negative charge under the infinitesimal $U(1)_m$ isometry generated by $m_\R$ contribute to the morphism spaces $\text{Hom}(\CB,\CB')$. This is a consequence of wall-crossing transformations. As $m_\R\to\infty$, the solitons with strictly positive charge decouple completely, leaving behind ungraded morphism spaces.

We can also attempt to describe this process geometrically, from the perspective of a sigma-model. In the B-model with target $\CM_H$ we start with the subcategory of $D^b\text{Coh}(\CM_H)$ generated by sheaves with with vanishing Chern classes and holomorphic Lagrangian support. At generic nonzero $m_\R$, we should consider an even smaller subcategory, generated by sheaves $\CB$ that are equivariant for the isometry $U(1)_m$ associated to $m_\R$, and are such that the real moment map $h_m=m_\R\cdot \mu_{H,\R}$ is bounded from below on the support $\text{Supp}(\CB)$. 
Then morphism spaces $\text{Hom}(\CB,\CB')$ will have \emph{non-negative} grading under $U(1)_m$. Subsequently sending $m_\R\to\infty$ should have the effect of quotienting $\text{Hom}(\CB,\CB')$ by the subspace of morphism with strictly positive charge
\be \text{Hom}(\CB,\CB')\;\leadsto\; \text{Hom}(\CB,\CB')/\text{Hom}(\CB,\CB')_{>0}\,. \label{Hom-0} \ee
The resulting quotient is neutral under the whole torus of the Higgs-branch isometry group that commutes with $U(1)_m$. It would be interesting to study this procedure in greater detail.

Similarly, in the B-model with target $\CM_C$, there are solitons charged under Coulomb-branch isometries, which endow morphisms spaces with an additional grading. 
As $t_\R\to\infty$, all the charged solitons decouple from the spectrum, leaving behind neutral morphism spaces.

As discussed in Section \ref{sec:3d2d}, the result of sending $m_\R\to\infty$ in the $\CM_H$ sigma-model should agree with the result of sending $t_\R\to\infty$ in the $\CM_C$ sigma-model. Both limits lead to the fully reduced theory $\CT_{2d}$. Correspondingly, the B-models with targets $\CM_H$ and $\CM_C$ should both reduce to the $Q^{++}$ twist of $\CT_{2d}$.
In the limit, the only remaining solitons that contribute to morphisms spaces are those coming from domain walls in 3d, whose mass is of order $R\,m_\R t_\R$. 

\subsubsection{Relation to derived categories $\CO$}
\label{sec:derO}

So far, we have argued that the $m_\R\to\infty$ limit of the B-model with target $\CM_H$ is equivalent to the $t_\R\to \infty$ limit of the B-model with target $\CM_C$, since they both coincide with the $Q^{++}$ twist of $\CT_{2d}$.
Let us denote the complex structures in which these B-models are defined as $I_{\zeta=0}^H$ and $I_{\zeta'=0}^C$, respectively, as in Appendix \ref{app:twist}.
While these B-models have many of the right properties for symplectic duality, they look very little like derived categories of modules $D^b\CO_H$ or~$D^b\CO_C$.

The category $D^b\CO_H$ does appear naturally in the $Q_{1,0}$ twist of the 2d Higgs-branch sigma-model. As explained in Appendix \ref{app:twist} and summarized in Figure \ref{fig:twist2}, this twist defines an A-model to $\CM_H$ in complex structure $I_{\zeta=1}^H$.
Kapustin and Witten \cite{KapustinWitten} defined a functor (generalized by Gukov and Witten \cite{GW-branes})
\be  \label{Dfunctor}
 \begin{array}{lccc}\CI\hspace{-.18cm}\CD\,: & \text{Fuk}(\CM_H) & \to & D^b \big(\hat \C[\CM_H]\text{-mod}\big) \\[.1cm]
 & \CB & \mapsto & \text{Hom}(\CB_{cc},\CB) \end{array} \ee
that sends any Lagrangian A-brane $\CB$ to the (derived) space of morphisms between a canonical coisotropic brane $\CB_{cc}$ and $\CB$. The brane $\CB_{cc}$ is such that its endomorphism algebra $\text{Hom}(\CB_{cc},\CB_{cc})=\hat \C[\CM_H]$ is a deformation quantization for the ring of functions on $\CM_H$ in complex structure $I_0^H$ \cite{KapustinOrlov}. Since the algebra $\text{Hom}(\CB_{cc},\CB_{cc})$ acts on 
$\text{Hom}(\CB_{cc},\CB)$, the latter space acquires the structure of a $\hat \C[\CM_H]$-module.

When $\CM_H$ is a cotangent bundle, Nadler and Zaslow proved that the functor $\CI\hspace{-.18cm}\CD$ provides an equivalence of categories \cite{NadlerZaslow}. This statement is expected to be true more generally, and we will assume here that it holds for the fully resolved Higgs and Coulomb branches of 3d $\CN=4$ gauge theories.

In a similar way, the $Q_{0,1}$ twist of a 2d Coulomb-branch sigma-model defines an A-model in complex structure $I_{\zeta'=1}^C$. Its category of boundary conditions is expected to be equivalent to a derived category of $\hat \C[\CM_C]$ modules.

In the presence of real mass and FI parameters, these (conjectural) equivalences are slightly modified. As explained in Appendix \ref{app:mt}, a real mass $m_\R$ induces a superpotential 
(up to signs and factors of 2)
\be W_H = m_\R\cdot \mu_{H,\C}^{\zeta=1} = m_\R\cdot(\mu_{H,\R} + i\, \text{Im}\,\mu_{H,\C}) \ee
in the 2d sigma-model to the Higgs branch in complex structure $\zeta=1$, viewed as a 2d $(2,2)$ theory. The real part of this superpotential is our familiar Morse function
\be  \text{Re}\,W_H = m_\R\cdot \mu_{H,\R} = h_m\,.\ee
The resulting A-model category will be a Fukaya-Seidel category $FS(\CM_H,W_H)$ rather than a Fukaya category, generated by Lefschetz thimbles for $h_m=\text{Re}\,W_H$ \cite{Seidel-Fukaya}.
(The physics of such massive A-models was developed in \cite{HIV}, and their categories of branes were the subject of \cite{GMW}.) The branes in the Fukaya-Seidel category are supported on Lagrangian cycles with $h_m$ bounded from below, and, correspondingly, the functor $\CI\hspace{-.18cm}\CD$ in \eqref{Dfunctor} maps them to $m_\R$-lowest-weight modules for $\hat C[\CM_H]$. We thus have
\be \CI\hspace{-.18cm}\CD\;:\; FS(\CM_H,W_H)\to D^b \CO_H^{k_t,m_\R}\,, \label{DFS} \ee
which we expect to be an equivalence.%
\footnote{The quantization parameter $k_t$ appearing on the RHS of \eqref{DFS} enters as a parameter of the canonical coisotropic brane $\CB_{cc}$ in the definition of $\CI\hspace{-.18cm}\CD$. The precise value is unimportant, since we know (Section \ref{sec:shuffle}) that categories $\CO_H^{k_t,m_\R}$ for different $k_t$ are derived equivalent. However, a particularly natural choice is to align $k_t\sim t_\R$. In this case simple, compact Lagrangian branes will map to ordinary modules, with no additional homological structure.}

Similarly, in the A-twisted sigma-model to the Coulomb branch, the FI parameter $t_\R$ induces a superpotential with real part $h_t=t_\R\cdot \mu_{C,\R}$, such that the image of $\CI\hspace{-.18cm}\CD$ becomes precisely $D^b\CO_C$.

\begin{figure}[htb]
\centering
\includegraphics[width=4in]{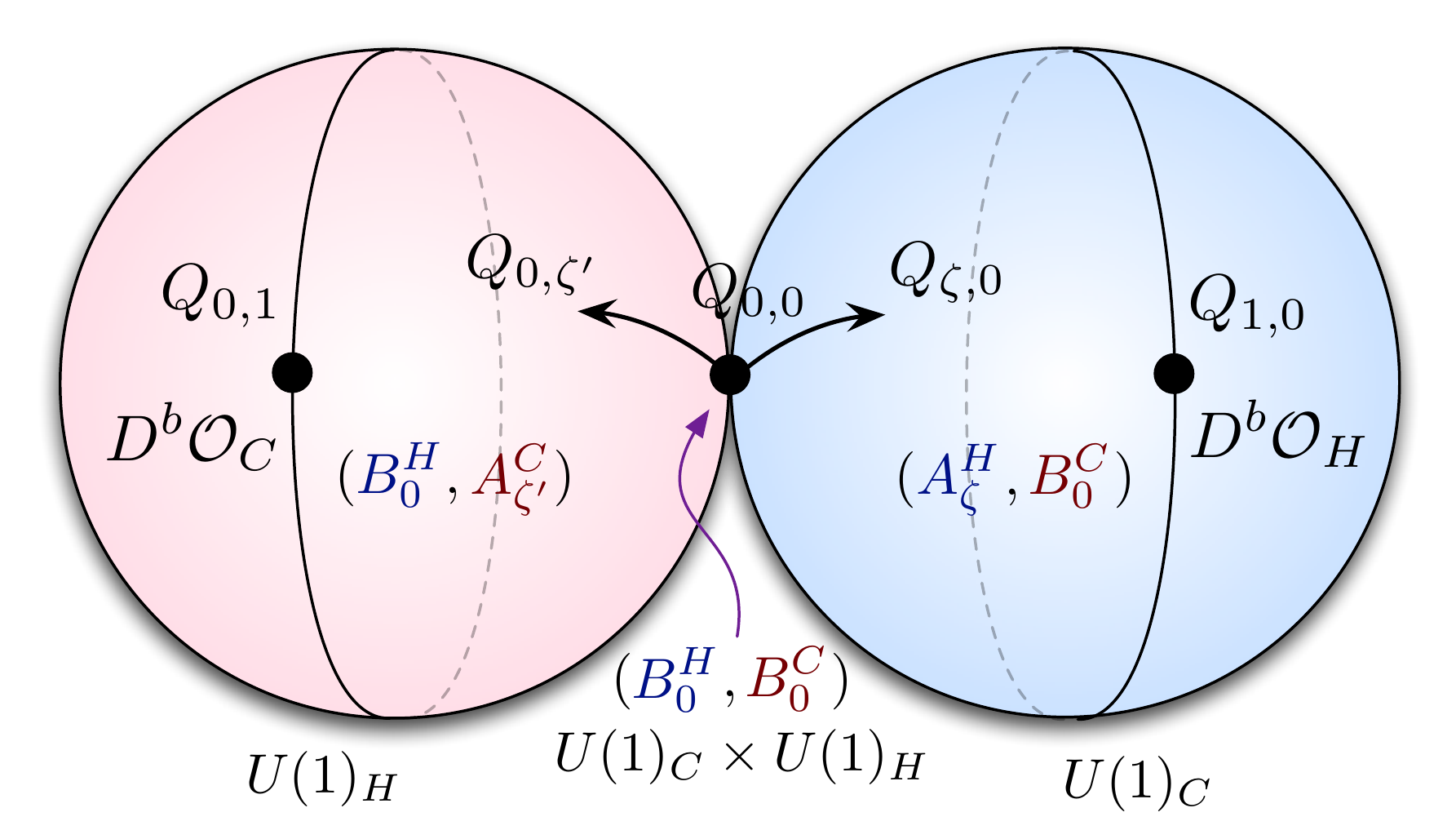}
\caption{2d topological twists that preserve at least one R-symmetry, repeated from Figure \ref{fig:twists} (page \pageref{fig:twists}); each twist can be identified as either an A-model or B-model twist in the Coulomb-branch or Higgs-branch sigma-models.}
\label{fig:twist2}
\end{figure}

We would like to propose that the A-model categories associated to the $Q_{1,0}$ twist of the $\CM_H$ sigma-model and the $Q_{0,1}$ twist of the $\CM_C$ sigma-model are both equivalent to the (half-BPS) category associated to the $Q_{0,0}=Q^{++}$ twist of $\CT_{2d}$.

To justify this, we proceed in two steps. First, we note that the A-model to (say) $\CM_H$ in complex structure $I_{\zeta=1}^H$ is independent of mass parameters $m_\R$, as long as they are generic and nonzero. This is a standard result, following from the fact that $m_\R$ is a chiral deformation of the 2d $(2,2)$ theory that we twist to get the A-model. Alternatively, we may use the fact that in an A-model the morphism spaces $\text{Hom}(\CB,\CB')$ are never graded under target-space isometries; in terms of \cite{GMW}, charged solitons never contribute to them.
Therefore, the A-model to $\CM_H$ at finite $m_\R$ is equivalent to the $Q_{1,0}$ twist of $\CT_{2d}$, obtained in the $m_\R\to \infty$ limit.

Second, we claim (conjecturally) that in the fully reduced theory $\CT_{2d}$, we can deform the twist $Q_{0,0}$ to any $Q_{\zeta,0}$ or $Q_{0,\zeta'}$ without changing the category of half-BPS boundary conditions. A B-model to $\CM_H$ \emph{would} jump discontinuously as $\zeta$ is deformed away from zero, because solitons charged under Higgs-branch isometries contribute to B-model morphisms but not to A-model morphisms at $\zeta\neq 0$. However, the fully reduced theory $\CT_{2d}$ avoids this problem precisely because it has no charged solitons. Thus, it is plausible that the categories of boundary conditions for $\CT_{2d}$ remain constant.

Putting everything together, we arrive at a chain of conjectural dualities that finally relate $D^b\CO_C$ and $D^b\CO_H$:
\be \raisebox{-.7in}{$\includegraphics[height=1.6in]{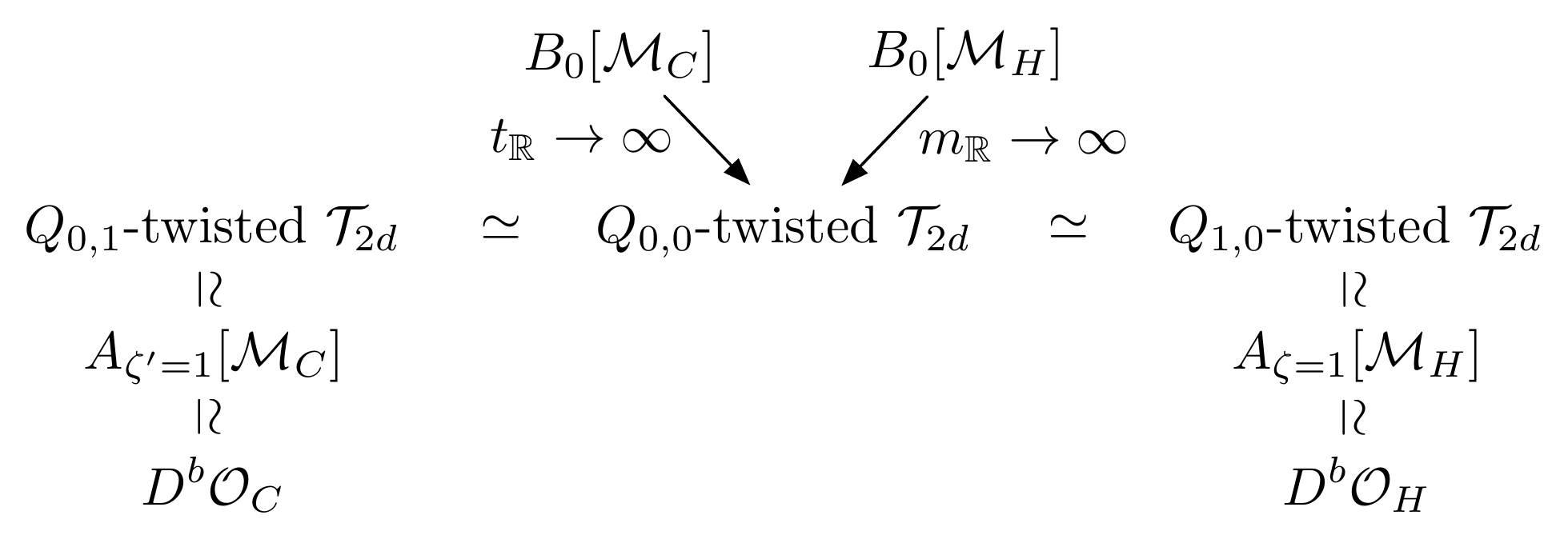}$}  \label{chain}\ee
It is worth emphasizing again that the homological and internal gradings are only manifest in the B-models and in the $Q_{0,0}$-twist of $\CT_{2d}$. They must be transported via the chain of equivalences to $D^b\CO_C$ and $D^b\CO_H$. 

\subsubsection{Relation to 3d Omega backgrounds}
\label{sec:Omega2d}

We may also connect the current discussion of topological twists directly to the collections of $\hat\C[\CM_H]$ and $\hat \C[\CM_C]$-modules that we found in three dimensions by turning on Omega backgrounds. The basic idea follows from work of Nekrasov and Witten \cite{NekWitten} and is discussed in Appendix \ref{app:RW}.

Consider the twisted $\wt\Omega$-background that quantizes the algebra of operators on the Higgs branch. The $\wt\Omega$-background supercharge $Q_{\wt \Omega}$ is a deformation of the Rozansky-Witten supercharge $Q_H = Q_{0,1}$. Rather than being nilpotent, $Q_{\wt \Omega}$ squares to an $\epsilon$-rotation of 3d spacetime in the $(x^0,x^3)$-plane parallel to a putative boundary. Following \cite{NekWitten}, we may deform the $(x^0,x^3)$-plane into a cigar $D$, whose asymptotic region is a cylinder of constant radius $R$. In the asymptotic region, let us define $x^3\sim x^3+R$ to be the coordinate along the cigar circle; so spacetime looks approximately like $S^1_R\times \R_{x^1}\times \R_{x^0}$. Asymptotically, can identify $Q_{\wt\Omega} = Q_{\zeta=R\epsilon,\zeta'=1}$. Compactifying fully to two dimensions (sending $R\to 0$ while holding $\epsilon'=R\epsilon$ fixed) leads to a theory on $\R_{x^1}\times \R_{x^0}^+$ with an A-type twist corresponding to $Q_{\epsilon',1}$. At $x^0=0$ (the tip of the cigar) lies a canonical coisotropic brane $\CB_{cc}^H$, whose algebra of operators is the same quantum algebra $\hat\C[\CM_H]$ appeared in 3d.
The supercharge $Q_{\epsilon',1}$ preserves no R-symmetries, consistent with the fact that the modules in 3d had no derived structure, and no internal grading.

\begin{figure}[htb]
\centering
\includegraphics[width=5.5in]{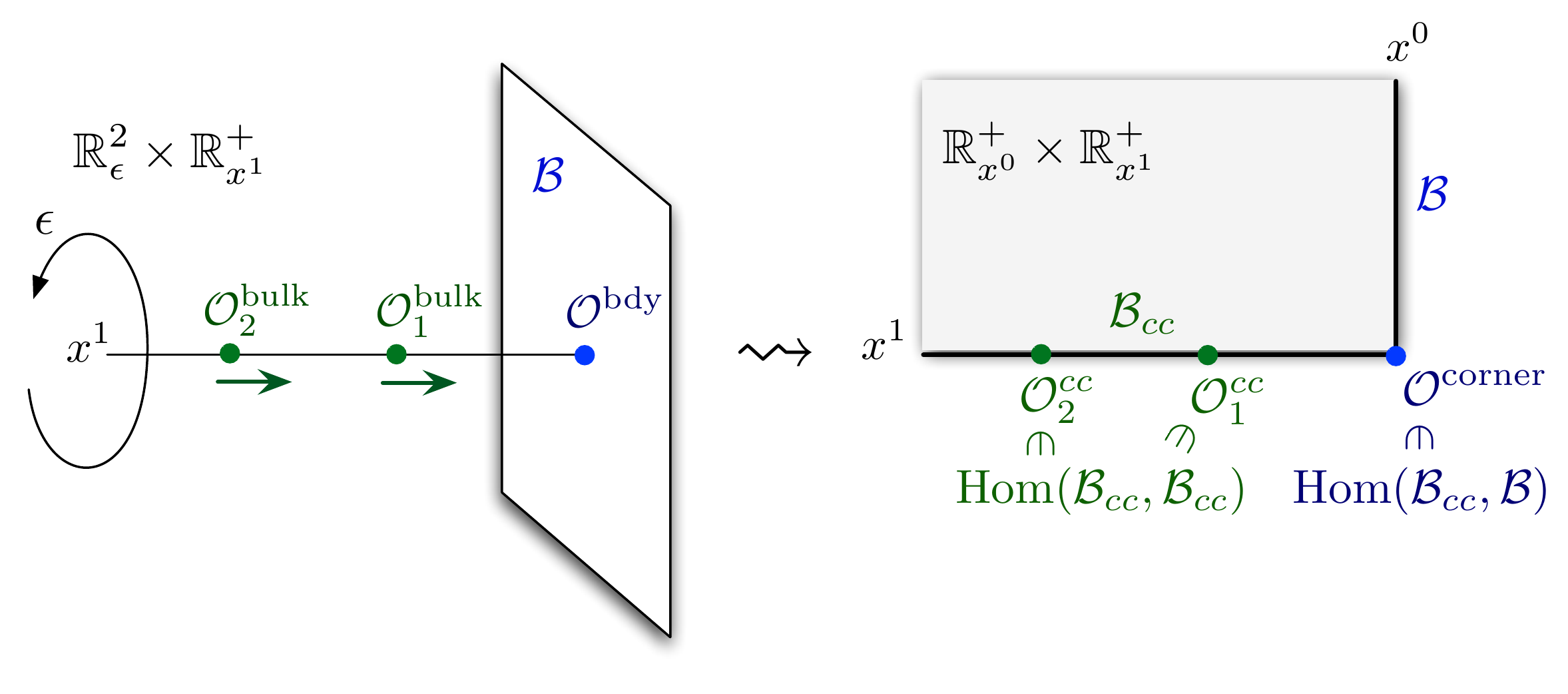}
\caption{Reduction to two dimensions of a 3d system in the Omega background times a half-space.}
\label{fig:omega2d}
\end{figure}

If we add a half-BPS boundary condition $\CB$ at $x^1=0$ and then along the cigar, we arrive at a 2d theory on a quadrant, as in Figure \ref{fig:omega2d}. The space of BPS local operators at the (corner) junction of the $\CB_{cc}^H$ and $\CB$ boundary conditions is identified with the $\hat \C[\CM_H]$-module that we found from from a three-dimensional analysis.

Similarly, the $\Omega$-background that quantizes the Coulomb-branch algebra reduces to a $Q_{1,\epsilon'}$ twist after cigar compactification to two dimensions, with a different canonical coisotropic brane $\CB_{cc}^C$.

Notice that at $\epsilon'=1$ both $\wt\Omega$- and $\Omega$-background supercharges reduce to the same 2d topological supercharge $Q_{1,1}$. There are two distinct types of canonical coisotropic branes $\CB_{cc}^H$, $\CB_{cc}^C$ in this theory. 
Given any half-BPS boundary condition $\CB$, we can compute the space of local operators sitting at a junction of $\CB$ and either $\CB_{cc}$ brane, obtaining two maps
\be    \hat \C[\CM_C]\text{-mod} \;\overset{\CI\hspace{-.15cm}\CD_C}{\longleftarrow}\; \text{(b.c. for the $Q_{1,1}$-twist)} \;\overset{\CI\hspace{-.15cm}\CD_H}\longrightarrow\; \hat\C[\CM_H]\text{-mod}\,.\ee
This is simply a two-dimensional reformulation of the ``symplectic correspondence'' of modules from Section \ref{sec:corresp}.

\subsection{Wall crossing revisited}
\label{sec:WC}

The identification of the module categories $D^b\CO_C$ and $D^b\CO_H$ with the category of boundary conditions in a B-type twist of the fully reduced theory $\CT_{2d}$ leads to several interesting predictions about their structure. We discuss one such prediction here, concerning the special collections of modules from Section \ref{sec:cast} (simples, standards, costandards, tiltings,...) that generate $D^b\CO_C$ and $D^b\CO_H$. Namely, we argue that every one of these collections appears as an \emph{exceptional} collection in a suitably generalized sense, and that the functors that relate the collections (twisting, shuffling, and even Koszul duality) can all be understood as wall-crossing transformations.

\subsubsection{Exceptional collections in 2d $\CN=(2,2)$ theories}
\label{sec:22excep}

We begin by reviewing in slightly more detail how the category of boundary conditions in a massive $(2,2)$ theory (with an R-symmetry) is built up from the spectrum of solitons, following \cite{GMW}, and how this is affected by the presence of additional flavor symmetry and twisted masses.

In the absence of extra flavor symmetries, the main conclusion of \cite{GMW} is that the category of half-BPS boundary conditions in a massive theory is generated by an exceptional collection  
$\mathfrak{Vac}$ whose objects are labelled by vacua $\nu$ of the theory. In physics terms, these objects 
represent ``thimble'' boundary conditions and general branes are built as bound states of elementary thimbles. 

The term exceptional collection means that the only morphism between an object $E_\nu$ in the collection and itself is the identity and 
that morphisms between different objects $E_\nu$, $E_{\nu'}$ only go in a specific direction, determined by the sign of the difference 
between the real part of the central charges $Z_\nu$ and $Z_{\nu'}$ attached to the corresponding vacua:
\be \begin{array}{rl} \text{Ext}^n(E_\nu,E_{\nu}) &= \C\,\delta_{n,0} \\[.1cm]
 \text{Ext}^n(E_\nu,E_{\nu'}) &= 0 \quad\text{if}\; \text{Re}(Z_\nu)<\text{Re}(Z_{\nu'})\,, \end{array} \label{excep} \ee
just as in \eqref{std-order} or \eqref{costd-order}.%
\footnote{The morphism spaces are cohomologies of a supercharge, and are always derived. We thus write ``Ext'' rather than ``Hom'' in \eqref{excep} to avoid confusion with standard homomorphisms of modules. Often one would simply write ``Hom.''} %
Concretely, the morphisms of the $\mathfrak{Vac}$ category are built from the spaces of BPS solitons of the theory. 
Each soliton is associated to two vacua $\nu, \nu'$ and has a central charge $Z_\nu - Z_{\nu'}$. The morphisms 
consist of sequences of BPS solitons with increasing argument of their central charge, from $-\pi/2$ to $\pi/2$. 

As the parameters of the theory are varied, the exceptional collection will jump in a specific way every time the central charge of a 
BPS soliton crosses the imaginary axis. The jumps across the positive and negative imaginary axis coincide with the standard notion of 
mutations of an exceptional collection.

It is interesting to consider an extreme situation where all central charges $Z_\nu$ have the same phase. If we then start varying this phase, we encounter a sequence of exceptional collections $E^{(n)}$, with jumps each time the phase of $Z_\nu$ passes $\pi/2$. The exceptional collections $E^{(n)}$ will be upper or lower triangular, depending on the parity of $n$. There is a sequence of collections, rather than only two, because there may be non-trivial monodromy as we parallel transport boundary conditions in parameter space (the point where all $Z_\nu \equiv 0$ is singular, since the theory becomes massless there). The categories of boundary conditions built from consecutive collections are related by the action of a $\pi$-rotation functor $\CR^\pi$, whose square is a Serre functor.

In the case of $\CT_{2d}$, the only solitons present are those descending from half-BPS domain walls in three dimensions. In the $(2,2)$ subalgebra containing $Q^{++}$ as a B-type supercharge, the central-charge function is  a complexification of the 3d central charge $h_\nu$ from Section \ref{sec:order}.
To be more explicit, recall that the $m_\R$ and $t_\R$ get complexified when putting the 3d theory on a circle, and that $\CT_{2d}$ is obtained by taking the $R\to 0$ limit while keeping $\hat m = \sqrt R\, m_\R$ and $\hat t=\sqrt R\, t_\R$ fixed. Then 
\be Z_\nu  \approx R\, h_\nu(m_\R,t_\R) = h_\nu(\hat m,\hat t)\,. \label{ZT2d} \ee

If we keep all $m_\R,t_\R$ real, then by comparing \eqref{excep} to \eqref{std-order} we find that the exceptional collection $\mathfrak{Vac}$ matches the structure of standard modules. More precisely, we expect there to exist an identification of our category with (say) $D^b\CO_H$ such that the exceptional collection is built from the standard modules. By applying the rotation functor $\CR^\pi$, we then find an exceptional collection corresponding to the costandard modules.
The rotation functor can be implemented in several equivalent ways: for example, by rotating the phase of all $t_\R$ to send $t_\R\to -t_\R$; or by varying the phase of all $m_\R$ to send $m_\R\to - m_\R$.

In terms of (say) the module category $D^b\CO_H$, wall-crossing transformations that come from varying $m_\R$ are implemented by shuffling functors, while transformations that come from varying $t_\R$ are implemented by twisting functors. The current analysis of massive (2,2) theories justified the assertion from Section \ref{sec:shuffle} that both kinds of transformations are manifestations of a single set of wall-crossing functors, controlled by the central charges \eqref{ZT2d}.

\subsubsection{Twisted masses and positive collections}
\label{sec:22pos}

Now, if a given 2d (2,2) theory has a global symmetry $G_F$ that leaves the topological supercharge invariant, the morphism spaces in the category of boundary conditions will be graded by the global symmetry. In addition, one may turn on twisted mass deformations $\wt m$ (valued in the complexified Cartan $\mathfrak t_F^{\C}$), which modify the central charges of BPS solitons by an amount proportional to their global charge.

When the twisted masses are set to zero, the conclusions of \cite{GMW} are unchanged, aside from the presence of the extra grading. The formalism of \cite{GMW} can also be adapted to the presence of twisted mass deformations, with one major modification: the generating collection $\mathfrak{Vac}$
will not no longer be an exceptional collection. Instead, morphisms of charge $q\in \mathfrak t_F^*$ will exist from $E_\nu$ to $E_{\nu'}$ 
only if $Z_\nu- Z_{\nu'} + q\cdot \wt m$ has positive  real part. We could call this a ``graded exceptional collection.'' 

Again, it is interesting to consider an extreme situation where all central charges $Z_\nu$ have the same phase \emph{and} all BPS solitons carry non-zero 
global charge. The latter condition is actually not restrictive at all: the global charge of solitons can always be re-defined as $q \to q + n_\nu - n_{\nu'}$. 
We can easily pick some (possibly fractional) $n_\nu$ shifts to make all charges of solitons non-zero. 

If we turn on an infinitesimal real twisted mass $\wt m$, the walls associated with solitons of positive and negative charge $q\cdot \wt m$ will separate from each other. 
In particular, at $\arg Z_\lambda = \pm \pi/2$ only solitons with either positive or negative charge will contribute to the spaces of morphisms. 
The collection of thimbles $\mathfrak{Vac}$ will not be an exceptional collection anymore, but rather a \emph{positive} (\emph{negative}) collection: except for identity morphisms, all morphisms 
have positive (negative) charge $q\cdot \wt m$. 

\begin{figure}[htb]
\centering
\includegraphics[width=5in]{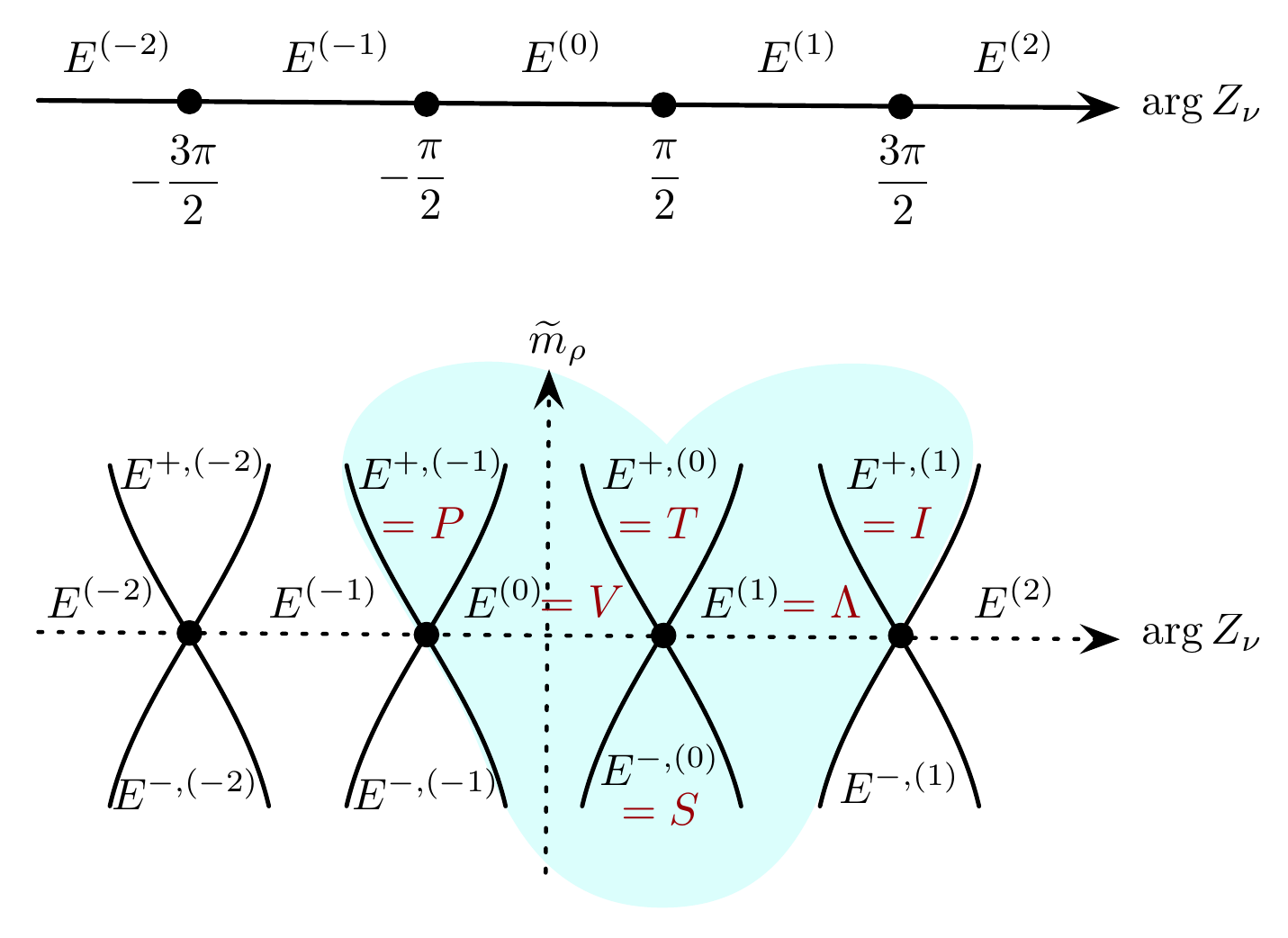}
\caption{Top: chamber structure in a slice of parameter space parameterized by a common phase of all the central charges $\arg Z_\nu$ at $\wt m_\rho=0$. Bottom: as the twisted mass $\wt m_\rho$ is turned on, new chambers open up, containing positive and negative collections of objects.}
\label{fig:wc}
\end{figure}

Thus we arrive at the following picture in the $\arg Z_\nu,\, \Re \, \wt m$ plane, depicted in Figure~\ref{fig:wc}. Along the $\arg Z_\nu$ axis, at $\wt m=0$, we will have the usual sequence of chambers with exceptional collections 
$E^{(n)}$. Above $\arg Z_\nu = n \pi - \pi/2$ we will find positive bases $E^{+,(n)}$ for positive $\Re \,\wt m$ and negative bases $E^{-,(n)}$ for negative $\Re \, \wt m$.
These chambers will be separated by bundles of walls associated to solitons with definite sign of the global charge and direction along the sequence of vacua.

The rotation functor $\CR^\pi$ now changes both the phase of $Z_\nu$ and of $\wt m$. We can decompose it into the product of two commuting functors,
\be \CR^\pi = \CR_Z^\pi \circ \tilde \CR\,,\ee
where $\CR_Z^\pi$ implements parallel transport in the space of central charges $Z_\nu$ at fixed $\wt m$ and functor $\tilde \CR$ reflects the sign of $\wt m$.
Nothing special happens at $\wt m =0$, so we do not need to worry about monodromy there.

There are two applications of these ideas to 2d compactifications of 3d $\CN=4$ theories. The first we have already encountered: if reduce the 3d theory to a 2d sigma-model, as on the two sides of Figure \ref{fig:2dlimits}, the B-type $Q^{++}$ twist of the theory will have morphism spaces graded by target-space isometries $G_F=G_H$ or $G_F=G_C$. For (say) the Higgs-branch sigma-model, the 3d real masses $m_\R$ enter as twisted masses. The above analysis tells us that at large $m_\R$, all morphism spaces in the category will have non-negative charge under the associated symmetry $U(1)_m$, as claimed in Section \ref{sec:massiveB}. Then, as $m_\R$ is sent to infinity, the morphism spaces simplify precisely as in \eqref{Hom-0}.

The second application is more interesting. The $Q^{++}$ twist of the fully reduced theory $\CT_{2d}$ still has a global symmetry, the anti-diagonal combination of the two vectorial R-symmetries $U(1)_H$ and $U(1)_C$.  Its charge was denoted $\rho$ in \eqref{erHC}; it corresponds to the internal grading in category $D^b\CO_H$, and the negative of the internal grading in category $D^b\CO_C$. We may therefore introduce an associated twisted mass $\wt m_\rho$, bringing us to the situation analyzed above. We find that the B-model category for $\CT_{2d}$ has \emph{three} infinite series of generating collections $E^{(n)}$, $E^{\pm,(n)}$, all related by triangular wall-crossing transformations that depend on the order of the vacua.

The $E^{(n)}$ are exceptional collections with respect to the order of the vacua (or its inverse), which we already identified above with standard/costandard modules and their translates by the Serre functor. In contrast, the $E^{+,(n)}$ are positive collections. As discussed in Section \ref{sec:Koszul}, a famous positive collection in category $D^b\CO_H$ is given by the tilting modules (or their translates under $\CR_Z^\pi$: the projective and injective modules). The tilting modules are related to standards/costandards by triangular transformations, precisely the way that $E^{+,(n)}$ are related to $E^{(n)}$. We thus propose that $E^{+,(n)}$ are precisely the tilting modules and their $\CR_Z^\pi$ translates.

Similarly, the $E^{-,(n)}$ are negative collections, related to $E^{(n)}$ by triangular transformations. By comparison to the discussion of Section \ref{sec:Koszul}, we are led to identify them with collections of simple modules and their translates under $\CR^\pi_Z$ translates.

Altogether, we find that the wall-crossing picture in Figure \ref{fig:wc} matches perfectly the picture of category $D^b\CO_H$ in \eqref{rels2}, with its various special collections related by triangular transformations! In terms of the category of B-type boundary conditions for $\CT_{2d}$, every single transformation appears as wall crossing. The reflection functor $\tilde \CR$ behaves precisely like the version of Koszul duality at the top of Figure \ref{fig:Koszul}, while the full rotation functor $\CR^\pi$ behaves like the modified Koszul duality at the bottom of Figure \ref{fig:Koszul}.

\subsection{The $\CN=2^*$ deformation and Landau-Ginzburg models}
\label{sec:LG}

The twisted mass $\wt m_\rho$ introduced just above in $\CT_{2d}$ breaks supersymmetry from 2d $\CN=(4,4)$ to 2d $\CN=(2,2)$. It has a well-known three-dimensional origin: it descends from the canonical real mass deformation of a 3d $\CN=4$ theory that breaks supersymmetry to 3d $\CN=2^*$.

We used $\wt m_\rho$ above to find positive collections of objects in the B-twist of $\CT_{2d}$, and to interpret Koszul duality as a (sequence of) wall-crossing transformations.
Another major advantage of turning on $\wt m_\rho$ is that it allows us to use 2d mirror symmetry to give a very concrete dual description of $\CT_{2d}$, as a Landau-Ginzburg model $\wt \CT_{2d}$. The category of boundary conditions in the B-twist of $\CT_{2d}$ then maps to a category of boundary conditions in the A-twist of the Landau-Ginzburg model.

When the original 3d $\CN=4$ theory is an $A_n$-type quiver gauge theory, the dual Landau-Ginzburg superpotential was derived in \cite{GK-3d}, and was shown to reproduce the Yang-Yang functional for a rational Gaudin model.  (This is a particular instance of Nekrasov-Shatashvili duality \cite{NS-I}.) Notably, the same superpotential appeared in \cite{GW-Jones}, in the study of M2/M5 brane systems. The physical reason for this is fairly clear: the 3d $A_n$-type quiver gauge theory can be engineered from a system of intersecting D3-NS5 branes, and both M2-M5 and D3-NS5 systems are dual to a common D2-NS5 system. The D2-NS5 system engineers our theory $\CT_{2d}$, and the Landau-Ginzburg superpotential is capturing its physics.

The paper \cite{GW-Jones} studied braid actions in the A-twisted Landau-Ginzburg theory coming from varying mass parameters. One claim of that paper was that these actions, at the categorical level, should provide a physical construction of Khovanov knot homology. (It is related to many other physical constructions of categorical braid actions and knot homology, \eg\ \cite{GukovWalcher, GW-surface, GukovStosic, GNSSS, CGR-quantumgroup}, \cf\ the basic idea in \cite{Gukov-gauge}, all ultimately tracing back to the physics of M2-M5 and related M5-M5 brane systems from \cite{OV, GSV}.) In the mathematics literature, braid actions on categories $D^b\CO_H$ and $D^b\CO_C$ have also been used to construct knot homology \cite{Webster-quivercat, Webster-HRT}, \cf\ the related \cite{SeidelSmith, CautisKamnitzer-sl2, CautisKamnitzer-sln}. One expects these various braid actions to all be equivalent. This provided a vital clue in our original identification of categories $D^b\CO_H$ and $D^b\CO_C$ with the B-twist of $\CT_{2d}$.

If we start from a general 3d $\CN=4$ gauge theory with gauge group $G$ and quaternionic representation $\CR=R\oplus R^*$, the Landau-Ginzburg theory dual to $\CT_{2d}$ has a superpotential of the form
\be \wt W(\sigma;m,t,\wt m_\rho) = \frac{\wt m_\rho}{2\pi i} \Big[ \sum_{\text{weights $\lambda\in R$}} \hspace{-.2cm} \log( \lambda\cdot (\sigma+m)) \;- \hspace{-.3cm} \sum_{\text{roots $\alpha_j\neq 0$}} \hspace{-.2cm} \log ( \alpha_j\cdot \sigma) \Big] + t\cdot \sigma \,.
\label{W-LG} \ee
It depends on dynamical fields $\sigma \in \mathfrak t_\C$, which are the complexifications of the real Coulomb-branch scalars $\sigma_\R$ in the original 3d theory; as well as on the usual rescaled, complexified mass and FI parameters $m,t$, and the twisted $\CN=2^*$ mass $\wt m_\rho$. In the case of an A-type quiver gauge theory, this superpotential should be compared with the Yang-Yang function \cite[Eqn. 3.52]{GW-Jones} the Bethe equations $\pd \wt W/\pd \sigma=0$ in \cite[Eqn 4.13]{GK-3d}.
The special scaling limit used to derive this potential in \cite{GK-3d} coincides with the scaling limit that defined $\CT_{2d}$ in Section \ref{sec:3d2d}.

We make a few brief comments on the structure of the superpotential \eqref{W-LG}, deferring further study of this Landau-Ginzburg theory to a forthcoming publication.

Though it is not entirely obvious, the critical points $\sigma_\nu$ of $\wt W(\sigma)$ are in 1--1 correspondence with the vacua $\nu$ of the original 3d $\CN=4$ theory. Indeed, in the limit $\wt m_\rho\to 0$, the critical values $\wt W(\sigma_\nu)$ are precisely the 2d central charges $Z_\nu$ (complexifications of $h_\nu$) that we encountered in Sections \ref{sec:order}, \ref{sec:shuffle}, and \ref{sec:22excep}. In this limit, the critical values are bilinear in $m$ and $t$, matching the structure from earlier discussions.

At finite $\wt m_\rho$, the function $\wt W(\sigma_\nu)$ becomes multivalued.
 In particular, each critical point $\sigma_\nu$ is associated with infinitely many critical values, differing by integer multiples of $\wt m_\rho$. This ambiguity reflects the internal $U(1)_\rho$ grading in the category of boundary conditions for $\CT_{2d}$; an extended discussion of such a phenomenon can be found in \cite[Section 4.1.4]{GW-Jones}.
Similarly, the difference of critical values $\wt W(\sigma_\nu)-\wt W(\sigma_{\nu'})$ is modified by $q\,\wt m_\rho$ for some $q\in \Z$, reflecting the structure of central charges in a theory with flavor symmetry that we described abstractly in Section \ref{sec:22pos}.

\subsection{A string-theory interpretation for $\CT_{2d}$}
\label{sec:M2M5}

There is a neat string-theory interpretation of the $\CT_{2d}$ theories derived from A-type quiver gauge theories in three dimensions. 

Consider a system of M2 branes stretched between two sets of M5 branes, which we can denote as M5$_H$ and M5$_C$, which share two common directions $01$,
are orthogonal in the $3456$ and $789\,10$ directions and well-separated along the $2$ direction. This is a system with $(4,4)$ supersymmetry. 

We can make the $x^3$ and $x^{10}$ directions compact, with radii $r_H$ and $r_C$, without changing the supersymmetry of the system. If the compactification radii are sufficiently small, 
the system has a dual description as a D3-D5-NS5 system in IIB string theory, engineering the A-type quiver gauge theory compactified on a large circle of inverse radius $R^{-1} \sim T_{M2} r_H r_C$. Indeed, the KK momentum corresponds to 
the charge of M2 branes wrapping both circles. 

The data of the quiver is encoded in the number of D3 branes ending on each five-brane \cite{Hanany:1996ie}. 
The separation between the fivebranes controls the masses and FI parameters of the theory. 
Notice that the corresponding central charges are associated to F1 and D1 strings stretched between the fivebranes, \ie\ M2 branes wrapping a single compactification circle. 
Thus we can identify the 3d masses and FI parameters with $r_H d_C T_{M2}$ and $r_C d_H T_{M2}$, where $d_{H,C}$ are the M5 brane separations in M-theory.
Finally, the domain walls tension is $R^{-1} T_{M2} d_C d_H$ and the corresponding soliton mass is $T_{M2} d_C d_H$.

We have thus identified in the M-theory geometry all the central charges that control the various scaling limits 
we are interested in. Clearly, the scaling limit that leads to $\CT_{2d}$ introduces a separation between the scale set by the 
M5 brane separations and the compactification radii, effectively focussing on the dynamics of the original uncompactified M2-M5$_H$-M5$_C$ system.
On the other hand, the naive 2d limits makes one compactification radius much smaller than the other, mapping the system to the D2-D4-NS5 IIA 
configuration that engineers the appropriate 2d gauge theory. 

This construction establishes an explicit physical link between the braid group actions that appear in the context of three-dimensional gauge theory and symplectic duality 
and the braid group actions that appear in M-theory contexts. 

\appendix

\section{Rewriting 3d $\CN=4$ as 2d $\CN=(2,2)$}
\label{app:2d}

In this appendix, we describe in some detail how to rewrite a 3d $\CN=4$ gauge theory as a 2d $\CN=(2,2)$ theory with infinitely many fields. We take 3d spacetime to have signature $(-,+,+)$ and coordinates $x^0,x^1,x^3$; this is convenient because it corresponds to a choice of gamma-matrices
\be \sigma_0 = \begin{pmatrix} 1 & 0  \\ 0 & 1 \end{pmatrix}\,,\qquad \sigma_1 = \begin{pmatrix} 0 & 1\\ 1& 0 \end{pmatrix}\,,\qquad \sigma_3 = \begin{pmatrix}1 & 0 \\ 0 & -1 \end{pmatrix}\,, \label{gamma} \ee
which are manifestly real. We want to view the 3d theory as a 2d theory on $\R^2$ with coordinates $x^0,x^3$, whose fields are valued in maps from $\R$ (parametrized by $x^1$) to the original 3d target.

As discussed in the main text, we want to choose a 2d $\CN=(2,2)$ subalgebra of the 3d $\CN=4$ algebra, in such a way that anti-commutators of supercharges do not generate translations in the $x^2$ direction. Such subalgebras are parametrized by the broken R-symmetry $[\text{R-symmetry of 3d $\CN=4$}]/[\text{R-symmetry of 2d $\CN=(2,2)$}]$, \ie\
\be \big(SU(2)_C\times SU(2)_H\big) /\big(U(1)_A\times U(1)_V\big) \simeq \cp^1\times \cp^1\,.\ee
The choice of subalgebra is equivalent to a choice of complex structure on the Higgs and Coulomb branches. Indeed, the vevs of chiral (respectively, twisted-chiral) operators with respect to a $(2,2)$ subalgebra are holomorphic functions on the Higgs (respectively, Coulomb) branches, in the corresponding complex structure. The subgroups $U(1)_C\subset SU(2)_C$ and $U(1)_H\subset SU(2)_H$ that preserve a given complex structure become the axial $U(1)_A$ and the vector $U(1)_V$ R-symmetries from the 2d perspective.

Our conventions for 2d $\CN=(2,2)$ supersymmetry and superspace are the same as in \cite{Witten-phases}, aside from scalings by $\sqrt{2}$ for some of the fermions. (One rather nice benefit of $\CN=(2,2)$ supersymmetry, in contrast with 3d $\CN=4$, is that all fields and interactions can be written in superspace.)
 Supercharges are $Q_\pm,\ol Q_\pm$, labelled by their eigenvalues under the 2d chirality matrix $\sigma_3$. Corresponding coordinates on superspace are $\theta^\pm,\bar\theta^\pm$. We set $x^\pm = \frac12(x^0\pm x^3)$, $\pd_\pm = \pd_0\pm \pd_3$, and in general for a 2d vector $A_a$,
\be A_\pm  = A_0 \pm A_3\,.\ee

To keep things simple, we'll focus on abelian gauge theories.

\subsection{Vectormultiplet}
\label{app:vector}

Having fixed a complex structure, the 3d $\CN=4$ abelian vectormultiplet contains a 3d gauge connection $A_\mu$, real and complex scalars $\sigma,\varphi$ (an $SU(2)_C$ triplet), and two complex fermions $(\lambda_\alpha,\eta_\alpha)$ that transform in the bifundamental of $SU(2)_C\times SU(2)_H$. The charges of these fields under 2d R-symmetry must be
\be  \begin{array}{c|cccccrr}
 & A_\mu & \sigma & \varphi & \lambda_\pm & \bar\lambda_\pm & \eta_\pm & \bar \eta_\pm \\\hline
 U(1)_A = U(1)_C & 0 & 0 & 2 & 1 & -1\, & 1\, & -1 \\
 U(1)_V = U(1)_H & 0 & 0 & 0 & 1 & -1\, & -1\, & 1 \end{array}
 \ee
They  can be grouped into a twisted-chiral superfield $\Sigma$ (the standard 2d field-strength multiplet) and a chiral superfield $S$:
\be \label{2dSigmaS}
\begin{array}{rl}
\Sigma = & \varphi + 2i\theta^+\eta_++2i\bar\theta^-\lambda_- -2\theta^+\bar\theta^-(D-iF_{03})-i\theta^+\bar\theta^+\pd_+\varphi+i\theta^-\bar\theta^-\pd_-\varphi \\[.1cm] &-2\theta^-\bar\theta^-\theta^+\pd_-\eta_++2\theta^+\bar\theta^+\bar\theta^-\pd_+\lambda_- +\theta^+\bar\theta^+\theta^-\bar\theta^-\pd_+\pd_-\varphi\,, \\[.2cm]
S = & A_1 -i\sigma + 2i\theta^+\bar\lambda_+ + 2i\theta^-\eta_- + 2\theta^+\theta^-F_\varphi - (\theta^+\bar\theta^+\pd_++\theta^-\bar\theta^-\pd_-)(\sigma+iA_1) \\[.1cm] &  + 2\theta^-\bar\theta^-\pd_-\bar\lambda_+ +2\theta^-\theta^+\bar\theta^+\pd_+\eta_- -\theta^+\bar\theta^+\theta^-\bar\theta^- \pd_+\pd_-(A_1-i\sigma)\,,
\end{array}
\ee
where $D$ and $F_\varphi$ are new real and complex auxiliary fields. The gauge-invariant twisted-chiral $\Sigma$ originates from an abelian vector superfield, $\Sigma = -\ol D_+D_-V$, where in Wess-Zumino gauge
\begin{align} \label{2dV}
 V = &\,\, \theta^+\bar\theta^+A_+ + \theta^-\bar\theta^-A_-  +\theta^-\bar\theta^+\varphi+\theta^+\bar\theta^-\ol\varphi +2\theta^+\bar\theta^+\theta^-\bar\theta^-D  \\
 & -2i(\theta^-\bar\theta^-\bar\theta^+\lambda_-+\theta^+\bar\theta^+\theta^-\eta_+ + c.c.)\,. \notag \end{align}
The standard supersymmetrized gauge transformation for $V$ is
\be V \to V - \Im \,\Lambda\,,\ee
with $\Lambda$ a chiral superfield.

Note that the 2d gauge connection $A_\mu$ has split into a 2d connection $A_\pm$ and a third component $A_1$ that combines with $\sigma$ to form a complex scalar. The chiral superfield $S$ that contains $A_1$ cannot be gauge invariant, but rather transforms as
\be S \to S + \pd_1\Lambda\,. \ee
A gauge-invariant Lagrangian density can then be constructed as
\be \label{2dLgauge}
 L_{\rm gauge} = \int dx^1 \int d^4\theta\, \frac{1}{4g^2}\Big[ -\frac12\Sigma\Sigma^\dagger +(\Im\, S+\pd_1V)^2 \Big]\,. \ee
This contains standard 2d kinetic terms (containing $\pd_\pm$ derivatives) of all the fields, as well as gauge-kinetic terms $(F_{1\pm})^2$ involving $\pd_1$ derivatives, and
a 2d ``scalar potential''
\be \int dx^1\,\frac1{2g^2}\big[ D^2+|F_\varphi|^2-2\sigma \pd_1D-|\pd_1\varphi|^2\big]\,.\ee
The kinetic terms involving $\pd_1$ derivatives for $\varphi$ are manifest, but for $\sigma$ they appear only after solving for the $D$-term, $D=-\pd_1\sigma$.

\subsection{Chern-Simons and FI terms}
\label{app:CS}

Twisted vectormultiplets of 3d $\CN=4$ (whose charges under $SU(2)_C$ and $SU(2)_H$ are swapped) can similarly be regrouped 2d chiral field strength $\wt \Sigma = -\ol D_+\ol D_-\wt V$ and a twisted-chiral $\wt S$. In three dimensions, such twisted vectormultiplets couple to ordinary vectormultiplets in mixed Chern-Simons interactions. Notably, the FI terms of a 3d $\CN=4$ theory are scalars of a background twisted vectormultiplet that couples in just this way.

A mixed Chern-Simons coupling at level $k$ between a twisted vectormultiplet and an ordinary vectormultiplet can be written in $\CN=(2,2)$ superspace as
\begin{align} \label{2dLFI}
 & \int dx^1\int d^4\theta \,\frac k2 \big[ \wt V\pd_1 V + \wt V\,\Im\, S - V\,\Im\, \wt S\big] \\
&\hspace{.5in} = \int dx^1 \int d^4\theta\, \frac k2\, \wt V\pd_1 V  -\int dx^1\Big[  \frac{k}{4i}\int d\theta^+d\theta^-\,\wt \Sigma S - \frac{k}{4i} \int d\theta^+ d\bar\theta^-\,  \Sigma \wt S + c.c.\Big]\,. \notag
\end{align}
The Lagrangian on the top line is manifestly invariant under ordinary gauge transformations $(S,V)\to (S+\pd_1\Lambda,V-\Im\,\Lambda)$, and is also invariant under twisted gauge transformations $(\wt S,\wt V)\to (\wt S+\pd_1\wt \Lambda,\wt V-\Im\,\wt \Lambda)$ after integrating by parts. On the bottom line we see that the second and third terms in the Lagrangian can be written succinctly as 2d ordinary and twisted superpotentials.

In order to include 3d FI terms, we just choose $k=1$ and set $\wt S,\wt \Sigma$ to constant values: the scalar in $\Im\,\wt S$ becomes a real FI parameter while the scalar in $\wt \Sigma$ becomes a complex FI parameter.

\subsection{Hypermultiplets}
\label{app:hyper}

Consider a single hypermultiplet. Having fixed a complex structure, we are accustomed to splitting it into a pair of 3d $\CN=2$ chiral multiplets $(X,Y)$, so that the scalars $X,\ol Y$ form a doublet of $SU(2)_H$. The complex fermions $\psi_\alpha^X$ and $\psi_\alpha^Y$ in the 3d multiplets organize into a doublet of $SU(2)_C$. Altogether, the R-charges are
\be \begin{array}{c|c@{\;\;}c@{\;\;}c}
& X,Y & \psi_+^{X},\psi_+^Y & \ol \psi_-^X,\ol\psi_-^Y \\\hline
U(1)_A= U(1)_C & 0 & -1 & 1  \\
U(1)_V = U(1)_H & 1 & 0 & 0
\end{array} \ee

In terms of 2d $\CN=(2,2)$ supersymmetry, we again find two chiral multiplets, with fermions reorganized as
\be \begin{array}{rl}
 \X = &  X +2\theta^+\psi_+^X +2\theta^-\ol\psi_-^Y -2\theta^+\theta^- F_Y \; -i(\theta^+\bar\theta^+\pd_++\theta^-\bar\theta^-\pd_-)X \\[.1cm] &-2i\theta^-\bar\theta^-\theta^+\pd_-\psi_+^X-2i\theta^+\bar\theta^+\theta^-\pd_+\ol\psi_-^Y-\theta^+\bar\theta^+\theta^-\bar\theta^-\pd_+\pd_-X \\[.2cm]
 \Y = & Y +2\theta^+\psi_+^Y -2\theta^-\ol\psi_-^X + 2\theta^+\theta^- F_X \; -i(\theta^+\bar\theta^+\pd_++\theta^-\bar\theta^-\pd_-)Y \\[.1cm] &-2i\theta^-\bar\theta^-\theta^+\pd_-\psi_+^Y+2i\theta^+\bar\theta^+\theta^-\pd_+\ol\psi_-^X-\theta^+\bar\theta^+\theta^-\bar\theta^-\pd_+\pd_-Y\,.
\end{array}\ee
Note that some signs in the definition of $\Y$ are flipped relative to those in $\X$. These signs are ultimately controlled by the holomorphic symplectic structure on the hypermultiplet moduli space $\R^4\simeq T^*\C$.

The 2d Lagrangian that encodes the 3d kinetic terms for the hypermultiplet is
\be L_{\text{hyper}} = \frac14\int dx^1 \int d^4\theta\, (\X\X^\dagger+\Y\Y^\dagger) + \Big[\frac{1}{2i}\int dx^1\int d\theta^+d\theta^-\, \X\pd_1\Y + c.c.\Big]\,. \label{Lhyper}\ee
This includes a scalar potential
\be |F_X|^2+|F_Y|^2 + iX\pd_1 F_X - i F_Y\pd_1 Y\,. \ee
Solving for auxiliary fields, we find $F_X = -i\pd_1 \ol X,\, F_Y = -i\pd_1\ol Y$, so that the $F$-term in $\X$ contains the $\pd_1$ derivative of $Y$ and vice versa.

If let the hypermultiplet transform with charge $n$ under a $U(1)$ gauge symmetry, and couple it to a vectormultiplet $(\Sigma,S)$, then the Lagrangian is modified:
\be L_{\text{hyper}} \to \frac14\int dx^1 \int d^4\theta (\X^\dagger e^{2nV}\X+\Y^\dagger e^{-2nV}\Y) +  \Big[\frac{1}{2i}\int dx^1\int d\theta^+d\theta^-\, \X(\pd_1-i n S)\Y + c.c.\Big]\,. \label{Lhyper-gauge}\ee
The total scalar potential of $L_{\text{vector}}$ and $L_{\text{hyper}}$ now takes the form
\begin{align} &\tfrac1{2g^2}\big(D^2+|F_\varphi|^2-2\sigma\,\pd_1D  -|\pd_1\varphi|^2\big) \notag \\
 &\hspace{.5in} + |F_X|^2+|F_Y|^2 + nD(|X|^2-|Y|^2)-n^2|\varphi|^2(|X|^2+|Y|^2) \\
&\hspace{.5in} +\big[iX(D_1-n\sigma)F_X -iF_Y(D_1-n\sigma)Y+nXYF_\varphi + c.c.\big]\,, \notag\end{align}
with covariant derivative $D_1=\pd_1-inA_1$. After solving for auxiliary fields, we recover the total scalar potential and kinetic energy (involving $\pd_1$ derivatives) of the original 3d $\CN=4$ theory, though in a somewhat nontrivial way. Note, in particular, that the D-term has become
\be -D = \pd_1\sigma+g^2\mu^\R\,,\qquad \mu^\R = n|X^2|-n|Y|^2\,, \label{22D} \ee
with $\mu^\R$ the real moment map of the $U(1)$ action. The F-terms are
\be \tfrac{-1}{2g^2}\ol F_\varphi  = n XY = \mu\,,\qquad \ol F_X = i(D_1+n\sigma)X\,,\qquad \ol F_Y = i(D_1-n\sigma)Y\,, \label{22F} \ee
and include the complex moment map for the $U(1)$ action. Altogether, after solving for auxiliary fields, we find the scalar potential
\be \tfrac1{2g^2}\big(|\pd_1\sigma+g^2\mu^\R|^2+|2g^2\mu|^2+|\pd_1\varphi|^2\big)+|(D_1+n\sigma)X|^2+|(D_1+n\sigma)Y|^2+n^2|\varphi|^2(|X|^2+|Y|^2)\,. \label{2dpot} \ee

\subsection{BPS equations: superpotential and Morse potential}
\label{app:hW}

There is a beautiful way to summarize the minima of the scalar potential \eqref{2dpot}, \ie\ half-BPS classical field configurations that are preserved by the supercharges in the 2d $\CN=(2,2)$ subalgebra of 3d $\CN=4$. In addition to the usual complex superpotential of 3d $\CN=4$ (viewed as a 3d $\CN=2$ theory),
\be W = \langle \varphi,\mu\rangle = n\varphi XY  \ee
(with $\mu$ the complex moment map for the gauge action), we introduce a ``Morse potential''
\be  h = \langle \sigma, \mu^\R\rangle = n\sigma (|X|^2-|Y|^2)\,, \ee
where $\mu^\R$ is the real moment map. Then the BPS equations are
\be dW = 0\,,\qquad D_1\Phi = - g^{\Phi\Phi'}\frac{\pd h}{\pd \Phi'} \label{BPS-app} \ee
for all fields $\Phi$, where $g^{\Phi\Phi'}$ is the inverse of the target-space metric. In other words, solutions of the BPS equations are gradient flows with respect to $h$.

This structure can be understood by writing the 3d $\CN=4$ theory as a 3d $\CN=1$ theory. Then modulo $dW=0$, $h$ is the real superpotential of the $\CN=1$ theory. In the $\CN=1$ theory, BPS configurations are Morse flows, just as in supersymmetric quantum mechanics. (A similar analysis for 2d theories appeared in \cite[Section 5.1.1]{Witten-path}. See also Appendix \ref{app:central}.)

\subsection{Sigma models}
\label{app:sigma}

Finally, we examine more closely the role of the holomorphic symplectic form that appeared, implicitly, in  superpotentials for hypermultiplets.
Suppose we have a 3d $\CN=4$ linear (ungauged) sigma model, whose hyperk\"ahler target has coordinates $\{X^i\}_{i=1}^{2n}$, with constant holomorphic symplectic form $\Omega=\Omega_{ij}dX^i\wedge dX^j$ and K\"ahler metric $g_{i\bar j}$. The natural generalization of \eqref{Lhyper} is
\be L_{\rm hyper} = \int dx^1\Big[\int d^4\theta \frac14 g_{i\bar j}\X^i(\X^{j})^\dagger + \int d\theta^+ d\theta^-\frac{1}{2i}\Omega_{ij} \X^i\pd_1\X^j + c.c.\Big]\,. \ee
The expression $\int dx^1\, \Omega_{ij}\X^i\pd_1\X^j$ can be understood geometrically as the pull-back from the target of a Liouville 1-form $\Lambda$, such that $d\Lambda=\Omega$. In other words,
\be  \int dx^1 \Omega_{ij}\X^i\pd_1\X^j \to \int_{\R(x^1)} \X^*(\Lambda)\,. \label{Liouville} \ee
This later expression makes sense for any sigma-model, linear or non-linear. The term \eqref{Liouville} played an important role in the study of boundary conditions for Rozansky-Witten theory \cite{KRS}.

\subsection{Boundary conditions for sigma models}
\label{app:Lag}

A key property of IR images of (2,2) boundary conditions is that they are supported on holomorphic Lagrangian submanifolds of the Higgs and Coulomb branches (\cf\ Section \ref{sec:genstruc}). We provide here a direct proof of this property.

Consider the effective IR description of a 3d $\CN=4$ gauge theory as a sigma-model to (say) the Higgs branch. At sufficiently low energy, we may focus on the neighborhood of a generic, smooth point in the Higgs branch (since the target-space metric has positive dimension).
Written as a 2d (2,2) theory, the effective sigma-model contains chiral fields $X^i:\R^2\to \text{Map}(\R_+,\CM_H)$ as above, such that for any fixed $x\in \R^2\times \R_+$ the $X^i$ are local complex coordinates on $\CM_H$. The bulk theory has a superpotential \eqref{Liouville},
\be W = \int_{\R_+} \X^*(\Lambda)\,,\ee
where $\Lambda$ is some choice of holomorphic Liouville one-form on $\CM_H$.

The most general (2,2) boundary condition for the sigma model can simply be constructed as a \emph{free} boundary condition for the superfields $\X^i$, coupled to some chiral boundary degrees of freedom $\Phi$ via a boundary superpotential $f(\X^i,\Phi)$. Given a variation of the action, we let the vanishing of a boundary variation determine the effective boundary condition for the $\X^i$.
The total superpotential becomes
\be W = f(\X^i\big|_\pd,\Phi) + \int_{\R_+}  \Phi^*(\Lambda) \,, \ee
and its variation includes boundary terms
\be \delta W = \Lambda_i \delta \X^i + \frac{\pd f}{\pd \X^i}\delta \X^i + \frac{\pd f}{\pd \Phi}\delta\Phi + \int_{\R_+}(...)\delta\X^i\,,\ee
which must vanish independently of the bulk part (...). This implies $\pd f/\pd \Phi=0$ (this is the boundary BPS equation for $\Phi$), and $\Lambda_i(\X) = \pd f/\pd \X^i$. We can express this succinctly as
\be  \Lambda\big|_\pd = df\,. \label{Ldf} \ee

Restricting to scalar fields, \eqref{Ldf} is precisely the condition that the boundary values of the $X^i$ lie on a holomorphic Lagrangian submanifold $\CL_H\subset \CM_H$ ``generated'' by $f$. Indeed, the equation puts at most $n=\frac12\text{dim}_\C \CM_H$ independent constraints on the boundary values of $X^i$, cutting out (locally) a holomorphic submanifold $\CL_H$ of dimension $\geq n$; and the holomorphic symplectic form must vanish at the boundary since $\Omega|_\pd = d\Lambda|_\pd = d^2f=0$, whence $\CL_H$ is Lagrangian.
(Conversely, any holomorphic Lagrangian looks locally like \eqref{Ldf} for a suitable function $f$.)

\section{Dirichlet boundary conditions and averaging} \label{app:Dir}
There is an alternative perspective on Dirichlet boundary conditions in abelian gauge theories, which 
may be useful in understanding and double-checking our prescription for their quantum Higgs and Coulomb branch images.

The idea is simple: replace Dirichlet boundary conditions with Neumann boundary conditions enriched by a $\C^*$-valued 2d chiral multiplet $\phi_a$
for each generator of the gauge group.
The expectation value of such chiral field will spontaneously break the gauge symmetry at the boundary. Boundary conditions 
for the hypermultiplets which explicitly break the boundary gauge symmetry can be incorporated by using $\phi_a$ as compensator fields
to promote them to gauge-invariant boundary conditions. Mathematically, this corresponds to applying the averaging functor $\text{Ind}_*$ from \cite[Section 3.7]{BernsteinLunts}.

In order to study such system, we may first add the compensator fields to a system of free hypermultiplets, and later add the gauge fields. 
Our first example is a basic $X=c$ boundary condition. Adding a compensator field $\phi$, we can replace it with an 
$X= e^\phi$ boundary condition. More precisely, we can start from an $X=0$ boundary condition and deform it by an $Y e^\phi$ superpotential.

Classically, the superpotential both sets $X= e^\phi$ and imposes the constraint $Y e^\phi=0$. Thus the classical Higgs branch image naively appears 
to be the $Y=0$, $X \neq 0$ sub-manifold. We will see that the actual image is likely closer to  
$X Y=0$, a direct sum of the $X=0$ and $Y=0$ manifolds with some extra 2d twisted chiral degrees of freedom along the 
$X=0$ sub-manifold, which break SUSY there in the absence of a mass deformation or twisted $\Omega$ background. (Mathematically, $XY=0$ is the singular support of the sheaf obtained from the averaging functor mentioned above.)

We now turn on the twisted $\Omega$ background. We can start from the space of operators of the form $Y^n e^{m \phi}|$ and set to zero combinations of the form 
\begin{equation}
\epsilon \partial_\phi P(\phi,Y) + Y e^\phi P(\phi,Y)
\end{equation}
The module action is given by the usual 
\begin{equation}
\hat Y P(\phi,Y)| = Y P(\phi,Y)| \qquad \qquad \hat X P(\phi,Y)| = \left(\epsilon \partial_Y P(\phi,Y) + e^\phi P(\phi,Y)\right)|
\end{equation}

We can pick generators $e^{n \phi}|$, with module action 
\begin{equation}
\hat Y e^{n \phi}| = - (n-1) \epsilon e^{(n-1) \phi}| \qquad \qquad \hat X e^{n \phi}| = e^{(n+1) \phi}|
\end{equation}
Thus $\hat X$ simply raises $n$ when acting on any generator, while $Y$ kills the $e^\phi|$ generator. 

We can give a simple, alternative description of the module: it is the quotient of the full algebra by the ideal generated by $\hat Y \hat X$:
$\hat X^n |$ maps to $e^{n \phi} |$ while $\hat Y^n |$ maps to $n! \epsilon^n e^{-n \phi} |$.
Another useful description is that of an extension built from the highest weight and lowest weight modules. 
If we had started from a $Y=c$ boundary condition, we would have obtained the opposite extension, the quotient of the full algebra by the ideal generated by $\hat X \hat Y$. (This opposite extension corresponds to applying an alternative averaging functor $\text{Ind}_!$ from \cite{BernsteinLunts}, rather than $\text{Ind}_*$.)

We could deform the setup further by turning on a mass $\wt m$ associated to the winding symmetry of the 2d chiral field. The superpotential becomes $Y e^\phi- \wt m \phi$.
The Higgs branch image is $X Y = \wt m$: the two branches have merged into a single manifold. 
We can pick generators $e^{n \phi}|$, with module action 
\begin{equation}
\hat Y e^{n \phi}| = \left[\wt m - (n-1) \epsilon \right] e^{(n-1) \phi}| \qquad \qquad \hat X e^{n \phi}| = e^{(n+1) \phi}|
\end{equation}
Thus $\hat X$ simply raises $n$ when acting on any generator, while $Y$ lowers $n$ and rescales the generator.
For generic $\wt m$ this is a natural, and rather unique quantization of the $X Y = \wt m$ manifold. 

If we had started from a $Y=1$ boundary condition, we would have obtained an isomorphic module, but 
the isomorphism would involve multiplication or division by polynomials in $\wt m$. Both modules can also be described as  
the quotient of the full algebra by $\hat Y \hat X-\wt m$, but again the isomorphism would involve multiplication or division by polynomials in $\wt m$.
The failure of the isomorphisms when $\wt m$ become certain multiples of $\epsilon$ should be a manifestation of the fact that the underlying boundary conditions are not 
equivalent. 

We can generalize this to a set of hypermultiplets with $X_i = c_i$ boundary conditions. Without loss of generality we can set $c_N=1$.
For example, we can add a compensator field for the diagonal symmetry 
acting on all hypers with charge $1$. Thus we start from $X_i=0$ b.c. and deform by $\sum_i c_i Y_i e^\phi$. Naively, we get 
$X_i = c_i e^\phi$, i.e. $X_i = c_i X_N$ and $X_N \neq 0$, and $\sum_i c_i Y_i e^\phi =0$, i.e. $\sum_i c_i Y_i =0$. 
The true answer is closer to $X_i = c_i X_N$ and $(\sum_i c_i Y_i) X_N =0$, with extra twisted degrees of freedom 
on the $X_N=0$ branch. 

The quantum Higgs h=branch module is spanned by the $\prod_i Y_i^{n_i} e^{n \phi}|$ monomials,
modulo expressions of the form 
\begin{equation}
\epsilon \partial_\phi P(\phi,Y_i)| + \sum_i c_i Y_i e^\phi P(\phi,Y_i)|
\end{equation}
The module action is given by the usual 
\begin{equation}
\hat Y_i P(\phi,Y_j)| = Y_i P(\phi,Y_i)| \qquad \qquad \hat X_i P(\phi,Y_j)| = \left(\epsilon \partial_{Y_i} P(\phi,Y_j) + c_i e^\phi P(\phi,Y_j)\right)|
\end{equation}

Without loss of generality we can set $c_N=1$ and pick generators $\prod_{i<N} Y_i^{n_i} e^{n \phi}|$, with module action 
\begin{align}
\hat Y_i \prod_{i<N} Y_i^{n_i} e^{n \phi}| &= Y_i \prod_{i<N} Y_i^{n_i} e^{n \phi}|  \cr \hat X_i \prod_{i<N} Y_i^{n_i} e^{n \phi}| &= \epsilon n_i Y_i^{-1} \prod_{j<N} Y_j^{n_j} e^{n \phi}| 
+ c_i \prod_{i<N} Y_i^{n_i} e^{(n+1) \phi}|\qquad i<N \cr
\hat Y_N \prod_{i<N}Y_i^{n_i} e^{n \phi}| &= - \sum_{i<N} c_i Y_i \prod_{i<N} Y_i^{n_i} e^{n \phi}|- (n-1) \epsilon \prod_{i<N} Y_i^{n_i} e^{(n-1) \phi}|\cr \hat X_N \prod_{i<N} Y_i^{n_i} e^{n \phi}|&= \prod_{i<N} Y_i^{n_i} e^{(n+1) \phi}|
\end{align}

We see the relations 
\begin{align}
&\hat X_i | = c_i \hat X_N |\qquad i<N \cr
&(\hat Y_N  + \sum_{i<N} c_i \hat Y_i) \hat X_N | = 0
\end{align}
We can identify the module as the quotient of the full algebra by that ideal. Indeed, we can identify the generators $\prod_{i<N} \hat Y_i^{n_i} \hat X_N^n|$ and 
$\prod_{i<N} \hat Y_i^{n_i} (\hat Y_N  + \sum_{i<N} c_i \hat Y_i)^{n}|$ respectively with $\prod_{i<N} Y_i^{n_i} e^{n \phi}|$ and $n! \epsilon^n \prod_{i<N} Y_i^{n_i} e^{-n \phi}|$.
We can also see the module as an extension built from the modules generated by the ideal $\hat X_i | = c_i \hat X_N$, $\hat Y_N  + \sum_{i<N} c_i \hat Y_i$ and 
the ideal $\hat X_i |=0$. 

As before, we can turn on a mass parameter for the winding number symmetry. This deforms the classical image to $X_i = c_i X_N$ and $(\sum_i c_i Y_i e^\phi)X_N =\wt m$.
The quantum ideal relations changes accordingly to 
\begin{align}
&\hat X_i | = c_i \hat X_N |\qquad i<N \cr
&(\hat Y_N  + \sum_{i<N} c_i \hat Y_i) \hat X_N | =\wt m |
\end{align}

Next, we can add the gauge fields. At first we can turn off the complex FI parameter, as typical for Neumann b.c., and include a Wilson line twist. 
The complex moment map acts as 
\begin{align}
(\hat Y_i \hat X_i+\hat Y_N \hat X_N) \prod_{i<N} Y_i^{n_i} e^{n \phi}| &= \epsilon \left(\sum_{i<N} n_i-n \right) \prod_{j<N} Y_j^{n_j} e^{n \phi}| 
\end{align}
Thus we can restrict the basis to $n = k_t + \sum_{i<N} n_i$. The generators are all monomials in the $Y_i$, $i<N$. The 
module action is the same as we computed in the main text, with $t_\C$ specialized to the appropriate integral values. 

We can turn on a generic value of $t_\C$ here, if we remember that the boundary anomaly of the topological symmetry can be cancelled 
by combining it with a 2d global symmetry with the same anomaly. Here we can use the winding number symmetry of $\phi$, which becomes anomalous as one gauges the 
translation symmetry. That means setting $t_\C = -\wt m + k_t \epsilon$. The complex moment map acts as 
\begin{align}
(\hat Y_i \hat X_i+\hat Y_N \hat X_N) \prod_{i<N} Y_i^{n_i} e^{n \phi}| &= \left( \wt m + \epsilon \left(\sum_{i<N} n_i-n \right) \right) \prod_{j<N} Y_j^{n_j} e^{n \phi}| 
\end{align}
We get the same constraint on $n$ as before and recover the module action in the main text for general $t_\C$. 

\section{Compactification to two dimensions}
\label{app:2dalgebra}

The purpose of this appendix is to collect several facts about the 3d $\CN=4$ super-Poincare algebra
and its relation to the 3d $\CN=(4,4)$ super-Poincare algebra. We begin by comparing central charges in the two algebras that control the masses of BPS objects. Then in \ref{app:twist} and \ref{app:mt} we describe families of topological twists in 2d $(4,4)$ theories that are relevant for symplectic duality.

Throughout this appendix, we will consider 3d theories on Minkowski spacetime with coordinates $x^0,x^1,x^3$ and (where needed) gamma-matrix conventions as in Appendix \ref{app:2d}. We compactify the theories to two dimensions along the $x^3$ direction, on a circle of radius $R$. Eventually we will add BPS boundary conditions at $x^1=0$, which descend to boundary conditions in 2d.

\subsection{Superalgebras and central charges}
\label{app:central}

\subsection*{$\CN=1$ supersymmetry}

Supersymmetry algebras often allow for a variety of central charges, which control the properties of BPS objects of various dimensions. 
The central charges are associated to conserved currents which appear as super-partners of the super-currents. 
They may include both scalar central charges associated to standard conserved currents and tensorial central charges 
associated to higher form conserved currents. 

A prototypical example in three dimension is an $\CN=1$ Landau-Ginzburg theory, defined by a set of real chiral multiplets $(\phi,\psi_\alpha)$ 
and some real superpotential $h(\phi)$. Classically, the theory has supersymmetric vacua labelled by critical points $\phi^*_i$ 
of $h$ and BPS domain walls interpolating between the vacua, preserving a 2d $\CN=(1,0)$ (or $\CN=(0,1)$ ) subalgebra of the 
3d $\CN=1$ symmetry algebra, given by solutions of ascending (or descending) gradient flow equations for $h$. 
The tension of domain walls is controlled by a central charge density proportional to the difference between critical values $h_i=h(\phi^*_i)$
at the vacua on the two sides of the wall. The corresponding conserved current is simply the two-form current $*dh$. 

The $\CN=1$ supersymmetry algebra, deformed by the corresponding vector central charge $C_{\alpha \beta}=C_{(\alpha\beta)}$, takes the form 
\begin{equation} \label{N13d}
\{ Q_\alpha,  Q_\beta \} =  P_{\alpha \beta} + C_{\alpha \beta}\,.
\end{equation}
If we compactly the 3d $\CN=1$ theory along the $x^3$ direction, restricting ourselves to domain walls that wrap the circle, the supersymmetry algebra reduces to a 2d $\CN=(1,1)$ subalgebra
\be \label{N12d} \begin{array}{rl}
\{ Q_L,  Q_L \} &=  P_L\,,  \\[.1cm]
\{ Q_L,  Q_R \} &=  P_3 + C_3\,, \\[.1cm]
\{ Q_R,  Q_R \} &=  P_R\,, \end{array}
\ee
where the KK momentum $P_3$ scales as $R^{-1}$ while the domain-wall central charge $C_3$ scales as $R$. 

Here the spinor indices $\alpha,\beta=\pm$ may be taken to indicate helicity in the $(x^0,x^3)$ plane (parallel to a potential boundary), just as in Appendix \ref{app:2d}. In contrast, the subscripts $L,R$ indicate left- and right-moving chiralities in the $(x^0,x^1)$ plane of a compactified 2d theory. The relation among spinors is
\be Q_\pm = \tfrac{1}{\sqrt{2}}(Q_L\pm Q_R)\,,\qquad Q_{L,R} = \tfrac1{\sqrt{2}}(Q_+\pm Q_-)\,.  \label{LRpm} \ee

Were we to reduce on a second circle to one dimension, we would find an $\CN=2$ super-quantum-mechanics, whose vacuum structure and instantons were related to Morse theory long ago \cite{Witten-Morse}. The real superpotential $h(\phi)$ plays the role of a Morse function on the target space of the quantum mechanics.

\subsection*{$\CN=2$ supersymmetry}

The story becomes more interesting already for $\CN=2$ theories. Forming complex combinations $Q_\alpha^\pm=Q_\alpha^1\pm iQ_\alpha^2$ of two real supercharges, we can write down the most general possible set of central charges:
\be \label{N23d}
\begin{array}{rl}
\{ Q^+_\alpha,  Q^+_\beta \} &= C^{++}_{\alpha \beta}\,, \\[.1cm]
\{ Q^+_\alpha,  Q^-_\beta \} &=  P_{\alpha \beta} + C_{\alpha \beta} + i Z \epsilon_{\alpha \beta}\,, \\[.1cm]
\{ Q^-_\alpha,  Q^-_\beta \} &=  C^{--}_{\alpha \beta}\,,
\end{array}
\ee
where $Z$ is real, $C_{\alpha\beta}$ is Hermitian, and $(C_{\alpha\beta}^{++})^\dagger = C_{\alpha\beta}^{--}$ are complex vectors. More compactly, in terms of the real supercharges,
\begin{align}
\{ Q^a_\alpha,  Q^b_\beta \} &=  \delta^{ab} P_{\alpha \beta} + C^{a b}_{\alpha \beta} + i Z \epsilon^{ab}\epsilon_{\alpha \beta}\,.
\end{align}
The superalgebra has a $U(1)_R$ symmetry that rotates $Q^\pm$ with charges $\pm 1$; it is preserved by $Z$ and $C_{\alpha\beta}$, but broken by any nonzero value of $C_{\alpha\beta}^{++}$.

We look at some concrete examples to see how the central charges may be realized. 

The $C^{++}_{\alpha \beta}$ vector charge may only occur in theories with no $U(1)_R$ symmetry. The prototypical example is a
3d  $\CN=2$ LG theory with a generic complex superpotential $W$ with non-degenerate critical points. Classically, the theory has BPS domain walls preserving 
a 2d $\CN=(1,1)$ subalgebra, associated to gradient flows between the critical points. The domain walls are associated to a central charge density 
proportional to the difference between critical values $W_i$ of the superpotential
at the vacua on the two sides of the soliton. The corresponding conserved current is simply the two-form current $*dW$. 

The central charges that are compatible with the $U(1)_R$ symmetry are more interesting. The scalar central charge 
$Z$ is a linear combination of the global charges of the theory, with coefficients that coincide, essentially by definition,
with the ``real masses'' $m_\R$, parameters that enter the theory as the scalar superpartners of background gauge multiplets.%
\footnote{In a Coulomb phase we should include gauge charges as well, but we are assuming the theory is massive}

In order to gain intuition on the $U(1)_R$-invariant vector supercharge, we can consider some generic 
$\CN=2$ gauge theory. If we focus on a 3d $\CN=1$ subalgebra generated by $Q^\zeta_\alpha = \Re \zeta^{-1/2} Q^+_\alpha$ and $(Q_\alpha^\zeta)^\dagger$, labelled by a phase $\zeta$, the corresponding real superpotential 
can be written as 
\begin{equation}
h_\zeta = \langle \sigma, \mu_\R \rangle + \Re\, \zeta^{-1} W 
\end{equation}
where $\sigma$ are the gauge multiplet scalars (including background real masses $m_\R$) and $\mu_\R$ the corresponding real moment maps. 

We thus recognize that $C_{\alpha \beta}$ is associated classically to the expectation values $c_i$ of $\langle \sigma, \mu_\R \rangle$ at the vacua of the theory. 
If the theory has an $U(1)_R$ symmetry, this is a rather special object. Classically, at a massive vacuum the gauge moment maps vanish and the 
flavor moment maps are typically linear combinations of the real FI parameters of the theory. Thus $c_i$ is typically a bilinear expression in the real masses 
and FI parameters. 

Quantum mechanically, the vector central charge is corrected in a very interesting fashion. We can gain further insight by compactifying the theory 
down to two dimensions as before. The super algebra reduces to 
\be \label{N22d}
\begin{array}{rl}
\{ Q_L^+,Q_R^+\} &= C_3^{++} \\[.1cm]
\{ Q_L^+, Q_R^-\} &= P_3+C_3+iZ \\[.1cm]
\{ Q_L^-, Q_R^+\} &= P_3+C_3-iZ \\[.1cm]
\{ Q_L^-,Q_R^-\} &= C_3^{--} 
\end{array}
\qquad
\begin{array}{rl} 
\{ Q_L^+, Q_L^-\} &= P_L \\[.1cm]
\{ Q_R^+, Q_R^-\} &= P_R
\end{array}
\qquad
\begin{array}{rl}
\{Q_L^+,Q_L^+\}&=0 \\[.1cm]
\{Q_L^-,Q_L^-\}&=0 \\[.1cm]
\{Q_R^+,Q_R^+\}&=0 \\[.1cm]
\{Q_R^-,Q_R^-\}&=0
\end{array}
\ee

As expected from the LG example, the complex vector charge $C^{++}_3$ goes to the vector central charge in the $(2,2)$ super-algebra. (In an LG model, the complex superpotential $W$ descends to a superpotential in 2d, whose critical values determine $C^{++}_3$.)
The real vector charge $C_3$, instead, combines with the KK momentum and scalar central charge into the \emph{axial} central charge of the $(2,2)$ superalgebra,
which is associated to the expectation values of an effective \emph{twisted} superpotential $\wt W$. In the large-radius limit, the effective twisted superpotential 
in a massive vacuum is known (see e.g. \cite{GGP-walls}) to be a quadratic form $K(m)$ in the complexified real masses $m=m_\R + \frac{i}{R}\oint A_{flavor}$, whose coefficients are the low-energy effective Chern-Simons couplings in that vacuum. We conclude that in a massive 3d $\CN=2$ theory with $U(1)_R$ symmetry, the real vector charge $C_3$ is controlled by the quadratic form $K(m)$. 

\subsection*{$\CN=4$ supersymmetry}

Finally, the $\CN=4$ 3d super-algebra takes the form 
\begin{equation}
\{ Q^{A\dot A}_\alpha,  Q^{B\dot B}_\beta \} = \epsilon^{AB} \epsilon^{\dot A \dot B} P_{\alpha \beta} + iZ^{AB}  \epsilon^{\dot A \dot B} \epsilon_{\alpha \beta}+ i \epsilon^{AB}  \wt Z^{\dot A \dot B}\epsilon_{\alpha \beta} + C^{AB;\dot A \dot B}_{\alpha \beta}\,,
\end{equation}
where $A, B,...$ are indices for a 
doublet of the $SU(2)_C$ R-symmetry and $\dot A, \dot B,...$ are indices for a 
doublet of the $SU(2)_H$ R-symmetry.
Here  $Z$ and $\wt Z$ are two types of scalar central charges transforming in vector representations of 
$SU(2)_C$  and $SU(2)_H$, respectively, and $C$ is a vector central charge that is carried by domain walls. 
The supercharges are complex linear combinations $Q_\alpha^{A\dot A}=(\sigma_E)_{a}^{A\dot A} Q_\alpha^a$ of four real spinors $Q_\alpha^a$, formed with Euclidean Pauli matrices $\sigma_E$, and therefore satisfy $(Q_\alpha^{A\dot A})^\dagger = \epsilon_{AB}\epsilon_{\dot A\dot B}Q_\alpha^{B\dot B}$\,.

The scalar central charges are well understood: $Z^{AB}=Z^{(AB)}$ is a linear combination of the conserved charges for Higgs-branch flavor symmetries, 
with coefficients given by the the mass parameters $m^{AB}$; while $\tilde Z^{\dot A \dot B}=\tilde Z^{(\dot A \dot B)}$ is a linear combination of the conserved charges for Coulomb-branch flavor symmetries, with coefficients given by the FI parameters $t^{\dot A \dot B}$. 

We can determine the properties of the vector central charge $C^{AB;\dot A \dot B}_{\alpha \beta}$ by extending 
our analysis of the $\CN=2$ case. We find that there is a bilinear pairing $K_i(\cdot, \cdot)$ associated to massive vacua such that the vector central charges are controlled by 
\begin{equation} \label{KCS-appendix}
K_i(m^{AB}, t^{\dot A \dot B})\,.
\end{equation}
 The pairing $K_i$ can be given a more physical interpretation by promoting masses and FI parameters to 
background vector and twisted-vector multiplets. It coincides with the value of the effective background Chern-Simons coupling pairing the two types of background 
vector multiplet.%
\footnote{Massive hypermultiplets and vectormultiplets do not contribute to the effective coupling, 
which is just a specialization of the bare coupling between topological $U(1)$ symmetries and gauge fields.} %
It would be nice to confirm this 
statement with an explicit analysis of the supercurrent multiplet in mass-deformed $\CN=4$ theories.  

Upon compactification to two dimensions, the 3d spinors split into a left-moving and a right-moving part. If we assume as above that no domain 
walls of the 3d theory wrap the whole 2d space-time, we should keep only the third component of the vector central charge. We could also allow some KK momentum. We find
\be
\begin{array}{rl}\{ Q^{A\dot A}_L,  Q^{B \dot B}_L \} &= \epsilon^{AB} \epsilon^{\dot A \dot B} P_L \\[.2cm]
\{ Q^{A \dot A}_L,  Q^{B \dot B}_R \} &= \epsilon^{AB} \epsilon^{\dot A \dot B} P_3 + C^{AB;\dot A \dot B}_3 +  i Z^{AB}  \epsilon^{\dot A \dot B} + i \epsilon^{AB}  \tilde Z^{\dot A \dot B}  \\[.2cm]
\{ Q^{A \dot A}_R,  Q^{B\dot B}_R \} &= \epsilon^{AB} \epsilon^{\dot A \dot B} P_R 
\end{array}
\ee
The result is a $(4,4)$ theory with a non-chiral $SU(2)_C \times SU(2)_H$ R-symmetry, possibly broken by nonvanishing central charges $Z,\wt Z$ and $C_3$, \ie\ by mass deformations.%

As the radius of the compactification circle tends to zero and KK modes decouple, the full chiral R-symmetry $SO(4)_L\times SO(4)_R$ of the $\CN=(4,4)$ superalgebra may be restored. However, BPS boundary conditions of the type considered in this paper again break the symmetry to a maximal torus of the diagonal $SU(2)_C\times SU(2)_H$, so that is all we shall discuss here.

If we further restrict ourselves to real mass and FI parameters, then the only nonzero components of the central charges will be (say) $Z:=Z^{+-}$, $\wt Z:=\wt Z^{\dot+\dot-}$, and $C_3 = C_3^{+-;\dot+\dot-}$. In this case, the algebra simplifies to
\be\label{N42d}\begin{array}{rl}
\{ Q^{++}_L, Q^{--}_R \} &= C_3+  P_3+ iZ +  i\tilde Z   \\[.1cm]
\{ Q^{+-}_L, Q^{-+}_R \} &=  C_3-  P_3 + iZ -  i\tilde Z   \\[.1cm]
\{ Q^{-+}_L, Q^{+-}_R \} &=  C_3-  P_3 -iZ +  i\tilde Z   \\[.1cm]
\{ Q^{--}_L,  Q^{++}_R \} &=  C_3+  P_3 -iZ -  i\tilde Z  
\end{array}\ee
along with $\{Q^{A\dot A}_L,Q^{B\dot B}_L\}= \epsilon^{AB}\epsilon^{\dot A\dot B} P_L$ and $\{Q^{A\dot A}_R,Q^{B\dot B}_R\}= \epsilon^{AB}\epsilon^{\dot A\dot B} P_R$ as usual.

Looking at the null space of the right hand side of the algebra, we can see that a half-BPS soliton state (annihilated by some half of the $Q$'s) may carry either KK momentum or 3d domain wall charge, 
but not both, and either Higgs of Coulomb branch flavor charge, but not both. Solitons that carry more complicated sets of charges can at most be quarter-BPS. 

A quarter-BPS soliton state may be annihilated, say, by linear combinations of the form 
\begin{equation} \label{q-BPS1}
Q^{+ -}_L - \zeta^{+-} Q^{+-}_R \qquad \qquad Q^{- +}_L - \zeta^{-+} Q^{-+}_R
\end{equation}
with phases $\zeta^{+-}$ and $\zeta^{-+}=(\zeta^{+-})^{-1}$ controlled by the ratio between $P_3-C_3$ and $Z - \tilde Z$;
or by linear combinations of the form 
\begin{equation} \label{q-BPS2}
Q^{+ +}_L - \zeta^{++} Q^{+ +}_R \qquad \qquad Q^{--}_L - \zeta^{--} Q^{--}_R
\end{equation}
with phases $\zeta^{+,+}$ and $\zeta^{--}=(\zeta^{++})^{-1}$ controlled by the ratio between $P_3+C_3$ and $Z+ \tilde Z$.
Half-BPS solitons are annihilated by both sets of supercharges. 

It is interesting to look at the relative scaling of the various contributions to the mass of a 2d soliton as a function of the compactification radius. 
The KK momentum contribution scales as $R^{-1}$. The contribution from Higgs of Coulomb branch flavor charges is independent of $R$, 
and proportional to $m_\R$ or $t_\R$. The 3d domain wall contribution scales as $R m_\R t_\R$. 

As a result, if we want to keep the theory massive as we send $R \to 0$ and also want to treat the Higgs and Coulomb branches democratically,
we will have to send real masses and FI parameters to infinity as $R^{-\frac{1}{2}}$. Then the only BPS particles that generically remain of 
finite mass carry only 3d domain-wall charge. This is an interesting limit for our purposes, studied in Section \ref{sec:3d2d}. Asymmetric limits that keep either real masses (or real FI parameters) 
fixed instead correspond to the naive dimensional reduction of a 3d gauge theory (or its mirror) to a 3d gauge theory.

\subsection{Boundary conditions and topological twists}
\label{app:twist}

The boundary conditions of type $(2,2)$ that we study throughout this paper preserve the four supercharges $Q_+^{++}$, $Q_+^{--}$, $Q_-^{+-}$, $Q_-^{-+}$ in the 3d $\CN=4$ superalgebra. From the perspective of a compactified 2d $\CN=(4,4)$ theory, these four supercharges become
\begin{equation} \label{Qbdy}
\begin{array}{l@{\qquad}l}Q^{++}_+=Q_L^{++} + Q_R^{++}\,, &   Q^{+-}_-=Q_L^{+-} - Q_R^{+-}\,, \\[.1cm]  Q^{-+}_-=Q_L^{-+} - Q_R^{-+}\,, & Q^{--}_+=Q_L^{--} + Q_R^{--}\,.
\end{array}
\end{equation}

The boundary conditions are compatible with a large family of topological twists of the compactified $(4,4)$ theory. By ``topological twist'' here we mean a choice of choice of supercharge $Q$ that 1) is nilpotent $Q^2=0$; and 2) generates all 2d spacetime translations by commutation with the rest of the bulk superalgebra, $P_0,P_1\in\{Q,*\}$, making all translations $Q$-exact.
This is slightly less than one usually requires for a topological twist (\cf\ similar discussions in \cite{GMW, Costello-Yangian}).
These properties ensure that correlation functions of $Q$-closed operators are independent of insertion points. They  also allow one to define a category of boundary conditions, for which morphism spaces $\text{Hom}(\CB,\CB')$ are defined to be $Q$-cohomology of the space of local operators at the junction of two boundary conditions.

However, these properties do not guarantee in general that the theory can be defined on curved backgrounds while preserving $Q$. This typically requires that $Q$ transform as a scalar under some mixture of Lorentz and unbroken (bulk) R-symmetry groups, which is an extra condition.
Thus properties (1) and (2) do not always lead to a TQFT in the standard sense \cite{Witten-TQFT, Witten-TSM}. 

Let us assume that only real mass and FI parameters are turned on, and that the compactification radius has been sent to zero, so that nontrivial KK modes decouple and $P_3=0$. Then, letting $Q = aQ^{++}_++bQ^{+-}_-+c Q^{-+}_-+dQ^{--}_+$ and using \eqref{N42d}, we find that
\be Q^2 = (ad-bc)(2P_L+2P_R+4C_3) = \det\left(\begin{smallmatrix} a & b\\ c & d \end{smallmatrix}\right) (2P_L+2P_R+4C_3)\,.\ee
Therefore, nilpotent supercharges are given by matrices $\left(\begin{smallmatrix} a & b\\ c & d \end{smallmatrix}\right)$ of rank one, up to overall rescaling. This space is $\cp^1\times\cp^1$. Letting $\zeta,\zeta'$ be affine parameters on $\cp^1$, we can parameterize the nilpotent charges as
\be Q_{\zeta,\zeta'} := Q^{++}_+ + \zeta' Q^{+-}_- + \zeta Q^{-+}_- + \zeta\zeta' Q^{--}_+\,. \label{Qzz}\ee
A short calculation shows that property (2) is satisfied (\ie\ translations are $Q$-exact) for all $\zeta$ and $\zeta'$.

\begin{figure}[htb]
\centering
\includegraphics[width=4in]{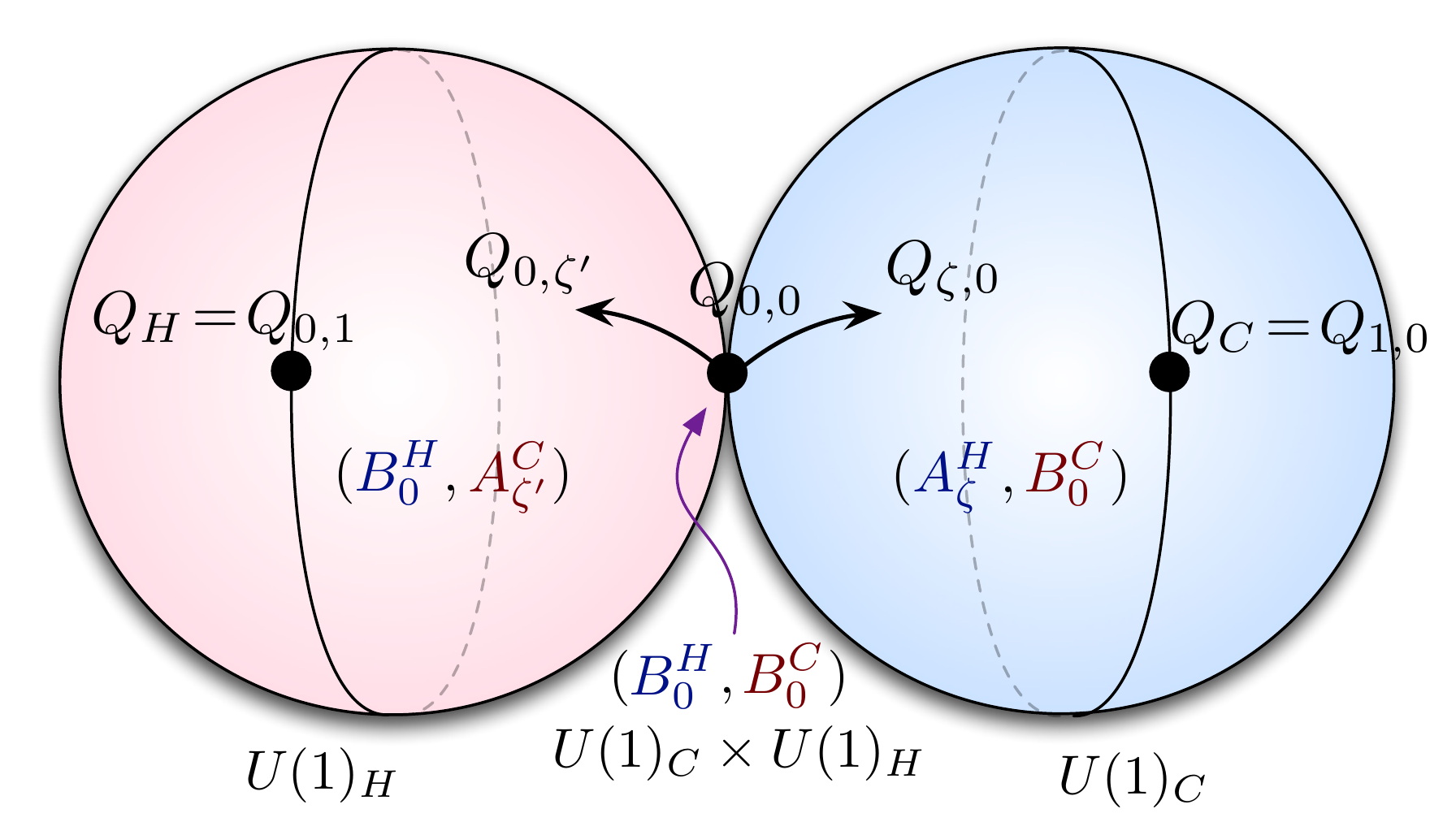}
\caption{The $\cp^1\cup \cp^1$ of topological twists in 2d $(4,4)$ theory that preserve at least one $U(1)$ R-symmetry, parameterized by $\zeta,\zeta'$ with $\zeta\zeta'=0$. The special supercharge $Q_{0,0}$ that preserves both R-symmetries reduces to a B-model on both Higgs and Coulomb branches; while $Q_{0,1}$ and $Q_{1,0}$ can be identified with reductions of Rozansky-Witten supercharges from~3d.}
\label{fig:twists}
\end{figure}

Recall that our BPS boundary conditions break $SU(2)_C\times SU(2)_H$ R-symmetry to the torus $U(1)_H\times U(1)_C$.
We are especially interested in supercharges that are invariant (up to rescaling) under at least one of these $U(1)$'s. Such supercharges lead to categories of boundary conditions with well-defined homological gradings (dg categories). As we explain momentarily, they can also be identified as A and/or B-model supercharges for standard topological twists. If \emph{both} R-symmetries are preserved, then the categories contain an extra internal (non-homological) grading, corresponding to the anti-diagonal of $U(1)_H\times U(1)_C$. It is easy to see that
\be \text{$Q_{\zeta,\zeta'}$ preserves}\quad  \begin{array}{ll} U(1)_H & \text{if $\zeta'=0$ or $\infty$} \\[.1cm]
  U(1)_C &\text{if $\zeta=0$ or $\infty$}\,. \end{array}
\ee
Thus, supercharges that preserve at least one R-symmetry live in a subspace $\cp^1\cup \cp^1\cup\cp^1\cup \cp^1\subset \cp^1\times \cp^1$, where at least one of $\zeta,\zeta'$ equals $0$ or $\infty$. Up to conjugation of $Q_{\zeta,\zeta'}$, which acts as the antipodal map (sending $(\zeta,\zeta')\mapsto(-1/\bar\zeta,-1/\bar\zeta')$), we can focus on the subspace $\cp^1\cup\cp^1 \subset \cp^1\times\cp^1$ where at least one of $\zeta,\zeta'$ equals zero. This space is depicted in Figure \ref{fig:twists}.

We can identify the $Q_{\zeta,\zeta'}$ that preserve an R-symmetry as A and/or B-model supercharges. The identification only makes sense if we 1) choose a 2d $\CN=(2,2)$ subalgebra of the 2d $\CN=(4,4)$ SUSY algebra, amounting to a choice of complex structures on the 3d Higgs and Coulomb branches; and 2) choose which operators to call ``chiral'' vs ``twisted-chiral'' with respect to the 2d (2,2) algebra, amounting to a choice of 3d Higgs branch vs. Coulomb branch. (This is a choice of mirror frame for the 2d (2,2) theory.)

It is fairly clear what sort of answer to expect due to the fact that our BPS boundary conditions define holomorphic Lagrangian submanifolds on both the Higgs and Coulomb branches. Namely, if we denote the complex structures on the respective branches as $(I_\zeta^H,I_{\zeta'}^C)$, such that the images of boundary conditions are holomorphic at $\zeta=\zeta'=0$ (or $\infty$), then we can only get a B-model on the Higgs branch at $I^H_0$ or $I^H_\infty$ (and otherwise an A-model), and a B-model on the Coulomb branch at $I^C_0$ or $I^C_\infty$ (otherwise an A-model). A natural guess would be that the twist parameters $(\zeta,\zeta')$ and the complex-structure parameters $(\zeta,\zeta')$ get correlated, so that $Q_{0,0}$ defines B-models $B_0^H$, $B_0^C$ on both branches in complex structures $(I_0^H,I_0^C)$; $Q_{\zeta,0}$ defines an A-model $A_\zeta^H$ on the Higgs branch in complex structure $I_\zeta^H$ and a B-model $B_0^C$ on the Coulomb branch in complex structure $I_0^C$, etc. This is summarized in Figure \ref{fig:twists}.

Our guess is easy to verify directly. Consider, say, $Q_{0,0}=Q^{++}$. If we use the 2d reduction to a Higgs-branch sigma model as on the RHS of Figure \ref{fig:2dlimits} (page \pageref{fig:2dlimits}), the local coordinates $X^A$ on the Higgs branch transform as doublets of $SU(2)_H$ and singlets of $SU(2)_C$. In a flat approximation, the supercharges act as $Q_\alpha^{A\dot A}X^B\sim \epsilon^{AB}\psi^{\dot A}_\alpha$.  The holomorphic coordinates in complex structure $\zeta=0$ are $X^+,(X^-)^\dagger$, and they are both annihilated by $Q^{++}_+$. In fact, they are both annihilated by both the left- and right-moving parts of the supercharge $(Q_{0,0})_L\sim Q^{++}_L$ and $(Q_{0,0})_R \sim Q^{++}_R$. Therefore, in a (2,2) sigma-model to the Higgs branch in complex structure $\zeta=0$, we identify $X^+,(X^-)^\dagger$ as chiral fields and $Q_L^+:=(Q_{0,0})_L$, $Q_R^+:=(Q_{0,0})_R$ as chiral supercharges, \cf~\eqref{N22d}. The combination
\be Q_{0,0} = Q_L^+ + Q_R^+  \label{B-model} \ee
is the standard form of the B-model supercharge \cite{Witten-TSM}. From the perspective of the a Coulomb-branch sigma-model, the chiral fields in complex structure $\zeta'=0$ are $\wt X^{\dot +}$, $(\wt X^{\dot -})^\dagger$, where $\wt X^{\dot A}$ is a doublet of $SU(2)_C$. An identical analysis shows that $Q_{0,0}$ is also a B-model supercharge for the Coulomb-branch sigma-model.

We next consider $Q_{\zeta,0}$ for $\zeta\neq 0$. Its left- and right-moving parts are $(Q_{\zeta,0})_L\sim Q^{++}_L + \zeta Q^{-+}_L$ and $(Q_{\zeta,0})_R \sim Q^{++}_R - \zeta Q^{-+}_R$. On the Coulomb branch, both left- and right-moving charges annihilate the chiral fields $\wt X^{\dot +}, (\wt X^{\dot -})^\dagger$ in complex structure $\zeta'=0$, so we again get a B-model in this complex structure. However, on the Higgs branch, there is no complex structure for which both $(Q_{\zeta,0})_L$ and $(Q_{\zeta,0})_R$ annihilate the chiral/holomorphic fields. In contrast, the holomorphic functions $X^+ + \zeta (X^-)^\dagger$ and $X^--\zeta^{-1} (X^+)^\dagger$ in complex structure $\zeta$ are both annihilated by $Q_L^+:=(Q_{\zeta,0})_L$ and the \emph{conjugate} $Q_R^+:= (Q_{\zeta,0})_R^\dagger = Q^{--}_R+\bar \zeta Q^{+-}_R$ so long as $\zeta$ lies on the unit circle (so that $\bar\zeta=\zeta^{-1}$).  In this case we have
\be Q_{\zeta,0} = Q_L^+ + (Q_R^+)^\dagger =  Q_L^+ + Q_R^- \,, \label{A-model} \ee
which is the standard form of an A-model supercharge. We conclude that, from the perspective of the Higgs branch, we get an A-model $A_\zeta^H$ in complex structure $\zeta$.

When $\zeta$ is not on the unit circle, the relation \eqref{A-model} is deformed. One finds precisely the generalized A-models of the type considered in \cite{Kapustin-A, KapustinWitten, NekWitten}, defined in terms of generalized complex geometry. In this sense, the entire family of charges $Q_{\zeta,0}$ for $\zeta\in \C^*$ define A-models $A_\zeta^H$ on the Higgs branch in complex structure $\zeta$.

An identical analysis shows that $Q_{0,\zeta'}$ defines a B-model $B_0^H$ on the Higgs branch and an A-model $A_{\zeta'}^C$ on the Coulomb branch in complex structure $\zeta'$. Similarly, we can identify a generic $Q_{\zeta,\zeta'}$ as an A-model-like supercharge on \emph{both} branches --- with the caveat that these A-models are missing homological gradings and also can never be promoted to full TQFT's (because a generic $Q_{\zeta,\zeta'}$ preserves no R-symmetry).

\subsection{Superpotentials and twisted masses}
\label{app:mt}

We want to turn on real mass and FI terms in a 3d $\CN=4$ theory. After compactification to two dimension, they can deform the A- and B-models of the previous section in various ways. In particular, in a reduction to a 2d sigma-model (as on the LHS or RHS of Figure \ref{fig:2dlimits} on page \pageref{fig:2dlimits}), viewed as an $\CN=(2,2)$ theory, these parameters can show up either as complex mass terms in superpotentials, or as twisted masses.

There are several ways to analyze this. In a given $(2,2)$ algebra, we may take commutators of left- and right-moving charges and compare with \eqref{N22d}. Then a nonzero contribution to $\{Q_L^+,Q_R^+\}$ can be identified with a central charge for solitons, coming from critical values of a superpotential; while a nonzero contribution to $\{Q_L^+,Q_R^-\}$ can be identified as a twisted mass. Alternatively, we may work directly at the level of Lagrangians: once we choose complex structures $\zeta,\zeta'$, we can first write the 3d $\CN=4$ theory (or its mirror) as a 3d $\CN=2$ theory, identifying twisted masses and superpotential terms; these descend to the correspond terms in 2d.

For example, let us choose $(\zeta,\zeta')=(0,0)$ and consider a Higgs-branch sigma-model. The fact that $\zeta=0$ tells us immediately that the real FI parameter $t_\R$ (suitably rescaled, as in Figure \ref{fig:2dlimits}) will appear as a resolution/K\"ahler parameter for the target $\CM_H$; we just need to find the role of $m_\R$.
The charges $Q^+_L$, $Q^+_R$ in \eqref{B-model} anti-commute with each other, so there is no superpotential. However, using $Q_R^-\sim(Q_R^+)^\dagger$, we find
\be \{Q_L^+,Q_R^-\} = \{Q_L^{++},Q_R^{--}\} \overset{\eqref{N42d}}{=} -C_3+iZ+i\tilde Z\,. \label{QQ-Bmodel} \ee
The contribution of $C_3$ on the LHS suggests the existence of a 2d twisted mass $m_\R$ (complexified by a flavor holonomy, and suitably rescaled). This result is also obtained directly from reduction of the 3d $\CN=4$ Lagrangian to two dimensions: as a 3d $\CN=2$ theory, $m_\R$ enters as a 3d twisted (or ``real'') mass, and descends to a 2d twisted mass. Again, there is no superpotential.

The analysis on the Coulomb branch at $(\zeta,\zeta')=(0,0)$ is identical: $m_\R$ is a resolution/K\"ahler parameter, $t_\R$ enters as a twisted mass, and there is no superpotential.

Next, let us consider $\zeta'=0$ and $\zeta\neq 0$. The Coulomb branch is still in complex structure $\zeta'=0$, with K\"ahler parameter $m_\R$. As explained above \eqref{A-model}, the B-model supercharge is $Q_{\zeta,0}=Q_L^++Q_R^+$ with $Q_L^+ = \frac{1}{\sqrt{1+|\zeta|^2}}(Q_L^{++}+\zeta Q_L^{-+})$ and $Q_R^+=\frac{1}{\sqrt{1+|\zeta|^2}}(Q_R^{++}-\zeta Q_R^{-+})$, where we now include the correct normalization factor. Since $\{Q_L^+,Q_R^+\}=0$ there is no superpotential. Moreover, just as in \eqref{QQ-Bmodel}, $\{Q_L^+,Q_R^-\} = -C_3+...$, indicating that $t_\R$ still enters as a twisted mass.

In contrast, at $\zeta'=0$ and $\zeta\neq 0$, the Higgs branch is in complex structure $\zeta$. Specializing to $|\zeta|^2=1$, we find that $t_\R$ plays the role of a complex deformation parameter.
The A-model supercharge is now $Q_{\zeta,0}=Q_L^++Q_R^-$ with $Q_L^+ = \frac{1}{\sqrt{2}}(Q_L^{++}+\zeta Q_L^{-+})$ and $Q_R^- = \frac{1}{\sqrt{2}}(Q_R^{++}-\zeta Q_R^{-+})$, whereas $Q_R^+ = (Q_R^-)^\dagger = \frac{1}{\sqrt{2}}(Q_R^{--}+\zeta^{-1} Q_R^{+-})$. Since $\{Q_L^+,Q_R^-\}=0$ there is no twisted mass; but $\{Q_L^+,Q_R^+\} = -C_3+...$, indicating the presence of a superpotential whose real part is $\langle m_\R,\mu_\R\rangle$, where $\mu_\R$ is the real moment map (in the original complex structure $I^H_0$) for the the flavor symmetry on the Higgs branch. A more natural way to write this superpotential is
\be W_\zeta = \langle m_\R, \mu_\C^\zeta\rangle\,, \label{W-Amodel}\ee
where
\be \mu_\C^\zeta = \frac{i}\zeta \mu_\C + \mu_\R -i\zeta\,\ol{\mu_\C} \ee
is the complex moment map in complex structure $\zeta$. Notice that for $|\zeta|=1$ we simply have $\Re(\mu_\C^\zeta) = \mu_\R$ as desired, but the expression \eqref{W-Amodel} should continue to be valid for general $\zeta\in \C^*$.

Conversely, if we take $\zeta=0$ and $\zeta'\neq 0$, we get a B-model $B_0^H$ on the Higgs branch with vanishing superpotential and twisted mass $m_\R$; and we get an A-model $A_{\zeta'}^C$ on the Coulomb branch with
\be W_{\zeta'} = \langle t_\R,\mu_\C^{\zeta'}\rangle\,, \ee
where $\mu_\C^\zeta$ is the moment map for the Coulomb-branch isometry group.

For generic $\zeta,\zeta'\in \C^*$, we obtain an A-model on either branch, with a superpotential
\be W_\zeta^{H} = \langle m_\R^{\zeta'},\mu_\C^\zeta\rangle\qquad\text{or}\qquad
 W_{\zeta'}^C = \langle t_\R^\zeta,\mu_\C^{\zeta'}\rangle\,. \label{Wzz}\ee

\subsection{Rozansky-Witten twists and Omega backgrounds}
\label{app:RW}

A 3d $\CN=4$ gauge theory with $SU(2)_H\times SU(2)_C$ R-symmetry admits two families of fully topological twists, corresponding to supercharges
\be Q_H^{(\gamma)} =  \delta^\alpha_{\dot A} (Q^{+\dot A}_\alpha+ \gamma\, Q^{-\dot A}_\alpha)\,,\qquad Q_C^{(\gamma')} = \delta^\alpha_A (Q^{A\dot+}_\alpha+\gamma'\,Q^{A\dot-}_\alpha)\,. \label{RW-charges} \ee
At generic $t_\R\neq 0$ and $m_\R=0$ (resp. $m_\R\neq 0$ and $t_\R=0$), the $Q_H^{(\gamma)}$-twisted (resp. $Q_C^{(\gamma')}$-twisted) theory is equivalent to Rozansky-Witten \cite{RW} theory on the Higgs (resp. Coulomb) branch in complex structure $\gamma$ (resp. $\gamma'$). 
If both $t_\R$ and $m_\R$ are nonzero, then the R-symmetry is broken to $U(1)_H\times U(1)_C$, and neither of the supercharges \eqref{RW-charges} can be preserved on generic curved backgrounds. However, the supercharges still give ``topological twists'' in the sense described at the beginning of Appendix \ref{app:twist}. In particular, the bosonic operators in the cohomology of $Q_H^{(\gamma)}$ (resp. $Q_C^{(\gamma')}$) provide holomorphic functions on the Higgs (resp Coulomb) branches in complex structure $I_\gamma^H$ ($I_{\gamma'}^C$); we have used the ring structure of such operators extensively throughout this paper.

The particular supercharges
\be Q_H = Q_H^{(0)} = Q^{++}_+ + Q^{+-}_-\,,\qquad Q_C=Q_C^{(0)} = Q^{++}_+ + Q^{-+}_- \ee
are compatible with the half-BPS boundary conditions that we study in this paper, whose images are holomorphic in complex structures $I_0^H$ and $I_0^C$. These supercharges are readily identified as distinguished twists of a compactified 2d $(4,4)$ theory. Comparing to \eqref{Qzz}, we find
\be Q_H = Q_{\zeta=0,\zeta'=1}\,,\qquad Q_C = Q_{\zeta=1,\zeta'=0}\,.\ee
Thus, we quickly recover the fact that the 2d reduction of (say) Rozansky-Witten theory on the $t_\R$-resolved Higgs branch is a B-model \cite{KRS}. However, we also see from Appendices \ref{app:twist} and \ref{app:mt} that the same theory can be viewed as an A-twisted Landau-Ginzburg model on the Coulomb branch. 

\begin{figure}[htb]
\centering
\includegraphics[width=3.2in]{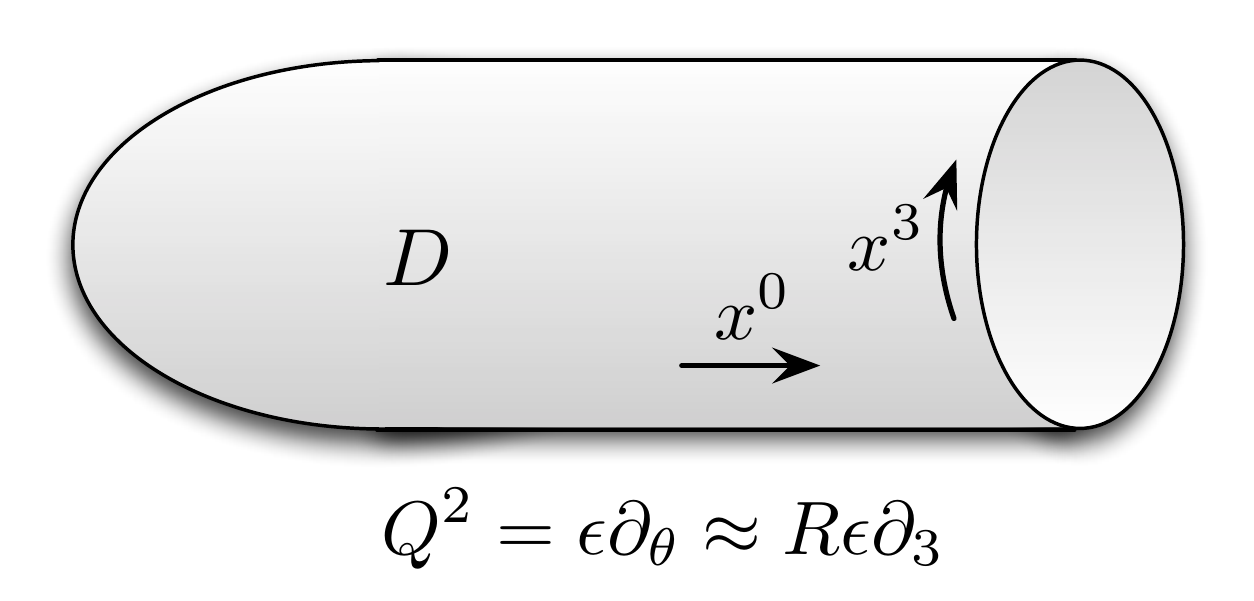}
\caption{Reducing an Omega-deformed theory to 2d, by compactifying on a cigar $D\times \R_{x^1}$.}
\label{fig:cigar}
\end{figure}

The Omega-backgrounds that quantize the algebra of functions on the Higgs and Coulomb branches correspond to deformations of $Q_H$ and $Q_C$, respectively. If we compactify the 3d Omega-deformed theory on a cigar as in \cite{NekWitten} (Figure \ref{fig:cigar}), we arrive at a topologically twisted 2d theory. The topological charges can be identified as $Q_{\zeta,\zeta'}$ with
\be \text{$\wt\Omega$-background:} \quad (\zeta,\zeta') = (R\epsilon,1)\,,\qquad  \text{$\wt\Omega$-background:}\quad (\zeta,\zeta') = (1,R\epsilon)\,, \label{Omega-2d} \ee
where $R$ is the asymptotic radius of the cigar.
To see this, note that the Omega-background supercharges are determined by the properties that 1) they reduce to Rozansky-Witten supercharges when $\epsilon=0$; and 2) in the asymptotic region of the cigar, where we can take the cigar circle to be the $x^3$ direction, the supercharge should satisfy $Q^2=R\epsilon\pd_3$. Property (1) is obvious in \eqref{Omega-2d}, and property (2) follows from the fact that $(Q_{\zeta,\zeta'})^2 = 2\zeta\zeta'P_3$ (using \eqref{Qzz} and \eqref{N42d}). If we keep $\epsilon':=R\epsilon$ fixed while taking the 2d limit $R\to 0$, the two Omega-background supercharges reduce to $Q_{\epsilon',1}$ and $Q_{1,\epsilon'}$ in the 2d $(4,4)$ theory. In either case, they lead to A-models on both Higgs and Coulomb branches, with no homological grading.
Alternatively, if we keep $\epsilon$ fixed, then we simply recover the Rozansky-Witten supercharges $Q_{0,1}$ and $Q_{1,0}$.

\acknowledgments{We are grateful to many friends and colleagues for insightful discussions and explanations during the preparation of this paper, including C. Beem, D. Ben-Zvi, T. Braden, A. Braverman, K. Costello, P. Deligne, S. Gukov, J. Kamnitzer, I. Losev, G. Moore, D. Nadler, H. Nakajima, T. Nevins, N. Proudfoot, L. Rozansky, N. Seiberg, C. Teleman, B. Webster, and E. Witten. We especially want to thank B. Webster for introducing us to symplectic duality and explaining much of the mathematical background for this paper. The research of MB, TD, and DG was supported by the Perimeter Institute for Theoretical Physics.
Research at Perimeter Institute is supported by the Government of Canada through Industry Canada and by the Province of Ontario through the Ministry of Economic Development \& Innovation.
MB gratefully acknowledges support from ERC Starting Grant no. 306260  ``Dualities in Supersymmetric Gauge Theories, String Theory and Conformal Field Theories" and IAS Princeton through the Martin A. and Helen Choolijan Membership.
TD was also supported by DOE grant DE-SC0009988 at IAS, and in part by ERC Starting Grant no. 335739 ``Quantum Fields and Knot Homologies,'' funded by the European Research Council under the European Union's Seventh Framework Programme.
JH would like to thank his advisor N. Proudfoot for his guidance and support.}

\appendix

\bibliographystyle{JHEP}

\bibliography{coulomb}

\end{document}